%% file: LIM_2017StatusReport.tex
\documentclass[titlepage]{tcibook}
\usepackage{fancyhea}
\usepackage[sort&compress,numbers]{natbib}  
\usepackage{work}
\usepackage{bm}       
\usepackage{graphicx}
\usepackage{hyperref}      
\usepackage[usenames,dvipsnames]{color} 
\usepackage{amssymb}
\usepackage{amsmath}
\usepackage{chngcntr}
\usepackage{booktabs}
\usepackage{multirow}
\usepackage{multicol}
\usepackage[mediumspace,Gray,squaren]{SIunits}  
\counterwithout{figure}{chapter}
\usepackage{titlesec}
\usepackage{wrapfig}
\usepackage[utf8x]{inputenc}

\def\hi    {H{\sc i}~}

\newcommand{\pixiefwhm}{$1.65^\circ$\, FWHM}

\definecolor{orange}{rgb}{1,0.3,0}

 \def\gtsima{$\; \buildrel > \over \sim \;$} 
\def\simgt{\lower.5ex\hbox{\gtsima}} 
\font\BFd=cmmib10
\font\BFt=cmmib10
\font\BFs=cmmib10 scaled 700
\font\BFss=cmmib10 scaled 500

\def\bbox#1{%
\relax\ifmmode
\mathchoice
{{\hbox{\BFd #1}}}
{{\hbox{\BFt #1}}}
{{\hbox{\BFs #1}}}
{{\hbox{\BFss #1}}}
\else \mbox{#1} \fi }

\newcommand{\avg}[1]{\ensuremath{\langle #1 \rangle}}
\def\x{{\bbox{x}}}
\def\k{{\bbox{k}}}

\def\lsim{\raise-.75ex\hbox{$\buildrel<\over\sim$}}

\DeclareUrlCommand\email{\urlstyle{rm}}

\begin{document}


\def\bibname{References}

\bibliographystyle{unsrtnat}

\raggedbottom


\parindent=0pt
\parskip=8pt
\setlength{\evensidemargin}{0pt}
\setlength{\oddsidemargin}{0pt}
\setlength{\marginparsep}{0.0in}
\setlength{\marginparwidth}{0.0in}
\marginparpush=0pt


\renewcommand{\chapname}{chap:intro_}
\renewcommand{\chapterdir}{.}
\renewcommand{\arraystretch}{1.25}
\addtolength{\arraycolsep}{-3pt}

\pagenumbering{roman} 
\chapter*{\huge Line-Intensity Mapping: 2017 Status Report
}
\vskip -9.5pt
\hbox to\headwidth{%
       \leaders\hrule height0.5pt\hfil}

 \input author_list_indexed.tex

\bigskip

\begin{center}
{ \large \bf Abstract}
\end{center}

\indent Following the first two annual intensity mapping workshops at Stanford in March 2016 and Johns Hopkins in June 2017, we report on the recent advances in theory, instrumentation and observation that were presented in these meetings and some of the opportunities and challenges that were identified looking forward. With preliminary detections of CO, [CII], Ly$\alpha$ and low-redshift 21cm, and a host of experiments set to go online in the next few years, the field is rapidly progressing on all fronts, with great anticipation for a flood of new exciting results. This current snapshot provides an efficient reference for experts in related fields and a useful resource for nonspecialists. We begin by introducing the concept of line-intensity mapping and then discuss the broad array of science goals that will be enabled, ranging from the history of star formation, reionization and galaxy evolution to measuring baryon acoustic oscillations at high redshift and constraining theories of dark matter, modified gravity and dark energy. After reviewing the first detections reported to date, we survey the experimental landscape, presenting the parameters and capabilities of relevant instruments such as COMAP, mmIMe, AIM-CO, CCAT-p, TIME, CONCERTO, CHIME, HIRAX, HERA, STARFIRE, MeerKAT/SKA and SPHEREx. Finally, we describe recent theoretical advances: different approaches to modeling line luminosity functions, several techniques to separate the desired signal from foregrounds, statistical methods to analyze the data, and frameworks to generate realistic intensity map simulations. 

 \pagebreak
 \smallskip
 
 \begin{center}
{ \large \bf List of Endorsers}
\end{center}

 \input endorser_list_indexed.tex






\eject

\begin{center}
  {\Large \bf Preface}
\end{center}
\bigskip

This Status Report presents the product of a growing global community of scientists who are involved in different aspects of line-intensity mapping research, an emerging field which promises new insights into the evolution of the Universe at low redshifts and into the Epoch of Reionization and Cosmic Dawn at higher redshifts. 
The line-intensity mapping field is still in a nascent phase, with only a handful of detected signals using a small list of instruments/datasets and a limited body of theoretical work to support it. This is set to dramatically change before the end of the current decade, as multiple experiments are coming online and considerable effort is devoted to plan observations, develop methods to analyze their data, and to study the implications for astrophysics and cosmology.  Currently dozens of papers refer to ``intensity mapping''.  While there are reviews of the study of the epoch of reionization with 21-cm fluctuations, there is no single document that provides an introduction to the broader field of intensity mapping, which includes growing attention to other atomic/molecular emission lines, a newer focus on cosmology at lower redshifts, and emerging prospects to study star formation and galaxy evolution up to high redshifts.  This article is intended to provide this introduction, present the current state of affairs, and lay out its opportunities and challenges looking forward.

Starting in early 2016, the line-intensity mapping community began a series of annual workshops to help coalesce and advance the  field. The pioneering meeting was a workshop titled {``Opportunities and Challenges in Intensity Mapping''} which took place at Stanford University (SLAC) March 21-23, 2016 and was attended by over 40 scientists. The second workshop, {``IM@Hopkins''}, was held at Johns Hopkins University, June 12-14, 2017, with over 50 participants. In between, during and after these two workshops, efforts were dedicated to developing a document presenting a status report of the line-intensity mapping field. Through the workshops, presentations, personal writing assignments and feedback on drafts, over 45 scientists (from over 25 institutions) have contributed to this 2017 Status Report, while a small writing group was responsible for editing, combining and integrating the individual contributions. The next community workshop will take place in February 2018, at the Aspen Center of Physics.

\newpage


\begin{center}
  {\LARGE \bf Executive Summary}
\bigskip
\end{center}

The aims of cosmology are to characterize the Universe on the
largest observable distance scales and to understand its origin
and evolution.  Efforts to address these questions are
intimately connected with parallel aims in extragalactic
astronomy to characterize and understand the origin and
evolution of galaxies and their interplay with the intergalactic
medium.  Progress in these questions are advanced today
primarily by cosmic microwave background (CMB) experiments and
by galaxy surveys. 
However, to anticipate the 2020 decadal survey planning process, 
it is important to identify and highlight new opportunities that may 
be fruitful in the advancement of cosmology and extragalactic astronomy. 

Line-intensity mapping is an emerging technique with potential for 
dramatic scientific payoff. Unlike galaxy surveys,
which determine the large-scale distribution of mass by
locating huge numbers of galaxies, intensity mapping measures
the integrated emission of spectral lines from galaxies and
the intergalactic medium (IGM) with (smaller) low-aperture instruments.  
Information about the line-of-sight distribution is obtained through the frequency
dependence. In this way, the cosmic luminosity density from a
variety of spectral lines can be mapped over potentially a huge
three-dimensional volume of the Universe and also at
redshifts not easily accessed with traditional galaxy surveys.
Line-intensity mapping can be uniquely applied to study
large-scale structure, the epoch of reionization (EoR), and
star/galaxy formation and has generated a tremendous flurry of 
activity over the past five years, in both theoretical research and the planning of dedicated experiments. 

While most of the line-intensity-mapping effort has been on 
21-cm emission from the neutral IGM, with several significant 
experimental efforts now afoot, there has more recently been 
rapid growth in the study of intensity mapping 
of line emission from galaxies, including the 21-cm line as well 
as those associated with rotational carbon-monoxide (CO)
transitions, the [CII] fine-structure line, and the Ly$\alpha$
line, among others.  Several groups are now pursuing
ground-based measurements of CO and [CII] fluctuations, and NASA
has recently selected SPHEREx for a MIDEX phase A study, 
a mission which has the potential for line-intensity mapping 
measurements of the Ha, Hb, OIII and Lya emission lines. 
The purpose of this status report is to describe the recent advances and 
prospects in line-intensity mapping. Given the significant prior
attention on 21-cm, this report focuses primarily,
though not exclusively, on intensity mapping with these other lines.

At large distance scales, line-intensity mapping probes large scale
structure much like a galaxy
survey, and thus addresses the growth of density perturbations,
the primordial power spectrum, and from these follow tests of
inflation and dark energy and possibly constraints to neutrino
masses. Since every photon is measured, including those from 
unresolvable faint sources, huge volumes can be probed 
at high redshift, 
allowing unique tests of the standard cosmological model and its possible extensions.
The intensity-mapping power spectrum can also be used to probe the 
luminosity functions of emitting galaxies, including luminosities too faint
to be accessed through traditional measurements.  Such
measurements, as well as cross-correlations between different
lines, offer the prospect to study the {\it universal} evolution
of star/galaxy formation at high redshift, rather than the
evolution of star formation in the high-redshift galaxies bright
enough to be imaged individually. In particular, components such as 
diffuse emission, dwarf galaxies, Ly$\alpha$ scattering and the 
IGM will be very difficult to see in a point-source survey, but may contribute 
significantly to the overall luminosity probed by line-intensity mapping. 
Combining [CII]/CO/Ly$\alpha$ with 21 cm can provide a wealth of new information 
on the EoR, such as evolution of bubble size, ionization state, and metal production. 
Given the statistical nature of the HI data, this information is not practically 
feasible with catalog-based surveys.

The intensity-mapping community has recently achieved several
breakthrough detections. 21-cm emission was first detected 
at $z \approx 0.8$ using GBT data in
cross-correlation with the DEEP2 galaxy survey, and followed up
in a deeper survey in cross-correlation with WiggleZ. Data from
the Sunyaev-Zel'dovich Array have provided a first detection of
the CO intensity auto-power. [CII] emission has been tentatively detected
in cross-correlation between Planck and SDSS
quasars. Ly$\alpha$ emission has been detected in
cross-correlation between BOSS cleaned spectra and the BOSS
quasars. These detections have provided new views of neutral
gas, molecular gas, and line transport and excitation. The  recent
BOSS measurement, to provide one example, implies a much higher overall Ly$\alpha$ luminosity 
than can be explained by the quasars themselves.

Intensity mapping instruments are now targeting a variety of interesting 
redshift ranges for the purposes of understanding the epoch of reionization, 
star formation, galaxy assembly, large scale structure, and Dark Energy. 
Multiple experiments anticipate having sensitivity to detect signatures 
of EoR, including HI experiments that target a power spectrum detection (HERA) 
and imaging (SKA-LOW), and experiments aiming for a detection in other 
emission lines (such as [CII] with TIME, CCAT-prime and CONCERTO). 
CO and [CII] lines have long been recognized as tracers of star formation, 
although much remains to be understood about the history of star formation and 
galaxy assembly. Intensity mapping experiments focusing on measurements 
of these lines plan to better understand these processes at high redshift 
around the peak of star formation, measuring star-formation properties in 
aggregate to improve constraints which are currently based on small sample 
sizes. The current generation of experiments range from tentative detections 
of CO (COPSS I, II) to pathfinders seeking a detection in CO (COMAP, AIM-CO) 
and [CII] (STARFIRE), and those targeting multiple lines (e.g.\ SPHEREx). 
Finally, neutral hydrogen traces the distribution of galaxies at lower redshift, 
and can be used for intensity mapping in radio frequencies at low spatial resolution 
(CHIME and HIRAX, optimized for a measurement of Dark Energy) or at higher resolution (SKA-MID). 

A range of analysis challenges must be overcome to best extract
science from upcoming line-intensity mapping surveys. 
Challenges include the presence of strong
continuum foreground contamination, confusion from
interloping emission lines and the non-Gaussian nature
of the signal. Finally, for many emission lines of interest, a
vast dynamic range in spatial scale is relevant, 
from the parsec scale of individual molecular clouds out to
cosmological lengths of several giga-parsecs. The
intensity-mapping community has started to attack these
challenges head-on. Work is underway to model and simulate
intensity-mapping signals reliably and to determine optimal
analysis strategies, maximizing the scientific return from the
upcoming surveys. 

Several statistical approaches can be used to confront the challenges
to intensity mapping. The cross power spectrum between
the redshifted 21-cm signal during the epoch of reionization and
line-intensity maps in other emission lines can help confirm
initial detections, even in the presence of strong foreground contamination, 
and is sensitive to quantities such as the typical size of ionized regions at
different stages of reionization. In the post-reionization era,
the 21-cm line-intensity mapping signal can be cross-correlated
with optical galaxy surveys, providing a powerful probe of
galaxy evolution. Analysts have established regimes in which line-intensity mapping is more 
powerful than traditional catalog-based surveys, and proposed novel methods 
to explore and realize its tremendous potential: 
The one-point probability distribution of voxel intensities can exploit
non-Gaussian information from upcoming  surveys; 
The ``multi-tracer" method can be applied profitably to multiple
line-intensity mapping surveys to extract key quantities without
the limitations of sample variance; 
A hybrid simulation
approach combining high resolution hydrodynamic simulations to
capture small scale physics, with the mass-peak-patch method to
identify dark matter halos across cosmological volumes, 
will allow rapid, yet accurate mock survey
realizations, important for anticipating future observations,
interpreting upcoming data and testing analysis pipelines; Finally, several
promising approaches for mitigating foreground interloper
contamination have been developed, including blind bright-voxel
and targeted masking techniques, as well as an anisotropic 
power-spectrum fitting methodology.

To close, there are great prospects for intensity-mapping
surveys to uniquely address a large range of science goals, 
expecting early results in the near future.  An array of small 
ground-based experiments, that capitalize upon existing hardware
and infrastructure with small budgets, are rapidly making headway.  They are guided
by the experience of larger and more mature 21-cm
projects that share some of the same science goals and
techniques.  There are good prospects for satellite missions
with extraordinary capabilities.  These experimental
developments are motivating theorists to identify
new science goals and opportunities and to address modeling and analysis
issues, thus further advancing the promise of multi-line intensity mapping.


\newpage

\setcounter{tocdepth}{1}
\tableofcontents



\eject
\pagenumbering{arabic} 
\setcounter{page}{1}


\chapter{Introduction}
\label{chap:intro}

\bigskip

\section{What is Line-Intensity Mapping?}
\label{sec:whatisIM}
\vspace{-0.15in}
Line-intensity mapping represents an exciting and rapidly
emerging new frontier in physical cosmology.   It uses the
integrated emission from spectral lines in galaxies and/or the
diffuse intergalactic medium to track the growth and evolution
of cosmic structure. The essential idea is to measure the
spatial fluctuations in the line emission from many individually
unresolved galaxies, rather than targeting galaxies one by
one.  The emission fluctuations trace the underlying large scale
structure of the Universe, with the frequency dependence
providing information about the distribution of emission along
the line of sight.  Unlike traditional galaxy surveys, which target only
discrete objects whose emission lies above some flux limit,
defined within a narrow aperture, intensity mapping is sensitive to {\em all
sources of emission in the line}. It is therefore advantageous
in studying faint and/or extended emission sources, and has prospects to 
further the {\it universal} study of galaxy formation/evolution 
(as opposed to the study of only the galaxies brightest enough to be 
imaged directly), in addition to probing the cosmological model in 
unexplored regimes. Since high angular resolution is not required, 
line-intensity mapping is also more economical than traditional galaxy surveys. 

\begin{figure}[h!]
\begin{center}
\includegraphics[width=0.85\linewidth]{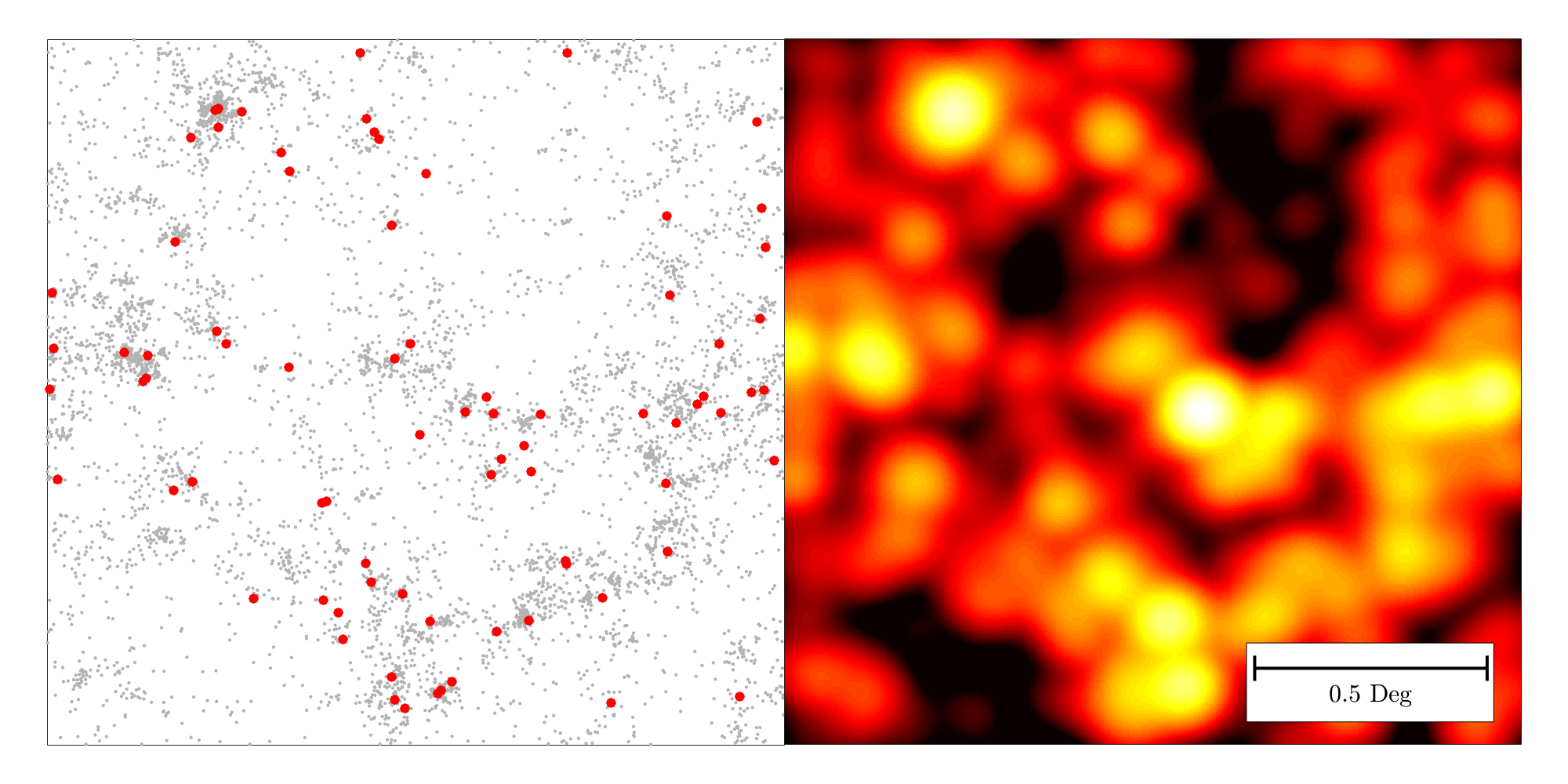}
\caption{A simulated $2.5\, {\rm deg}^2$ field with galaxy positions ({\it Left}) 
and the corresponding CO intensity map ({\it Right}). Luminosities were drawn 
from a Schechter function model (Breysse et al. 2016). Sources bright enough 
to detect with 1hr of VLA time are marked in red (see Li et al. 2016). 
({\it Figure: Patrick Breysse})}
\label{fig:IMmerit}
\end{center}
\end{figure}

Fig.~\ref{fig:IMmerit} provides a powerful demonstration of the potential gain. 
It compares the Very Large Array (VLA), an advanced radio telescope 
observatory consisting of 27 dishes, with a single-dish carbon-monoxide 
(CO) intensity mapping instrument (COMAP), in terms of their ability to 
observe a $2.5\, {\rm deg}^2$ sky patch. COMAP plans to spend $\sim$1500 
hours observing a field of this size, whereas the VLA would take $\sim$4500 
hours to cover the same area. While the VLA would detect only $\sim\!1\%$ 
of the total number of CO-emitting galaxies, COMAP will produce a map of 
the intensity fluctuations sensitive to emission throughout the field.
\vspace{-0.1in}

\section{Targets for Line-Intensity Mapping}

\begin{figure}[h!]
\begin{center}
\includegraphics[width=\linewidth]{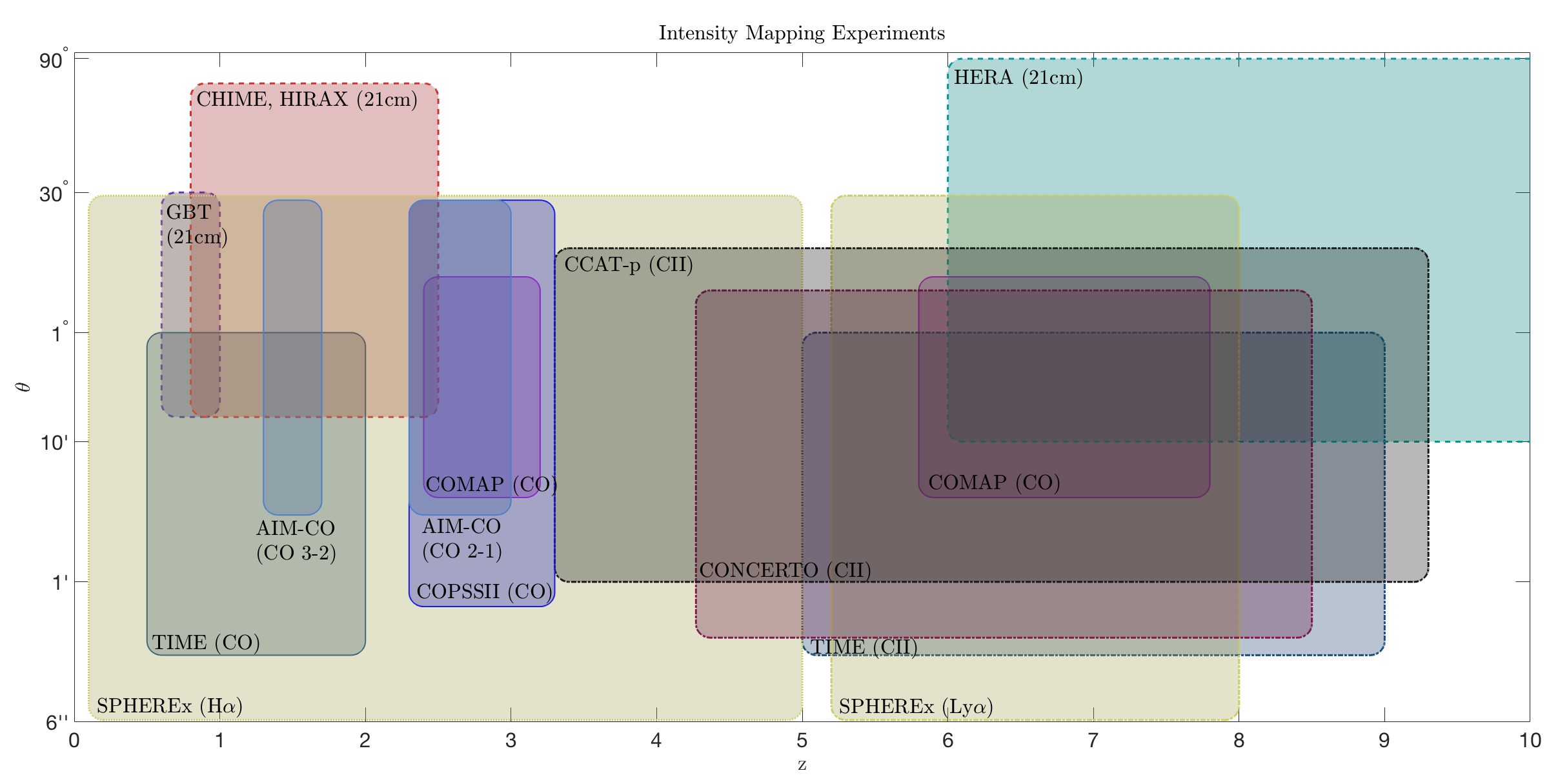}
\end{center}
\caption{A representative list of current and proposed intensity mapping experiments. 
The horizontal axis shows the redshift range of each experiment and the vertical axis 
indicates the range between the maximum resolution of the instrument and the total 
sky coverage. These include COPSSII, AIM-CO and COMAP which will target CO at 
medium redshifts, CONCERTO, CCAT-p and TIME which target [CII] at EoR 
redshifts, GBT, CHIME, HIRAX and HERA which target 21cm and SPHEREx which can 
measure H$\alpha$ and Ly$\alpha$ over a wide range of redshifts at high-resolution. 
({\it Courtesy of Ely Kovetz and Patrick Breysse})}
\label{fig:targets}
\end{figure}

A wide range of spectral lines have been considered for intensity mapping 
studies \cite{Visbal:2011ee}. While most work has focussed on the 21-cm line, 
there is growing interest in the [CII] fine-structure line \cite{Gong:2011mf,Crites:2014,Yue:2015sua}, 
the Ly$\alpha$ line \cite{Silva2013,Pullen:2013dir,Gong:2013xda,Comaschi:2015waa}, 
and rotational CO lines
\cite{Righi:2008br,Lidz:2011dx,Pullen:2012su,Mashian:2015his,Breysse:2014uia,Breysse:2015baa}.
Much of the initial motivation was to probe the epoch of reionization \cite{Suginohara:1998ti} 
(EoR) at redshift $z\sim10$, but increasingly the focus has extended to 
large-scale structure at lower redshifts. The experimental front has been 
evolving rapidly with several preliminary detections and a host of new 
experimental projects, which include an array of suborbital instruments, 
and at least two major satellite mission concepts \cite{Dore:2014cca,Cooray:2016hro}. 
Fig.~\ref{fig:targets} shows the accessible scales and redshifts of some of these 
upcoming line-intensity mapping experiments.

\begin{figure}[h!]
\begin{center}
\includegraphics[width=0.93\linewidth]{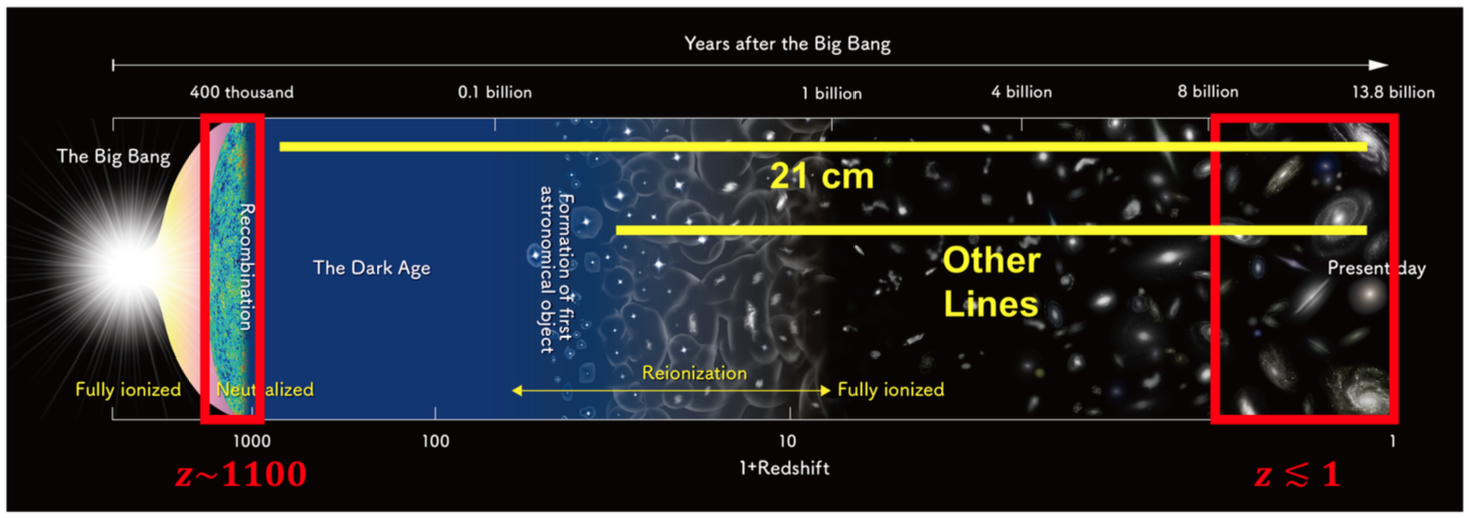}
\end{center}
\caption{Schematic view of the history of the universe. Red frames show the periods 
observed by the CMB (left) and by large galaxy surveys (right). The gap in between includes 
some of the most important periods in cosmic history, including the birth of the first 
galaxies during the dark ages, the epoch of reionization, and the growth of galaxies 
into the forms we see today. Yellow lines show the time periods accessible to intensity 
mapping surveys targeting the 21-cm line (top) as well as other lines (bottom), including 
CO, [CII], Ly$\alpha$, and many others. (Original image credit: NAOJ)}
\label{fig:IMUniverse}
\end{figure}

One or more of the lines above is observable from redshifts of order unity to redshifts 
potentially as high as 20 or more. Under the prevailing $\Lambda$CDM cosmological 
model, this will make it possible to track the growth of the first structures, the reionization 
of the Universe and the emergence of dark energy  (see Fig.~\ref{fig:IMUniverse}).  
The accessible cosmic volume is so large that it may be possible to identify even 
small deviations from $\Lambda$CDM. Meanwhile, measurements of line-emission 
over large volumes of space at high redshift may provide a unique window into 
astrophysical properties such as the star-formation rate and the density of molecular clouds.

\section{Basic Formalism}

\begin{figure}[h!]
\begin{center}
\includegraphics[width=0.5\linewidth]{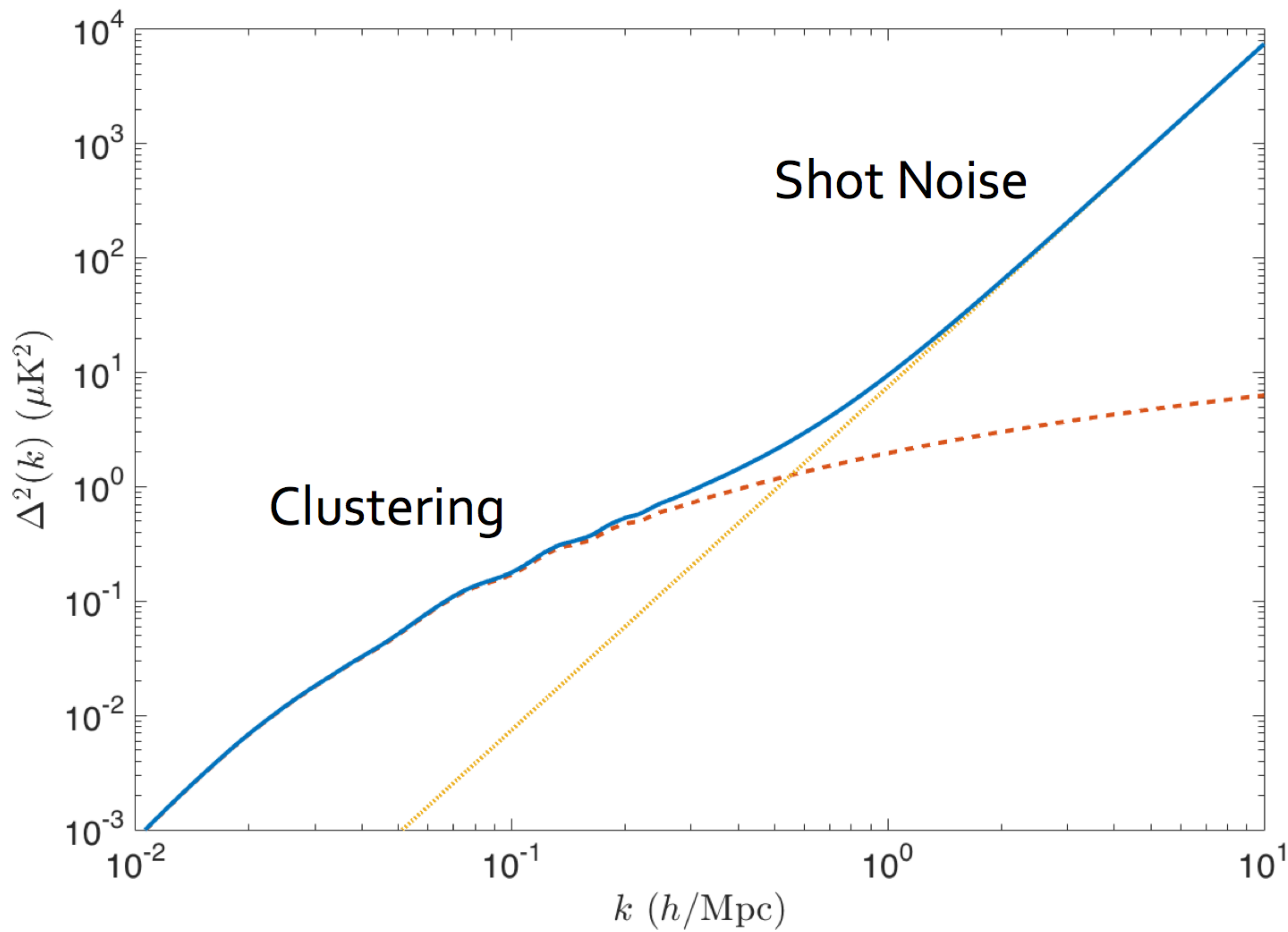}
\end{center}
\caption{The clustering and shot noise contributions to the line-intensity mapping 
power spectrum (same model as Fig.~\ref{fig:IMmerit}).}
\label{fig:PowerSpectrum}
\end{figure}

Before setting out to report on recent advances in the field and preview various efforts 
going forward, it is useful to present some of the basic  formalism. We focus on the 
power spectrum of fluctuations in a line-intensity map of emission from galaxies at some 
redshift $z$. As the line-emitting galaxies are a discrete, biased tracer of the underlying 
dark matter fluctuations, the intensity mapping power spectrum will consist of clustering 
and shot noise components (shown in Fig.~\ref{fig:PowerSpectrum})
\begin{equation}
P_k(z) = \langle I(z)\rangle^2b^2(z)P_m(k,z)+P_{\rm shot}(z)
\label{eq:PowSpec}
\end{equation}
where $b(z)$ is the redshift-dependent bias and the average intensity $\langle I(z)\rangle$ 
and shot noise power spectrum $P_{\rm shot}(z)$ are determined by the first and second 
moments of the line luminosity function $\Phi(L,z)\equiv dn(z)/dL$
\begin{equation}
\langle I(z)\rangle \propto \int\limits_0^\infty L\Phi(L,z)dL,~~~~~~~~~~P_{\rm shot}(z) 
\propto \int\limits_0^\infty L^2\Phi(L,z)dL.
\end{equation}

Examining the different quantities in Eq.~(\ref{eq:PowSpec}), one can get an idea 
of the different aspects of intensity mapping research. It is clear that the power 
spectrum above can ultimately be used to test the standard cosmological model 
in different epochs on various scales, through the dependence on the spectrum 
of dark matter fluctuations, as well as to study the astrophysical processes that 
take place in galaxies at different redshifts, as these determine the luminosity 
function of the target emission line. 

\section{The Scope of this Report}

The rest of this document presents a snapshot of the line-intensity mapping field, 
based on contributions from dozens of scientists. While currently some of the most 
advanced stages of research involve mapping of the 21-cm line, with several major 
projects currently getting underway, we pay special attention to the more newly 
emerging areas of intensity mapping with other atomic and molecular 
lines\footnote{Our discussion of 21-cm studies of the epoch of reionization is brief, 
given several prior reviews \cite{Furlanetto:2006jb,Morales:2009gs,2012RPPh...75h6901P} on the subject.  
We also refer readers to Refs.~\cite{Dore:2014cca,Dore:2016tfs} for more detailed 
discussions than provided here of intensity mapping with the Ly$\alpha$ line.}.
In the first chapter, we describe the main science targets of line-intensity mapping 
research, which include the study of large-scale structure (cosmological parameters, 
inflation, dark energy), the star-formation history, and the study of galaxy 
evolution\footnote{Line-intensity mapping measurements may also have important 
implications for a variety of other areas of astrophysics---including asteroids, the 
interstellar medium, stars, and nearby galaxies (for more detail on these applications, 
see \cite{Dore:2016tfs}).}. The next chapter describes the handful of preliminary 
detections to date. The multi-thronged experimental frontier, which consists of several 
collaborations targeting various lines with several ground-based instruments and at 
least two major satellite mission concepts, is reviewed in a separate chapter. On the 
theory side, several challenges need to be faced in order to fully exploit this technique 
and realize its ultimate potential as an astrophysical and cosmological probe. As we 
elaborate in dedicated sections below, one such challenge is how to model the intensity 
mapping signal; another is devising techniques to disentangle the signal from foregrounds 
(including line-interlopers in particular); and a third is how to best interpret the measured 
signal and extract efficient statistical constraints from it on quantities of interest ranging 
from the expansion rate of the Universe to the evolution of the star formation rate density.  As can be seen, a great deal of progress has already been made in this newly emerging field, and the potential for new discoveries is quite promising.

\chapter{Science Goals}
\label{chap:goals}

\bigskip

Line-intensity mapping has the potential to provide important information in a variety 
of areas of astrophysics.  The initial focus of intensity-mapping efforts was on the 
epoch of reionization \citep{Furlanetto:2006jb,Morales:2009gs}, primarily with 21-cm 
fluctuations.  It was soon realized, though, that 21-cm intensity mapping can, with 
fewer experimental challenges, map large-scale structure at lower redshifts.  It has 
then more recently been realized that relatively modest efforts can begin to probe 
the epoch of reionization and map large-scale structure at intermeditate redshifts 
with CO rotational lines, the [CII] line, and Ly$\alpha$, and perhaps other lines.  
Recent theoretical work and initial experimental efforts have demonstrated that line 
intensity mapping has much to offer for the study of star formation, galaxy 
formation/evolution, and the intergalactic medium at high redshift.  The field is rapidly 
growing and evolving, and new ideas for science with intensity mapping (e.g., new 
avenues to seek radiatively decaying dark matter \cite{Creque:2017} are now emerging 
at an accelerated pace).  The science case for a space mission to do Ly$\alpha$ 
intensity mapping addresses a number of questions in Galactic, stellar, 
and planetary astronomy \cite{Dore:2016tfs} that we do not address here.  It is also 
important to note---as the case of fast-radio-burst studies with CHIME \cite{Amiri:2017qtx} 
demonstrates---that observational capabilities developed for intensity mapping may 
ultimately be useful in other ways that we cannot anticipate now.

\section{Cosmology}

\subsection{Large-scale structure}

Precision observational cosmology is becoming an increasingly crowded field, 
with a host of large-scale structure techniques now reaching a sufficient level of 
maturity to seriously compete with (and complement) the precision of CMB experiments. 
Notable examples include baryon acoustic oscillation (BAO) surveys, which have 
obtained $\sim 1\%$ precision on distance measurements out to $z \sim 1$ (with 
galaxies \citep{Kazin:2014qga, Ross:2016gvb}) and $z \sim 2.4$ (with the Ly$\alpha$ 
forest) \citep{Delubac:2014aqe}; weak gravitational lensing surveys, now covering 
areas as large as $5,000$ deg$^2$ to appreciable depths \citep{Heymans:2013fya, 
Jarvis:2015hvz, Kohlinger:2017sxk, Troxel:2017xyo}; and redshift-space distortions 
from spectroscopic galaxy surveys, which have achieved better than $10\%$-level 
constraints on the growth rate of large-scale structure out to $z \sim 1$ (see the 
compilation in \citep{Macaulay:2013swa}). Taken together, these observables have 
pinned down the parameters of the standard flat $\Lambda$CDM model at around 
the 1\% level or better \citep{Ade:2015xua}, as well as uncovering several possible 
anomalies that, if confirmed, may point to new physics \citep{Battye:2013xqa, 
Raveri:2015maa, DiValentino:2016hlg, Nesseris:2017vor, Renk:2017rzu, Freedman:2017yms}.

\begin{figure}[h!]
\centering
\includegraphics[width=0.66\textwidth]{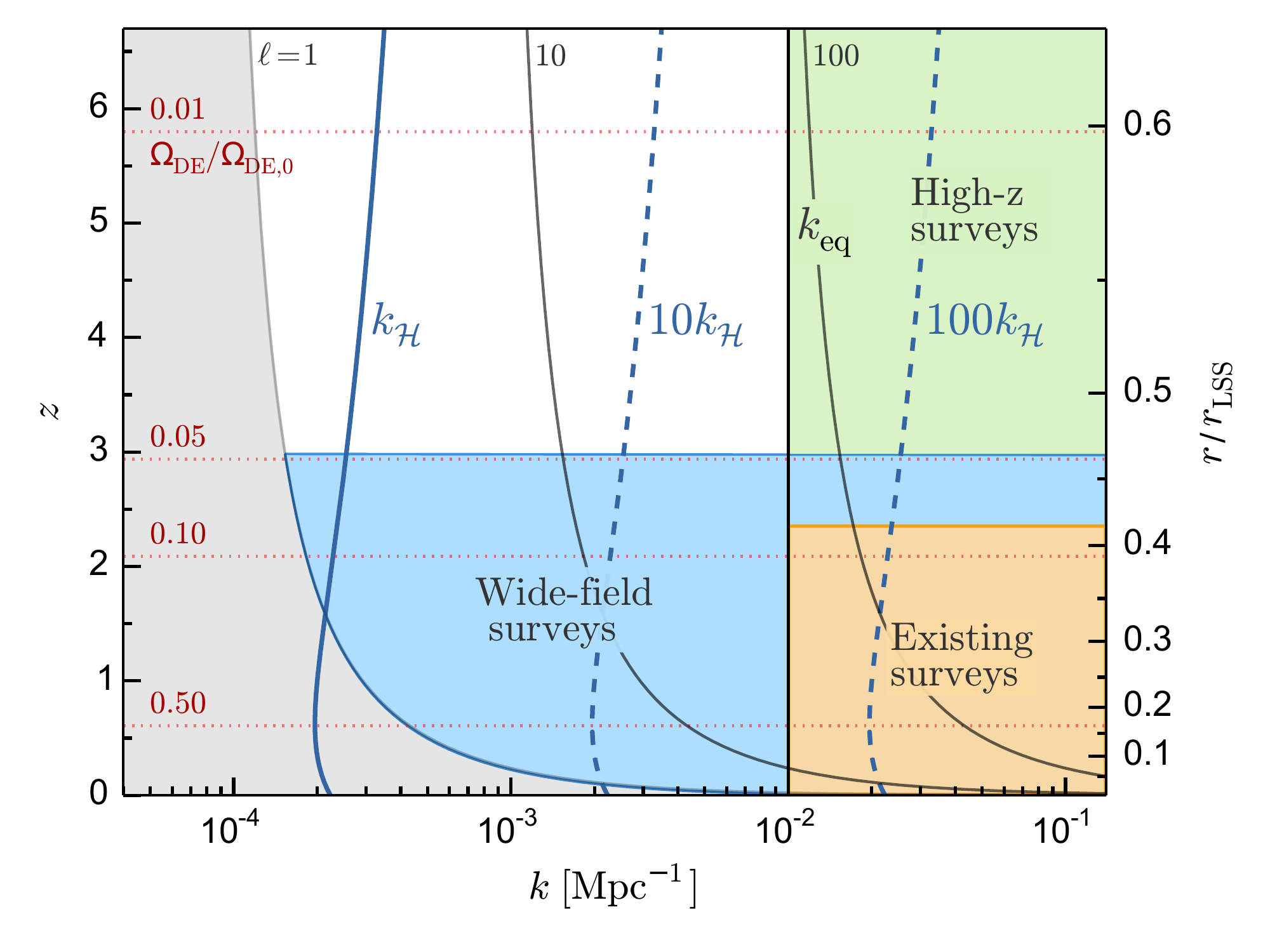}
\caption{Distance scales that are nominally reachable by current and future 
large-scale structure surveys, as a function of redshift (left vertical axis) and 
comoving distance to last scattering (right vertical axis). The red dotted lines 
show how the dark energy density is expected to change with redshift (assuming 
$w \simeq -1$), while the blue solid/dashed lines correspond to multiples of the 
Hubble scale, $k_\mathcal{H} \sim \mathcal{H}$, as a function of redshift. The 
black curved lines show the Limber-approximated Fourier wavenumbers 
corresponding to several spherical harmonic wavenumbers, $\ell$, as a function 
of redshift. The shaded gray region shows modes with wavelengths larger than 
half the sky, and the thick black line shows the matter-radiation equality scale, 
$k_{\rm eq}$, where the matter power spectrum turns over. 
({\it Courtesy of Phil Bull})}
\label{fig:scales}
\end{figure}

The question is whether intensity mapping, as a relative newcomer, has much 
to offer beyond what is already being provided by more mature methods like 
spectroscopic and photometric galaxy surveys. To answer this, it is instructive 
to look at the range of distance scales and redshifts that have been probed by 
existing methods, and that will be probed in the near future by planned surveys 
such as DESI \citep{Aghamousa:2016zmz}, Euclid \citep{Amendola:2012ys}, 
and LSST \citep{Abate:2012za}. An illustration is shown in Fig.~\ref{fig:scales}, 
which focuses on linear scales (those with $k \lesssim 0.2$ Mpc$^{-1}$) out to 
$z \approx 6$; i.e. the post-reionization era.

Existing surveys (orange region in Fig.~\ref{fig:scales}) have covered only a 
small corner of the space of accessible linear modes. They have mostly been 
restricted to redshifts of unity or less, with some limited coverage out to 
$z \approx 2.4$ from the Ly$\alpha$ forest. Arguably, no existing survey 
has reached scales larger than the matter-radiation equality scale 
$k_{\rm eq} \approx 10^{-2}$ Mpc$^{-1}$, which marks the turnover in the 
matter power spectrum (the closest so far is BOSS \cite{Gil-Marin:2014sta}).

Forthcoming wide-field galaxy surveys (blue region) will be able to extend our 
reach to much larger scales, beyond the matter-radiation equality scale, and 
perhaps even out to the Hubble scale ($k \sim H_0$). This regime is particularly 
interesting for performing tests of general relativity and detecting signatures 
of primordial non-Gaussianity, as discussed below. Intensity mapping has a 
definite role to play here, as while the planned galaxy surveys will be able to 
reach the low-$k$ regime in principle, they are not optimized for the task, and 
will likely struggle to mitigate a variety of large-scale systematic effects. Intensity 
mapping surveys tend to be cheaper and enjoy much faster survey 
speeds than optical galaxy surveys, as so it is plausible to design specialized 
surveys to study the very largest scales. An example of such a survey is SPHEREx, 
a space-based mission that will be surveying the entire sky with low resolution 
spectroscopic observations adequate for intensity mapping \citep{Dore:2014cca}. 
While not strictly optimized for the task, 21cm intensity-mapping experiments with large 
fields of view (e.g. SKA1-MID, CHIME and BINGO) should also be better suited 
to recovering large angular scales, especially at higher redshifts.

Intensity mapping is also invaluable for reaching higher redshifts (green region). 
Beyond $z \sim 3$, optical and near-IR galaxy surveys become much more 
difficult due to a combination of the large distances involved and the redshifting 
of the emission of the galaxies. The only currently planned galaxy surveys in the 
$z \gtrsim 3$ regime will cover comparatively small areas (e.g., HETDEX 
\cite{Hill:2008mv}). Intensity mapping is better suited for studying large-scale 
structure at high redshift, as the dilution of the signal---the aggregate emission 
from many galaxies, instead of just one---is less severe with distance/redshift.

Most current and near-future surveys focus on the $k \gtrsim k_{\rm eq}$, 
$z \lesssim 2$ regime because it corresponds to the period of dark energy domination, 
and contains a large number of linear Fourier modes (the total number of Fourier 
modes in a 3D survey naively scales like $k_{\rm max}^3$). There are a number 
of good reasons to extend our observations to larger scales and higher redshift 
though, especially if the goal is to perform precision tests of fundamental physics.

\begin{figure}[h!]
\centering
\includegraphics[width=7.6cm]{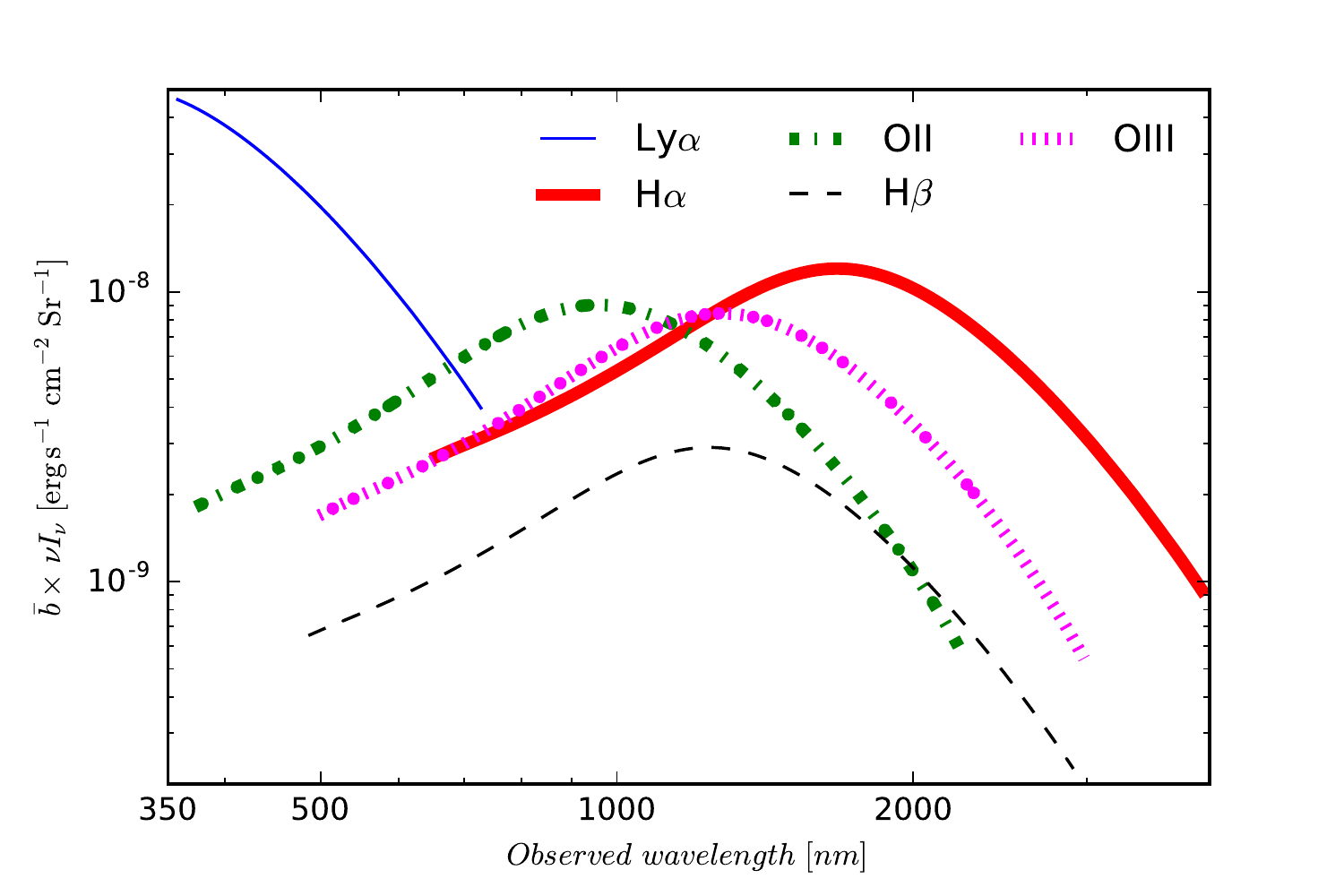}
\includegraphics[width=7.6cm]{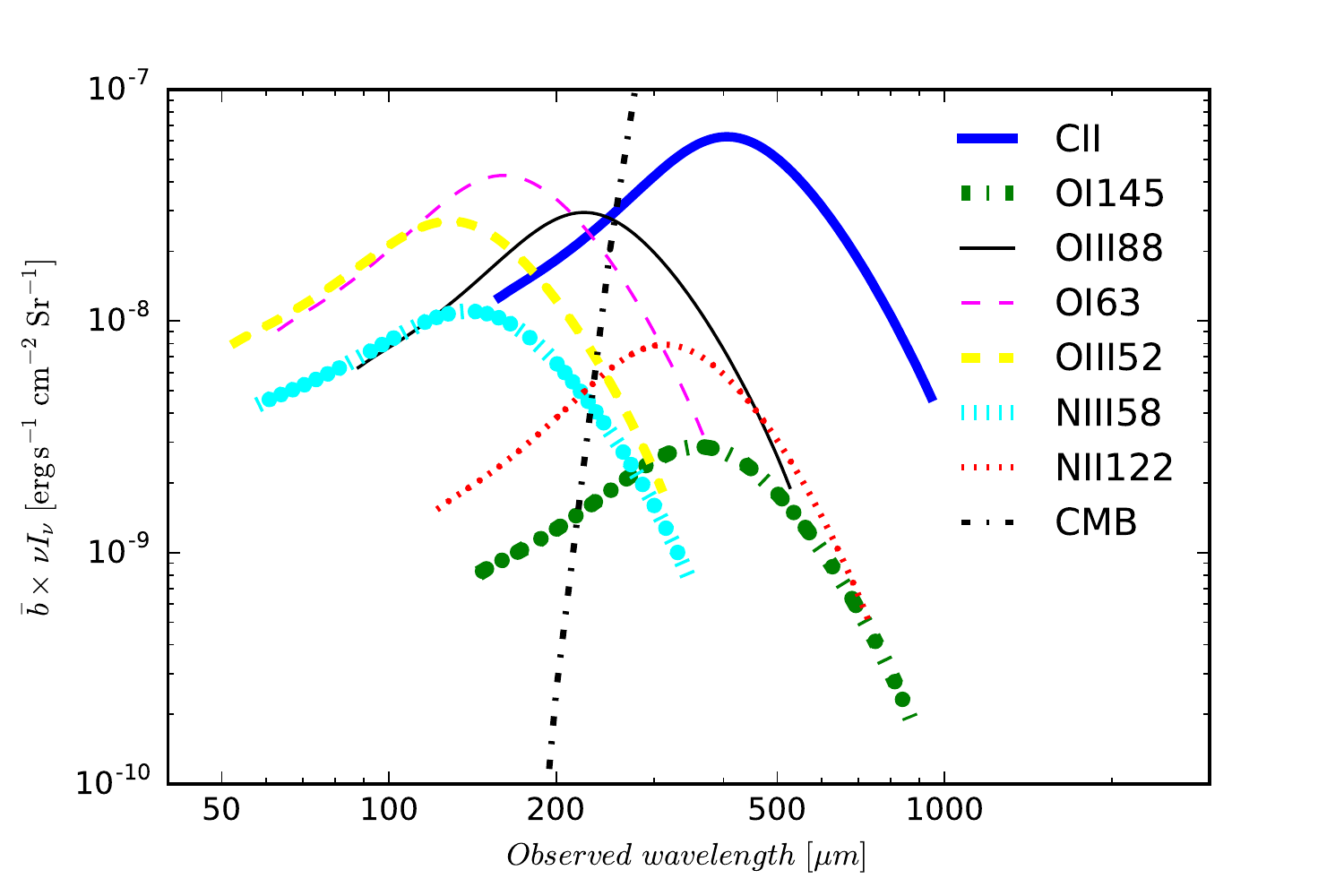}
\includegraphics[width=7.6cm]{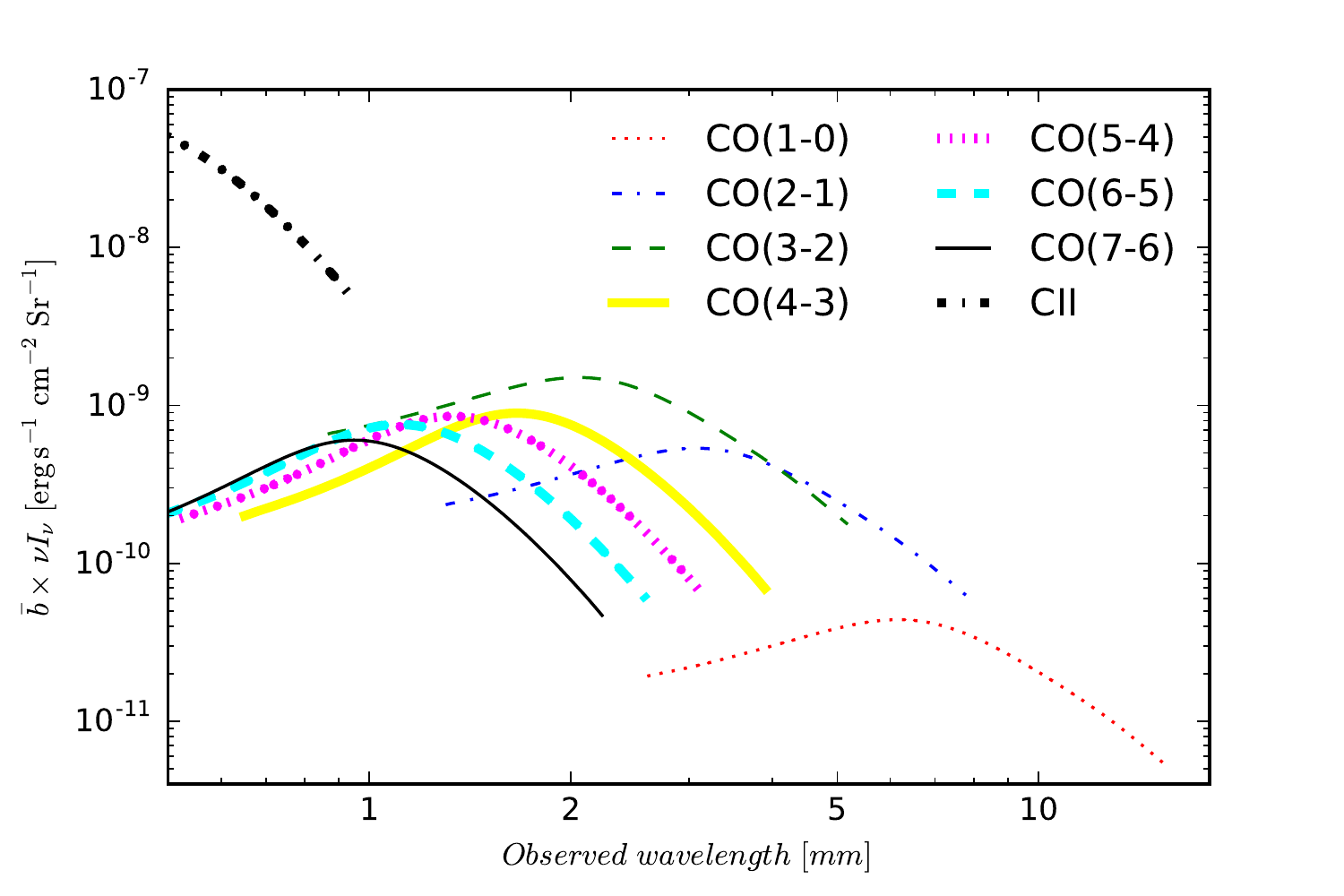}
\includegraphics[width=7.6cm]{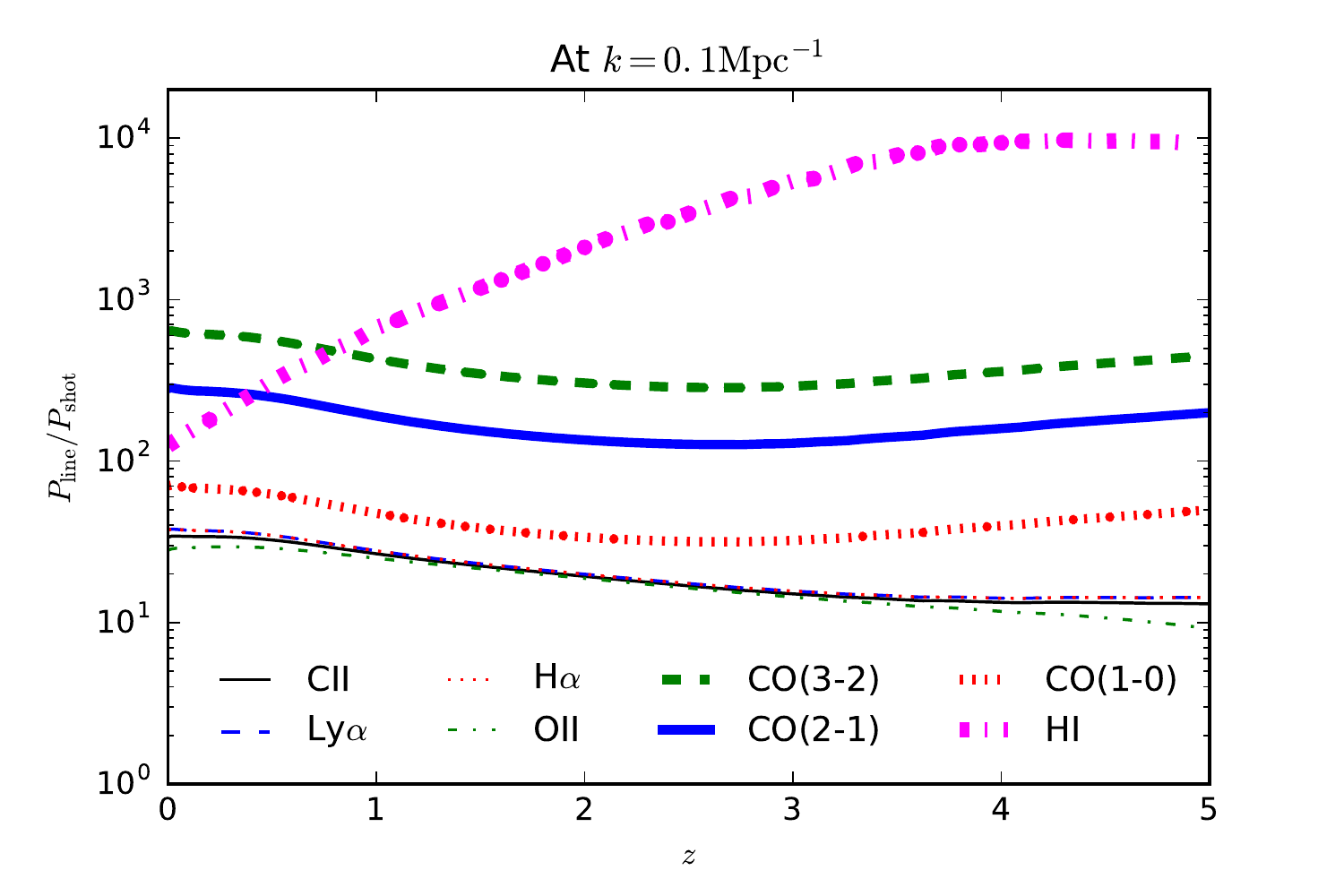}
\caption{{\bf Comparison among atomic and molecular emission lines}. \emph{Top left:} 
Optical and Near Infra-red emission lines. \emph{Top right:} Far Infra-red emission lines. 
\emph{Bottom left:} Far Infra-red to milimeter emission lines. \emph{Bottom right:} 
Comparison between power spectrum of a given line at $k=0.1$ Mpc$^{-1}$ and its 
shot-noise assuming an instrumental noiseless experiment.  
({\it Courtesy of Jos\'e Fonseca})}
\label{fig:comp_lines}
\end{figure}

Below we consider the prospects of different line-intensity mapping measurements 
in placing constraints on $\Lambda$CDM through power spectrum measurements. 
As described in the Introduction, the amplitude of the intensity mapping power spectrum 
is given, at first order, by the power in dark matter density fluctuations times the square 
of the product of the bias and intensity of the line. Therefore, when comparing atomic 
and molecular emission lines it is not enough to compare their brightness; one has to 
also include the bias. In the top panels and bottom left of Figure \ref{fig:comp_lines} 
one can see how a selection of the most intense lines compare to each other in different 
regions of the electromagnetic spectrum \cite{Fonseca:2016qqw}. These comparisons 
are highly dependent on the assumed model and the assumed gastrophysics and are 
only intended to suggest which lines might be targeted with intensity mapping experiments. 
The bottom right panel of Figure \ref{fig:comp_lines} gives the ratio between the power 
spectrum of a given line at a fixed comoving scale and its shot-noise power. In the future, 
one hopes that intensity mapping experiments become ever more sensitive to the point 
where shot noise becomes the dominant noise term. In that limit one can see that the 
radio to submilimeter lines are the ones that can perform better. It is also in that limit 
one should look for emission-line information encoded in the shot noise.

\subsection{Probing the $\Lambda$CDM Cosmological Model} 

It is worth asking whether any current, planned or future intensity mapping instrument 
can provide robust measurements of the power spectrum at redshifts beyond the reach 
of current galaxy surveys, irrespective of the models one uses to estimate the cosmological 
average intensity of an emission line (see the Modeling Section for more details on how 
this can be done). An extensive study of this has been done in \cite{Fonseca:2016qqw}, 
from which we present a short summary in Table \ref{tab:expsum}, indicating the lines 
used and the assumptions for the instrument specifications and observational strategies. 
While some lines still have low signal-to-noise ratios (SNRs), others have a great potential 
to measure not only the power spectrum, but the baryon acoustic oscillations (BAO) with great resolution. 

\begin{table*}[h!]
\centering
\caption{Survey details in estimating $P(k)$ for the different lines in consideration.}
\begin{tabular}{l c c c c c c c c c}        
\hline\hline                 
 & Line & Area & $z$  & $\delta z$ & $\delta \Omega$ & $\Delta z$ & $P_N$ &$k$ range & SNR \\
&  &  [$\deg^2$] &   &  &  [$10^{-9}$ Sr]&   & [erg$^2$s$^{-2}$cm$^{-4}$Sr$^{-2}$Mpc$^3$] & [Mpc$^{-1}$]& \\
\hline
HETDEX& ${\rm Ly\alpha}$& 300& 2.1 & 0.005 & $0.213$& 0.4 &7.24 $\times 10^{-16}$&0.009-0.3& 489\\
SPHEREx& ${\rm H\alpha} $& 90 & 1.9 & 0.07  & $0.903$& 0.4 & 2.59 $\times 10^{-12}$&0.009-0.3& 105 \\
&  OII& 90 &  1.2 & 0.05 & $0.903$& 0.4 &  1.34 $\times 10^{-11}$ & 0.02-0.1& 11\\
TIME-like& [CII] &  100  & 2.2 &0.002 & $13.5$& 0.4& 4.59 $\times 10^{-13}$ & 0.01-0.3& 294\\
& &  2 (50h) &  &&&& 3.67 $\times 10^{-13}$  & 0.05-0.3& 42\\
& CO(3-2) &   250 & 2.0 &0.01 & $85$& 0.4& 3.31 $\times 10^{-16}$ & 0.01-0.3 & 471 \\
&  &   2 (50h) & & & & & 1.06 $\times 10^{-16}$ & 0.05-0.3 & 45 \\
\hline                                  
\end{tabular}
\label{tab:expsum}     
\end{table*}

For instance, the forthcoming Hobby-Eberly Telescope 
Dark Energy Experiment (HETDEX \cite{Hill:2008mv}) can be a promising 
probe of the BAO signal, using Ly$\alpha$ intensity mapping. 
HETDEX was originally planned to analyze the large-scale clustering of Ly$\alpha$ emitters (LAE) 
detected with high S/N over 400 ${\rm deg}^{2}$ (300 ${\rm deg}^{2}$ for 
the spring and 100 ${\rm deg}^{2}$ for the fall field) at $1.9 < z < 3.5$ 
(corresponding to $350-550\,{\rm nm}$ wavelengths or the $545-856\,{\rm THz}$ 
frequency range). However, the unbiased nature of HETDEX, which uses Integral Field Units, 
will also provide a Ly$\alpha$ intensity map, and its cross-correlation with the detected LAEs 
can be a powerful probe of the BAO signal \cite{Saito:2017xx}. 

\begin{figure}[h!]
\begin{center}
\includegraphics[width=0.6\textwidth]{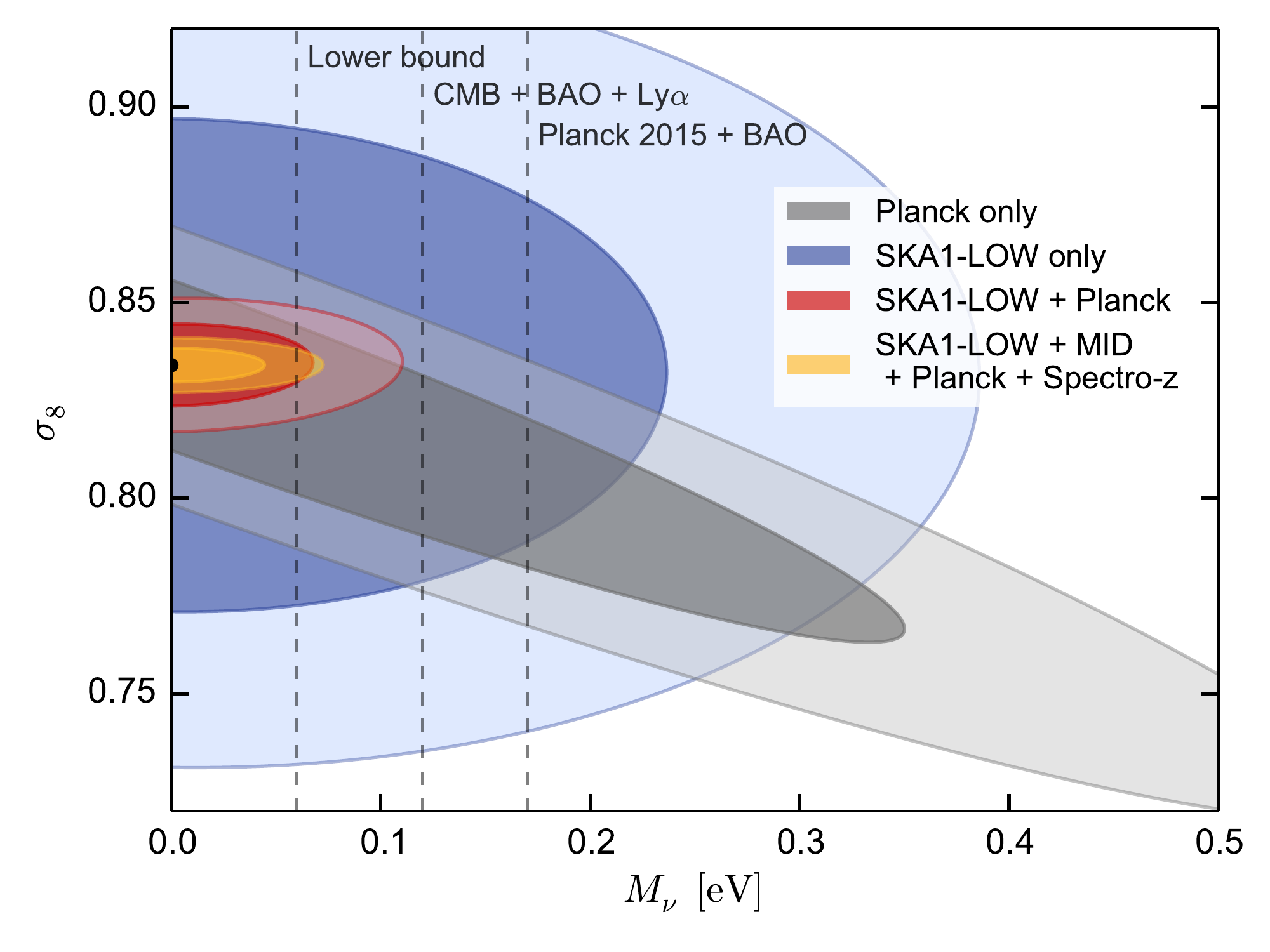}
\caption{1- and 2$\sigma$ constraints on the sum of the neutrino masses, 
and its degeneracies with $\sigma_8$, from Planck (grey), SKA1-LOW (blue), 
SKA1-LOW + Planck (red) and SKA1-LOW + SKA1-MID + Planck + Euclid (yellow). 
The lower limit from neutrino oscillations, together with the bounds from 
CMB+BAO+Ly$\alpha$, and Planck 2015 95\% limits are shown as vertical 
dashed lines from left to right respectively. ({\it Courtesy of Francisco Villaescusa-Navarro})}
\label{fig:mnu_sig8}
\end{center}
\end{figure}

One important goal of modern cosmology is a measurement of neutrino masses. 
Knowing their masses may also lead us to determine their hierarchy, allowing 
us to rule out a whole class of theoretical models aiming at explaining their masses. 
Neutrinos leave characteristic signatures on cosmological observables mostly 
due to the fact that they have very large thermal velocities, that make their 
dynamics very different to those of cold dark matter or baryons. In order to 
weigh neutrinos with cosmological observables, data from both large scales, 
e.g.\ CMB, and small scales, e.g.\ the Ly$\alpha$-forest, is needed. This is 
precisely the regime where 21-cm intensity mapping holds promise since very large 
cosmological volumes can be probed down to relatively small scales. Besides, 
since intensity mapping can probe a very wide redshift range, the limitations on 
their bounds, arising from the degeneracies between neutrinos and other 
cosmological parameters, can be somewhat relaxed.  \citet{FVN2015} performed a detailed study 
through hydrodynamic simulations of the constraints that a combination of data 
from SKA1-MID, SKA1-LOW, Planck and a Euclid-like spectroscopic galaxy 
survey can place on neutrino masses. They find that the neutrino masses can 
be constrained (see Fig. \ref{fig:mnu_sig8}) with a very competitive error of 
$34$ meV ($1\sigma$).

\begin{figure}[!]
	\centering
	\includegraphics[width=\textwidth]{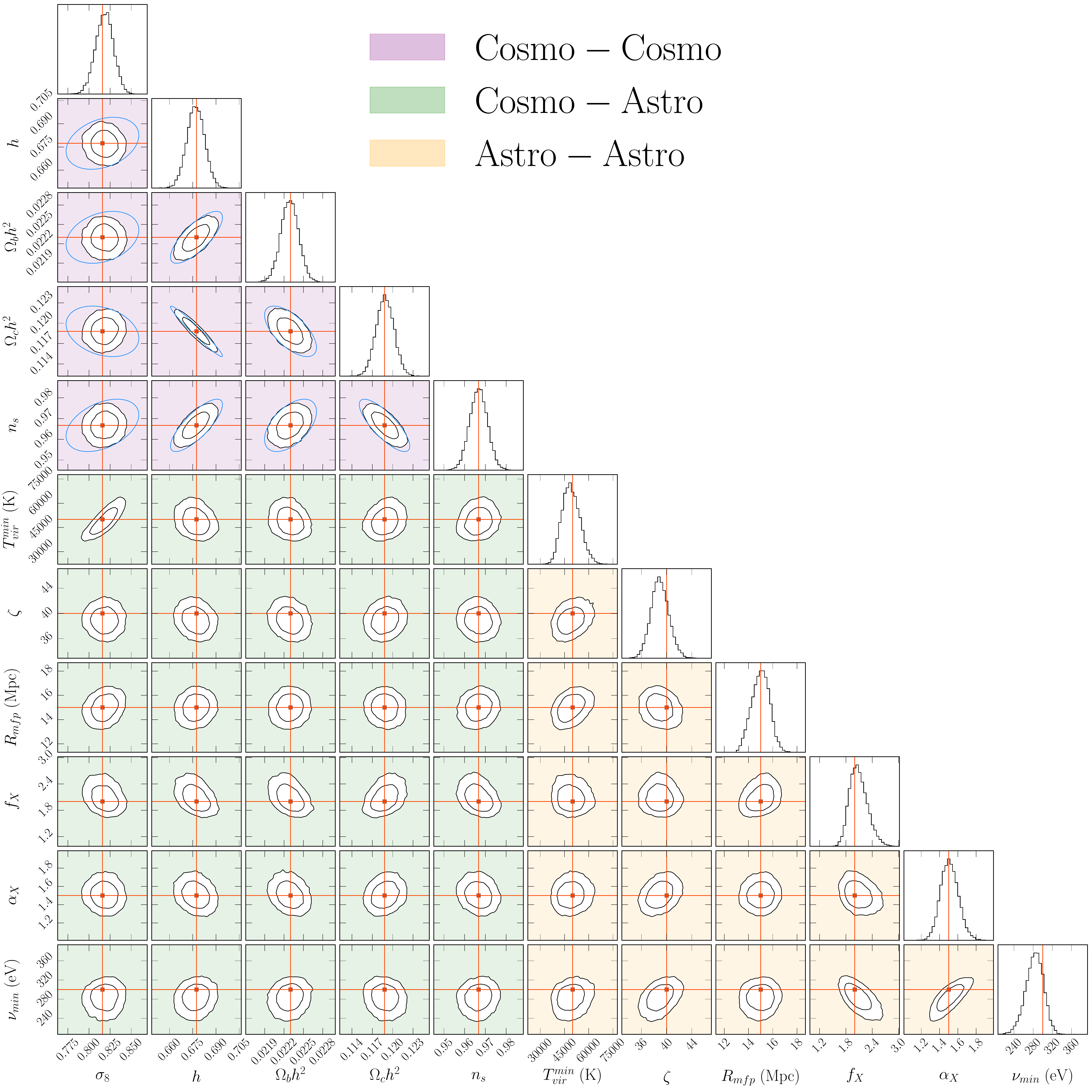}
	\caption{Forecasted $68\%$ and $95\%$ confidence regions for HERA, 
	using a reionization model parameterized by $\sigma_8$, $h$, $\Omega_c h^2$, 
	$n_s$, the minimum virial temperature of ionizing halos $T_{\rm vir}^{\rm min}$, 
	the ionizing efficiency of UV photons $\zeta$, the mean free path of ionizing 
	photons $R_{\rm mfp}$, the X-ray flux amplitude $f_X$, the X-ray spectral index 
	$\alpha_X$, and the X-ray cut-off frequency $\nu_{\rm min}$. Red squares 
	indicate the fiducial values of this forecast, while the blue contours show the 
	$95\%$ confidence regions from \emph{Planck} alone. Reproduced from 
	\cite{kern_2017}. ({\it Courtesy of Adrian Liu})}
	\label{fig:tri}
\end{figure}

As mentioned above, there exist degeneracies between cosmological parameters 
and astrophysical parameters at the epoch of reionization \cite{liu_2016b}. For 
example,  Figure \ref{fig:tri}  shows forecasted parameter constraints from a 
HERA power spectrum measurement \cite{kern_2017}, where one sees a degeneracy 
between $T_{\rm vir}^{\rm min}$, the minimum virial temperature of ionizing halos, 
and $\sigma_8$. This arises because a higher $\sigma_8$ accelerates structure 
formation, and thus moves reionization to an earlier epoch. One can then compensate 
for this by raising $T_{\rm vir}^{\rm min}$, which limits the ionizing influence of 
galaxies to only the most massive galaxies, thus delaying reionization. From this, 
we see that even with the \emph{Planck}-level errors on cosmological parameters 
assumed in Figure \ref{fig:tri}, uncertainties in cosmological parameters cannot be 
neglected even if the goal is on constrain astrophysics.

While the necessity of including cosmological parameters in the analyses may 
seem like a burden, it is also an opportunity, for it means that high-redshift 
$21\,{\rm cm}$ experiments can in principle place competitive constraints on 
cosmological parameters. Conceptually, this is unsurprising, given their vast 
($\sim$hundreds of Gpc$^3$) survey volumes. The challenge, however, is in 
the extraction of this cosmological information from amidst the messy astrophysics 
of reionization. There are several approaches to this:
\begin{itemize}
\item The most direct approach is simply to treat the astrophysical and cosmological 
parameters on an equal footing, varying both in simulations of the $21\,{\rm cm}$ 
power spectrum \cite{liu_2016b,kern_2017}. This then allows a simultaneous fit 
of all parameters to the observed power spectrum. The disadvantage of this is that 
reionization simulations are computationally expensive, and therefore this approach 
has only been explored for semi-analytic simulations (and even then the problem 
is only barely tractable \cite{kern_2017}).
\item An alternative approach has been to assume that the power spectrum of the 
ionization field and the cross-power spectrum between the ionization and density 
fields are both proportional to the matter power spectrum. The proportionality 
constants are then parameterized by a functional form motivated by fits to radiative 
transfer simulations. The free parameters in the functional form are treated as nuisance 
parameters in the data analysis and marginalized out \cite{mao_2008,clesse_2012}. 
For this approach to be robust, however, early observations must confirm the 
parametrized forms. Additionally, typical treatments of this approach have assumed 
that the nuisance parameters are not themselves dependent on cosmological parameters.
\item A final approach is to move away from the model dependence of the previous 
approaches, and to instead use redshift space distortions. The core idea is that redshift 
space distortions are sourced by the underlying density field, rather than ionization 
fluctuations. Building on this idea, \citet{barkana_2005} showed that to linear order, 
a separation of the $21\,{\rm cm}$ power spectrum into powers of $\mu$ (where $\mu$ 
is the cosine of the angle of the Fourier mode with respect to the line of sight) yields a 
$\mu^4$ component that depends only on the density fluctuations. By extracting this 
component from the data, one  evades astrophysical parameters and directly 
constrains cosmological ones. However, in doing so, one discards information content 
(since the portions of the signal that are intertwined with astrophysical parameters 
\emph{do} contain cosmological information, albeit information that is difficult to access). 
Forecasts therefore predict less stringent parameter constraints with this method 
\cite{mao_2008}. Moreover, one must be cognizant of the possibility that nonlinear 
fluctuations could spoil the separation of astrophysics and cosmology, with simulations 
suggesting that the separation works only early in reionization ($\lesssim 40\%$ 
ionized; \cite{shapiro_2013}).
\end{itemize}
In practice, a robust estimation of cosmological parameters from high redshift 
$21\,{\rm cm}$ surveys may involve a combination of all these approaches.

Lastly, we note that instead of directly constraining cosmological parameters, 
a possible alternative is to use high redshift $21\,{\rm cm}$ surveys to concentrate 
on measuring astrophysical parameters, which can then be used to forward model 
any reionization nuisance parameters in other cosmological probes. For example, 
CMB studies need to fit for the optical depth $\tau$, which arises due to the Thomson 
scattering of CMB photons by electrons in the IGM. Since these electrons are the 
result of the reionization process, a successful $21\,{\rm cm}$ measurement can---through 
modeling---be converted into a prediction for $\tau$ \cite{liu_2016a}. Though such 
a measurement would be model-dependent (recall that the $21\,{\rm cm}$ brightness 
temperature probes a specific mixture of ionization, temperature, velocity, and density 
fluctuations, rather than each of these individually), it would  nonetheless be 
valuable, since a $\tau$ prediction would break $A_s$-$\tau$ degeneracies in the CMB.

\subsection{Going beyond $\Lambda$CDM}

\begin{figure}[h!]
\centering
\includegraphics[width=0.8\textwidth]{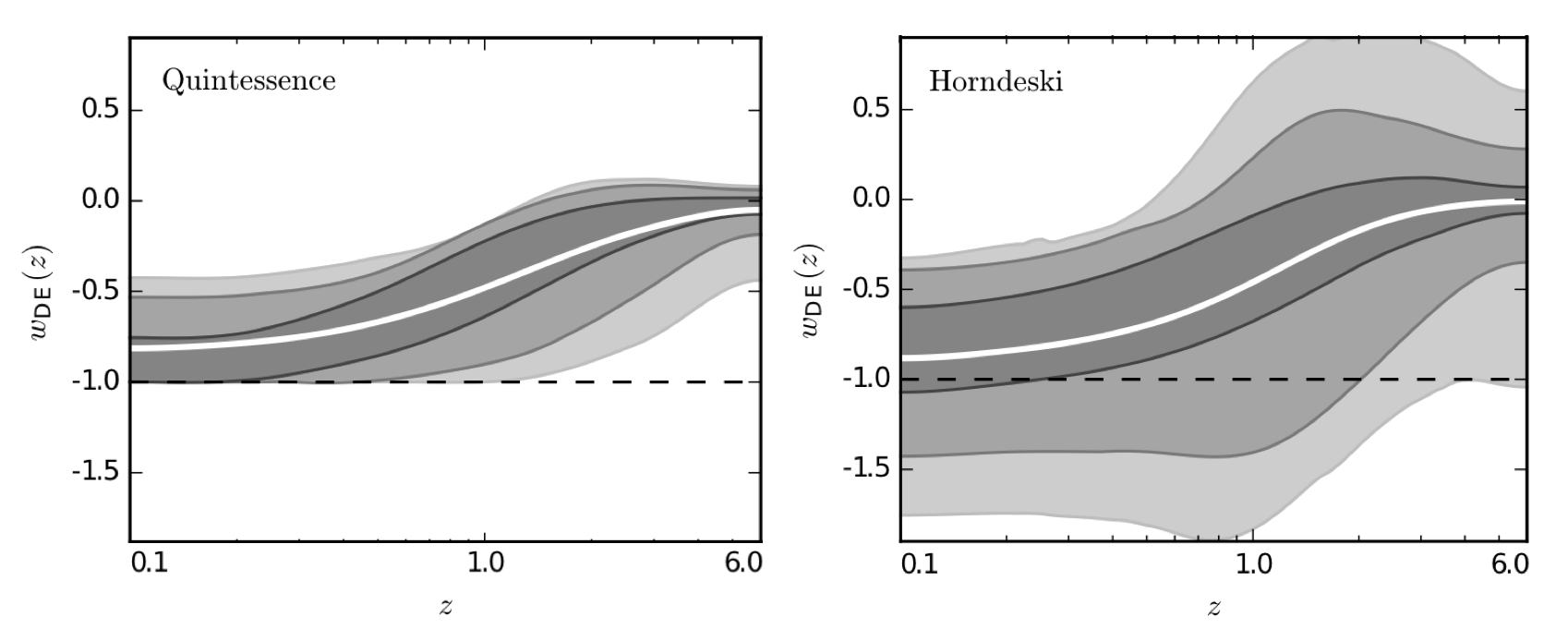}
\caption{Prior distributions of $w(z)$ for the Horndeski class of scalar field models, 
adapted from the Monte Carlo study of \protect\cite{Raveri:2017qvt}. The left panel 
shows the results for minimally-coupled (quintessence) subclass, while the right is 
for the general class. The contours show 68\%, 95\%, and 99.7\% intervals, while 
the white line shows the median. There is a clear tendency in these models to deviate 
from $w=-1$ at $z \gtrsim 1$. ({\it Courtesy of Phil Bull})}
\label{fig:deevol}
\end{figure}

Intensity mapping is ideally suited to provide constraints on the possible time 
variation of the equation of state $w(z)$ of dark energy which is one of the major 
focuses of observational cosmology. While dark energy dominates the cosmic 
energy density at the lowest redshifts, this is not necessarily the best regime in 
which to look for departures from a cosmological constant, $w = -1$. Most current 
surveys focus on a broad redshift range around $z \sim 1$, but significant deviations 
from a cosmological constant are more likely to occur at higher redshift, $z \gtrsim 2$, 
in a large class of models. Results from \cite{Raveri:2017qvt}---a Monte Carlo 
exploration of many millions of physically-viable Horndeski single-scalar field 
models---are shown in Fig.~\ref{fig:deevol}. These models have five arbitrary 
time-dependent coupling functions that were parametrized in several different ways, 
with coefficients chosen from broad prior distributions. The models were then 
evolved to find $w(z)$, and unphysical models (e.g. with ghost instabilities) discarded. 
The result is a prior probability distribution over the space of possible Horndeski 
$w(z)$ functions. As shown in Fig.~\ref{fig:deevol}, both the simplest Horndeski 
subclass (minimally-coupled quintessence models), and the most general, exhibit 
a typical `tracking' behavior, where $w \simeq -1$ at low redshift, but deviates 
significantly ($w \to 0$) at $z \gtrsim 2$. While there can be several different 
causes of this behavior, the most common is due to non-minimal couplings, 
which force the scalar field to track the evolution of the dominant component 
of the cosmic energy density. This seems to be quite generic, at least for 
scalar field theories. As per Fig.~\ref{fig:deevol}, the transition from matter 
domination ($z \gtrsim 2-3$) therefore seems to be a particularly promising place 
in which to search for deviations from a cosmological constant, making it an 
interesting target for intensity mapping surveys (see, e.g., \cite{Bull:2015lja}).

Beyond constraining modified gravity theories through their equation of state, 
intensity mapping experiments offer the promise of constraining screening mechanisms, 
\cite{Khoury:2003aq, Khoury:2003rn} through their environment dependent behavior. 
Astrophysical tests of gravity theories, including chameleon \cite{Khoury:2003rn} and Vainshtein 
\cite{Vainshtein:1972sx} screening depend on the distinct signatures of each theory 
and the environment under study.  Thus nuanced experiments can tease out their effects, 
in varied ways, including a comparison of the mass distribution inferred by different probes. 
Intensity mapping experiments, in concert with high resolution optical imaging and 
spectroscopy will allow us to study the different behavior of different components of 
galaxies and thus potentially constrain or rule out these theories. For a comprehensive 
discussion of such tests, see \cite{Jain:2013wgs} and references therein.

The 21cm signal from pre-stellar times is a sensitive probe of new physics, and dark matter
in particular, much in the same way of the CMB \cite{Dvorkin:2013cea,2015PhRvL.115g1304A,Slatyer:2016qyl}. 
For example, dark matter annihilation at early times could provide an early and uniform 
source of heat \citep{EvoliMesingerFerrara2014}.  This model is distinguishable from 
standard models by its spatial distribution and timing \citep{2014MNRAS.439.3262M}. 
The effect is generally one of de-emphasizing large fluctuations and increasing the pace 
of heating. The opposite is also possible, a non-zero cross-section between cold dark 
matter and baryons could drive the gas temperature below or above that expected by Hubble 
expansion, depending on the dissipation of kinetic energy into heat as a result of friction 
between the baryon and dark matter fluids, significantly increasing the emission/absorption 
signal \citep{2015PhRvD..92h3528M}. In both of these cases intensity mapping at $z>20$ 
becomes an exploration of basic physics.


Another strong motivation to access ultra-large scales with intensity mapping 
measurements is the possibility to detect hints of primordial non-Gaussianity, 
which occurs in certain types of multi-field inflation (see, e.g., the discussion 
in \cite{dePutter:2016trg}). If non-Gaussian correlations were imprinted in the 
curvature perturbations left over after inflation, then one would expect to see 
a scale-dependent bias in the two-point function of a biased tracer of the 
underlying density field \citep{Dalal:2007cu}. This effect is mostly limited to 
very large scales (the bias correction is $\propto k^{-2}$), and so as with  
relativistic corrections, a multi-tracer approach will be useful here too 
\citep{Seljak:2008xr, Alonso:2015sfa, Fonseca:2015laa} (see the Techniques 
Section for more details).

\subsection{The Physics of the Epoch of Reionization} 

\begin{figure}[h!]
\begin{center}
\includegraphics[width=0.66\textwidth]{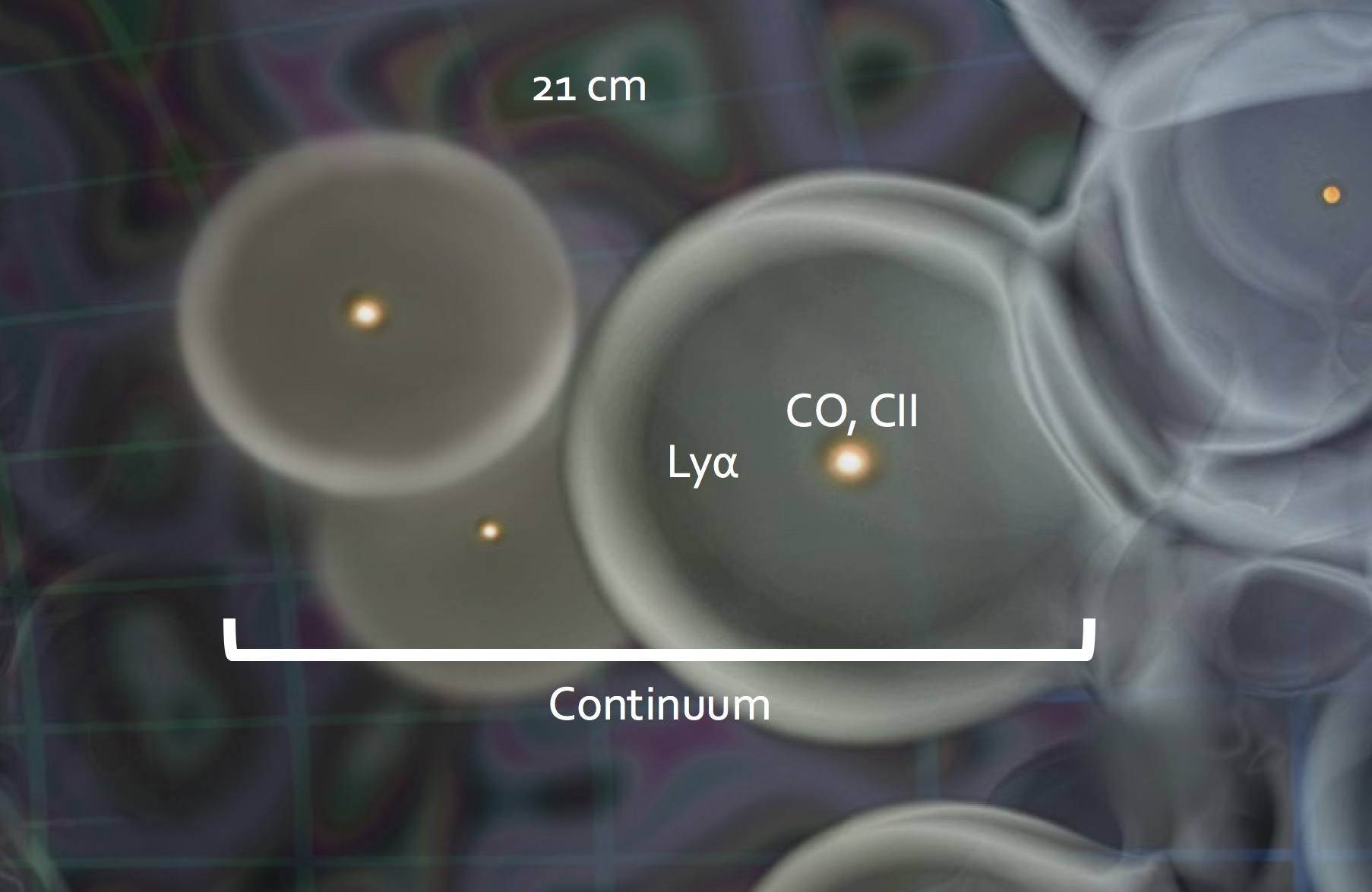}
\vspace{0.1in}
\caption{This is an illustration of the signals from CO and [CII] from within galaxies, 
Ly$\alpha$ from around galaxies, and 21-cm from neutral IGM. Fainter lines can be 
seen in cross-correlation, while the continuum is measured as the cosmic infrared background. (Background 
Image Credit: Scientific American - pending approval) 
({\it Courtesy of Patrick Breysse})} 
\end{center}
\label{fig:eorIM}
\end{figure}

 Galaxy surveys by the Hubble Space Telescope have dramatically improved 
our understanding of the Epoch of Reionization (EoR) \cite{2015ApJ...802L..19R}, 
but our knowledge of the detailed connection between the galaxy population 
and the evolving state of the intergalactic medium can be revolutionized through 
line-intensity mapping signals. This is especially true if one uses measurements 
of several different lines originating from this epoch in tandem, to obtain complementary 
information about the physical processes that are taking place \cite{Lidz:2011dx}. 
To appreciate the potential of line-intensity mapping in providing a unique picture 
of EoR, consider the illustration in Fig.~\ref{fig:eorIM}. The 21-cm line maps the neutral 
gas from outside of the ionized bubbles, while CO and [CII] lines trace the star-forming 
galaxies that create the ionizing photons, most of which are too faint to detect. Ly$\alpha$ 
probes the galaxies as well, along with the halos around them.  Together, these 
lines allow the study of the wide range of spatial scales and physical processes 
which contribute to reionization.  Cross-correlations between these lines will add 
additional information about their relationships, and correlations with fainter lines 
will probe astrophysical conditions in ever more detail.  

The prospects to probe the astrophysics of reionization with intensity mapping 
experiments have been considered in more detail for the $21\,{\rm cm}$ line. 
For instance, instruments such as the Hydrogen Epoch of Reionization Array 
(HERA; \cite{deboer_2017}) are forecasted to provide $\sim 10\%$ to $20\%$-level 
constraints on parameters such as the minimum virial temperature of ionizing halos, 
the ionizing efficiency of these halos, the UV and X-ray emissivities of ionizing sources, 
and the mean free path of ionizing photons in ionized regions \cite{pober_2014, greig_2015,greig_2017}. 
Such forecasts assume that the survey volume is subdivided 
into several redshift bins (say, bins of thickness $\Delta z = 0.5$), and that a series 
of power spectrum  measurements are made, one in every bin. Note that unlike with 
low-redshift 21-cm intensity mapping surveys, experiments targeting 
reionization are sky-noise dominated (owing to the brighter foreground sky at the 
lower observation frequencies). Thus power spectrum measurements generally lose 
signal to noise in the highest redshift (lowest-frequency) bins fairly quickly (depending on the 
particular survey and the foreground removal techniques used). However, these noisier 
measurements are still crucial for breaking degeneracies between reionization 
parameters \cite{pober_2014}. We also note that degeneracies between cosmological 
parameters and astrophysical parameters need to be taken into account when 
extracting astrophysical information from IM measurements, even with 
\emph{Planck}-level uncertainties on the former \cite{liu_2016b}.

\subsection{Cross-correlation of Intensity Mapping with Other Probes}

Large-area 21-cm intensity mapping surveys 
\cite{Battye:2012tg,Bandura:2014gwa,Newburgh:2016mwi,Fonseca:2016xvi,Santos:2015gra,Chen:2012xu} 
that trace low to medium redshift line emissions allow for the possibility of 
interesting cross-correlations with other probes. One possibility is to consider 
cross-correlation with CMB surveys. The southern hemisphere location of HIRAX 
\cite{Newburgh:2016mwi} (see Experiments Section), for example, makes it ideal 
to cross-correlate the measured 21cm brightness fluctuations with various CMB 
probes of the large-scale structure that will be measured by AdvACT \cite{Henderson:2015nzj}
and SPT3G \cite{Benson:2014qhw}.
Direct correlation with CMB surveys is challenging because of the loss of long-wavelength 
line-of-sight modes in the 21cm brightness field after foreground removal (see Techniques 
Section). However, higher order correlations of the 21cm brightness field \cite{Moodley:2017}
or tidal field reconstruction \cite{Zhu:2016esh} can provide an observable signal. 
Higher order cross-correlations with the reconstructed CMB lensing field \cite{Moodley:2017}
will allow us to constrain the high-redshift matter power spectrum and the change 
in the HI density and bias with cosmic time, independently of the optical galaxy bias 
that enters in the optical galaxies-21cm intensity mapping cross-correlation \cite{Masui:2012zc}.

An advantage of IM over conventional spectroscopic galaxy surveys lies in the 
ability to cover extremely large cosmological volumes in a relatively short time. 
Galaxy surveys must threshold far above the noise level to ensure that candidate 
sources are not simply noise fluctuations. This throws away a large fraction of the 
signal, and results in slow survey speeds, especially when spectra must be obtained. 
Thresholding is not necessary with intensity mapping, where the whole (noisy) 
signal can be kept in the analysis. The reduced resolution requirements of IM also 
allow for survey instruments with a wider instantaneous field of view. This, along 
with the high spectroscopic resolution of radio receivers, allows 21\,cm IM 
experiments to achieve extremely high survey speeds, translating into considerably 
wider survey areas at a given depth.
 
Increasing survey volumes is a straightforward way of improving precision on the BAO, 
as galaxy surveys tend to be limited by sample variance rather than other factors. 
Purpose-designed IM surveys are easily capable of surpassing the effective volume 
of almost any galaxy survey, but suffer from an array of additional systematic effects 
that make the sample variance limit harder to achieve. Mitigation of these issues, 
particularly foreground contamination and instrumental calibration uncertainty, has 
been discussed elsewhere, but remains a serious unknown in performance 
comparisons of IM and galaxy survey techniques.

One way around this is to consider synergistic analyses that combine intensity mapping 
data with overlapping galaxy surveys. Cross-correlation was already used to good effect 
to filter-out foreground contamination in the GBT $\times$ WiggleZ 21\,cm signal detection 
\cite{Masui:2012zc}. Reconstruction of the real-space density and velocity fields using a 
combination of 21\,cm IM and low-density galaxy samples has also been investigated as 
a way of correcting for foreground removal effects and sharpening the recovered BAO 
feature (\cite{Seo:2015aza,Cohn:2015ljb}). Another possible target was suggested in 
\citep{Wolz:2017rlw}, which demonstrated how cross-correlations with galaxy catalog 
subsamples can characterize gas in galaxies as a function of optical properties. 
Using tomography techniques, it may also be possible to use the high-resolution 
spectroscopic redshift information from low angular resolution IM surveys to improve 
photometric redshift estimates from galaxy surveys \cite{Alonso:2017dgh}. Most 
planned low-redshift 21\,cm IM surveys are large enough that there will generally 
be significant overlap with at least one large galaxy survey, although some level of 
coordination in survey design will be necessary to optimize returns from combined 
analyses with next generation galaxy surveys such as LSST. 

\section{Galaxy Assembly and Star-Formation/IGM Interplay}

Intensity mapping opens a new method to probe the physics of phenomena at high 
redshift that would otherwise be unreachable by classical methods.  Current 
measurements of the star-formation history, for example, indicate that the star 
formation rate density (SFRD) increased as the Universe evolved until reaching a peak 
around $z\sim2-3$, and has been declining ever since \cite{Madau:2014bja}. 
However, our knowledge of star formation in the distant universe comes from the 
relatively small population of high-redshift galaxies bright enough to be imaged 
directly (see, e.g. \citep{2017MNRAS.467.1222H}). The contribution from these 
highly obscured galaxies (which are detected in the rest-UV) is highly uncertain.

This situation therefore has the potential to greatly improve using intensity mapping, 
which grants access to the statistical properties of a large population of faint sources 
which cannot be individually detected. Through intensity mapping, one can trade object 
localization for an unbiased, highly sensitive measure of high-redshift molecular gas 
that provides a direct comparison to optical and near-infrared star-formation histories.  
Carbon-monoxide intensity mapping, for example, provides a particularly compelling avenue 
toward understanding high-redshift star formation.  Most of our current knowledge 
of star formation in high-redshift galaxies from UV/optical/IR observations comes 
from stellar light and emission lines from the hot ionized gas in the ISM.  
Star formation, however, takes place in the cold molecular gas which is traced by line emissions such as CO.

Targeted observations have probed cold gas in galaxies out to redshifts 
$z\simeq 7$ \cite{Walter:2003zh}, but only in rare extremely bright objects that are 
not characteristic of the high-redshift star-forming population \cite{Sargent:2013sxa}.  
Blind surveys have integrated on regions that are too narrow to provide the statistics 
required to constrain the high-redshift CO luminosity function \cite{2016MNRAS.462.3265D}.  
As a result, models for the CO luminosity density at high redshift vary by orders of 
magnitude \cite{Breysse:2015saa}. By collecting the light from thousands, if not millions, 
of unresolved galaxies---not just the few that are bright enough to be imaged 
directly---intensity mapping offers the prospect to pin down the cosmic CO luminosity 
density, and thus the star-forming-gas reservoir, at high redshifts.

\begin{figure}[h!]
\begin{center}
\includegraphics[width=0.8\textwidth]{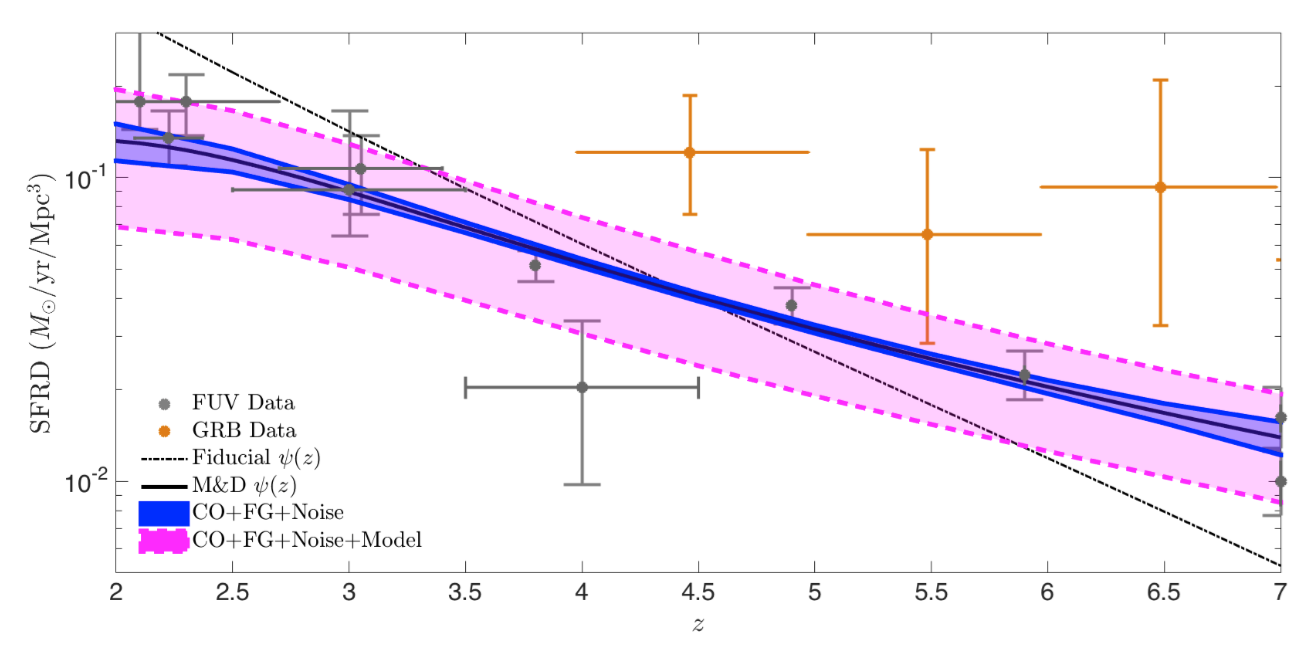}
\caption{Comparison between predicted constraints on star-formation rate density 
from CO intensity mapping and from existing FUV (grey points) \cite{Madau:2014bja} 
and GRB (orange points) data \cite{Kistler:2013jza}. Solid black curve $\psi(z)$ 
shows fit to FUV data \cite{Madau:2014bja}.  Blue curves show $\pm1\sigma$ SFRD 
uncertainty forecast with CO intensity mapping, taking into account foregrounds and 
noise. Dashed magenta lines include a 10\% model uncertainty in the adopted CO-FIR 
and FIR-SFR relations. From \citet{Breysse:2015saa}. ({\it Courtesy of Ely Kovetz})}
\label{fig:sfrd}
\end{center}
\end{figure}

To get an idea of the prospects of line-intensity mapping to access astrophysical information 
at high redshifts, we consider the use of CO intensity mapping at  $2 < z < 3$. 
As elaborated in different Sections of the report below, current (i.e. COPSS; 
\cite{2015ApJ...814..140K}) and future CO intensity mapping experiments 
(i.e., COMAP; \cite{Li:2015gqa}) will provide direct constraints on the CO luminosity 
function at these redshifts. These can then be used to infer constraints on key 
astrophysical quantities, such as the SFRD across cosmic times.
In \cite{Breysse:2015saa}, it was demonstrated that the cosmic star-formation 
history can be effectively measured with one-point statistics of CO maps, 
using a $P(D)$ analysis to infer the luminosity function of CO-emitting sources 
from the measured voxel intensity distribution (see Section 5 for more details). 
This constraint hinges on our ability to model the relation between the CO luminosity and the SFRD. 
This can be done through a series of empirical relations, namely the relation between 
the CO luminosity and the far infrared luminosity (FIR) of a galaxy, and subsequently 
the relation between FIR luminosity and the star-formation rate (more in Section 5). 
In an ideal setting, with no foregrounds, noise or modeling 
uncertainty,  an experiment such as full COMAP (see Section on Experiments) will 
be able to constrain the SFRD to $\sim1\%$ accuracy at redshift $z\lesssim3$. 
Fig.~\ref{fig:sfrd} demonstrates the potential of this technique when line foregrounds, 
instrumental noise and modeling uncertainties are included. 
Comparisons between intensity maps and galaxies using targeted 
observations in small patches of sky with advanced instruments such as ALMA, 
could substantially reduce the latter. 

An additional line which has been proposed as an interesting high-redshift 
probe is the He\,{\sc II} 1640 \AA{} recombination line \cite{2015MNRAS.450.2506V}. 
This line is thought to be a signature of metal-free Population III (Pop III) 
stars formed from pristine gas left over from the Big Bang. This is because 
massive Pop III stars produce many more He\,{\sc II} ionizing 
(i.e. $>54.4~{\rm eV}$) photos than metal enriched stellar populations 
\cite{2002A&A...382...28S}. One of the main challenges in observing Pop III 
stars is that they are generally expected to form in very small galaxies which 
will be extremely difficult to detect directly even with powerful telescopes 
such as JWST. Intensity mapping is capable of measuring the cumulative 
signal from all of these very faint sources and is therefore one of the most 
promising probes of Pop III stars. In \cite{2015MNRAS.450.2506V}, the 
He II intensity mapping signal was estimated as a function of redshift. 
It was shown that the Pop III signal may dominate 
over the contribution from quasars and Wolf-Rayet stars, and could potentially 
be measured at $z\approx 10$ with high signal-to-noise by a space-based 
instrument that could be built in the relatively near future. Another desirable 
feature of the He\,{\sc II} 1640 \AA{} line is that (much like Ly$\alpha$) it 
requires no metal enrichment. In principle its intensity maps could be 
observable at very high-redshift before large quantities of metals have 
been produced. Thus, it may be a promising line to cross correlate with 
21-cm emission during the EoR.

Beyond single-line cross-correlations, molecular line physics enables cross-correlation 
between lines of isotopologues, for instance $^{12}$CO and $^{13}$CO, providing 
probes into the density of cosmic molecular gas, on top of the inherent ratio between 
the two isotopologues as a function of redshift \cite{Breysse:2016opl}. The spectral 
proximity of the two emission lines (rest $\Delta \nu = 5$ GHz) allows current and 
planned experiments to measure both lines simultaneously, minimizing noise that 
may arise from systematics between different experiments. While the $^{13}$CO 
is subdominant to the $^{12}$CO at any given frequency, the cross-correlation 
between the two lines can expose this information, which is a function solely of the 
molecular physics and demographics. \citet{Breysse:2016opl} demonstrate that 
constraints of this cross-correlation from currently planned experiments will provide 
information on the molecular gas distribution, as well as the isotope ratio of $^{12}$CO 
and $^{13}$CO; the former directly related to the mechanism of star formation, the 
latter to star-formation history. This probe can potentially provide direct 
insight into the physics of star formation at a redshift previously inaccessible.

Finally, intensity mapping surveys are well-suited to measuring the integrated 
emission from faint sources of emission. A substantial fraction of the baryons in the 
Universe reside in diffuse, faintly-glowing phases of the IGM, and have thus proven 
difficult to detect and characterize \cite{Fukugita:1997bi,Cen:1998hc}. A recent 
Ly$\alpha$ intensity mapping analysis was able to measure some properties of the 
faint IGM \cite{Croft2016}. Similar techniques can be used to further constrain the 
density, evolution, and clustering behavior of the IGM, and will complement other 
probes such as the kinetic Sunyaev-Zeldovich effect \cite{Schaan:2015uaa}.


\chapter{First Detections}
\label{chap:firstdec}

\bigskip

There have been several detections of line-intensity fluctuations with existing 
instruments and novel observational approaches and analyses.  These already 
provide unique constraints to the reservoir of cold molecular gas at high redshift.  
We describe these detections in order of decreasing wavelength.

\section{Detection of 21cm in Cross Correlation}

The first detection of the redshifted 21-cm emission in the intensity mapping 
regime was demonstrated by \citet{Chang:2010jp}. The authors conducted a 
$21$\,cm survey spanning the redshift range of $0.53 < z < 1.12$, corresponding 
to a comoving distance of $1400-2600$\,Mpc/h, by utilizing the $800$\,MHz 
receiver at the Green Bank Telescope (GBT). The survey fields overlap with two 
of the DEEP2 optical galaxy redshift survey fields \citep{Davis:2000vr}, each 
$120' \times 30'$ in angular size out to $z=1.4$, and contain $10,000$ DEEP2 
galaxies in total. The FWHM of the GBT beam of $15'$ corresponds to 
$9$\,Mpc/h (comoving) at $z=0.8$, and the high intrinsic spectral resolution is 
binned to $430$\,kHz, or $2$\,Mpc/h. 

The dominant sources present in the data are the radio frequency interference 
(RFI) and continuum foreground emission from astrophysical sources, notably 
the Galactic and extragalactic synchrotron radiation. The RFI is polarized and 
excised using cross-correlation of the two linearly polarized radio signals. The 
astrophysical continuum sources present a fluctuation in  brightness temperature 
of $\sim 125$\,mK, about $1000$ times brighter than the sought-after $21$\,cm 
signals. However, they are spectrally smooth and distinct from the $21$\,cm fluctuations 
which have redshift and thus frequency structures, and can be identified as the 
dominant spectral eigenmodes using the singular-value decomposition technique 
and subtracted from the data.

\begin{figure}[h!]
\begin{center}
\includegraphics[width=0.49\textwidth]{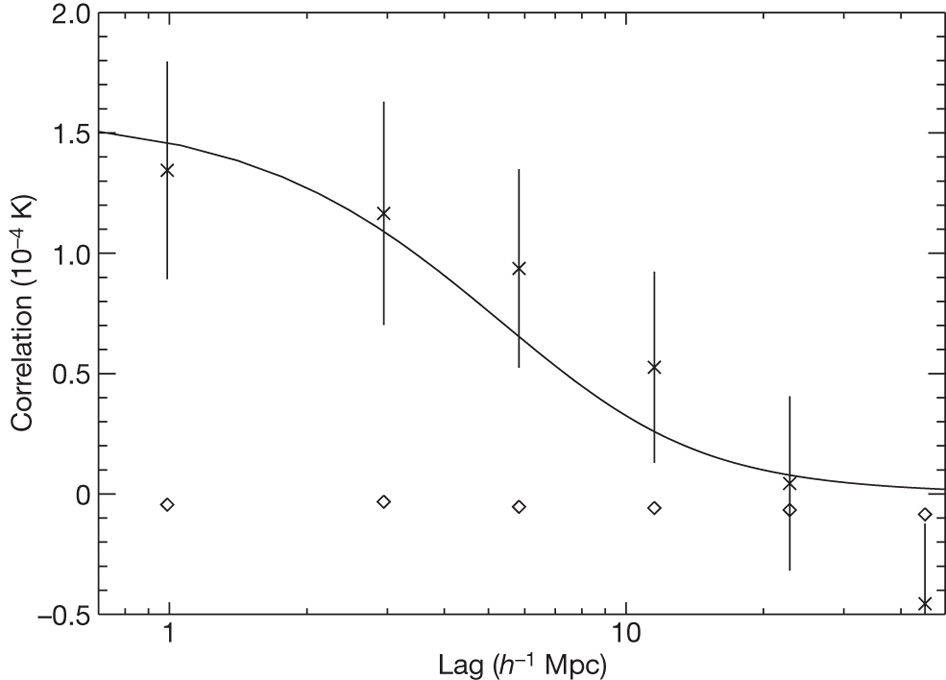}
\includegraphics[width=0.49\textwidth]{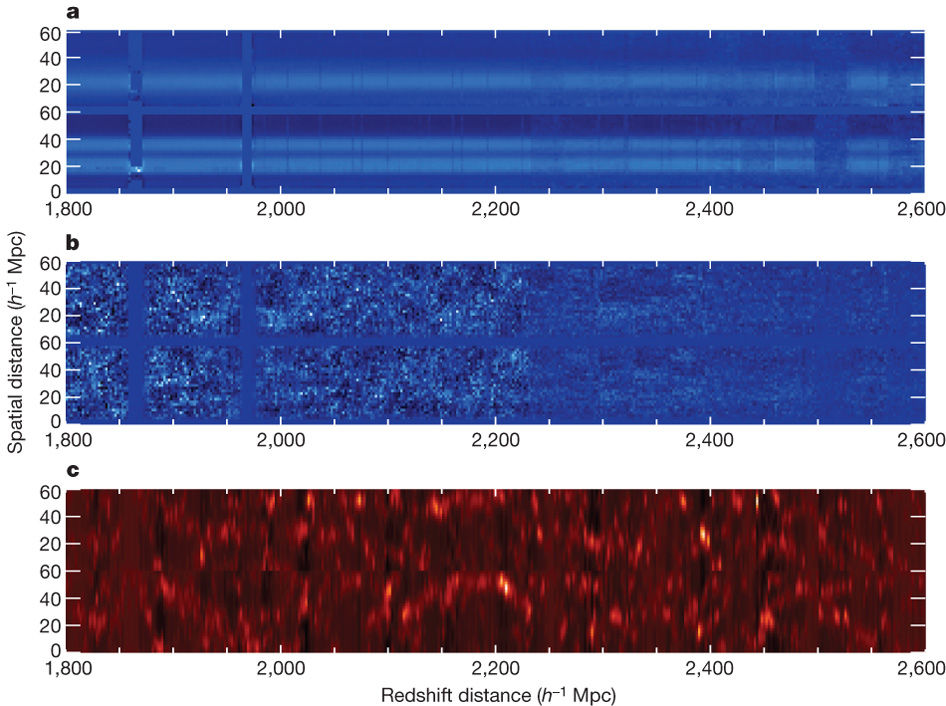}
\caption{\textit{Left:} The cross-correlation between the $21$\,cm brightness 
temperature and the DEEP2 galaxy density field. Crosses are the measured 
cross-correlation amplitudes, and error bars are the $1\sigma$ bootstrap errors 
generated using randomized optical data. The diamonds are the mean null-test 
values over $1000$ randomizations of the optical source position. The solid line 
indicates a power law DEEP2 galaxy correlation model \citep{Coil:2004rj}, 
convolved with the GBT beam pattern and velocity distortions, with an amplitude 
from the best-fit value of the cross-correlation. \textit{Right:} Spectra of the DEEP2 
field. Panels a, b, c show the radio flux after RFI excision, a proxy for the 21-cm 
intensity field after foreground subtraction, and the density field as traced by the 
DEEP2 galaxies, respectively. The fields are arranged with redshift horizontal and 
spatial vertical. ({\it Courtesy of Tzu-Ching Chang})}
\label{fig_gbt}
\end{center}
\end{figure}

After the above analysis, the authors measure the residual intensity field and 
reported an upper limit to the $21$\,cm brightness  temperature fluctuation of 
$464 \pm 277\,\mu{\rm K}$, on a pixel scale of $(2 {\rm Mpc}/h)^3$, at a mean 
redshift of $0.8$. The authors further performed cross-correlation analysis of the 
$21$\,cm intensity field with the underlying cosmic density field as traced by the 
DEEP2 galaxies \citep{Coil:2004rj}. They reported a positive cross-correlation on 
a scale of $\sim 10$\,Mpc/h, with an amplitude of $157\pm 42\,\mu{\rm K}$ at zero 
lag Fig.\ref{fig_gbt}. The amplitude constrains a combination of HI abundance, bias 
parameters and stochasticity, and the authors infer a value for HI abundance, 
$\Omega_{\rm HI} = (5.5 \pm 1.5) \times 10^{-4} (1/rb)$ at $z\sim 0.8$, where 
$r$ is the stochasty between the $21$\,cm and optical galaxy tracers, and $b$ 
the bias parameter of HI. The respective intensity fields of the radio, $21$\,cm 
and optical galaxies are shown in Fig.\ref{fig_gbt}.

The cross-correlation detection is significant as it verifies that the 21cm intensity 
field correlates with and thus traces the distribution of optical galaxies, which are 
known tracers of the underlying matter distribution. This serves as a proof of 
concept for the line-intensity mapping technique, and signifies line-intensity mapping 
as a viable tool for large-scale structure studies. Subsequent, deeper observations with the GBT have measured the  $21$\,cm signal in cross-correlation \citep{Masui:2012zc} 
with WiggleZ \citep{Drinkwater:2009sd} and used the auto- and cross-correlation 
to bound the $21$\,cm signal \citep{Switzer:2013ewa}. \citep{Switzer:2015ria} 
describes the methods used in analyzing the GBT maps.

\section{Detection of CO Fluctuations} 

Intensity mapping experiments are well-suited for data sets with large survey volumes, 
requiring only modest point-source sensitivity to detect an aggregate signal. As such, 
the Sunyaev-Zel'dovich Array (SZA) -- a $3.5\,{\rm m} \times 8$-element subset of the 
Combined Array for Research in Millimeter-wave Astronomy (CARMA) -- was an optimal 
choice of instrument for such an experiment. This is due in part to the frequency coverage 
($27-35 $\,GHz, covering CO(1-0) from $z=2.3-3.3$), relatively large field of view 
(140\,arcmin$^2$) and compact configuration of the SZA. 

Starting in 2011, the SZA was used in an intensity mapping experiment focused on 
measuring the abundance and evolution of molecular gas in the era leading up to the 
peak of cosmic star formation. This project -- known as the CO Power Spectrum 
Survey (COPSS) -- made use of both archival data and new observations with the SZA. 
There were two primary goals for COPSS: placing the first-ever limits of the CO power 
spectrum at high redshift, and exploration of the astrophysical and systematic contaminants 
that might impede future efforts. With a total survey volume of more than 10 million cubic 
megaparsecs, it is one of the largest blind surveys to date targetting molecular gas emission 
at high redshift, more than an order of magnitude larger than similar efforts with ALMA, 
VLA, and PdBI. 

\begin{figure}[h!]
\begin{center}
\includegraphics[width=0.49\textwidth]{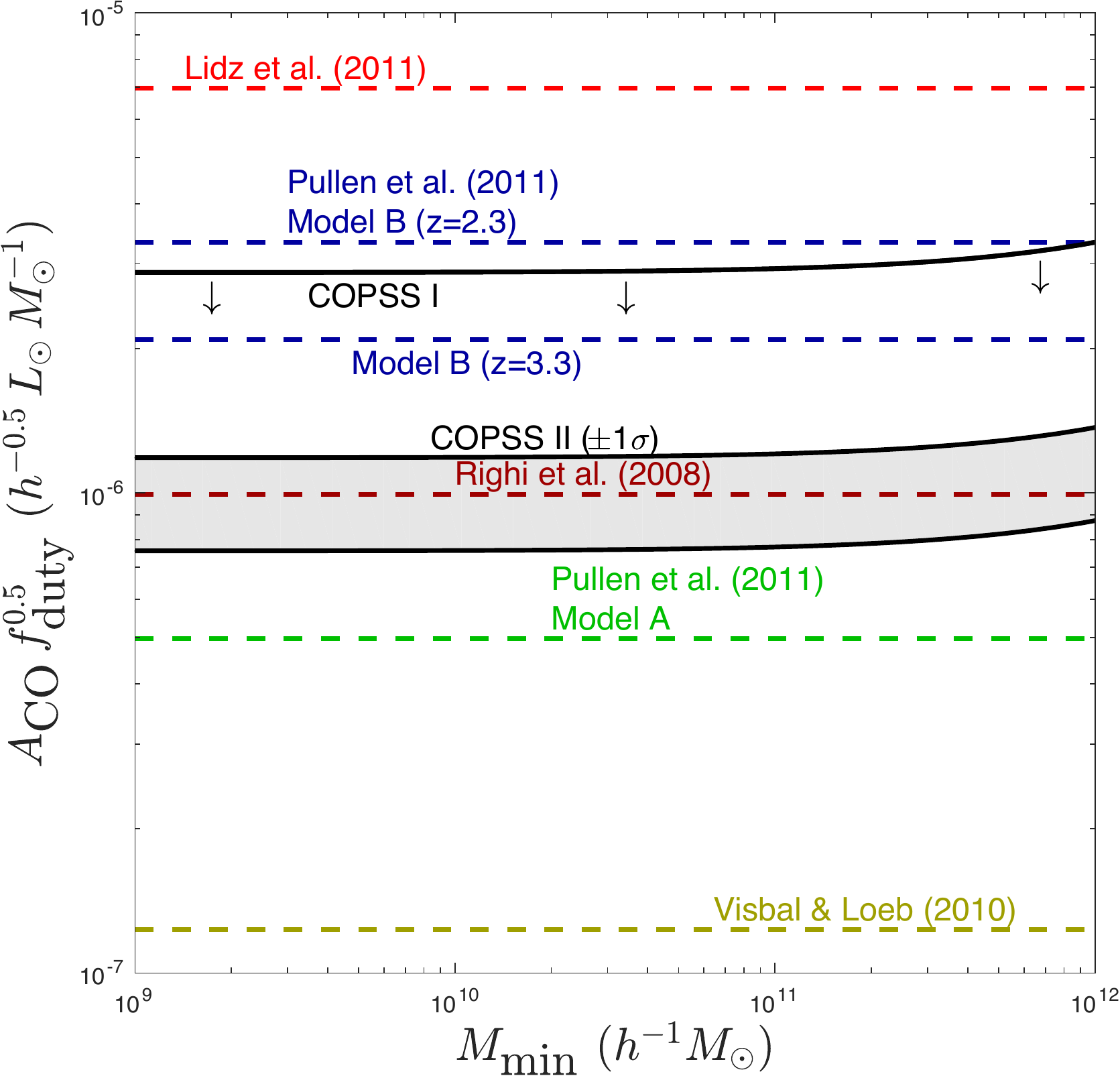}
\includegraphics[width=0.49\textwidth]{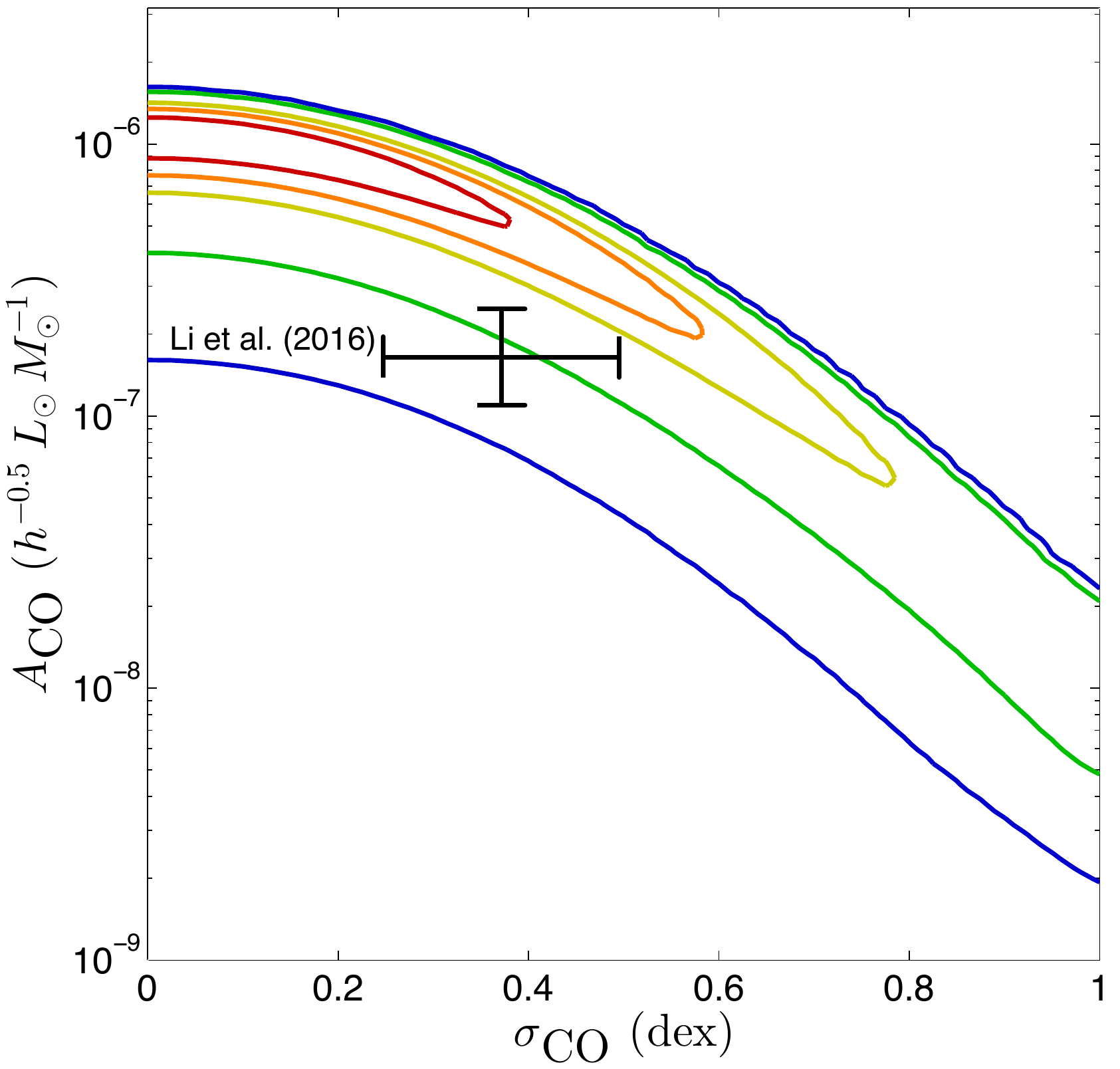}
\caption{\textit{Left}: Constraints on $A_{\textrm{CO}}$ versus minimum halo mass for 
CO emission ($M_{\textrm{min}}$, as derived from the COPSS experiment \cite{2015ApJ...814..140K,Keating2016}. 
These constraints are shown against the theoretical expectations from a number of models. 
\textit{Right}: Constraints on $A_{\textrm{CO}}$ versus the scatter in the halo mass to CO 
luminosity relationship ($\sigma_{\textrm{CO}}$), with the 95.4\% confidence limits shown in blue.  
Also shown are theoretical estimates from Li et al. (2016). 
({\it Courtesy of Karto Keating})}
\label{fig_aco}
\end{center}
\end{figure}

At the conclusion of the project, COPSS yielded a tentative ($\sim 3 \sigma$) detection of 
bulk CO emission, constraining the CO power spectrum to $P_{\textrm{CO}} = 3.0 \pm 
1.3 \times 10^{3} \mu\textrm{K}^{2} \ h^{-3}\,\textrm{Mpc}^{-3}$ at $z \sim 3$. As shown in 
Fig.~\ref{fig_aco}, these constraints excluded several theoretical models, and placed 
significant constraints on both the CO luminosity function and the cosmic molecular gas 
density at high redshift.

\section{Tentative Detection of [CII] at Medium Redshifts} 

The CII line emission typically comprises 0.1--1\% of the far-infrared luminosity 
in low-redshift galaxies, making it an ideal candidate for detection using 
cross-correlations.  At redshifts $2<z<3.2$, the [CII] emission line appears in the 
$545$\,GHz Planck band; thus its intensity can be measured by cross-correlating 
the Planck map with $z\sim 2.5$ quasars.  In \citet{Pullen:2017ogs}, the authors 
cross-correlate the 545 GHz Planck map with an overdensity map constructed 
from the SDSS BOSS DR12 quasar catalog (P\^{a}ris et al., in prep.).  The $545\,{\rm GHz}$ 
map is dominated by dust emission from the Galaxy \citep{Adam:2015wua} and 
cosmic infrared background (CIB) emission from young stars \citep{Ade:2013zsi}.  
The dust emission cancels out in the cross-correlation, though it contributes to the 
errors.  The CIB emission correlates with the quasars, so the authors use 
cross-correlations with the 353 and 857 GHz maps from Planck, as well as 
cross-correlations between the Planck maps and CMASS galaxies from $z=0.57$ 
\citep{Alam:2016hwk,Reid:2015gra,2015ApJS..219...12A} to perform a MCMC fit 
over all the cross-power spectra to estimate parameters in a CIB halo model 
\citep{2012MNRAS.421.2832S}.  This procedure breaks the degeneracy between 
the [CII] and CIB emission in the 545 GHz map to isolate the [CII] signal.

\begin{figure}
\begin{center}
\includegraphics[width=0.5\textwidth]{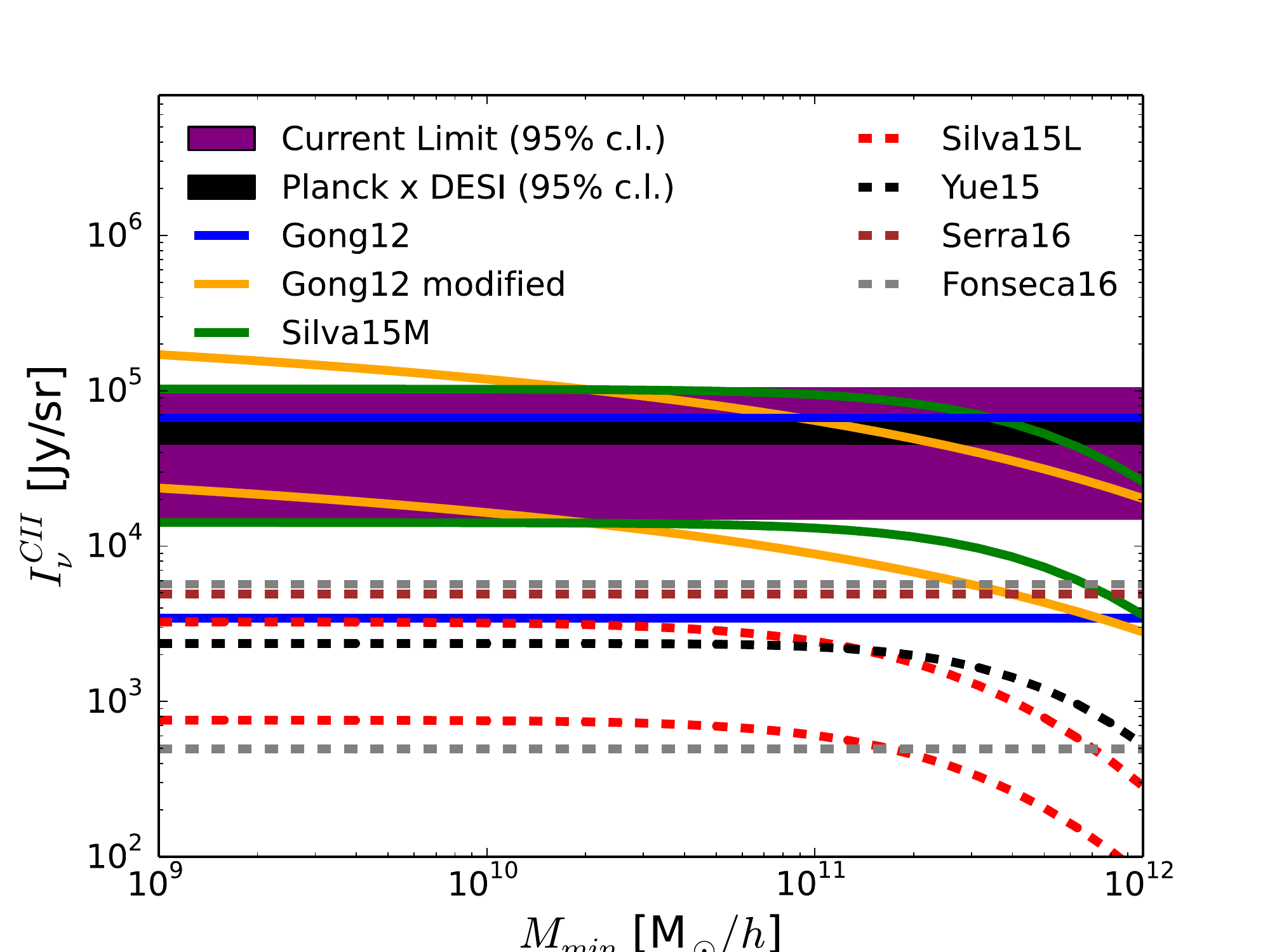}
\vspace{0.025in}
\caption{Measurement of the [CII] intensity with 95\% confidence limits. 
Also shown are the range of predictions for several [CII] intensity models, including 
collisional excitation models (solid lines) and scaling relations (dashed lines), as 
functions of minimum halo mass $M_{\rm min}$.  The measurement favors the 
collisional excitation models which appear at the high end of the range of models, 
although no models are ruled out by 3$\sigma$. ({\it Courtesy of Anthony Pullen})}
\label{F:acii} 
\end{center}
\end{figure}

In Fig.~\ref{F:acii}  the constraint on the intensity of [CII] emission from \citet{Pullen:2017ogs}, 
$\mathrm{I_{CII}}=5.7^{+4.8}_{-4.2}\times10^4$ $\mathrm{Jy/sr}$ (95\% c.l.) is plotted, along 
with predictions from various models
 \citep{2012ApJ...745...49G,Silva:2014ira,Yue:2015sua,Serra:2016jzs,Fonseca:2016qqw}. 
 Comparing the likelihoods of the CIB \& [CII] model and the CIB only model, the Bayesian 
 Information Criteria, 96.5 and 96.0 respectively, are similar, implying that both interpretations 
 are plausible and more data is needed to confirm the [CII] detection.  The results favor 
 the models where [CII] ions emit due to collisional excitations from electrons 
 \citep{2012ApJ...745...49G,Silva:2014ira} over the scaling relations constructed from 
 luminosity function measurements, though none of the models are ruled out.  Also 
 shown in the figure are forecasts where the BOSS quasar maps are replaced with quasars 
 from the Dark Energy Spectroscopic Instrument (DESI) \citep{Levi:2013gra}, finding 
 that a similar method could yield a signal-to-noise ratio of 10 for the [CII] emission.

\section{Cross-Correlation between Ly$\alpha$ Emission and Quasars}

Flux from the Ly$\alpha$ line occurs in many environments, and is produced by 
many sources, including young stars, quasars and the ultraviolet background 
(see e.g., \citet{Pullen:2013dir}). Because of the high cross section for scattering of 
neutral hydrogen, much of this Ly$\alpha$ emission is expected to be extended, and 
intensity mapping techniques are therefore useful. As with other optical emission, 
intensity mapping with Ly$\alpha$ is affected by contamination from other lines, but 
also potentially instrumental and other systematic effects. Cross correlation with objects 
with known redshift such as galaxies, known Ly$\alpha$ emitters, quasars, the Ly$\alpha$ 
forest, metal and other absorption lines all offer in principle a route to measure an intensity 
mapping signal without contamination. One would like to take integral field spectra of as 
much of the sky as possible (e.g., HETDEX \citet{Hill:2008mv}). At present, the largest 
single dataset available is the Luminous Red Galaxy (LRG) sample from SDSS/BOSS 
(total fiber area of 0.4 sq. deg. for DR12). \citet{Croft:2015nna} 
made a first attempt at Ly$\alpha$ intensity mapping. The authors used BOSS spectra by 
subtracting the best fit LRG model and cross-correlating the residual flux with SDSS quasars.

During the \citet{Croft:2015nna} analysis it became evident that even when using cross-correlation techniques achieving robust measurements from line-intensity mapping data is complex. The low surface brightness of targeted IM flux means that light can leak into nearby spectral columns 
on a fiber-fed spectrograph CCD. The authors eliminated pairs of fibers separated 
by 5 columns or less because of this. The work of \citet{Croft:2015nna} should be seen as a trial of this kind of observational analysis, and a tentative first detection (see below). Insights for the future include the desirability of selecting cross-correlation centers from entirely different datasets from the intensity map, and also of tracking light scattering in the instrument.

\begin{figure}
\centering
\includegraphics[height=0.5\textwidth]{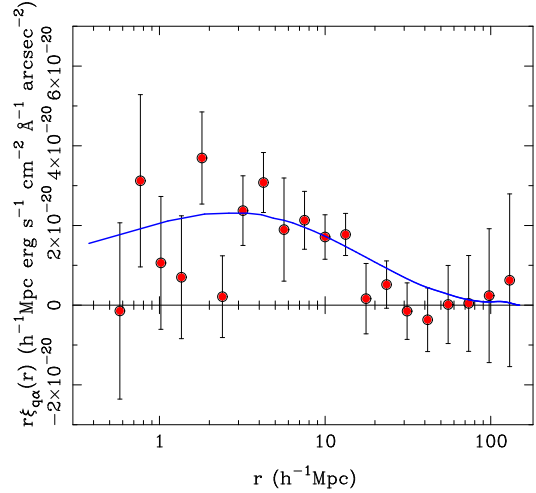}
\caption{
The quasar-Ly$\alpha$ emission cross-correlation function from \citet{Croft:2015nna}. 
The points represent results from SDSS BOSS over the redshift range $z=2-3.5$,
with jackknife error bars. The smooth curve is a best fit linear CDM correlation function. 
({\it Courtesy of Rupert Croft})}
\label{xiqe} 
\end{figure}

After eliminating all contaminants found, the \citet{Croft:2015nna} result was at face value an $8\sigma$ 
detection of Ly$\alpha$ emission with large-scale clustering ($1 -15{\;h^{-1}{\rm Mpc}}$) 
consistent with the CDM shape (see Figure \ref{xiqe}). The amplitude (proportional to the 
mean Ly$\alpha$ surface brightness) was extremely high however, equivalent to thirty 
times the Ly$\alpha$ emission from previously known Ly$\alpha$ emitters (but consistent 
with extinction corrected SFR). The signal to noise in the cross-correlation is nonexistent 
beyond $15{\;h^{-1}{\rm Mpc}}$, meaning that the measurement  only comes from the 
$\sim 3\%$ of the volume closest to quasars. It is energetically possible that quasar HeII 
reionization or jet heating is responsible for the signal rather than star formation. In this 
case, most of volume of space (far from quasars) would have much lower Ly$\alpha$ 
surface brightness.  This picture can be tested by cross-correlating with a more space-filling 
tracer (such as the Ly$\alpha$ forest). At time of writing a trial measurement of the Ly$\alpha$ 
forest-Ly$\alpha$ emission cross-correlation has not led to a detection. It therefore seems 
that likely that the \citet{Croft:2015nna} signal is local to quasars, or else due to unknown contamination. 
Current data is almost at the level required to detect a Ly$\alpha$ emission- Ly$\alpha$ 
forest cross-correlation signal from {\it known} Ly$\alpha$ emitters.  Many firm detections 
therefore seem certain from upcoming larger datasets such as  HETDEX \citet{Hill:2008mv}, 
PAU \citep{2012SPIE.8446E..6DC}, and J-PAS \citep{Benitez:2014ibt}.

\chapter{Experimental Landscape}
\label{chap:experiment}

\bigskip

As described in Section~\ref{chap:intro}, low resolution maps of various 
emission lines have the potential to efficiently measure 
bulk properties of the Universe across the past 14 billion years of cosmic 
history. In this section, we describe current and upcoming instruments dedicated 
to intensity mapping of various lines in different redshift ranges for the purposes 
of probing the epoch of reionization, star formation, and large scale structure 
and dark energy. A table of these experiments and their salient parameters is 
given in Table~\ref{tab:experiments}. 

\begin{table}[h]
\centering
\footnotesize
\begin{tabular}{| l l l l l | }
\hline
Experiment & Line & Frequency & Redshift range & Location \\ \hline 
HERA & HI & $50-250 \,{\rm MHz}$ & $5-27$ & South Africa \\
SKA-LOW & HI & $50-350 \,{\rm MHz}$ & $3-7 $& Australia  \\
CCAT-prime & [CII] & $185-440\, {\rm GHz}$ & $3.3-9.3$ & Chile \\
TIME & [CII] & $200-300\,{\rm GHz}$ & $5.3-8.5$ & North America  \\
CONCERTO & [CII] & $200-360\,{\rm GHz}$ & $4.3-8.5$ & Chile  \\
COPSS & CO & $27-35 \,{\rm GHz}$ & $2.3-3.3$ & North America \\
mmIME & CO, [CII] & $300, 100, 30\,{\rm GHz}$ & $1-5$ & various \\
AIM-CO & CO & $86-102\,{\rm GHz}$ & $1.2-1.7$, $2.4-3.0$ & China  \\
COMAP & CO & $26-34\,{\rm GHz}$ & $2.4-3.4$, $5.8-7.8$ & North America \\ 
STARFIRE & [CII], NII & $714-1250\,{\rm GHz}$ & $0.5 - 1.5$ & Sub-orbit (balloon) \\ 
SPHEREx & H$\alpha$ (H$\beta$, [OII], [OIII]), Ly$\alpha$ & $60-400\,{\rm THz}$ & $0.1-5$, $5.2-8$ & Space \\
CHIME & HI & $400-800\,{\rm MHz}$ & $0.8-2.5$ & North America  \\
HIRAX & HI & $400-800\,{\rm MHz}$ & $0.8-2.5$ & South Africa  \\ 
SKA-MID & HI & $350 \,{\rm MHz} - 14\,{\rm GHz}$ & $0-3$ & South Africa \\
BINGO & HI & $939-1238 \,{\rm MHz}$ & $0.13-0.48$ & South America  \\ \hline
\end{tabular}
\caption{Parameters for various intensity mapping instruments described below.}
\label{tab:experiments}
\end{table} 

\section{Epoch of Reionization Science at z=5--27}

A description of the science goals related to the Epoch of Reionization (EoR) 
was presented in Section~\ref{chap:goals}, here we describe the current 
experiments dedicated to measuring this signal. Measurements from the 
CMB constrain the EOR to have occurred around $z_{re}\sim9$, and so 
21\,cm experiments seeking to make a measurement with \hi mapping build 
instrumentation at radio frequencies: $\sim140\,{\rm MHz}$, while experiments targeting 
the dusty star-forming component with [CII] will target frequencies near $200\,{\rm GHz}$.

\subsection*{OVRO-LWA}
The Owens Valley Long Wavelength Array (OVRO-LWA) is a low-frequency interferometer located at the Owens Valley Radio Observatory (OVRO) near Big Pine, California. The array currently consists of 288 dual-polarization broadband dipole antennas operating between $27\,{\rm MHz}$ and $85\,{\rm  MHz}$. 251 antennas are located within a 200 m diameter core for surface brightness sensitivity, 32 antennas extend to longer baselines (up to $1.5\,{\rm km}$), and the remaining 5 antennas are equipped with noise-switched front end electronics for total power radiometry.  The full array will eventually extend to 352 antennas and baselines of $2.5\,{\rm km}$. The 5 total power antennas are used by the Large Aperture Experiment to Detect the Dark Age (LEDA) project to measure the sky-averaged signature of HI absorption from the Cosmic Dawn \citep{price2017}. The LEDA project also provides a 512-input correlator with 58 MHz instantaneous bandwidth \citep{2015JAI.....450003K}. As well as the LEDA total power experiment, the OVRO-LWA will target the spatial power spectrum of 21-cm fluctuations from the Cosmic Dawn era.

The OVRO-LWA surveys the sky north of -30$^\circ$ at 10 arcmin resolution ($k_\perp < 0.3\,h\,\text{Mpc}^{-1}$ at $z=20$) with $24\,{\rm kHz}$ channelization ($k_\parallel < 10\,h\,\text{Mpc}^{-1}$ at $z=20$). The OVRO-LWA will place upper limits on the spatial power spectrum of 21-cm fluctuations between $30 > z > 16$. These fluctuations are primarily sourced by inhomogeneous star formation and heating of the early universe \citep{2014MNRAS.437L..36F}. At $80\,{\rm MHz}$, the foreground brightness temperature can be an order of magnitude larger than at $200\,{\rm MHz}$. As a first step on the path to characterization of 21-cm fluctuations, the OVRO-LWA has been used to produce 8 foreground maps evenly spaced between $36.528\,{\rm MHz}$ and $73.152\,{\rm MHz}$ \citep{eastwood2017}. The OVRO-LWA also responds to LIGO and Swift triggers, monitors nearby stars for stellar flares and magnetospheric radio emission from exoplanets, as well as monitoring the radio emissions of the Sun and Jovian system.

\subsection*{HERA and its predecessors}
The Hydrogen Epoch of Reionization Array (HERA; \url{http://reionization.org} 
\citep{deboer17}) is a dedicated radio interferometer optimized to deliver high 
signal-to-noise measurements of redshifted 21\,cm \hi emission to detect and 
characterize the EoR and Cosmic Dawn.  Operating over the the frequency 
range $50 < \nu <  250$ MHz ($27 < z < 5$), HERA covers the period of the 
formation of the first stars and black holes $\sim 0.1$Gyr after the Big Bang 
($z \sim 30$) past the full reionization of the intergalactic medium (IGM) 
$\sim1$ Gyr later ($z \sim 6$). HERA will 
enable high precision measurement of the reionization history \citep{liu_2016a}; 
constraints on the physics of reionization \citep{greig15}; cosmological constraints 
using combined 21\,cm power spectrum and global signal measurements 
\citep{liu_2016b}; a robust statistical characterization of the reionization and X-ray 
heating power spectra \citep{ewall-wice16xray}; and has the sensitivity to enable 
first images of large scale \hi structure \citep{carilli_and_sims16} and perform 
cross-correlation analyses with galaxy surveys \citep{malloy13,beardsley15}. 

HERA is a second-generation instrument which combines efforts and lessons 
learned from the Murchison Widefield Array (MWA) and the Donald C. Backer 
Precision Array for Probing the Epoch of Reionization (PAPER), as well as the 
MIT EoR experiment (MITEoR) and the Experiment to Detect the Global EoR 
Step (EDGES).  The HERA array, currently under construction with completion 
in late 2019, will be $\sim350$ 14-meter diameter non-tracking dishes.  High 
surface brightness sensitivity and redundant calibration is achieved with a 
close-packed hexagonal array 300\,m across \citep{dillon16}, with imaging 
outriggers out to $\sim1$\,km .  Its substantial collecting area provides an order 
of magnitude more sensitivity than first generation instruments.  Careful 
electromagnetic simulation and measurement of the antenna, feed and 
electronics aim to control systematic effects due to spectral non-smoothness 
\citep{neben16, ewall-wice16dish,patra17,thyagarajan16}.  

HERA is fully funded from the NSF Mid-Scale Instrumentation Program and 
the Betty and Gordon Moore Foundation, and is currently commissioning the 
first 37 antennas. 

\begin{figure}[h!]
\centering
\includegraphics[height=1.6in]{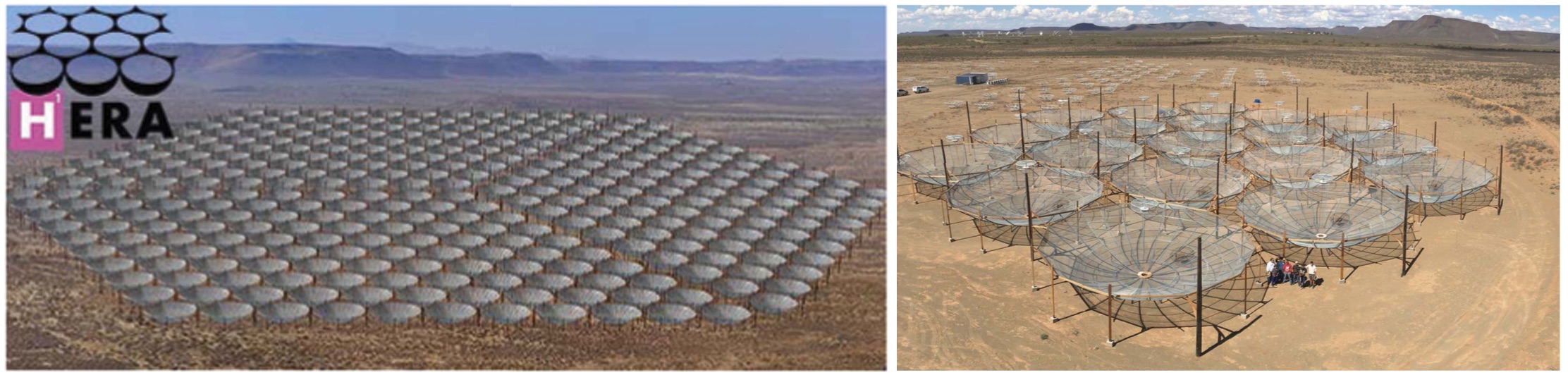}
\vspace{5pt}
\caption{{\it Left:} Rendering of the 320-element core of the full HERA-350 array.  
{\it Right:} Picture of 19 HERA 14 m, zenith-pointing dishes (with PAPER elements 
in the background) currently deployed in South Africa. 
({\it Courtesy of James Aguirre})}
\label{fig:HERA}
\end{figure}

\subsection*{SKA-LOW}

The Square Kilometre Array (SKA) is a large, international, general-purpose radio 
telescope facility, split between two sites in the Karoo desert in the Western Cape 
region of South Africa, and the Murchison region in Western Australia. Three 
pathfinder telescopes (ASKAP, KAT7, and MeerKAT) have already been constructed 
on these sites, with initial construction of Phase 1 of the array (SKA1) slated to 
begin around 2018, and an early science phase expected in 2020. This will be 
followed by full operations starting around 2023, and an eventual upgrade to 
SKA2 (with $10\times$ the projected sensitivity of SKA1) to begin operations around 2030.

The Phase 1 design incorporates two sub-arrays. The first, SKA1-LOW, is a low 
frequency array consisting of $\sim 500$ stations of 256 dipole antennas each, 
covering the band from $50-350$\,MHz, and will be constructed on the Australian 
site. The primary science goal of LOW is to study the EoR using intensity maps of 
the 21cm line, in particular by the direct imaging of ionization bubbles (seen in negative 
against the neutral hydrogen emission) over angular scales of arcminute to degrees 
\citep{Koopmans:2015sua}. This adds imaging capabilities that are not possible with 
earlier-generation EoR experiments aiming for a statistical detection of the 21cm power 
spectrum at the relevant redshifts. The sheer number of receivers, coupled with an 
advanced beamforming capability, will make SKA1-LOW an extremely sensitive and 
flexible intensity mapping experiment. 

\subsection*{CCAT-prime}

CCAT-prime (CCAT-p) will be a $6$ meter aperture telescope located at $5600$ 
meter site on Cerro Chajnantor in the Atacama Desert in northern Chile.  The CCAT-p 
design strives to optimize surface brightness sensitivity and mapping speed through 
the telluric windows at wavelengths from $200\,\mu m$ to $3.3\,{\rm mm}$.  The 
instantaneous field of view (FoV)  is maximized through the use of a crossed-Dragone 
optical arrangement where an off-axis concave primary delivers the beam to an 
off-axis and concave secondary which forms a flat focal plane with a FoV of  
$\sim2^{\circ}$, $4^{\circ}$, and $8^{\circ}$ diameter at $350\,\mu m$, $1.1\,{\rm mm}$, 
and $3.3\,{\rm mm}$ wavelength respectively.  The unobscured optical path with 
minimized telescope panel spacing minimizes both emissivity and the effects of 
low-level optical side-lobes.  The total wave-front error requirement is $<11\,\mu m$ 
rms ($7\,\mu m$ goal) so that CCAT-p operates very efficiently in the short submillimeter 
bands.  CCAT-p will be located at $5600$ meter elevation on Cerro 
Chajnantor---about $500\,{\rm m}$ above the nearby ALMA array and APEX telescopes 
in northern Chile where the water vapor is sufficiently low to enable routine operation 
in the $350\,\mu m$ window, frequent operations in the $200\,\mu m$ telluric window, 
and improved performance in the longer wavelengths.

\begin{figure}[h!]
\begin{center}
\includegraphics[width=0.7\columnwidth]{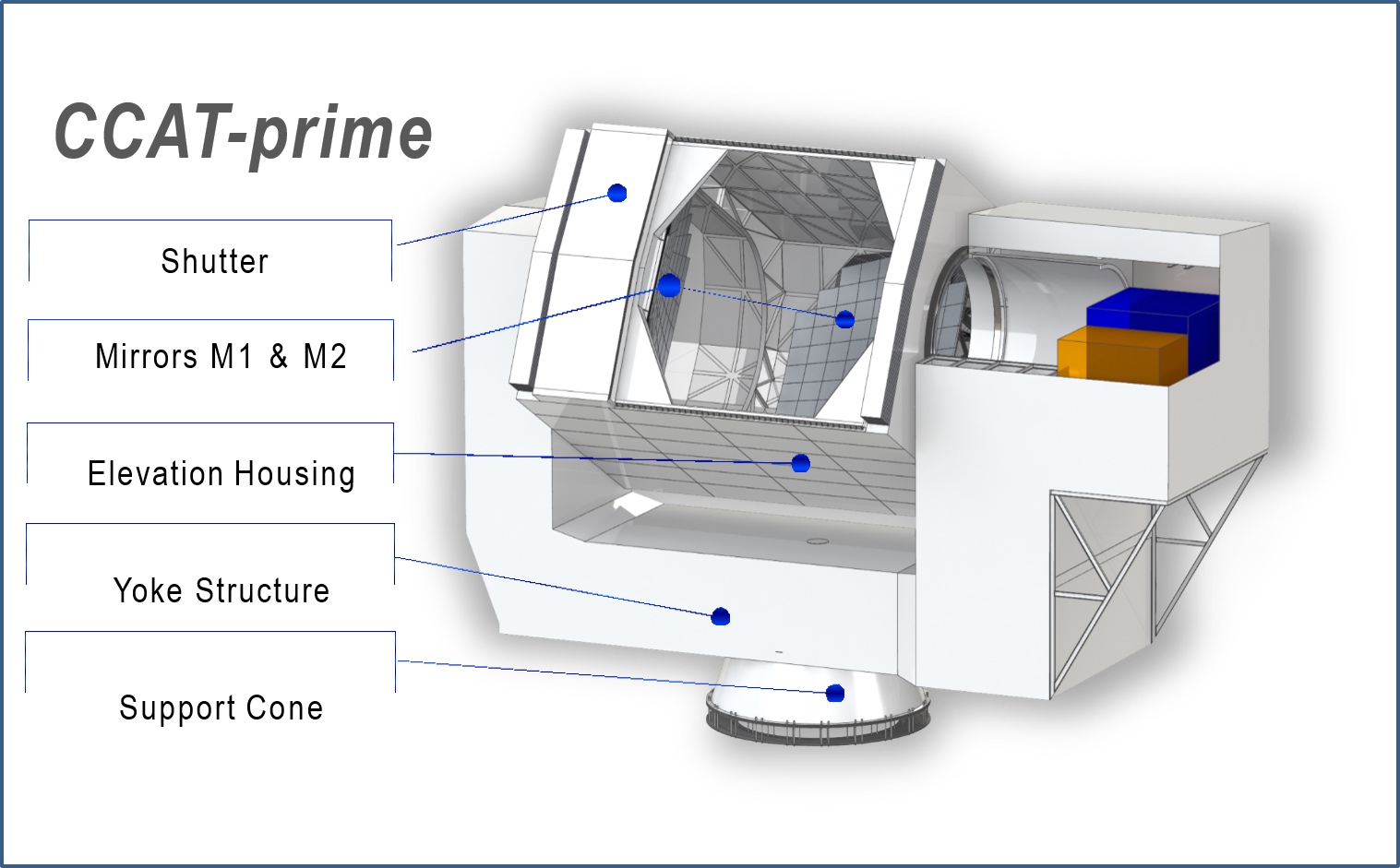}
\caption{The CCAT-prime design from Vertex Antennentechnik, GmbH. 
({\it Courtesy of Gordon Stacey})}
\end{center}
\end{figure}

CCAT-p offers an excellent platform from which to pursue LIM science, particularly 
in the [CII] $158\,\mu m$ line at redshifts from roughly $5<z<9$.  At these redshifts 
[CII] IM traces out the process of reionization and, quite fortunately, the signal is 
transmitted through the very clean 190 to 315 GHz portion of the mm-wave telluric 
transmission spectrum.  The expected aggregate clustering signal size-scale is of 
the order $1-2\,{\rm arcmin}$, which is a very good match to the CCAT-p diffraction 
limited beam ($\sim64$ to $40\,{\rm arcsec}$ at $190$ to $300\,{\rm GHz}$). The 
intensity mapping signal needs to be mapped at spectral resolving powers 
$\sim500\,{\rm  km/s}$ over approximately $16$ square degree region to noise levels 
below $8 \times 10^{-14}\,{\rm W/m^2/sr}$ (cf. \citet{Gong:2011mf}). These sort of 
noise levels over such a broad field require large numbers of mm-wave bolometer 
pixels which can be employed within spectrometers that either spatially, or spectrally 
multiplex.  For CCAT-p a spatially multiplexing spectrometer is planned based on a 
wide-field Fabry-Perot Interferometer.  This system should be able to deliver the 
requisite sensitivity with an imaging array of $\sim 4000$ pixels in about $4000$ 
hours of integration time.  These array formats are well within the capabilities of 
today's TES bolometer technology.  

CCAT-p is being constructed under the original CCAT framework, and is a partnership 
of Cornell University, the Universities of Bonn and Cologne, and CATC, a consortium 
of Canadian academic institutions.  CCAT-p is funded for construction and will begin 
detailed design in July 2017. First light is anticipated in June 2021.

\subsection*{TIME}


 The TIME (Tomographic Ionized-carbon Mapping Experiment \cite{Crites:2014}) is a novel high-throughput millimeter-wave imaging spectrometer array designed to make pioneering measurements of the redshifted $157.7\, \mu {\rm m}$ line of singly ionized carbon from the EoR  at $5.3 < z < 8.5$.  [CII] is the most energetic emission line in galaxies at wavelengths longer than $40\,\mu {\rm m}$ and is a bolometric tracer of total star-formation.  TIME will also produce high significance measurements of the molecular gas density through the epoch of peak start formation via detections of CO clustering fluctuations in multiple rotations transitions from redshifts $0.5 < z < 2$.  Finally TIME will measure the kinetic Sunyaev-Zeldovich effect in galaxy clusters, using spectral subtraction of atmospheric noise to improve mapping speed by a factor of $\sim5$ over previous surveys.

\begin{figure}[h!]
\begin{center}
\includegraphics[width=0.75\textwidth]{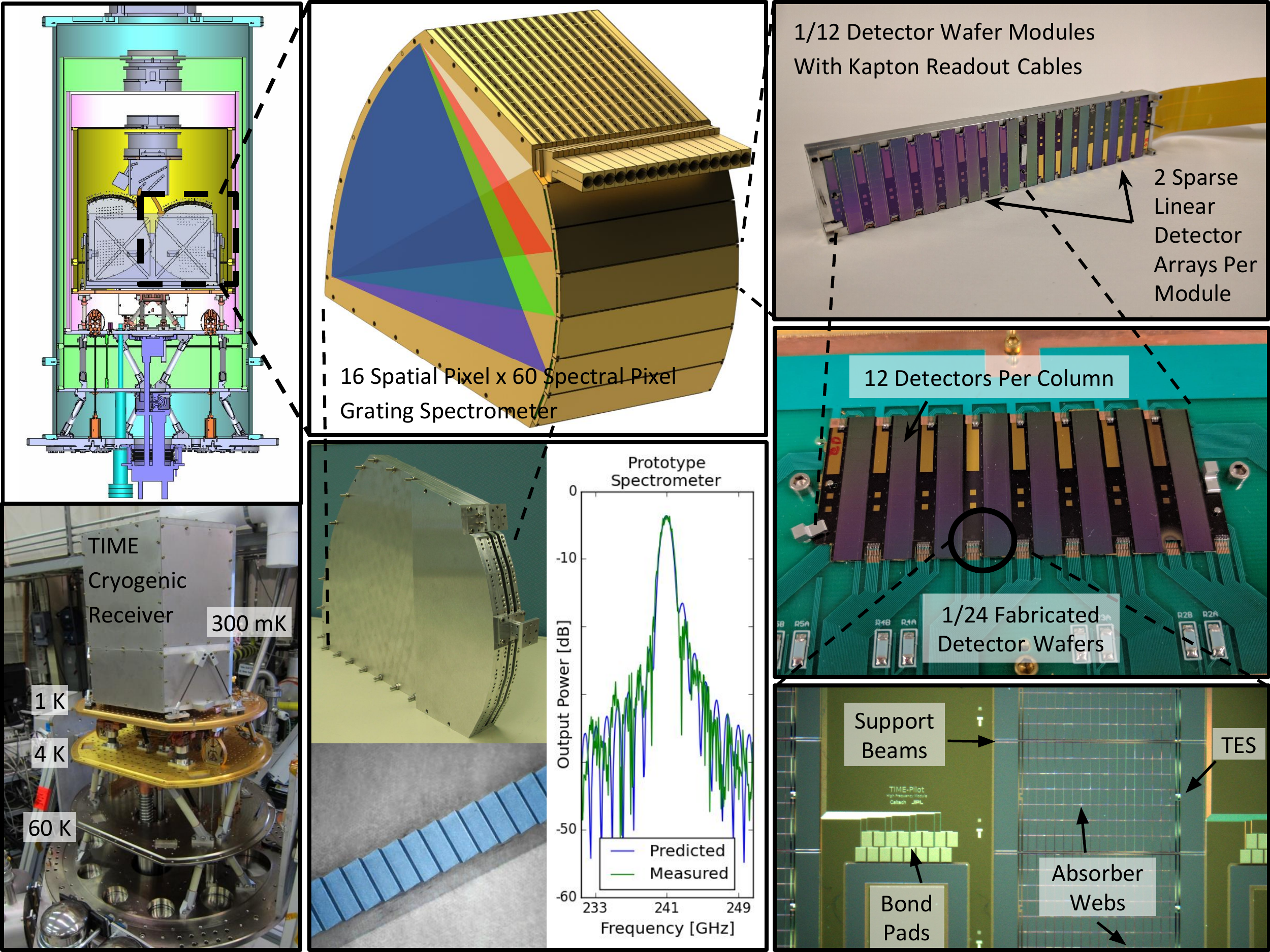}
\vspace{0.025in}
\caption{TIME instrument overview. The instrument is housed in an existing closed-cycle 4K-1K-300mK cryostat (bottom left) with a large cryogenic volume for the spectrometer stacks and optics (top left). 32 waveguide grating spectrometers (top center) are assembled into two stacks of 16; they couple the same 1-D linear field on the sky via an array of feedhorns and single-polarization waveguide feeds illuminated through a polarizing grid. Each grating spectrometer is similar to that used in Z-SPEC, but at lower resolving power. The dispersed light is detected with twelve 2-D arrays of TES bolometers which span the spectrometer stacks (right) with a total of 1920 detectors. The TES detectors (lower right) are similar to those built at JPL but with mesh absorbers.  A linear array of 11 $150\,{\rm GHz}$ broadband channels view the same sky as the spectrometers via a dichroic filter, and will be used in surveys of the kSZ effect.  Prototype TIME gratings in a ``mini-stack", a shortened version of one of the TIME spectrometer stacks, have been produced (bottom center). Each grating has 190 facets and provides resolving power in excess of $150$ over the full $183-326\,{\rm GHz}$ range.  Their spectral profiles have been measured using a coherent source and diode detector (bottom center).
({\it Courtesy of Abby Crites and Jamie Bock})}
\label{fig:TIME1}
\end{center}
\end{figure}

               TIME uses a linear array of 32 two-dimensional waveguide spectrometers with a spectral resolving power of $\sim100$, a lower resolution version of the spectrometer technology first developed for the Z-SPEC instrument.  The spectrometers are arranged in two stacks of 16, covering the frequency range of $183-326\,{\rm GHz}$, and view the sky in two polarizations off a beam splitter to maximize sensitivity.  Arrays of sensitive TES bolometers, read out by time-domain SQUID amplifiers, detect light from the spectrometers.  TIME couples to the telescope using relay optics that form an image of the primary mirror inside the cryostat, and a 4K cold stop to reduce stray light.  As shown in Fig.~\ref{fig:TIME1}, much of the cryogenic hardware for the instrument has already been assembled. First versions of the spectrometers, feeds and detectors have all been prototyped, and 300 nights of winter observing time on the APA 12-m telescope is secured for TIME observations.  TIME will carry out its first CII survey starting in late 2018.                              
                               
\begin{figure}[h!]
\begin{center}
\includegraphics[width=\textwidth]{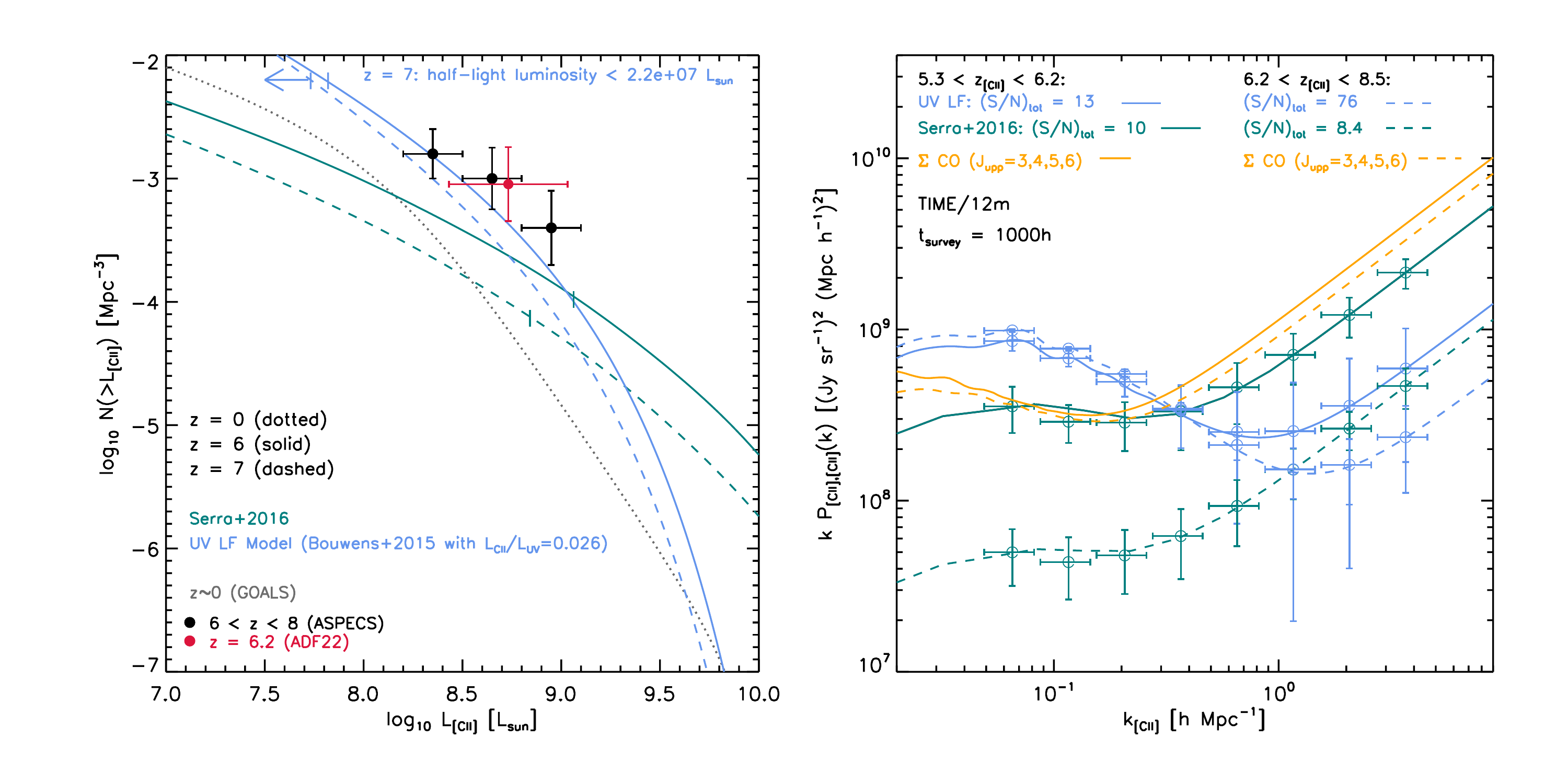}
\caption{{\it Left}: [CII] luminosity function models for EoR, plotted as cumulative number counts per ${\rm Mpc}^3$.  The \citet{Serra:2016jzs} model is shown in teal for $z=6$ (solid) and $z=7$ (dashed). Blue curves show the UV LF model based on the high-z UV luminosity function measured by HST and recent [CII] systems detected by ALMA. Vertical ticks indicate the depth in $L_{\rm [CII]}$ that recovers half of the total [CII] intensity (much fainter for the UV LFs; at $z=7$ UV LF, the half-[CII] light depth is unconstrained). Data with error bars (not including cosmic variance) are shown for results from ALMA deep field experiments ASPECS (\citet{2016ApJ...833...71A}) and ADF22 (\citet{2017PASJ...69...45H}). (While \citet{2016ApJ...833...71A} have corrected their data for incompleteness and false detections, the [CII] detections have yet to be spectroscopically confirmed.) For reference, the $z\sim0$ [CII] number counts are shown in dotted dark gray, as measured for local luminous infrared galaxies (\citet{0004-637X-834-1-36}). {\it Right}: 3-D power spectra (in $kP(k)$ units) of EoR [CII] per the UV LF and \citet{Serra:2016jzs} models at left. Two redshift ranges are shown. Orange curves show the signal from low-redshift CO fluctuations, when cast into the [CII] comoving frame.
({\it Courtesy of Jamie Bock})}
\label{fig:TIME2}
\end{center}
\end{figure}

               TIME targets a first detection of the [CII] signal (in both the clustering and Poisson regimes), spanning the spatial scales corresponding to $0.1 < k\,{\rm [h/Mpc]} < 1$.  TIME uses a linear survey strategy to maximize sensitivity on large scales, with a survey area of $0.3' \times 1 \,{\rm deg}$, or a comoving volume of $2 \times 10^5\, {\rm (Mpc/h)^3}$.  TIME uses spectral information to remove atmospheric fluctuations on large scales, a method previously demonstrated by observations with Z-SPEC. While forecasts for the [CII] power spectrum vary between models, as shown in Fig.~\ref{fig:TIME2}, TIME is able to detect [CII] clustering fluctuations in a 1000 hour survey on the APA 12-m telescope at $S/N \geq 10$ at $z = 6-7$.  The [CII] luminosity functions associated with these models are consistent with recent ALMA measurements of [CII] in galaxies at $z=6-7$.  A first detection of [CII] clustering fluctuations would constrain the total star formation rate during the epoch of reionization, while detection of Poisson power would constrain the integrated [CII] luminosity function, and set up the case for a new generation of [CII] mapping measurements with higher sensitivity.
               
\begin{figure}[h!]
\begin{center}
\includegraphics[width=0.5\textwidth]{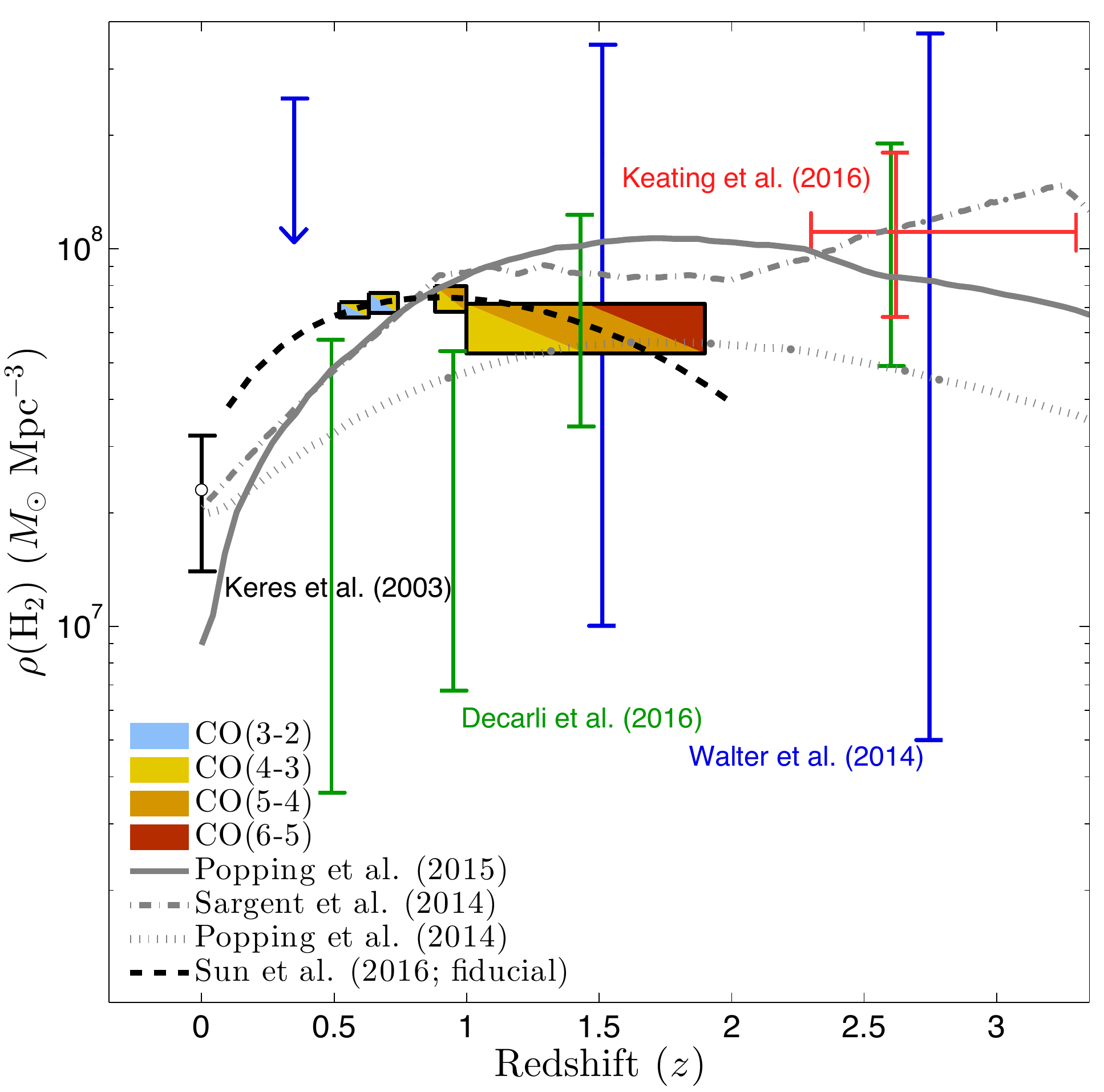}
\caption{Current theoretical predictions of the evolution of H2 density from several groups are shown, as well as the state of current measurements. The constraints from TIME are shown as boxes, colored by the pair of CO transitions that will be cross-correlated within the data to uniquely identify power at each redshift.  The TIME measurement is subject to systematic uncertainty in the conversion of CO to H2, though this uncertainty applies almost identically to all of the measurements shown. Outside of a small gap around $z\sim0.75$, TIME will chart the evolution in H2 density across 5 Gyr of cosmic time, starting from the period of peak star formation activity when depletion of molecular gas may have led to the subsequent rapid decline in the cosmic star formation rate density.  
({\it Courtesy of Jamie Bock})}
\label{fig:TIME3}
\end{center}
\end{figure}

               TIME will also make high signal to noise detections of CO rotational line emission at lower redshifts.  These can be used to estimate the density of molecular gas that fuels star formation, as shown in Fig.~\ref{fig:TIME3}.  The CO detections are shown assuming spectral cross-correlations between rotational levels, which are more robust against sources of systematic errors compared with auto-correlations.  The CO emitting galaxies must be removed from the [CII] signal, and two methods have been developed to surmount this problem, either using the strong anisotropy of the CO power spectrum in co-moving [CII] coordinates or masking the CO galaxies spectrally and spatially using an external catalogue, noting that modestly over-masking galaxies does not substantially degrade the [CII] signal (see more description of these techniques in Section 5).

\subsubsection*{CONCERTO}
\begin{figure}[h!]
\begin{center}
\includegraphics[scale=0.24]{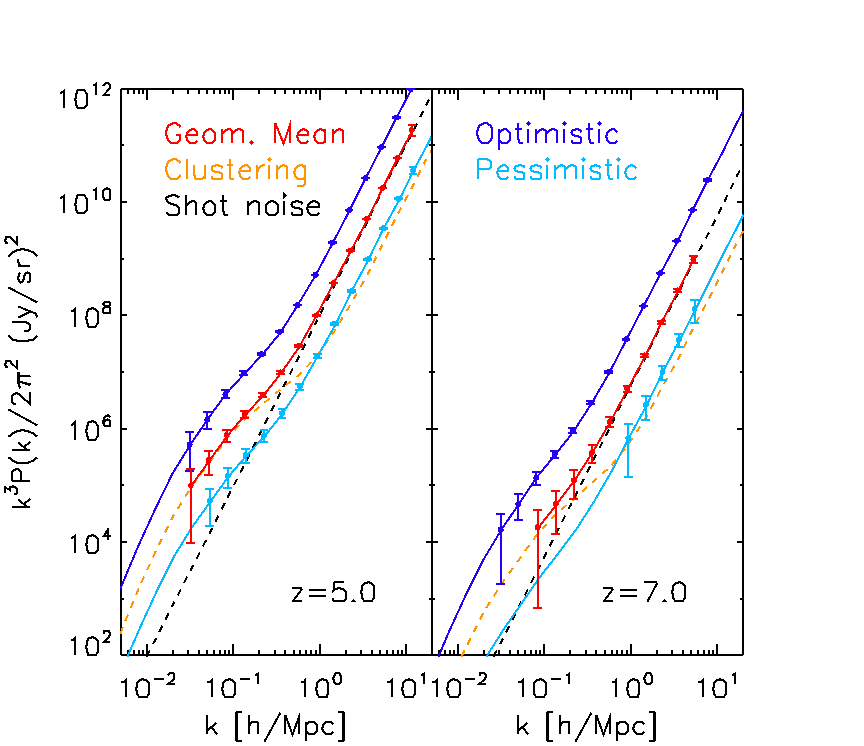}
\includegraphics[scale=.46]{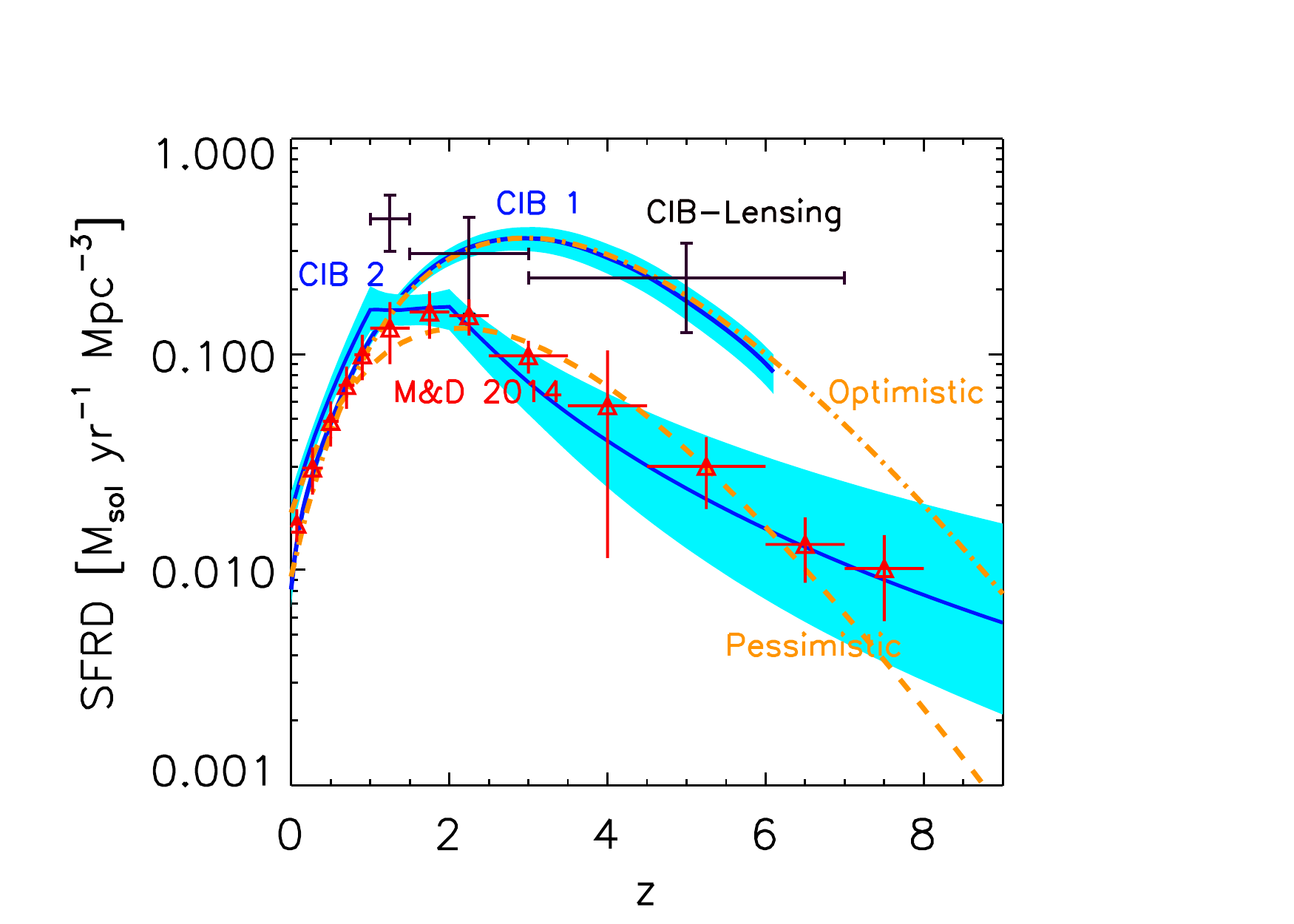}
\caption{{\it Left:} Predicted [CII] power spectrum (for three scenarios: optimistic, 
pessimistic and the geometrical mean) at redshifts $z=5$ and $z=7$. For the geometrical 
mean, both the clustering and shot-noise terms are shown. Error bars have been computed 
using the spectrometer and survey characteristics of the CONCERTO experiment. 
Only points with $S/N>1$ are shown. {\it Right:} Figure illustrating how different the 
SFRDs derived from galaxies (red points, which come exclusively 
 from UV-selected galaxies \citet{Madau:2014bja}) and CIB anisotropies 
are. Different models of CIB anisotropies, that fit equally well the measured power 
spectra, give very different SFRD (cf CIB1 and CIB2 models: dark blue curves and 
$1\sigma$ in light blue). Also shown are the measurements derived from the 
cross-correlation between the lensing map of the CMB and the CIB (black points, 
\citet{Ade:2013aro}). SFRD derived from the two [CII] models that are  
considered in the right panel (optimistic and pessimistic cases) are shown in orange. At $z>5$, SFRD 
from the dusty star-formation is an ``incognitus mundus". ({\it Courtesy of Guilaine Lagache})}
\label{fig:concerto_fig} 
\end{center}
\end{figure}

CONCERTO proposes to measure [CII] at redshifts $4.5<z<8.5$ and CO 
intensity fluctuations arising from $0.3<z<2$ galaxies. The CONCERTO 
instrument will use Kinetic Inductance Detectors (KID; \cite{Doyle2008, Doyle2010}) 
following the successful development of the NIKA2 camera 
\cite{2017arXiv170700908A}. CONCERTO is planned to be deployed to the APEX telescope, 
which is a 12-m antenna located at a $5105\,{\rm m}$ altitude on the Llano de Chajnantor in 
Northern Chile. The instrument is based on a dilution cryostat, not requiring liquid helium 
or nitrogen, and able to assure continuous operation, i.e. no recycling or other dead time. 
The field of view is exceeding $100\,{\rm arcmin}^2$. Spectra are obtained using a Martin-Puplett 
Fourier-transform spectrometer with variable resolution (settable from $1\,{\rm GHz}$ to $5\,{\rm GHz}$) 
located at room temperature in front of the cryostat. Such a fast (compared to the atmospheric 
noise) spectrometer can now be used to get the spectra thanks to the very small time 
constants ($<0.1\,{\rm msec}$) of KIDS. Each of the $\sim3000$ pixels will acquire fast 
interferograms and will be able to extract a spectrum in $0.2-2$ second integration time. 
Typically, faster acquisitions relate to spectral resolution $R = \nu/\delta \nu\sim100$, 
slower ones apply to $R\sim300$. CONCERTO will produce, roughly once per second, 
a data cube containing the spectral image. The projections for the sensitivity and signal-to-noise ratios are 
shown in Fig.~\ref{fig:concerto_fig}.  

\vspace{-0.1in}

\begin{figure}[h!]
\begin{center}
\includegraphics[scale=0.25]{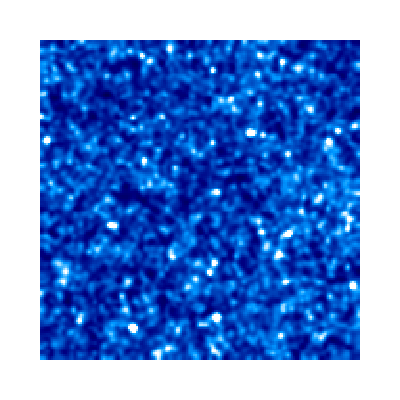}
\hspace{-0.6cm}\includegraphics[scale=0.25]{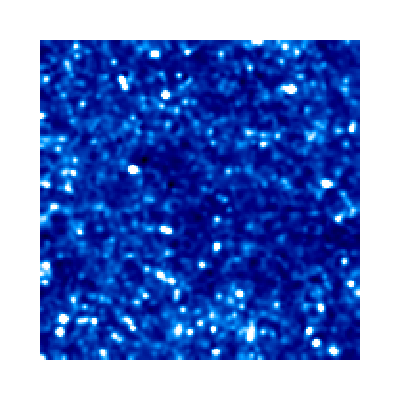}
\hspace{-0.6cm}\includegraphics[scale=0.25]{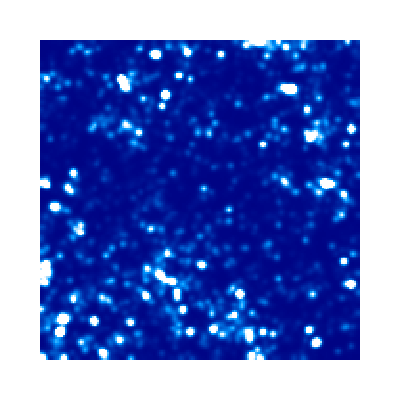}
\includegraphics[scale=0.25]{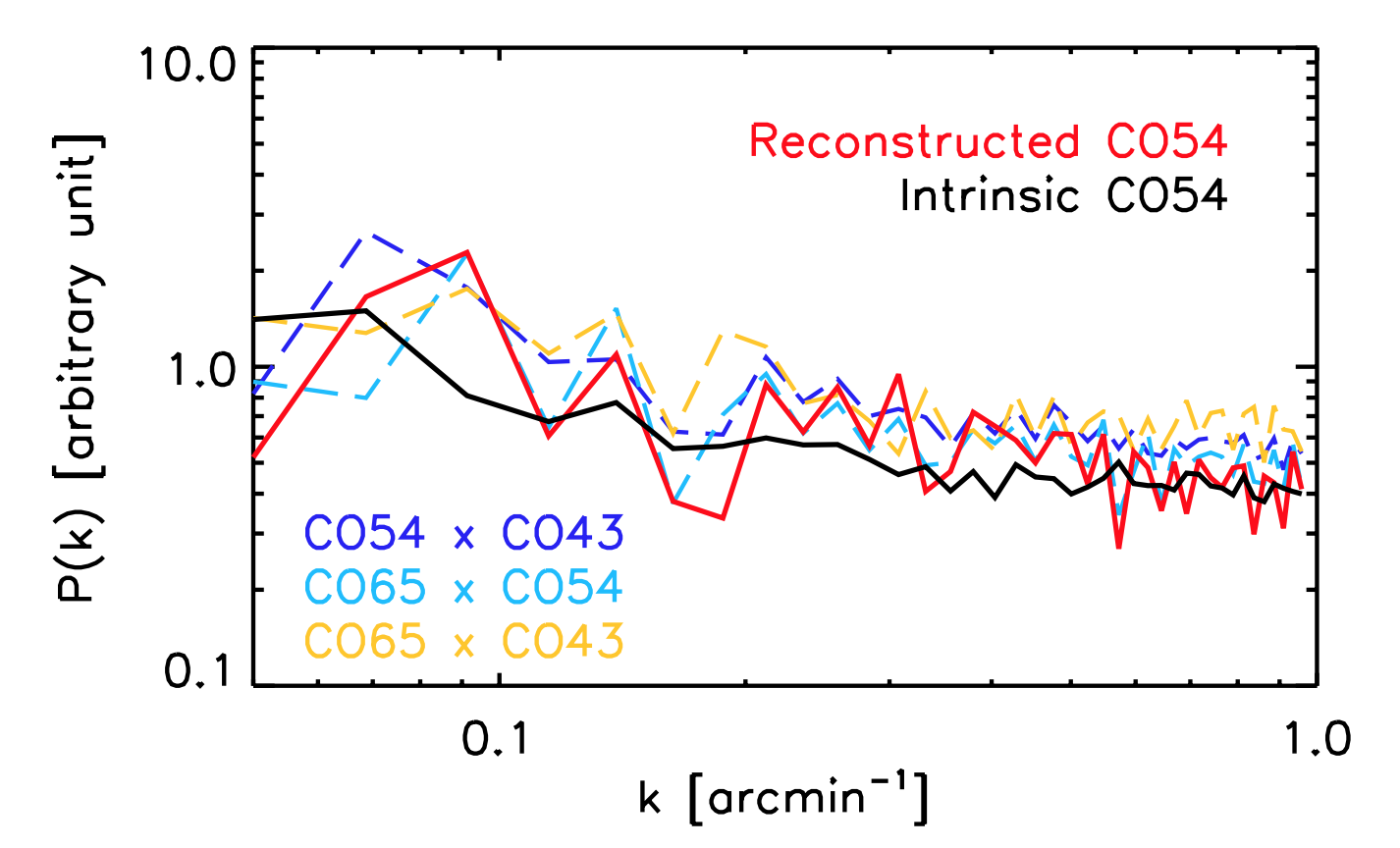}
\caption{\label{fig_simu_CO} First three images: our $1.4\times1.4\,\,{\rm deg}^2$ simulated sky maps at $z=5.5\pm0.1$. From left to right: total signal, CO+[CII] only, [CII] only. The display range has been adapted for each image (the standard deviation of the [CII] map being 30 times lower than that of the total signal map). Right figure: first comparison of the intrinsic and reconstructed CO(5-4) power spectrum at $z=[1.1-1.2]$ (a foreground for [CII] at $z=6$). It has been estimated using cross-correlations of CONCERTO maps at the frequencies of the CO(4-3), CO(5-4) and CO(6-5) lines. ({\it Courtesy of Guilaine Lagache})}
\label{fig:concerto_fig2} 
\end{center}
\end{figure}

To remove the CO contamination to the [CII] signal, CONCERTO will use the same 
methods mentioned for TIME and described in more detail in Section 5. 
But its wide frequency range will also permit to mitigate the CO contamination using 
the cross-correlation of multiple CO lines detected at each redshift \cite{Visbal2010}:  
between two and four CO lines are simultaneously observed at the same redshift for 
all $z>0.35$. This method is being successfully tested using realistic sky simulations 
(with [CII] and CO line emissions added to the SIDES simulations of \citet{Bethermin:2017ngy}, 
as shown in Fig.~\ref{fig:concerto_fig2}).

\vspace{-0.1in}

\section{Galaxy Assembly and Star Formation at z=2--10}

There are two primary lines targeted by intensity mapping experiments to explore star 
formation science: CO and [CII]. Experiments targeting the rotational transitions of 
carbon monoxide are divided into two primary classes: interferometric and single-dish 
with focal plane arrays. Interferometric experiments provide better control of systematics 
and make measurements directly in the Fourier regime. Fields of view, however, tend to 
be small putting these experiments primarily in the shot-noise regime. Single-dish focal 
plane array experiments, on the other hand, can map large areas efficiently, pushing 
sensitivity into the clustering regime where cross-correlation with other large-area 
intensity mapping experiments can be performed. Systematic errors, in particular 
bandpass ripples, may prove challenging to remove or calibrate. It is important to note 
that some of the instruments described in the previous subsection, such as TIME and 
CONCERTO, are also very much relevant to the study of star formation science.

\subsection*{COPSS and mmIME} 

{\it COPSS I and II --} The COPSS I and II surveys are two interferometric experiments 
carried out using the Sunyaev-Zeldovich Array (SZA; \cite{2015ApJ...814..140K,Keating2016}). 
The SZA is an array of 8 3.5-m diameter antennas located in Southern California with receivers 
at 1 cm wavelength (27 -- 35\,GHz).  The 1 cm receivers are sensitive to the J=1-0 transition 
from $2.3 < z < 3.3$ over k-modes from 0.5 to $10.0\rm \, h\,  Mpc^{-1}$.  The array is primarily 
configured in a compact configuration with minimum baselines of 4.5\,m. COPSS I consisted 
of an analysis of archival data from 44 fields, each observed for approximately 20 hours. 
COPSS II consisted of new observations of 22 fields for a total of nearly 3000 hours on the sky. 
Fields include GOODS-N and other deep fields with extensive optical spectroscopy, which are 
suitable for cross-correlation. Both experiments demonstrated the ability to detect and remove 
foregrounds through Fourier filtering as well as handle systematic errors. 

As described in the Chapter on First Detections, COPSS II detected a power 
$3.1^{+1.2}_{-1.3} \times 10^3\rm \, \mu K^2\, (h^{-1}\, Mpc)^3$ , which is non-zero at $99.2\%$
confidence, setting significant limits on the CO luminosity function and H$_2$ mass density.

{\it Millimeter Intensity Mapping Experiment (mmIME) -- } Motivated in part by the success of 
COPSS, and in advance of new instruments dedicated to intensity mapping experiments, a 
new intensity mapping-focused experiment was started in 2016, utilizing existing radio instruments. 
This new experiment, referred to as mmIME,  will focus on the multiplicity of bright transitions 
found in the cold gas of high redshift galaxies at millimeter wavelengths. The goal of these new 
observations is three-fold: confirm the tentative detection from COPSS, expand the redshift 
range beyond $z\sim3$, and explore other spectral lines (i.e.; higher-order rotational transitions of CO, [CII]). 

mmIME will seek to utilize both archival data and new observations, leveraging the combined 
power of the VLA, the Atacama Compact Array (ACA), and SMA at 1\,cm, 3\,mm, and 1\,mm 
respectively. mmIME will take advantage of the fact that all three instruments have similar 
primary beam sizes and sensitivity to similar spatial modes to cross-correlate measurements 
between these three instruments, separating the different line/redshift components at a 
given frequency to allow for the exploration of multiple line species over a broad redshift 
range ($z\sim1{-}5$).

\begin{figure}[!h]
\begin{center}
\includegraphics[width=0.93\textwidth]{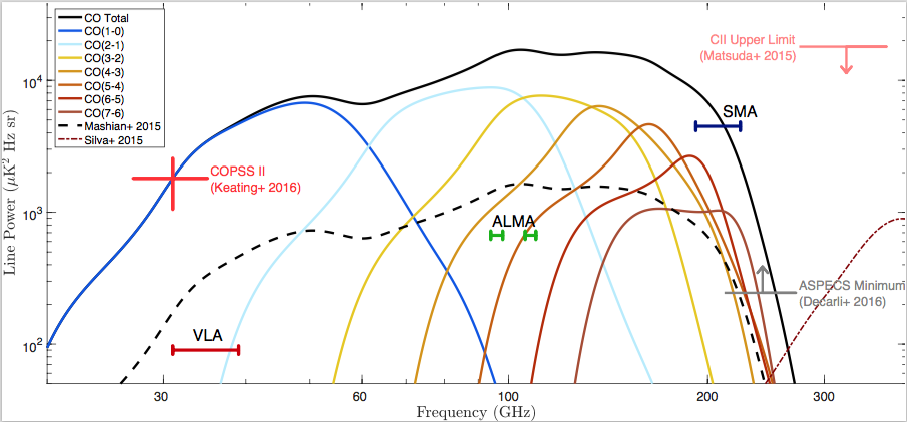}
\caption{Theoretical expectations for the line power of various transitions, versus the 
estimated sensitivity of the pilot survey of mmIME. The fiducial model takes estimates 
from \citet{Mashian:2015his}, scaled to match the power measured in COPSS. Also 
shown are the upper and lower limits set by blind 1mm surveys with ALMA 
\cite{2015MNRAS.451.1141M}. Additionally shown are estimates on the line power 
from [CII] from \citet{Silva:2014ira}, although recent results from ASPECS suggests at 
there may be more than 10 times more power than what is estimated. The full mmIME 
survey is expected to have improved sensitivity by factor of $\sim3$, with significantly 
greater spectral coverage between 30 GHz and 300 GHz.  ({\it Courtesy of Karto Keating})}
\label{fig:fig_linepow}
\end{center}
\end{figure}

As of mid-2017, pilot studies utilizing VLA, ALMA, and SMA are underway, and projections 
from this pilot survey are given in Figure~\ref{fig:fig_linepow} and show an improvement 
in the existing constraints on the CO power spectrum by approximately an order of magnitude, 
surveying an area approximately 10$\square^{\circ}$ in size and covering approximately an 
octave in frequency between 1\,cm and 1\,mm.

\subsection*{AIM-CO}

ASIAA Intensity Mapping for CO (AIM-CO) is an interferometric experiment using the 
Yuan-Tseh Lee Array (YTLA; \citet{2016AAS...22742604B}).  Located on Mauna Loa, 
the YTLA is an array of 13 1.2-m dishes on a hexagonal close-packed array with a 
minimum spacing between antennas of $1.4\,{\rm m}$.  Each dish has a 3\,mm receiver 
sensitive from 86--102\,GHz. A dual polarization 7-element correlator was delivered in 
2017 with an instananeous bandwidth of 4\,GHz with 4096 spectral channels.   The 
interferometric nature of the experiment was chosen in order to provide maximum control 
of telescope and foreground systematics and produce reliable detection of the expected 
signal. AIM-CO is sensitive to the J=3-2 transition from $2.4 < z < 3.0$ and $J=2-1$ from 
$1.2 < z < 1.5$ over k-modes from $0.5$ to $3\,{\rm h Mpc^{-1}}$

Following commissioning in 2017, a two-year observational campaign is planned. 
The observing campaign will match survey fields observed as part of with the COPSS 
SZA observations experiment \cite{2015ApJ...814..140K,Keating2016} to enable 
cross-correlations at (both radio and optical wavelengths).  Cross-correlations are 
critical for validation of the auto-correlation signal, disambiguation of different transients, 
and extraction of host galaxy-dependent effects.  The AIM-CO experiment will be a factor 
of 10 more sensitive than COPSS II, providing an opportunity to confirm the COPSS II 
detection and probe evolution of the CO power with redshift. 

\begin{figure}[!h]
\begin{center}
\includegraphics[width=0.7\columnwidth]{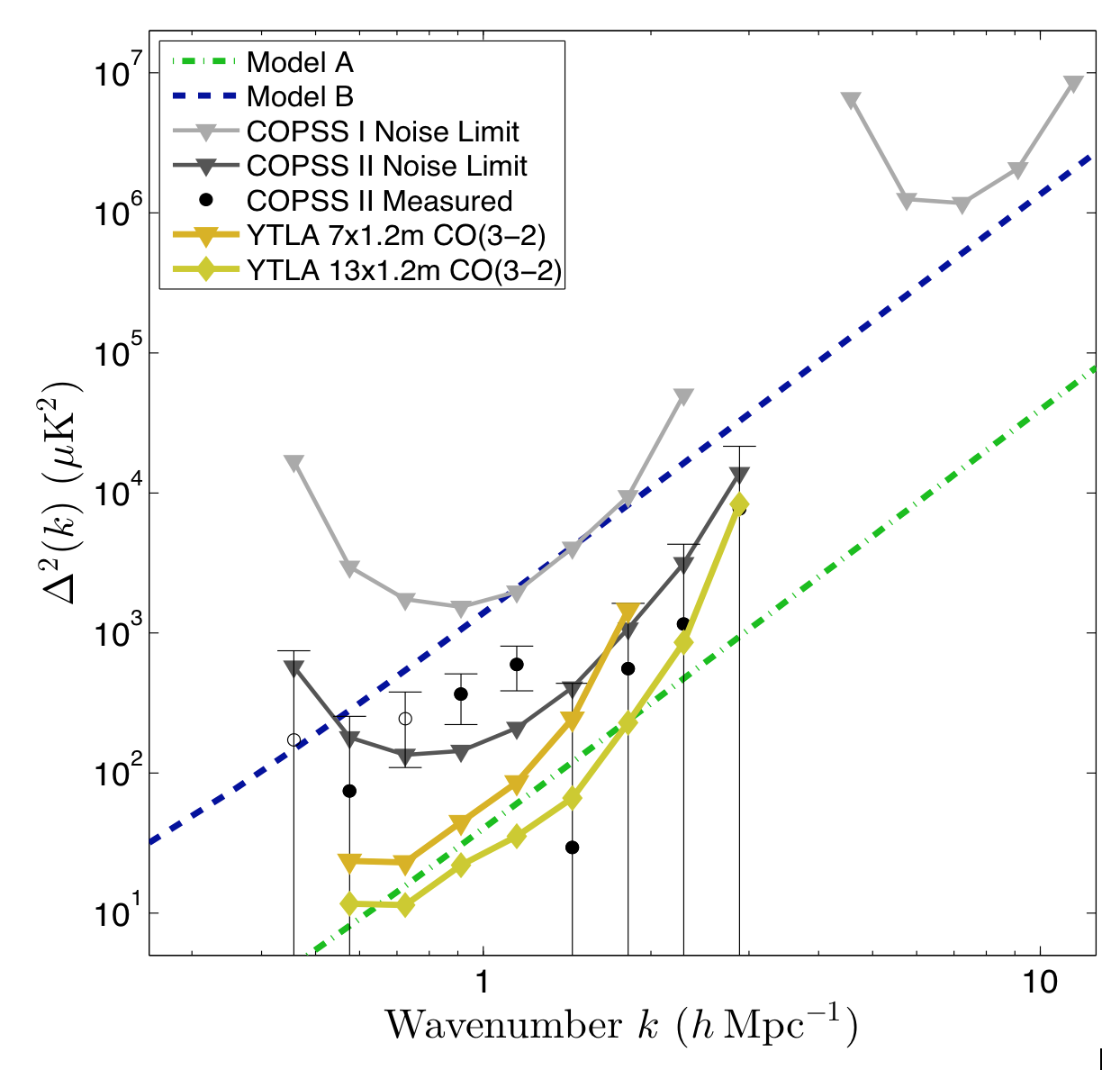}
\caption{AIM-CO sensitivity compared against COPSS I and II measurements and 
theoretical models from \citet{Pullen:2012su}. ({\it Courtesy of Geoff Bower})}
\end{center}
\end{figure}

\subsection*{COMAP}

The CO Mapping Array Pathfinder (COMAP; \cite{2016AAS...22742606C}) is part of a 
program aiming to trace the distribution of star-forming galaxies at the Epoch of Reionization 
(EoR). Constraining the CO power spectrum from the EoR will ultimately require measurements 
at multiple frequencies and focal-plane arrays with hundreds of elements. As a first step 
towards this goal, Phase I of COMAP comprises a 10.4-m telescope, located at the Owens 
Valley Radio Observatory (OVRO), equipped with a 19-pixel spectrometer array that will 
map a total of 10 square degrees of sky in the frequency range 26--34\,GHz, with spectral 
resolution R~800. This band will be sensitive to CO(1-0) in the redshift slice $z=2.4-3.4$ 
and to CO(2-1) in the redshift slice $z = 5.8-7.8$. A CASPER-based digital backend will 
process 8 GHz from each of the 19 pixels.

The aim of this pathfinder experiment is to i) demonstrate the feasibility and future potential 
of wide-field CO intensity mapping, and ii) provide a test-bed for the technology development 
and observational strategies. Phase I of COMAP will focus on constraining the CO power 
spectrum from the Epoch of Galaxy Assembly, at $z=2.4-3.4$.

The receiver and digital backend are currently under construction, with a two-year observing 
campaign due to begin in 2018. The COMAP collaboration includes Caltech, JPL, Stanford 
University, University of Maryland, University of Miami, University of Manchester, University 
of Oslo and ETH, Zurich. Phase I of the project is fully funded by NSF (AAG and MRI awards), 
the Keck Institute for Space Studies as well as contributions from institutional partners.

\begin{figure}[!h]
\begin{center}
\includegraphics[width=0.7\columnwidth]{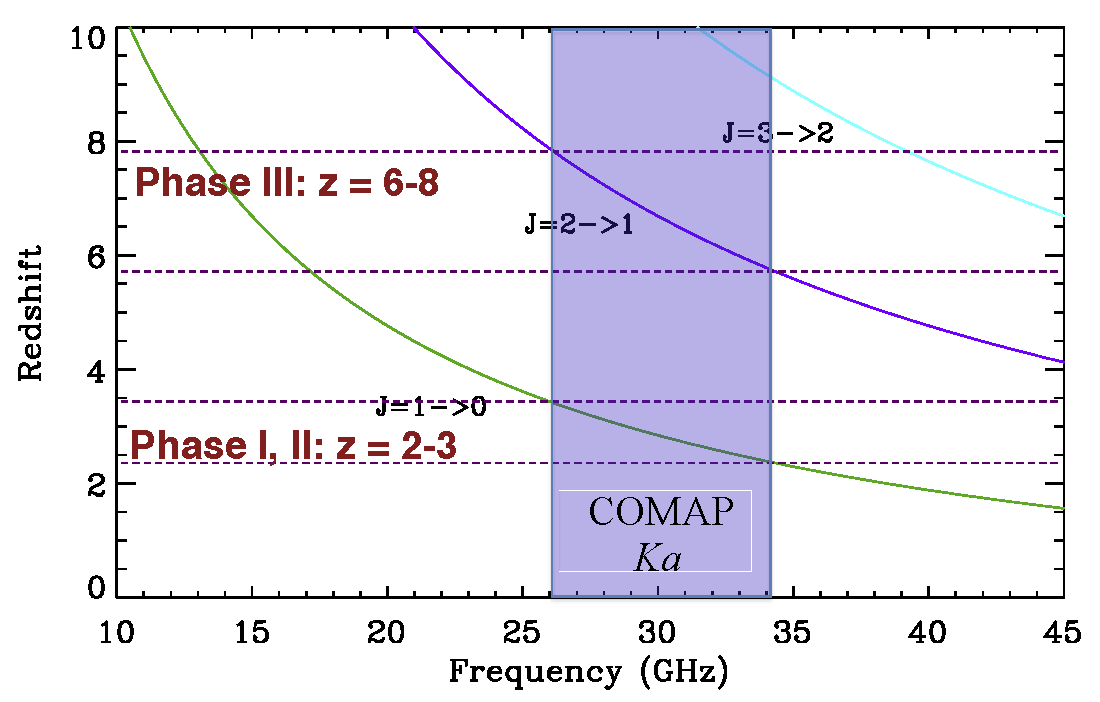}
\caption{The COMAP experiment \cite{2016AAS...22742606C}. ({\it Courtesy of Kieran Cleary})}
\end{center}
\end{figure}

\subsection*{STARFIRE}
 
The Spectroscopic Terahertz Airborne Receiver for Far-InfraRed Exploration ({\sc starfire}) 
is an integral-field spectrometer using kinetic inductance detectors, operating from 240 - 420 
$\mu$m and coupled to a 2.5 meter low-emissivity carbon-fiber balloon-borne telescope. 
Using dispersive spectroscopy and the stratospheric platform, {\sc starfire} can achieve
 better performance than SOFIA or Herschel-SPIRE FTS.  {\sc starfire} is designed to study 
 the ISM of galaxies from $0.5 < z < 1.5$, primarily in the [CII] (158 $\mu$m) line, and also 
 in cross-correlation with N{\sc ii} (122 \ensuremath{\mu\mathrm{m}}).   This offers a view 
 of the star-forming medium with minimal impact from dust extinction through the period of 
 peak cosmic star formation and into the current epoch where the star formation begins to 
 decline.  {\sc starfire} will be capable of making a high significance of the [CII] power spectrum 
 in at least 4 redshift bins and measuring the [CII] $\times$N{\sc ii}\ power spectrum at 
 $z \sim 1$.  The intensity mapped power spectra will be sensitive to the one- and two-halo 
 clustering, as well as the shot noise, and will relate the mean [CII] intensity as a function 
 of redshift (a proxy for star-formation rate density) to the large scale structure \citep{Uzgil2014}.  
 In addition, {\sc starfire} will measure $\sim50$ individual redshifts, but will also be able to 
 stack on optical galaxies below the SPIRE confusion limit to measure the [CII] luminosity of 
 more typical galaxies.  Detector development for {\sc starfire} is currently funded by NASA 
 \citep{barlis17}.

\begin{figure}[h!]
\centering
\includegraphics[height=4.5in]{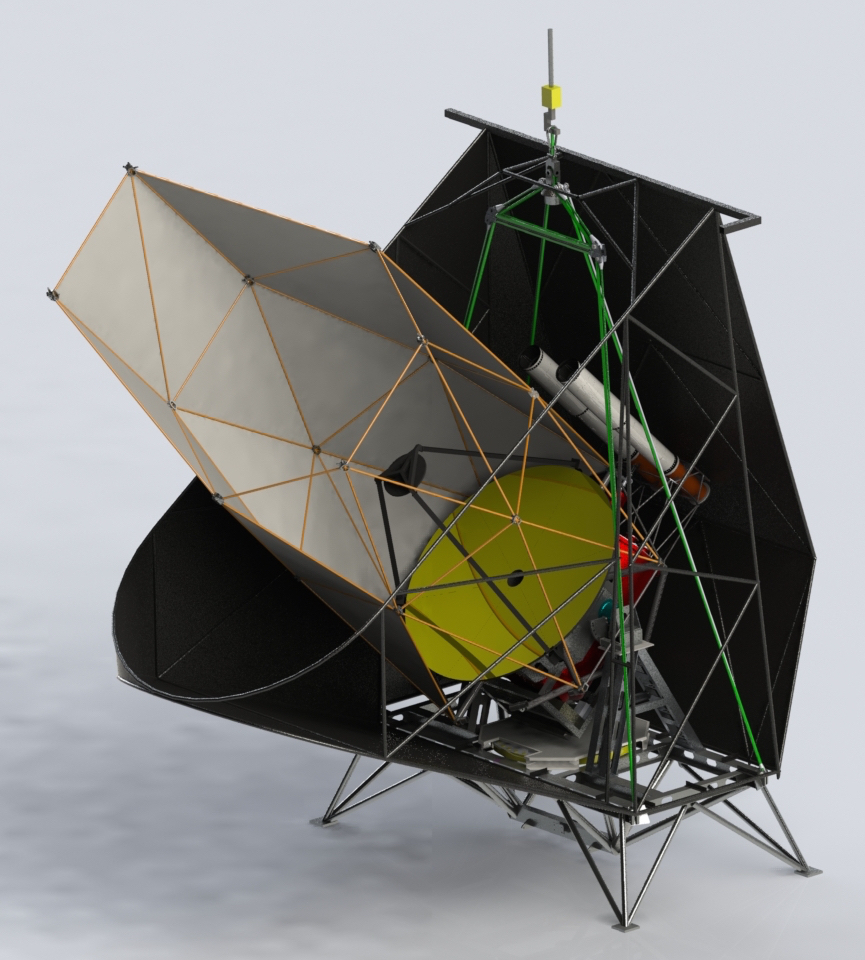}
\includegraphics[height=4.5in]{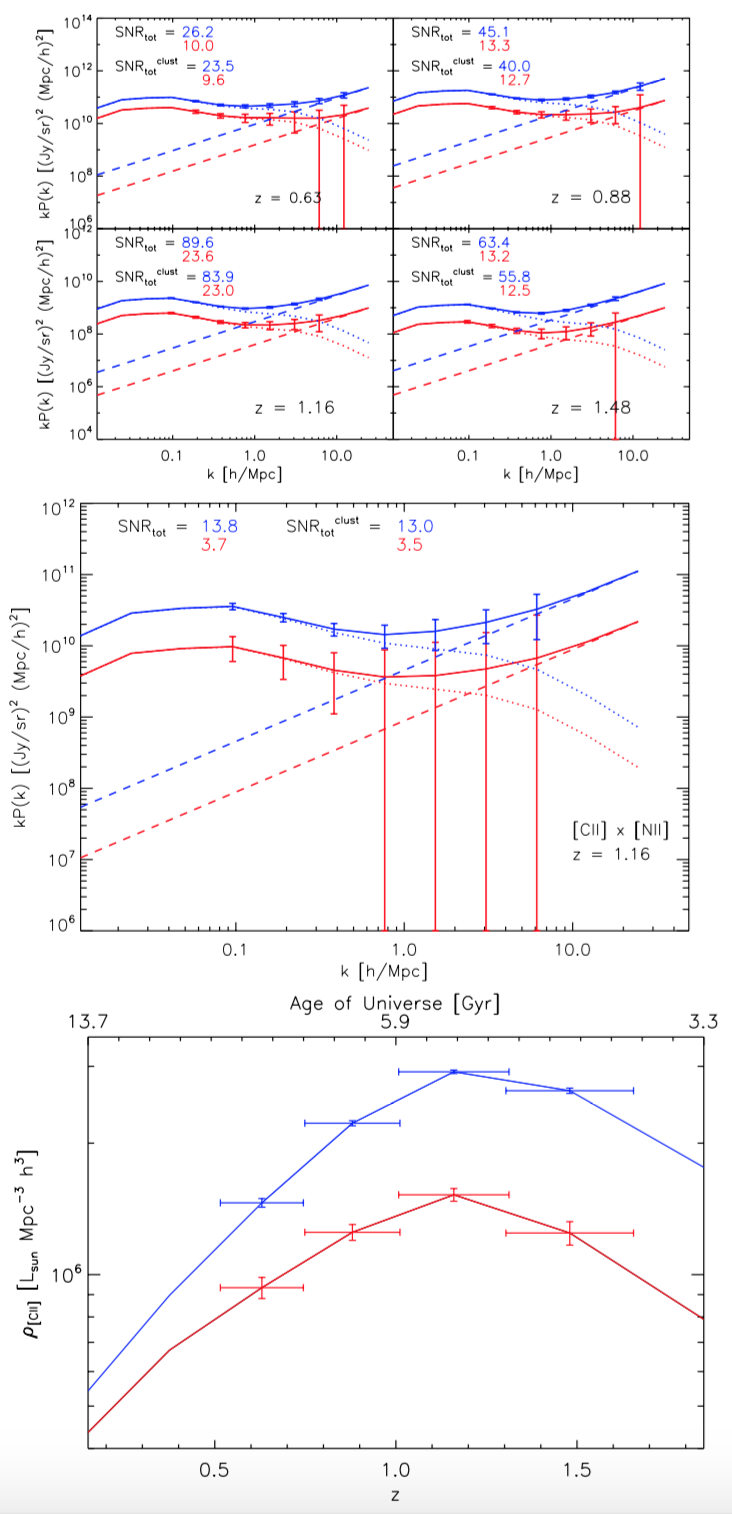}
\vspace{5pt}
\caption{Left: {\sc starfire}\ telescope and gondola in a cutaway view.  {\sc starfire}\ will 
re-use the design for the gondola, cryostat, readout electronics, and star camera design 
from BLAST-TNG.  Right: Top panel: Predicted total autocorrelation power spectra for the 
[CII] line at four redshifts. Solid curves are the total power, dotted lines show the sum 
of the 1- and 2-halo clustering terms, and dashed lines show the ``shot noise" or Poisson 
term due to the discrete nature of galaxies. Error bars are for a survey of 100 hours in 0.1 
deg$^2$. Power spectrum amplitudes were predicted using the same method described 
in \citet{Uzgil2014}, except for an updated IR luminosity function \citep{gruppioni13}.  Also 
used, in addition to the \citet{spinoglio12} [CII]-IR relation (red curve) used 
in \citet{Uzgil2014}, which under-predicts [CII] luminosities in high-redshift sources appropriate 
for this study, is a constant [CII]-IR relation of $L_{\rm [CII]}/L_{\rm FIR} = 3\times10^{-3}$ 
(blue curve). The two cases likely to bound the actual [CII] power spectrum are considered, 
and  the actual sensitivity (expressed in terms of the signal-to-noise ratio, SNR) is expected to lie 
between the red and blue power spectra. Middle: Predicted [CII]-[N{\sc ii}]122$\mu$m 
cross-power spectrum at $z = 1.16$. Color-coding and linestyles are the same as in the top 
panel.  A detection is possible even in the pessimistic case.  Bottom: [CII] luminosity 
density as a function of redshift.  Red and blue curves refer to the same respective [CII]-IR 
relations used in previous panels. The error bars in $z$ represent the redshift coverage for 
each {\sc starfire}\ band.  A high-significance measurement of the evolution of [CII]\ 
luminosity is possible with {\sc starfire}. ({\it Courtesy of James Aguirre})}
\label{fig:Starfire}
\end{figure}

\subsection*{Far-IR Spectrometers}

Far-infrared spectroscopy is uniquely well-suited to study the inner workings of galaxies throughout cosmic history.  The suite of rest-frame mid- and far-IR fine-structure transitions of elements such as carbon, oxygen, nitrogen, neon and iron in their various ionization states tracks total star formation activity, encodes the stellar and interstellar properties, and provides a census of heavy element contents, all without uncertainties due to dust obscuration.   Importantly, 3-D intensity mapping targeting the brightest of these lines  naturally overcomes source confusion to provide built-in look-back-time encoding, a key aspect for isolating the faint signals form the early universe from the much brighter `foregrounds'.    To date, sensitive far-IR capability has remained unrealized due to the technical challenges:  it requires a cryogenic telescope above the atmosphere, combined with sensitive detector array technology that must be built by the astrophysics / cosmology community.    However, large-format far-IR detectors have progressed greatly recently,  and NASA, ESA, and JAXA are studying options for cryogenic far-IR space missions featuring wideband spectroscopy which could begin in the next decade.

The exquisite surface brightness sensitivity of these facilities will make them excellent platforms for line intensity mapping in the far-IR (here taken to be 30 microns to 600 microns).   Initial 3-D power spectrum uncertainty estimates \cite{Visbal2010, Uzgil2014, Serra:2016jzs} based on expected instrument sensitivities for space-borne far-IR spectrometers (e.g. \cite{Nakagawa_17, Bradford_15}) indicate promise.  Clustering in the bright lines (both auto- and cross spectra) is readily detectable even at the epoch of reionzation with the ambitious space concepts.  These clustering signals constrain the total luminosity function integral in these lines, as well as their luminosity-averaged ratios, charting the total star formation history and probing the evolution of aggregate galaxy properties through cosmic time.

\subsection*{SPHEREx}

The Spectro-Photometer for the History of the Universe, Epoch of Reionization, 
and Ices Explorer (SPHEREx) is a NASA MIDEX-class mission currently undergoing 
Phase A study. SPHEREx is designed to conduct the first full sky spectrometric 
surveys in the near-infrared and will operate with $R \sim 42$ between $0.75$ and $4.18$ 
microns and $R \sim 135$ between $4.18$ microns and $5$ microns. It will perform 
4 all-sky surveys over the course of a 2-year mission. SPHEREx will chart the origin 
and history of galaxy formation through a deep survey centered near the ecliptic poles, 
allowing tomographic intensity mapping of large-scale structure at a complete set of near 
IR wavelengths. The deep spectro-imaging survey produces the ideal data set for full 
tomographic mapping of large-scale structure with $dz \sim 0.2$ resolution in the 
intensity mapping regime and higher when resolving individual sources. It will probe 
the inflationary history of the universe, the evolution of galaxies since the epoch of 
reionization, the origin of water in planetary systems, as well as creating a high-legacy 
spectral catalog over the entire sky [16]. At low redshifts SPHEREx will detect multiple 
lines with SNR $> 10$, the dominant lines being H$_\alpha$ for redshifts $0.1 < z < 5$, H$\beta$ 
for redshifts $0.5 < z < 2$, and [OIII] for redshifts $0.5 < z < 3$. At high redshifts $5.2 < z < 8$, 
SPHEREx accesses the Ly$_\alpha$ line, providing a crucial probe of the formation 
and evolution of EOR galaxies \cite{Silva2013,Pullen:2013dir}. Measurement of H$\alpha$ 
clustering measures cosmic star-formation rate as traced by bolometric line emission, 
integrated over all galaxy luminosities and including emission from any diffuse intra-halo 
component. Foreground line confusion from lower redshift [OIII] and H$\beta$ lines can be 
robustly removed by cross-correlating spectral lines in multiple bands. For example, $z = 3$ 
H$\alpha$ line fluctuations are detected in a band centered at $2.62\,\mu {\rm m}$, while at 
the same redshift [OIII] fluctuations are present in a band centered at $2.00\, \mu {\rm m}$. 
These secondary lines are useful, however: cross-correlating two independent line measurement 
traces the galaxies at z = 3 without masking, and naturally rejects any line contaminants or 
systematics that may be present in one of the two bands. In addition to the tomographic 
measurements and two main science cases, the SPHEREx all-sky archive will be a resource 
for numerous exciting and diverse astronomy investigations.

\begin{figure}
\begin{center}
\includegraphics[width=0.8\linewidth]{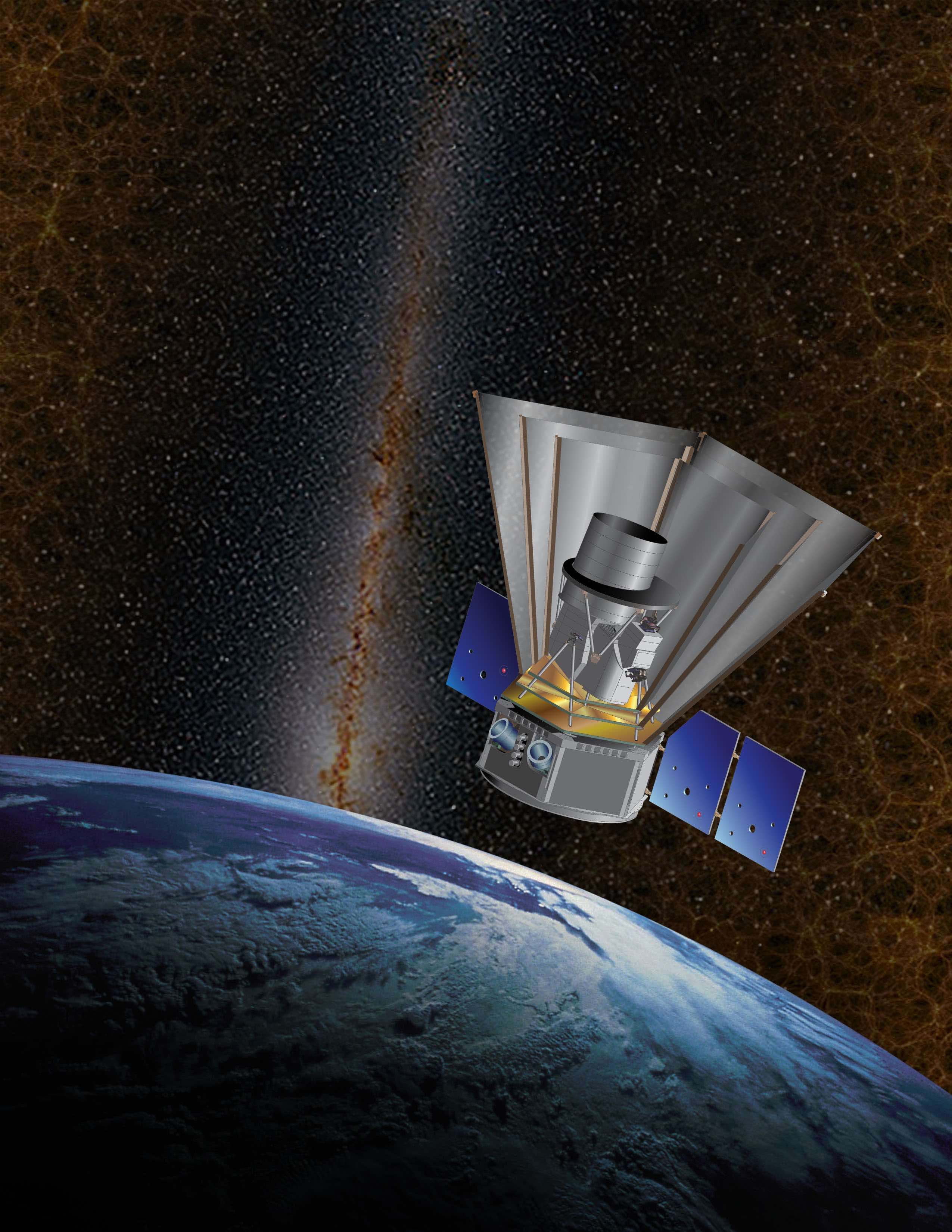}
\end{center}
\caption{SPHEREx is a NASA Medium Explorer mission designed to 1) constrain the physics 
of inflation by studying its imprints on the three-dimensional large-scale distribution of matter, 
2) trace the history of galactic light production through a deep multi-band measurement of 
large-scale clustering, and 3) investigate the abundance and composition of water and biogenic 
ices in the early phases of star and planetary disk formation. SPHEREx will measure near-infrared 
spectra from $0.75-5.0$ microns over the entire sky. It implements a simple instrument design with 
a single observing mode to map the entire sky four times during its nominal 25-month mission. 
The resulting rich legacy archive of spectra will bear on numerous scientific investigations. 
({\it Courtesy of Olivier Dor\'e})}
\label{fig:WFIRST}
\end{figure}

\subsection*{CDIM} 

The Cosmic Dawn Intensity Mapper (CDIM) is a NASA Probe-class Mission Study designed 
to be a survey instrument optimized for reionization studies, answering critical questions on 
how and when galaxies and quasars first formed, the history of metal build-up, and the history 
and topology of reionization, among other questions. CDIM will be a 1.0m-1.3\,m-class infrared 
telescope capable of three-dimensional spectro-imaging observations over the wavelength 
range of 0.75--7.5 $\mu$m, at a spectral resolving power $\Delta\lambda/\lambda \sim 500$. 
CDIM will provide spectroscopic imaging over ~10 $\square^{\circ}$ instantaneous FoV, at 1 
arcsecond/pixel. The depths, in equivalent broad-bands by combining narrow-band images, 
are comparable to depths reached by WFIRST (Figure~\ref{fig:WFIRST}). This will be achieved 
with linear variable filters (LVFs) sitting on top of a focal plane of 36 $2048^{2}$ detectors. 
The two-tiered wedding-cake survey, taking over three years, will consist of a shallow tier 
spanning close to 300 $\square^{\circ}$ and a deep tier of about 25 $\square^{\circ}$. CDIM 
will complement JWST with cosmological survey fields, as JWST will be limited to a handful 
of deep fields with a total area of several hundred $\square^{'}$, and provide spectroscopic 
data beyond $2<\,{\rm m}$ for the WFIRST surveys, allowing spectroscopic detection of H$\alpha$ 
at z$>$ 2 and galaxies at z $>$ 6. 

CDIM survey data will allow to (i) determine spectroscopic redshifts of WFIRST-detected 
Lyman-break galaxies out to $z\sim10$; (ii) establish the environmental dependence 
of star-formation during reionization through clustering and other environmental measurements; 
(iii) establish the metal abundance of first-light galaxies during reionization over two decades of 
stellar mass by spectrally separating NII from H$\alpha$ and detecting both H$\beta$ and OIII; iv) 
measure 3D intensity fluctuations during reionization in both Ly$\alpha$ and H$\alpha$; and (v) 
combine intensity fluctuations with 21-cm data to establish the topology of reionization bubbles.

\begin{figure}[h!]
\begin{center}
\includegraphics[width=0.66\linewidth]{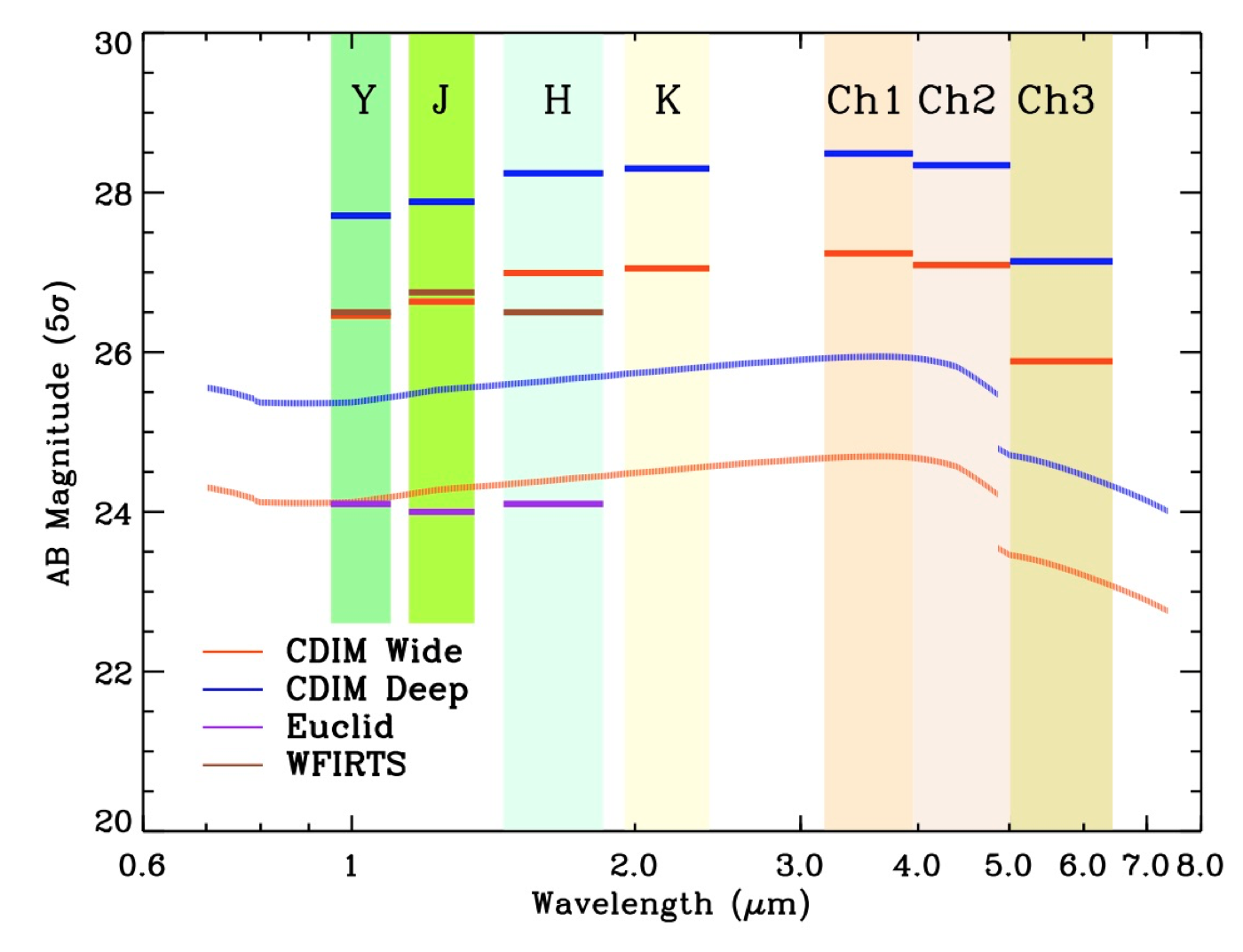}
\end{center}
\caption{Depth vs wavelength. CDIM covers 0.75 to 7.5 $\mu$m at R=500 (leading to 1360 
narrow-band images); 5$\sigma$ depth (AB mag) is shown in thin lines for the full R=500 spectrum. 
The 5$\sigma$ depths in broad bands (by combining appropriate narrow-band images) in 
z, Y, J, H, K, Ch1, Ch2, and Ch3 are shown as red and blue bars for both wide (300 $\square^{\circ}$) 
and deep (25 $\square^{\circ}$) surveys, respectively. The wide survey depths are roughly matched 
to WFIRST-HLS (brown bars in YJH) while the deep survey depth is matched to WFIRST 
medium-deep survey expected over an area of 25 $\square^{\circ}$ (depths not shown). 
Euclid depths are shown in purple bars. CDIM is best-suitable for wide-field R=500 spectro-imaging over areas $> 20 \square^{\circ}$. ({\it Courtesy of Tzu-Ching Chang})}
\label{fig:WFIRST}
\end{figure}

\subsection*{PIXIE}
PIXIE, proposed as a NASA Medium-class Explorer mission, is designed to measure the 
CMB polarization and absolute spectrum on large angular scales. It uses an FTS 
absolute spectro-polarimeter architecture similar to FIRAS, but with sensitivity 
$\sim 1000\times$ greater. This sensitivity is made possible largely through $100$\,mK 
cooling, which brings detectors to background-limited sensitivity. It will map the sky at 
\pixiefwhm\ (effective Gaussian) across frequencies from 
$\approx 45$\,GHz to $\sim 2$\,THz in $15$\,GHz bands. 

While it was not designed with intensity mapping in mind, PIXIE and similar architectures 
provide a unique constraint on the line emission monopole \citep{Mashian:2016bry, Serra:2016jzs, 2016ApJ...833...73C} 
and anisotropy \citep{Pullen:2017ogs, Switzer:2017kkz} on the largest scales on the sky. 
PIXIE's $15$\,GHz-wide, adjacent bands are well-matched to [CII] and higher-$J$ CO 
intensity mapping after reionization (PIXIE is not well-matched to CO $J=1-0$ observations 
because it would require twice the linear size, difficult to accommodate in the MIDEX cost cap). 
In addition, its narrow, adjacent bands with precise inter-band calibration aid in modelling and 
removing the galactic and extragalactic foreground emission \citep{Pullen:2017ogs, Switzer:2017kkz}. 
A spectrometer also contains more redshift information, which may be used to calibrate photometric 
redshifts in surveys such as LSST \citep{Alonso:2017dgh}.

\subsection*{LIME/CIBER}

The near-IR extragalactic background contains a host of lines which appear in the visible 
and near-IR originating from high-redshift structure responsible for reionizing the Universe 
\cite{Cooray2004, Kashlinsky2004}. LIME proposes to measure the power spectrum of these 
sources at various redshifts, focusing on the continuum emission between emission lines 
from 1--2\,$\mu$m. The proposal for LIME is based on successful measurements of extragalactic background light (EBL)
from the Cosmic Infrared Background Experiment (CIBER), designed to characterize the 
$1{-}2 \, \mu$m EBL \cite{Bock2006}, and over the course of four sounding rocket flights 
\cite{Zemcov2013} has successfully measured the amplitude of the near-IR background 
fluctuations on arcminute scales in two broad bands \cite{Zemcov2014}.  

CIBER detects 
an electromagnetic spectrum that is nearly Rayleigh-Jeans with an indication of a turn over 
at $\lambda=1.1 \, \mu$m, a spectrum that is significantly bluer than the integrated light from 
galaxies.  These fluctuations have been interpreted as arising from intra-halo light (IHL) from 
old, low mass stars residing in dwarf galaxies or dissociated from their parent galaxies during 
merging events over the history of the Universe.  This population has implications for large 
scale structure formation, implying the existence of a previously undetected population that 
may account for a non-negligible fraction of the missing baryons in the Universe. 

The need 
for further measurements of the fluctuating component of the near-IR EBL motivates CIBER-2, 
a second rocket-borne instrument designed to conduct comprehensive measurements of 
EBL anisotropy on arcsecond to degree angular scales in six broad bands covering 
$0.5 {-} 2.5 \, \mu$m \cite{Lanz2014}.  With an intensity mapping figure of merit an order of 
magnitude larger than its predecessor's, CIBER-2 is designed to measure the near-IR EBL 
anisotropy with the sensitivity, spectral range, and spectral resolution required to disentangle 
the contributions to the near-IR EBL from reionization, IHL, and local galaxies and foregrounds.  
The instrument is currently being fabricated and integrated, and first flight is expected in 2018.  

\begin{figure*}[t!]
\centering
\includegraphics*[width=6in]{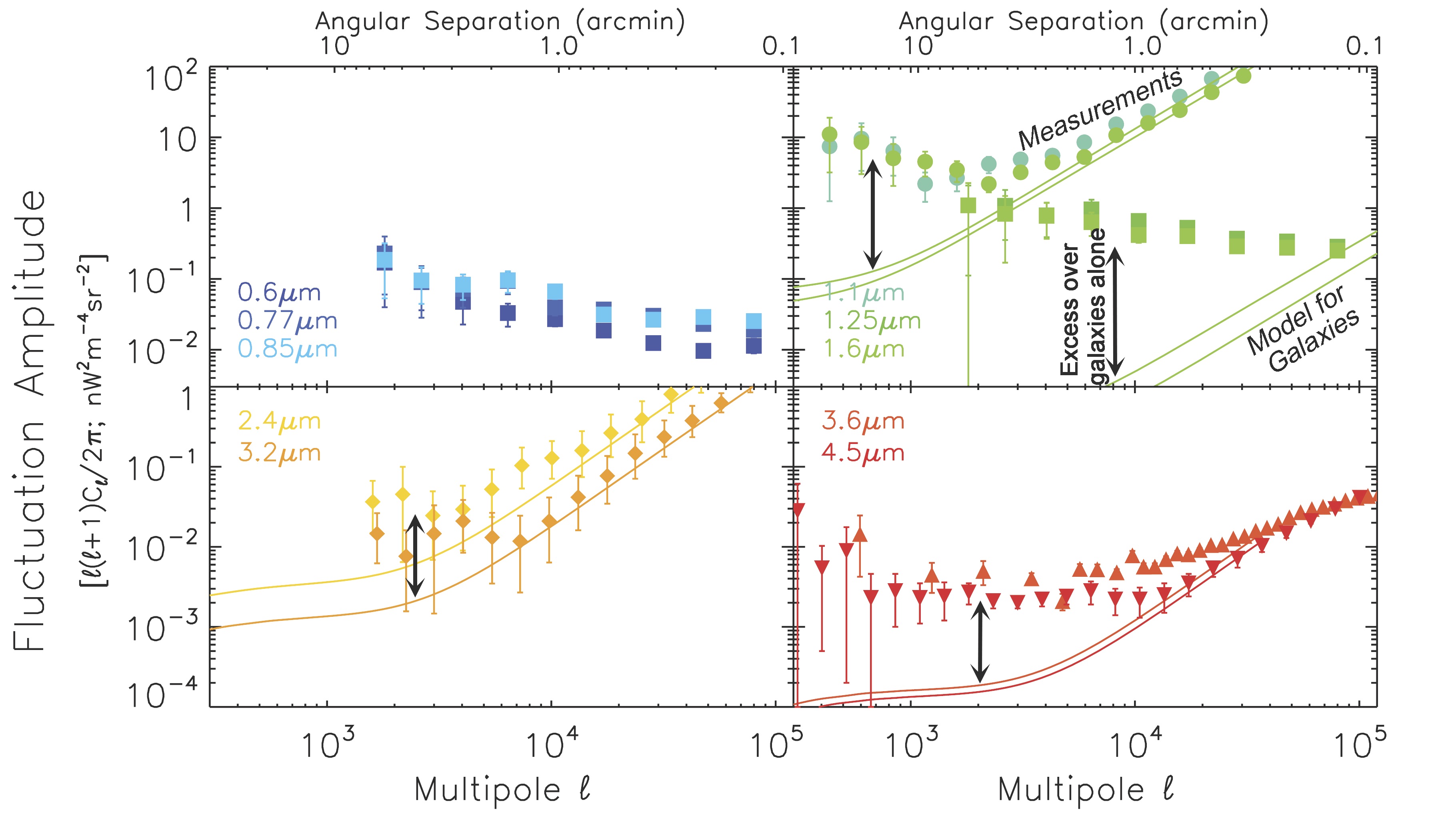}\\
\vspace{-0.1in}
\caption{Fluctuation power measurements
    compared to expectations from galaxies in the universe.  The
  current set of optical/IR measurements are shown by the data points,
  and corresponding models for the power spectrum of galaxies at the
  equivalent source masking depths \cite{Helgason2012} are shown as
  curves.  These data all show deviation from the galaxy-only model at
  large angles corresponding to scales of a few arcminutes and larger.
  \emph{The source of this excess emission is currently unknown.}  The
  electromagnetic spectrum of the excess fluctuation is Rayleigh-Jeans
  to about $1.2 \, \mu$m at which point it falls sharply into the
  optical \cite{Zemcov2014,Ketron2015}.  The measurements include: HST
  (squares - \cite{Ketron2015,Thompson2007}); \textit{Spitzer}
  (downward facing triangles - \cite{Kashlinsky2012}, upward facing
  triangles \cite{Cooray2012}); \textit{Matsumoto2011} (diamonds -
  \cite{Matsumoto2011}); and CIBER (circles - \cite{Zemcov2014}).
  Formal correlation between CIBER and \textit{Spitzer} data set is
  close to unity, which is strong evidence that the emission seen at
  all wavelengths is astrophysical in origin and arises from a common
  source. ({\it Courtesy of Michael Zemcov})}
  \label{fig:pssummary} 
\end{figure*}

\section{Large Scale Structure and Dark Energy at Redshifts z=0--2.5}

Observations of neutral hydrogen supported within galaxies promise to measure 
redshift dependence of Baryon Acoustic oscillations \cite{Bassett:2009mm}, resulting 
in a precise constraint of Dark Energy models at redshifts difficult to probe with optical 
galaxy surveys. Projections from 21\,cm measurements of BAO show promise for tight 
constraints on dynamical dark energy 
\cite{Peterson:2006bx,2010ApJ...721..164S,Shaw:2013wza,Shaw:2014khi}, and 21\,cm 
emission from galaxies at high redshift has been demonstrated through cross-correlation 
with deep optical galaxy surveys \cite{2010arXiv1007.3709C, 2013ApJ...763L..20M}, and 
a limit has been placed from the autocorrelation spectrum \cite{Switzer:2013ewa}. Upcoming 
instruments dedicated to measuring the matter power spectrum from 21-cm emission are 
described in this section, and focus on redshifts between 0.2--2.5 to target interesting 
regimes for Dark Energy probes. The SKA-MID will form a high-resolution galaxy survey 
in this redshift regime while other radio interferometers can be optimized to be sensitive to 
the scales of interest for BAO to form experiments dedicated to better understanding Dark Energy. 

\subsection*{CHIME} 
The Canadian Hydrogen Intensity Mapping Experiment (CHIME) is a new cylindrical 
transit radio interferometer located at the Dominion Radio Astrophysical Observatory 
(49$^{\circ}$19'15''N~119$^{\circ}$37'26''W). CHIME consists of 4 cylindrical reflectors 
with 1024 dual-polarization feeds operating between $400-800\,{\rm MHz}$ across four parabolic 
$f/0.25$ cylinders which are each 20\,m~wide~$\times$~100\,m long. It will have angular 
resolution 20' -- 40' and survey $f_{\mathrm{sky}}\sim3/4$ with 50\,K receiver noise across 
1024 frequency channels. The primary science goal of CHIME is to measure the expansion 
rate of the Universe and better understand the nature of Dark Energy with Baryon Acoustic 
Oscillations (BAO) via intensity maps of neutral hydrogen between a redshift range of 
$z\sim0.8-2.5$. 

As is true of all 21\,cm instruments, foreground emission from the Milky Way galaxy 
is far brighter than the cosmological signal of interest ($\sim$100\,$\mu$K signal compared 
to foreground emission as high as 700\,K). CHIME will use a Karhunen-Loeve transform 
method of foreground filtering, a technique that has been verified with simulations 
\cite{Shaw:2014khi} however it requires that the foregrounds appear as spectrally smooth, 
setting a stringent requirements for instrument calibration. The CHIME collaboration is 
exploring a broad array of calibration techniques for gain and beam calibration 
\cite{Newburgh:2014toa,Berger:2016ejd}. The wide sky coverage and high frequency 
resolution will provide maps of Large Scale Structure which could be cross-correlated 
with a variety of current and upcoming galaxy/quasar surveys (SDSS, DES, LSST, DESI, HSC, Hetdex). 

\begin{figure}[t!]
\begin{center}
\includegraphics[width=0.75\linewidth]{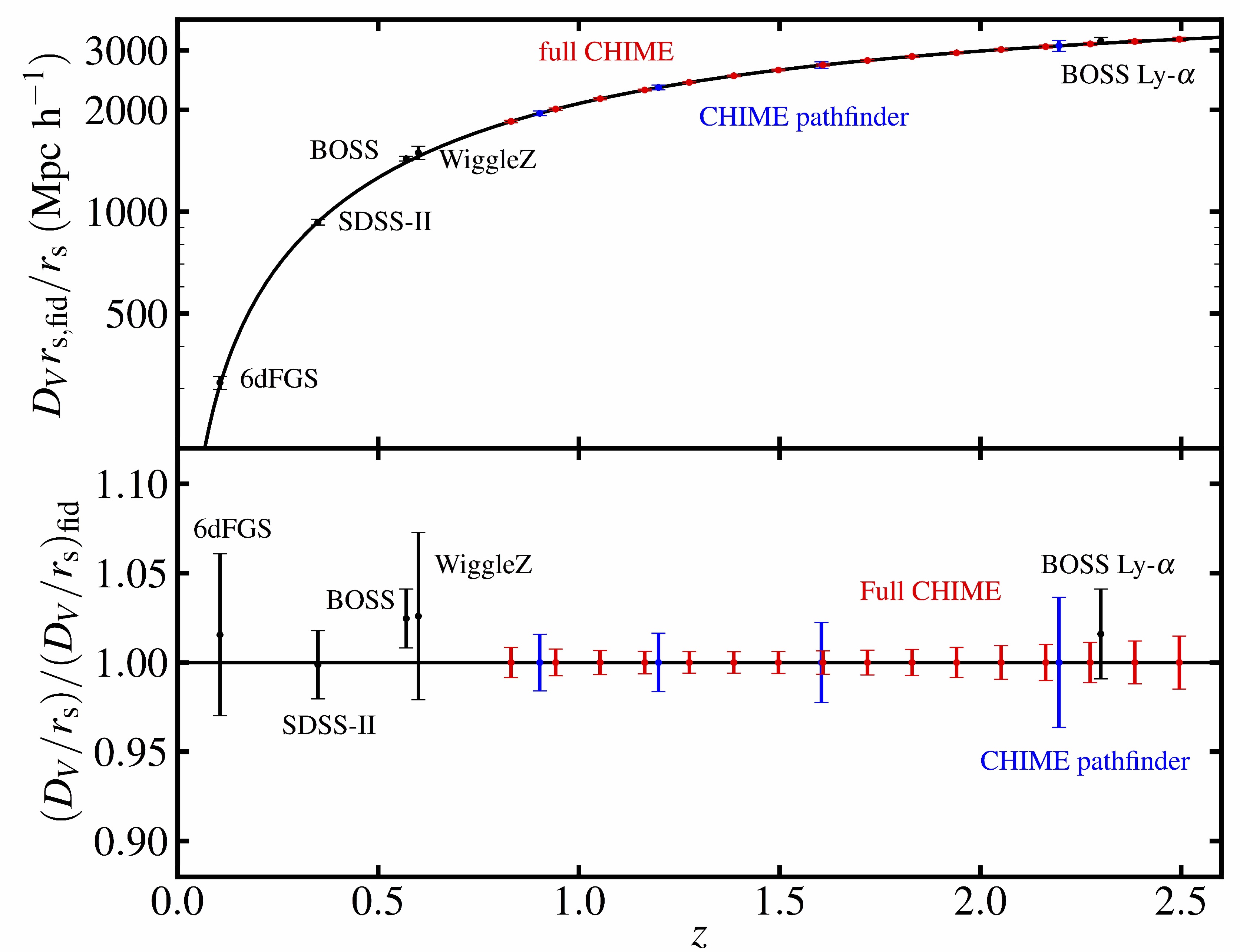}
\end{center}
\caption{Projections for CHIME and CHIME pathfinder error bars on the cosmological 
distance volume. CHIME and the CHIME pathfinder are expected to make sample variance 
limited measurements of BAO, improving constraints on Dark Energy. CHIME, as a 
physically larger interferometer, has improved error bars at higher redshift due to its higher 
resolution. ({\it Courtesy of J.R. Shaw})}
\label{fig:CHIME}
\end{figure}

CHIME has been built and will begin observations by the end of 2017, and will operate for 
five years to make sample variance limited measurements of BAO (the error bar projections 
are shown in Figure~\ref{fig:CHIME}). Prior to building CHIME, the collaboration built the 
CHIME pathfinder \cite{2014SPIE.9145E..22B}, a test-bed instrument with 128 dual-polarization 
feeds deployed on two 20\,m~wide~$\times$~36\,m long cylinders. The CHIME pathfinder 
started its two year survey in December 2015, and has published one constraint on the 
brightness of fast radio bursts (FRBs) \cite{Amiri:2017qtx}, and should be capable of interesting 
constraints on Dark Energy (see pathfinder projections in Fig.~\ref{fig:CHIME}).

\subsection*{HIRAX} 

The Hydrogen Intensity and Real-time Analysis eXperiment (HIRAX) \cite{Newburgh:2016mwi} 
is a planned radio telescope array that will consist of $\approx 1000$ close packed 6\,m dishes 
that will be deployed in South Africa. HIRAX will operate between 400--800\,MHz in 1024 linearly 
spaced frequency bins, corresponding to a redshift range of $0.8 < z < 2.5$ and a minimum 
$\delta z/z$ of $\approx 0.003$. It will employ first stage amplifiers directly embedded in the 
wide-band feed, and use radio-frequency-over-fiber to transmit the radio signals from the 
dishes to the correlator to reduce the signal loss and temperature variations. 

HIRAX will survey the majority of the southern sky to chart the expansion history 
of the universe and place competitive constraints on the dark energy equation of state and 
its time evolution. In addition to BAO cosmology, the large survey area and real-time analysis 
capabilities of the HIRAX array will make it a powerful tool for identifying pulsars and 
astrophysical transients such as fast radio bursts (FRBs) as well as providing an excellent 
platform for studying neutral hydrogen absorbers. The extensive overlap with other cosmological 
surveys in the Southern hemisphere should provide many opportunities for a variety of 
cross-correlation studies, including improved photometric redshift errors for LSST \cite{Alonso:2017dgh} 
and possibilities for understanding the \hi bias from cross-correlations with CMB surveys 
\cite{Moodley:2017}, and for joint constraints on cosmological parameters.
 
A drone based calibration system is being developed for HIRAX, which will allow the 
array to be calibrated in situ, at zenith pointing, or over the range of operational dish elevations.
The drone can easily be programmed to execute arbitrary beam mapping flight paths in 
the array far field. For a 6 m dish, the far field is at most $2 D^2/\lambda = 200$ m at 800 MHz.
Data taking and beam mapping are performed with a modified version of the ECHO 
software developed by Danny Jacobs \cite{jacobs16}.

As of the time of writing, an initial eight-element prototype array has been deployed at the 
Hartebeesthoek Radio Astronomy Observatory (HartRAO), providing the first end-to-end test
 of the HIRAX hardware. Build up to the full array will take place over an estimate period of 
 three years, beginning in 2019. Because of the inherently modular nature of the array, the 
 completed portions of the array will be able to operate throughout the array deployment.

\subsection*{Meerkat and SKA-MID}

The SKA will have a second sub-array, SKA1-MID, which is a mid-frequency dish array to be constructed 
on the South African site. In addition to $\sim 130$ new (15m diameter) SKA1 dishes, it will 
also incorporate the 64 (13.5m) dishes of the co-sited MeerKAT array. SKA1-MID dishes 
will eventually be fitted with 5-band receivers, covering bands from 350 MHz up to 14 GHz. 
Initial construction will likely prioritize Bands 2, 5, and 1, the most relevant for 21cm intensity 
mapping being Band 2 (covering $950-1760$ MHz), and Band 1 ($350-1050$ MHz). The 
MeerKAT receivers support 2 bands with different frequency ranges to the SKA1 receivers. 

\begin{figure}
\centering
\includegraphics[scale=0.33]{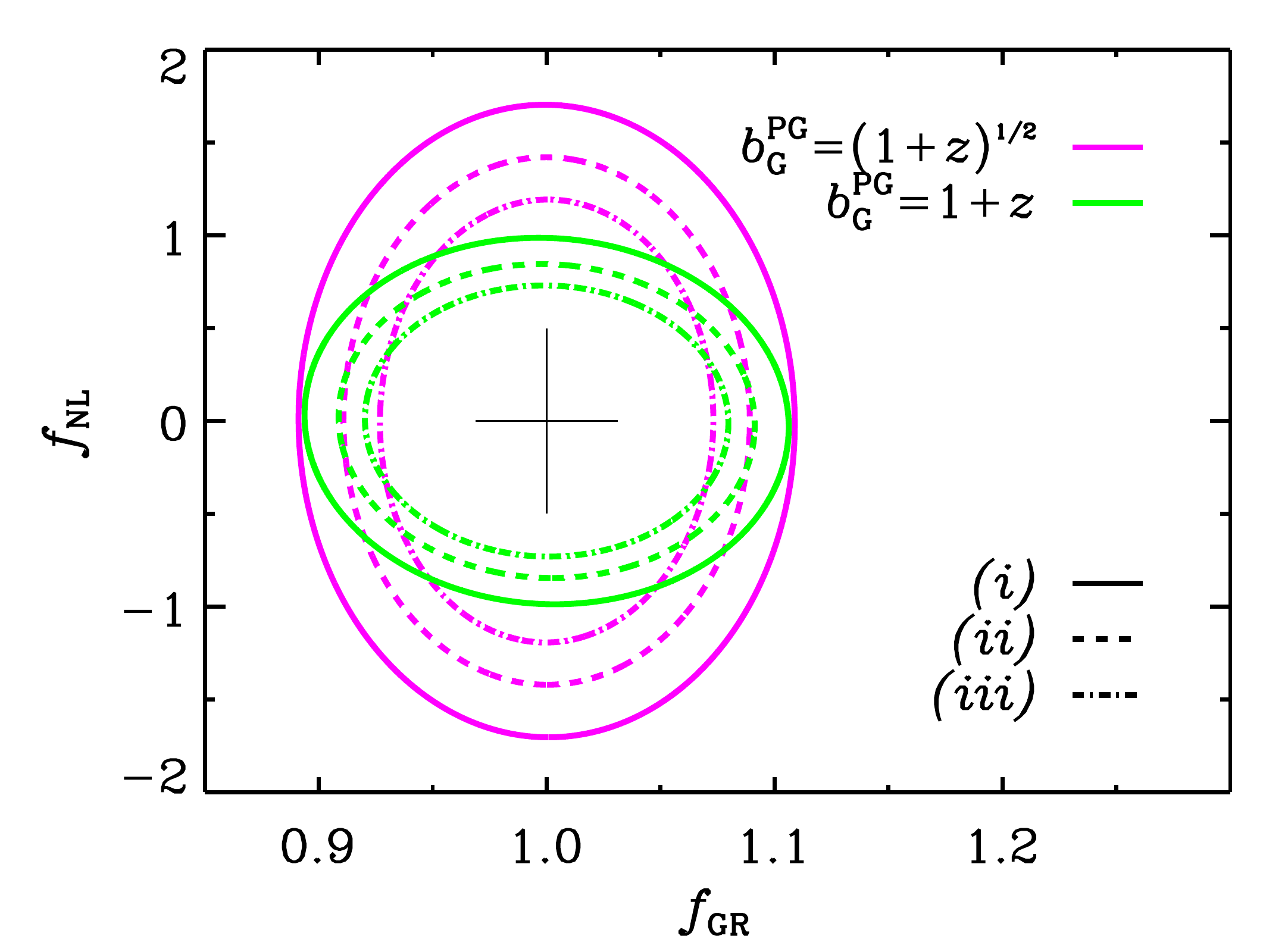}
\includegraphics[scale=0.5]{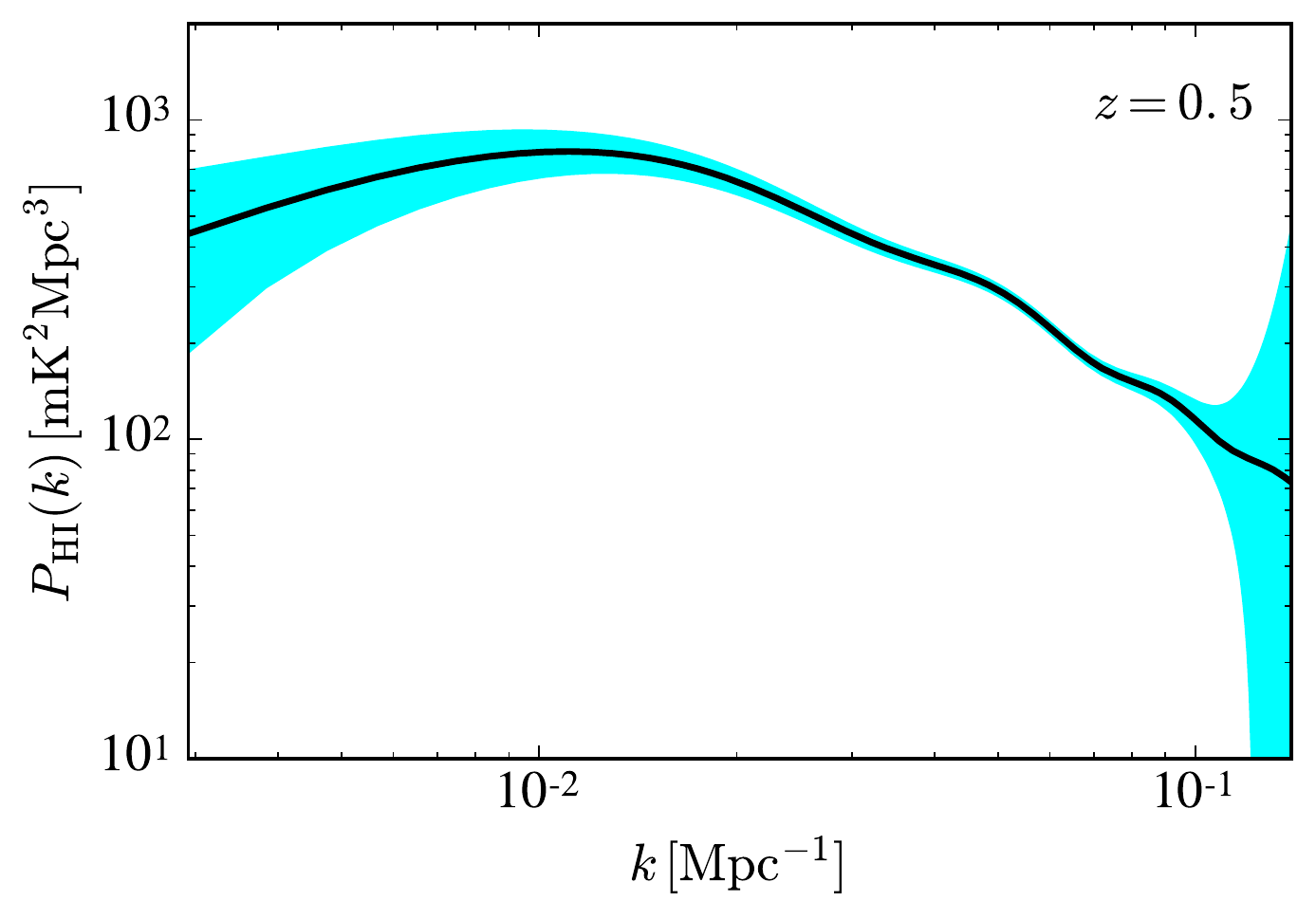}
\caption{Left: joint marginal error contours using the multi-tracer technique for primordial non-Gaussianity ($f_{NL}$) and GR corrections ($f_{GR}$) considering different scenarios for the bias and Euclid/LSST surveys \protect\citep{2015ApJ...812L..22F}. Right: HI power spectrum detection with MeerKLASS, showing the expected signal (black solid) and measurement errors (cyan) \protect\citep{MeerKLASS}. ({\it Courtesy of Mario Santos})}
\label{fig:SKA_Meerkat}
\end{figure}

Since its baselines are not small enough to probe the most relevant cosmological scales, the accepted plan is to use the auto-correlations from each dish which will be supported by calibration against noise diodes and an appropriate scan strategy to reduce $1/f$ noise. Assuming that residual foregrounds and calibration uncertainties can 
be controlled well enough, a 10,000 hour SKA1-MID IM survey over 30,000 deg$^2$ would 
rival large galaxy redshift surveys such as Euclid in terms of constraints on the dark energy 
equation of state \citep{Bull:2015lja} and provide powerful tests of dark energy models and modifications to General Relativity \citep{2014arXiv1405.1452B, 2016ApJ...817...26B}. Due to its low resolution in single dish mode, SKA1-MID will be particularly transformational on very large scales, where it can provide unique constraints on primordial non-Gaussianity and make the first detections of general relativistic corrections. This will be especially powerful when in combination with future optical/infrared surveys such as Euclid and LSST by using the multi-tracer technique \citep{2015ApJ...812L..22F,Alonso:2015sfa} (see Fig.~\ref{fig:SKA_Meerkat}, left).

In the immediate future, with 64, 13.5m dishes and two frequency bands covering the z=0 to 1.4 redshift range, MeerKAT, the SKA precursor in South Africa, will have the capability to produce high impact cosmological constraints using the same approach for HI intensity mapping (e.g. using the auto-correlations from each of its dishes). A wide area survey has therefore been proposed, known as MeerKLASS (MeerKAT Large Area Synoptic Survey \cite{MeerKLASS, Santos:2017qgq}) which aims to start observations in 2018. Covering an area of $\sim 4000 \, {\rm deg}^2$ for $\sim 4000$ hours, it will potentially provide the first ever measurements of the baryon acoustic oscillations using the 21cm intensity mapping technique \citep{2017MNRAS.470.4251P}, with enough accuracy to constrain the nature of dark energy (see Fig.~\ref{fig:SKA_Meerkat}, right). The combination with multi-wavelength data over the same area, will give unique additional information, such as stringent constraints on primordial non-Gaussianity using the multi-tracer technique \citep{2017MNRAS.466.2780F}. It will also be a crucial step on the road to using SKA1-MID for cosmological applications, as described in the top priority SKA key science projects.

\subsection*{BINGO}

BINGO \citep[Baryon acoustic oscillations In Neutral Gas Observations][]{2013MNRAS.434.1239B, 2016ASPC..502...41B} 
is a large (40m) transit telescope that is planned for construction in an open-cast mine in Minas Corrales, 
near the Uruguay-Brazil border. It will target baryon acoustic oscillations at $0.13 < z < 0.48$, with a 
drift-scan survey over a $15 \times 200$ degree patch of the sky. The telescope will have a large focal 
plane with at least 50 feeds operating in autocorrelation mode, with pseudo-correlation receivers to 
promote gain stability and thus reduce the impact of $1/f$ noise. The feed horns are large -- 4.5m 
long with 1.7m apertures -- and will be constructed from a lightweight foam wrapped in a conductor, 
to reduce mass. The instantaneous field-of-view is $15 \times 15$ degrees, with a $\sim 40$ arcmin 
resolution at 1 GHz. Being one of the few planned non-interferometric 21cm IM surveys, BINGO will 
be capable of recovering large scales in the HI distribution in addition to the BAOs. While foregrounds 
and instrumental effects remain important, they are expected to be less drastic than for lower-frequency 
experiments targeting the EoR \citep{2015MNRAS.454.3240B}. BINGO will be largely funded by a grant 
from FAPESP (Brazil), with contributions from the other partners in the UK, Uruguay, Switzerland, 
South Africa and China. First results are expected by 2020.


\chapter{Theoretical Backbone}
\label{chap:theory}





\bigskip

\section{Modeling}
\label{sec:model}

Line-intensity mapping has the potential to provide us with valuable insights 
into many physical quantities which play a role in the physics of the ISM, such as 
the strength of the radiation field, the number density of hydrogen and electrons in 
the neutral and ionized medium respectively, the gas metallicity and the mean star-formation rate. 
The observed signal in an intensity mapping survey depends on many physical processes 
governing the interplay between energetic sources, and gas and dust in galaxies. It is 
therefore very important to carefully consider how to model this signal, both when forecasting 
the signal-to-noise ratio of upcoming and future surveys, and when interpreting the results of these surveys.

Two main approaches are usually employed in modeling the strength of a given emission line: 
through interpretation of the clustering signal; or through its dependence on the local environment. 
The first method is phenomenological, and is usually built upon empirical relations between 
the intensity of a line and the measured far-infrared luminosity of a galaxy, which quantifies 
its total star-formation rate.  Using scaling relations between star-formation rate 
(or infrared luminosity) and halo mass (derived from abundance matching), it is then possible to 
express the line luminosity of a galaxy as a function of its host halo mass, and use a halo model approach to compute 
statistical quantities of interest such as the three-dimensional power spectrum 
\citep{Visbal:2010rz, Pullen2013, Breysse2014,Serra:2016jzs}.  This method, while providing rough 
predictions for the expected amplitude and shape of power spectra, relies on empirical relations 
based on a very limited set of observations of galaxies at some particular redshifts. Scaling 
relations based on different recipes result in quite different amplitudes of the power spectra 
\citep[often of an order of magnitude or more, see, e.g., discussion in][]{Lidz:2016lub,Cheng2016}. 
Moreover, the redshift evolution of these relations is often poorly known, which can be 
problematic when targeting, e.g., the epoch of reionization. Furthermore, the power of intensity 
mapping is to probe objects too faint for individual detections---and therefore by definition outside 
of the range of the measured scaling relations.

The second approach is based on numerical simulations and semi-analytic models 
of galaxy formation and evolution. These models capture in detail many processes 
governing the physics of the Interstellar Medium (ISM), and are able to predict the 
emission from multiple lines up to very high redshifts 
\citep[see, e.g.][]{Popping2016, Hopkins2014}. However, they also require numerous 
assumptions, which often makes them computationally demanding, inflexible, and hard 
to implement for quick computations.

Most of the modeling work up to now has been focused on forecasting results for 
future surveys, but the real challenge of modeling will be to translate intensity 
mapping measurements into useful astrophysical information. Scaling relation 
models are straightforward to work with, but they provide relatively little insight into 
the physical processes which govern line emission. Semi-analytic models, on the 
other hand, are tied intimately into detailed physics of the galaxy population, but this 
complexity makes them difficult to use and may obscure useful information amid 
a large number of assumptions and free parameters. 
Going forward, it will be important to consider intermediary approaches to model 
the line emission of atoms and molecules in galaxies combining phenomenology, 
numerical analysis, and when available, direct observations. 

\subsection{[CII] line-intensity mapping}\label{CII}

The [CII] 157.7$\, \mu$m fine-structure line arising from the 
$^2\rm P_{3/2} \rightarrow ^2 P_{1/2}$ transition is the brightest among all metal 
lines emitted by the interstellar medium (ISM) of  star-forming galaxies. It is associated 
with the star formation in galaxies \citep{Boselli2002,DeLooze2011,Wisnioski2015,Herrera2015} 
and plays a key role in the energy balance of galaxies, as it provides one of the most 
efficient cooling processes for the neutral ISM. A particularly attractive application of 
intensity mapping is the study of faint galaxies in the Epoch of Reionization. 
In fact, if the galaxy luminosity function has a sufficiently steep faint end (as deduced 
from the study of UV luminosity functions), the observed radiation is actually dominated 
by unresolved sources \citep{Uzgil2014}.  

The signal is however swamped by a combination of different foregrounds and 
contaminating lines that must be removed. In addition to the far-infrared (FIR) continuum 
foreground, other emission lines emitted from a range of redshifts fall at the same 
frequency of the [CII] signal; they act as \textit{contaminants}.  Among these are the OI 
line ($\lambda 145\, \mu$m), the two NII lines ($\lambda = 122, 205\, \mu$m), and a 
handful of CO rotational transition lines in the range 200-$2610\, \mu$m. Among these, 
the CO rotational transition lines are the most relevant here. For example, since the 
CO(4-3) line has a wavelength $651~\rm \mu m$, if emitted from $z = 0.45$ galaxies, 
it contaminates the [CII] emission from $z = 5$ galaxies. However, the luminosity distance 
from $z=0$ to 0.45 is only $\sim 5\%$ of that to $z=5$. As the flux is inversely proportional 
to the square of the luminosity distance, whereas the proper distance interval that 
corresponds to the same bandwidth is $\propto (1+z)^{-3/2}$, the CO flux can be higher 
than the [CII] one, even ignoring the cosmological evolution of the star-formation rate density.  

The relevant quantity for experiments is the power spectrum of the signal. 
This can be obtained analytically \citep{Gong:2011mf,Uzgil2014}, or numerically 
\citep{Yue:2015sua}. Although deducing the [CII] signal from simulation should be more 
robust, it implies a number of difficulties related to the large volumes required at the same 
time resolving galaxies contributing to the faint end of the luminosity function. For this 
reason, very often hybrid methods are used in which N-body simulations are used to provide 
the correct large scale structure correlation properties, and empirical relations between 
star formation, metallicity and [CII] emission are used. 

\citet{Yue:2015sua} used star-formation rates and metallicity derived small scale 
(box size 10 $h^{-1}\,\rm Mpc$) galaxy simulations including a sub-grid treatment of the 
interstellar medium \citep{Pallottini2014} with an empirical approach to compute the 
$L_{\rm CII}-M_{\rm h}$ relation taken from \citet{Vallini2015}. This relation is subsequently 
applied to halo catalogs built from the large-scale N-body simulation {\tt BolshoiP}, to 
generate mock maps of [CII] signal. The FIR continuum foreground, derived from 
abundance-matching techniques was added to the generated mock maps, along with 
contamination from CO lines. The latter was computed from intermediate 
$L^{\rm J}_{\rm CO} - L_{\rm IR}$ relations that were better fitted from measurements 
of both local and high-redshift samples. Mock maps were generated for FIR continuum 
and CO emission, in close analogy with the [CII] mock maps.

FIR foreground removal and CO masking experiments on the total mock maps was 
performed to recover the [CII] angular power spectrum. Efficient subtraction of the CO 
contamination can be achieved if the map has a sufficiently high spatial resolution 
(necessary to preserve a statistically meaningful number of pixels). In practice, a 
minimum resolution of $\sim40''$ is required to remove CO contamination by dropping 
all pixels containing galaxies brighter than $m_K =22$ in the relevant redshift range.

\begin{figure}
\begin{center}
\includegraphics[scale=0.46]{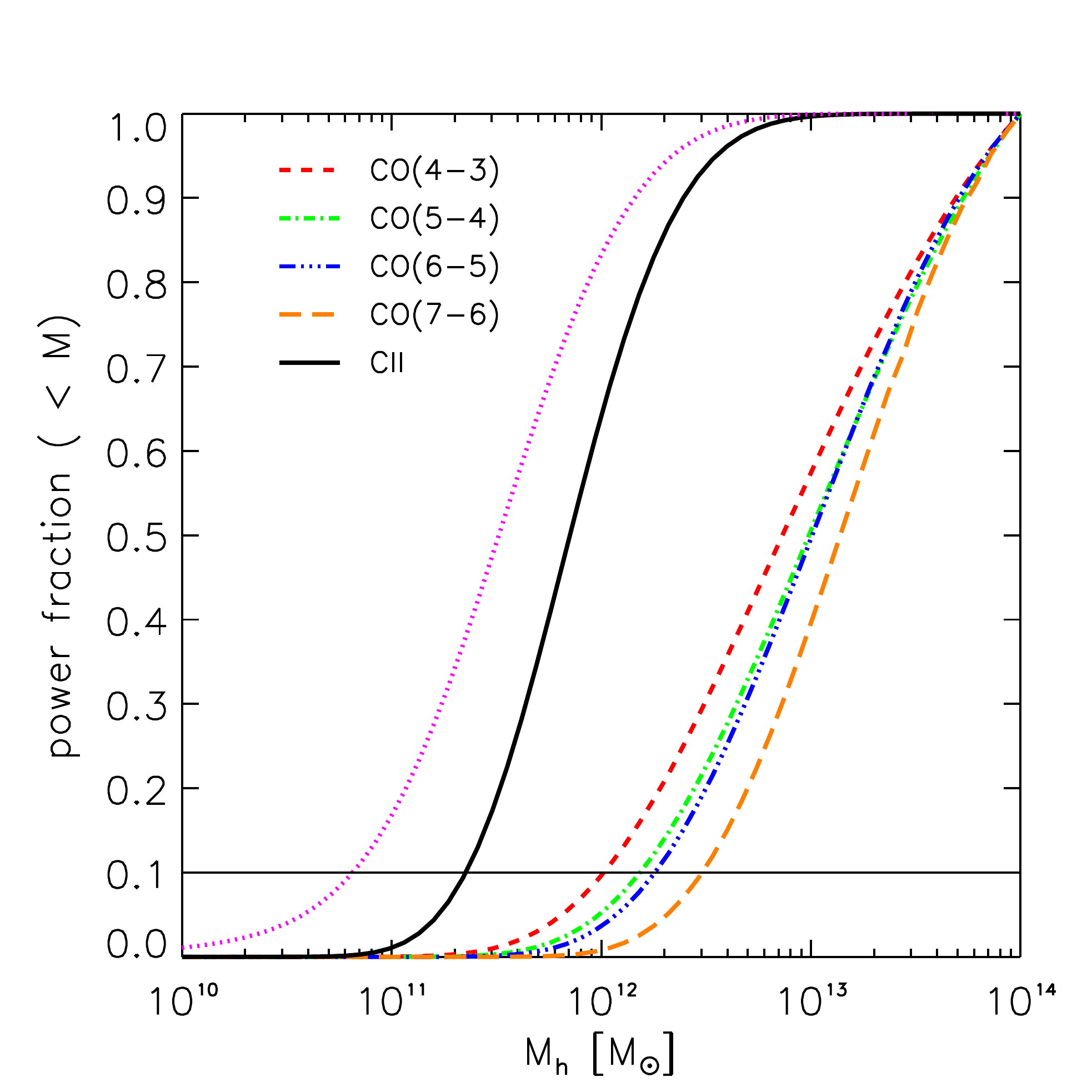}
\caption{Fractional contribution from halos below $M_{\rm h}$ to the clustering 
term of the angular power spectrum for the $316\pm0.65$\,GHz frequency bin for 
various emission lines. Note the contribution from several contaminating CO lines. 
The purple dotted line is the FIR continuum. The horizontal line guides the eye on 
the reference value 0.1. Adapted from \cite{Yue:2015sua}. 
({\it Courtesy of Andrea Ferrara})} 
\label{Ferrara1}
\end{center}
\end{figure}

The [CII] signal from $z_{\rm CII} \sim 5 - 6$ is detectable, for example, by a ground-based, 
noise-limited telescope with a 6~m aperture, $T_{\rm sys} = 150\, \rm K$, a FIR 
camera with $128\times128$ pixels in about $5000\, \rm hr$ total integration time.
Fig. \ref{Ferrara1} shows that at redshift 5, most of the [CII] fluctuation signal is from 
halos in the mass range $M_h =10^{11-12}\, M_{\odot}$ (H-band apparent magnitude 
$\sim 26.8 - 23.8$). According to this model, halos below $10^{11}\, M_{\odot}$ produce less than 1\% of the 
total [CII] fluctuations. The measured high-$z$ [CII] signal is particularly useful for studies 
of halos in the above narrow mass range. To access the fainter galaxies responsible for 
reionization, the [CII] flux from bright galaxies needs to be measured by a high-resolution 
interferometer array, and then subtracted in the relevant pixels. In practice, this operation 
would be challenging. For example, there are $\sim 7000$ halos in the light-cone used here whose 
[CII] emission appears in the $(316\pm0.65)$~GHz frequency bin and 
$\gtrsim 10^{-22}$~Wm$^{-2}$, corresponding a mass $M_{\rm h}\gtrsim 6\times10^{10}\, M_{\odot}$.
In order to resolve them with a signal-to-noise ratio of $> 5$, assuming a line width 
50~km s$^{-1}$, the required sensitivity is $4\times10^{-5}\, \rm Jy$. For comparison, 
the 10 hr ALMA r.m.s. noise  at $316\, \rm GHz \pm 50\, \rm km\, s^{-1}$ is $\sim 7\times10^{-5}\, \rm Jy$.  
To detect the [CII] signal from these sources, a space telescope with background-limited 
sensitivity, and an interferometer array that could measure radiation flux of all bright 
sources at relevant redshifts, are necessary.

\subsection{CO Line-Intensity Mapping, Scaling Relations, and Direct Measurements}\label{CO}

\begin{figure}[th!]
\begin{center}
\includegraphics[scale=0.850]{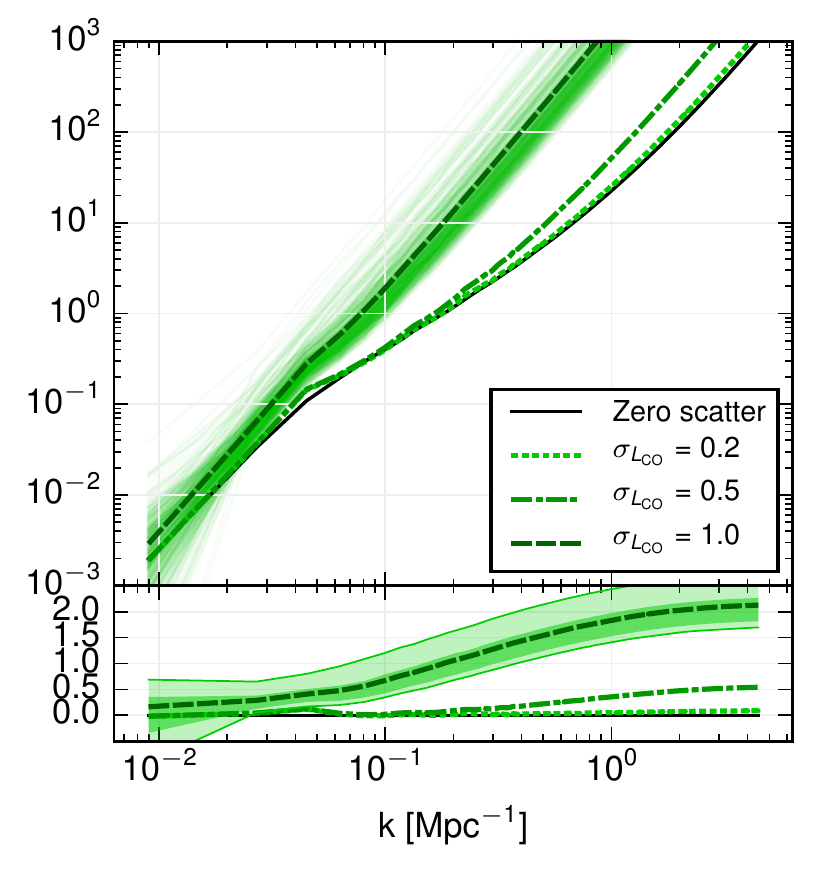}
\includegraphics[scale=0.850]{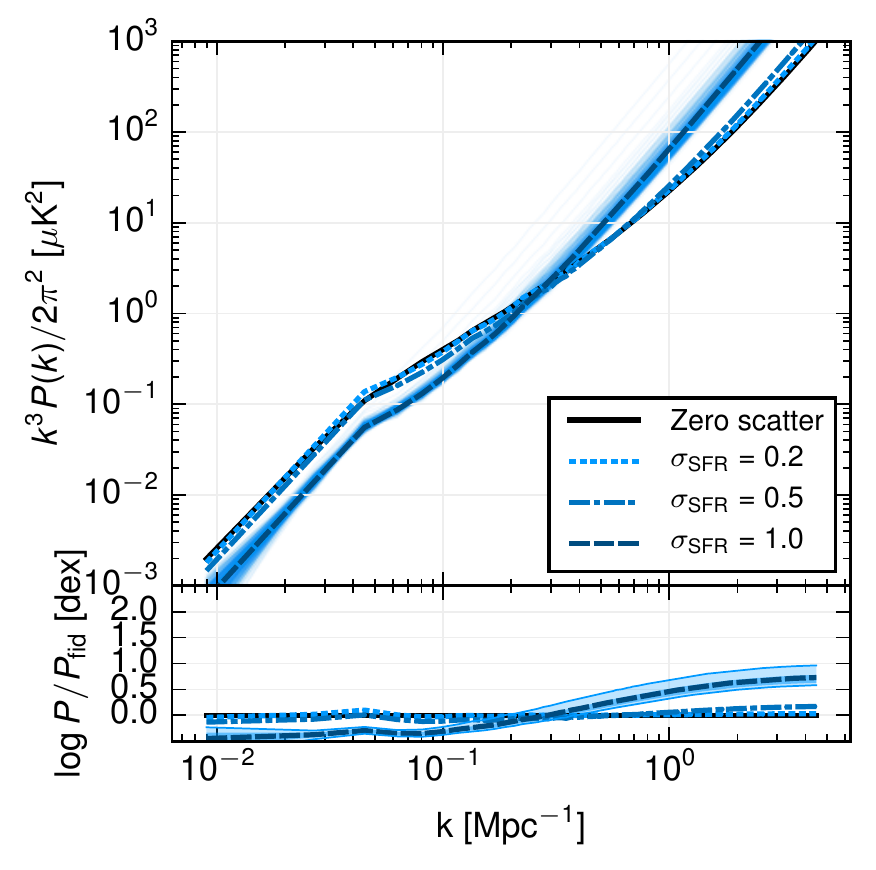}
\caption{Effect of scatter on the CO power spectrum, as parameterized by: SFR given 
$M_h (\sigma_{\rm SFR}, left)$ or $L_{\rm CO}$ given SFR $(\sigma_{L_{\rm CO}} , right)$. 
As scatter approaches 1 two things are evident: (1) the power spectrum begins to look 
like a pure shot noise spectrum, since any clustering signature is increasingly buried by 
the large halo-to-halo scatter, and (2) the scatter introduces significant variance into the 
power spectrum, shown as thin lines (shaded region in the lower panels) for 100 realizations. 
{\it (Courtesy of Tony Li)} }
\label{fig:Pk_li}
\end{center}
\end{figure}

Carbon monoxide (CO) is the most abundant molecular species after $\rm H_2$, 
whose emission lines arise from a \lq\lq ladder\rq\rq\ of rotational transitions ($J\to J-1$), 
with the ground-state CO(1–0) transition at $\nu_{1\to0}=115.27$ GHz ($\lambda=2.6\, \rm mm$) 
and higher transitions at $\nu_{J\to J-1}=J\nu_{1\to0}$. 
CO traces the metal-enriched cool to warm molecular ISM, where stars form efficiently, 
which motivates the empirical conversion between CO and 
$\rm H_2$ \citep[for a review, see][]{Bolatto2013}, and by extension star-formation. 

Several phenomenological models of the expected CO line-intensity mapping signal exist 
\citep[e.g.,][]{Righi2008,Visbal:2010rz,Carilli2011,Lidz:2011dx,Pullen2013,Silva:2014ira}.  Some are analytic 
halo models based on halo occupation distributions, some rely on semi-analytic models, 
while others directly populate light-cone catalogs generated from numerical dark matter halo simulations 
(\citealp[see Table 3 of][for a summary of the model differences]{Li:2015gqa}).  
A common thread among many of them is the use of star-formation rate as a proxy for CO luminosity, 
through the use of empirical scaling relations based on observables, for which bolometric infrared 
luminosity ($L_{IR}$) --- which traces the light from young stars as absorbed and re-radiated by dust 
grains, which are abundant in star-forming regions --- is the most direct.  $L_{IR}$, whose units are 
$\rm L_{\odot}$, refers commonly to the integral of the thermal spectral energy distribution 
\citep[effectively a modified blackbody, $\nu B(\nu)$; e.g.,][]{Blain2003} over the rest-frame 
8--1000$\, \micro$m region.

The SFR--$L_{IR}$ relation is typically taken from \citet{Kennicutt1998}: 
\begin{equation}
\frac{{\rm SFR}}{M_\odot\,\text{yr}^{-1}} = \delta_\text{MF}\times10^{-10}\left(\frac{L_\text{IR}}{L_\odot}\right),
\end{equation} 
where $\delta_{\rm MF} \approx 1$ when assuming an initial mass function (IMF) from \cite{Chabrier2003}. 

A power-law relation then connects the IR and CO luminosities:
\begin{equation}
\log{\left(\frac{L_\text{IR}}{L_\odot}\right)}=\alpha \log{\left(\frac{L'_\text{CO}}{\text{K km s}^{-1}\text{ pc}^2}\right)}+\beta.
\end{equation}
Values for galaxies in the redshift range $z \approx 1-3$ for $(\alpha,\beta)$ span 
(1.13, 0.53), (1.37, -1.74), (1.0, 2.0), (1.17, 0.28) taken from \citep{Daddi2010,Carilli2013,Dessauges2015,Greve2014}, 
highlighting the uncertainties that are still present in the scaling relations. 

\begin{figure}[h!]
  \begin{center}
    \includegraphics[width=0.4\textwidth]{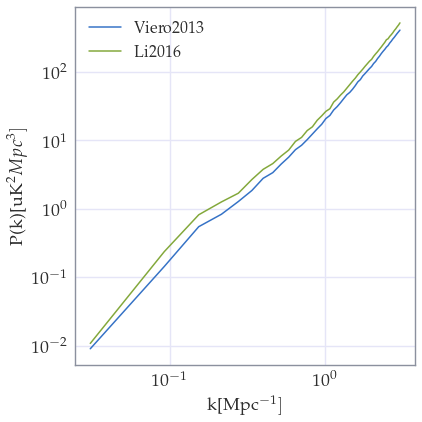}
    \caption{Comparing the \citet{Li:2015gqa} and \citet{Viero2013b} models applied to {\sc imapper2}$^1$.  
    \citet{Li:2015gqa} converts SFR to $L_{\rm IR}$ to $L_{\rm CO}$, while \citet{Viero2013b} uses fits 
    to direct measurements of the $L_{\rm IR}$ via {\sc simstack}.  Both use a minimum halo mass of 
    $10^{9}\, \rm M_{\odot}$ and $\sigma_{L_{\rm CO}} = 0.3$, and a scaling relation 
    $\alpha_{\rm CO} = 1.37$ \citep{Carilli2013}. {\it (Courtesy of Marco Viero)}}
    \label{fig:Pk}
  \end{center}
 \end{figure}

In practice the mapping from halo mass to stellar mass and/or star-formation rate, and 
then to $L_{\rm IR}$ and finally CO luminosity are stochastic, and the level of scatter 
introduced at each step to account for that has consequences on the predicted power 
spectrum (Fig.~\ref{fig:Pk}). \citet{Li:2015gqa} added log-scatter $\sigma_{\rm SFR}$ 
and $\sigma_{\rm L_{CO}}$ of both 0.3, motivated by the scatter in models of SFR at a 
given halo mass \citep{Behroozi2013a}. As shown in the following \lq\lq FIRE 
Simulations\rq\rq\ section, that may be an underestimate\footnote{{\sc imapper2} is a 
modeling code to populate dark matter halo catalogs 
with CO line luminosities, available on github at {https://github.com/tonyyli/imapper2/}}. Also 
notable is a recent model from \citet{Wu2017} which predicts a deficit from \citet{Kennicutt1998} 
of $L_{\rm IR}$ in low-mass ($\log{M/M_{\odot}} < 10 $) galaxies, which could impact the 
strength of the power spectrum signal.

Alternatively, models are under development that explore the dependence of the IR 
properties of galaxies on their physical properties --- including stellar mass, age, and 
extinction --- by fitting directly to measurements (Fig.~\ref{fig:Pk}).  This in turn would 
bypass the SFR to $L_{\rm IR}$ step, and more naturally account for stochasticity in the 
global sample.  Because measuring $L_{\rm IR}$ for large samples \citep[e.g., 1 million 
objects in the $\rm 2\,deg^2$ COSMOS field;][]{Laigle2016} with ALMA is prohibitively 
expensive, and submillimeter images (SPIRE, SCUBA2, etc.) are so source-confusion 
dominated as to be effectively continuum intensity maps, statistical methods like stacking 
\citep[e.g., \sc{simstack}][]{Viero2013b} are necessary.  The first published model to use 
this approach was \citet{Sun2016}, which found a scatter of $L_{\rm IR}$ around the main 
sequence of star formation of $\sigma_{\rm LIR} = 0.3$.

\subsection{Ly$\alpha$ Line-Intensity Mapping}\label{Lya}

The Ly$\alpha$~ line most likely represents the optimal feature for a line-intensity mapping 
experiment. Historically, it has been used for high-$z$ Ly$\alpha$-emitting (LAE) galaxy 
searches as it is the most luminous UV line \citep{Ouchi2008,Matthee2015}. 
These searches are often hampered by the fact that intergalactic \hbox{H~$\scriptstyle\rm I\ $} 
can scatter the bulk of Ly$\alpha$~ photons out of the line of sight, making systematic 
detections of LAE during the EoR very challenging \citep{Dayal2011}. Line-intensity mapping 
can overcome these problems, thanks to its sensitivity to even diffuse emission from the IGM. 
Therefore Ly$\alpha$~ intensity mapping seems a very promising tool to study the properties 
of early, faint and distributed EoR sources.

\citet{Silva2013} studied this problem with semi-numerical tools, focusing on the EoR emission 
at $z=7$. In a following work, \citet{Gong:2013xda} tackled the problem of line confusion, 
proposing the masking of contaminated pixels as a cleaning technique. These authors find 
that recombinations in the ISM of galaxies largely dominate the Ly$\alpha$~ intensity and 
power spectrum. \citet{Pullen2014} developed a simple analytical model to study the evolution 
of the Ly$\alpha$~ power spectrum at $z > 2$: their results are qualitatively different from 
\citet{Silva2013}, in that they conclude that diffuse IGM emission is the dominant source. 
Recently, \citet{Croft2016} attempted to observe the large scale clustering of Ly$\alpha$ 
emission at $z=2-3$ by cross-correlating the residuals in the SDSS spectra with QSOs. 
They claim a detection of a mean Ly$\alpha$~ surface brightness $\simgt 10$ times more 
intense than the one inferred from LAE surveys, but still compatible with the unobscured 
Ly$\alpha$ emission expected from LBGs. 

Finally, in a series of papers, \citet{Comaschi2016a} (see also \cite{Comaschi2016b, Comaschi:2015waa}) 
presented a complete study of Ly$\alpha$ intensity mapping including theoretical expectations 
and strategies to identify the optimal IM experiment. These authors find that Ly$\alpha$ 
intensity from the diffuse IGM emission is 1.3 (2.0) times more intense than the ISM emission 
at $z = 4(7)$; both components are fair tracers of the star-forming galaxy distribution. However, 
the power spectrum is dominated by ISM emission on small scales  ($k > 0.01\, h\,{\rm Mpc}^{-1}$) 
with shot noise being significant only above $k = 1 \,h\,{\rm Mpc}^{-1}$. At very large scales 
($k < 0.01\,h\,{\rm Mpc}^{-1}$) diffuse IGM emission becomes important. The comoving Ly$\alpha$ 
luminosity density from IGM and galaxies, 
$\dot \rho_\alpha^{\rm IGM}  = 8.73(6.51) \times 10^{40} {\rm erg~}{\rm s}^{-1}{\rm Mpc}^{-3}$ 
and $\dot \rho_\alpha^{\rm ISM}  = 6.62(3.21) \times 10^{40} {\rm erg~}{\rm s}^{-1}{\rm Mpc}^{-3}$ at $z = 4(7)$, 
is consistent with recent SDSS determinations.

The predicted power spectrum is 
$k^3 P^\alpha(k, z)/2\pi^2 = 9.76\times 10^{-4}(2.09\times 10^{-5}){\rm nW}^2{\rm m}^{-4}{\rm sr}^{-2}$ 
at $z = 4(7)$ for $k = 0.1\, h\, {\rm Mpc}^{-1}$.  Is such signal detectable? In principle the above 
Ly$\alpha$~ PS for $z<8$ is in reach of a small space telescope (40 cm in diameter, similar to the 
proposed SPHEREx, see \cite{Dore2014}); detections with low S/N are possible only in some optimistic 
cases up to $z\sim10$. However, the foreground interloper emission lines represent a serious source of confusion, and therefore must be removed. The host galaxies of these interloping lines can be resolved via an ancillary photometric galaxy survey in the NIR bands (Y, J, H, K). If the hosts are removed down to 
AB mag $\sim26$, then the Ly$\alpha$~ PS for $5 < z < 9$ can be recovered with good S/N. If [CII] 
intensity mapping data is available, by cross-correlating the two signals the required  depth of the 
ancillary galaxy survey could be is within reach of Euclid (AB mag $\sim24$). Alternatively, 
\cite{Comaschi2016b} suggested a promising method based on the cross-correlation between 
diffuse Ly$\alpha$~ emission and LAEs. Using this technique, they show that signal-to-noise of the 
observed cross-correlation power spectrum does not depend significantly if the variance of the random 
noise introduced by contaminating lines is $\sigma_{\rm N} \le 10 \sigma_\alpha$. In these conditions 
the mean line intensity, $I_\alpha$, can be precisely recovered. Even if 
$\sigma_{\rm N} = 100 \sigma_\alpha$, $I_\alpha$ can be constrained within a factor $2$. 
Since removing the contaminating lines in future IM surveys will not be an easy task, relying on 
a solid cross-correlation is crucial to extract information from these experiments.

\subsection{Modeling the HI Power Spectrum on Non-Linear Scales}

\begin{figure}\centering
\includegraphics[scale=0.9]{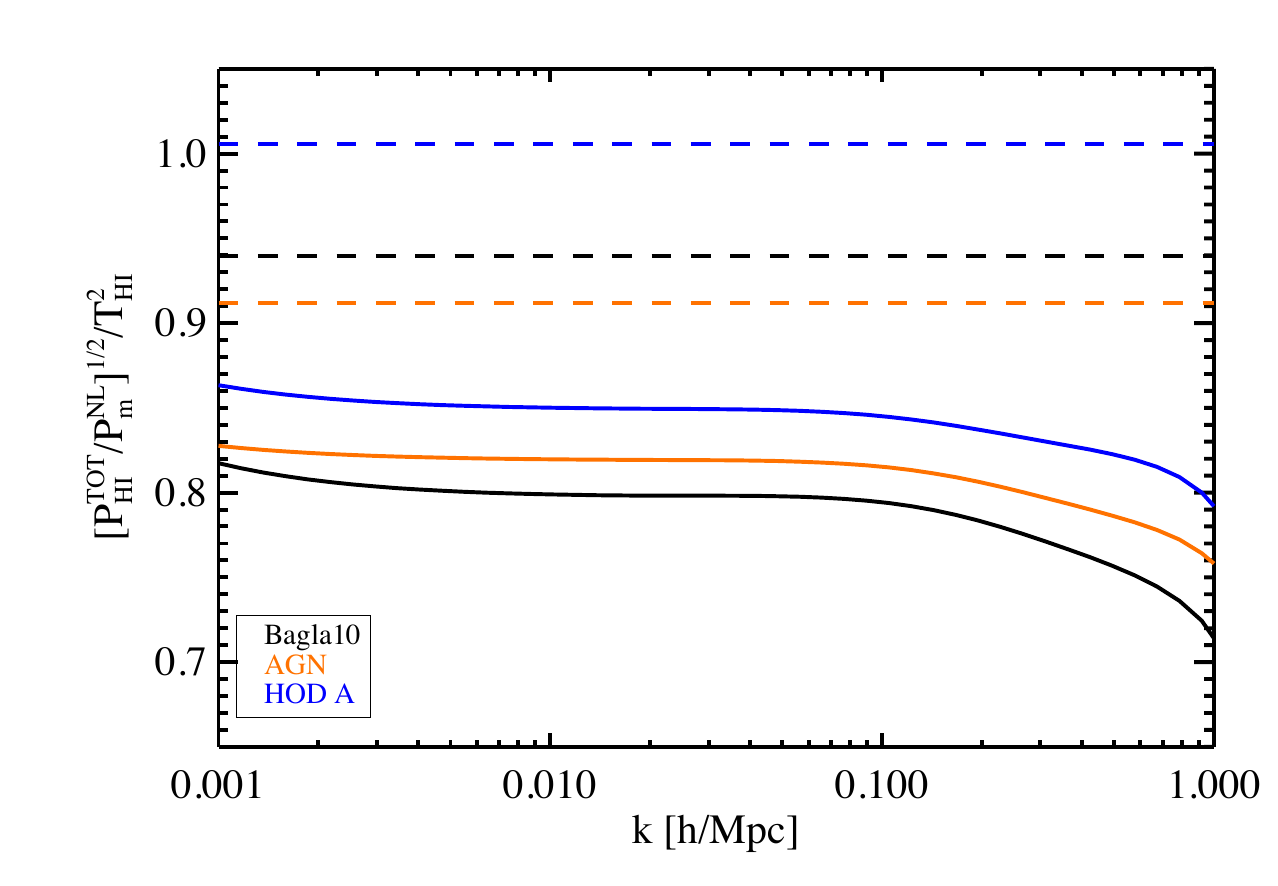} 
\caption{Scale dependence of the HI effective bias at $z=1$,defined as the ratio 
$\sqrt{P_\mathrm{HI}(k)/P_\mathrm{m}^\mathrm{NL}(k)}/\overline T_\mathrm{HI}^2$ (right) 
for several MHIMh prescriptions in real space. $P_\mathrm{m}^\mathrm{NL}$ is the non-linear 
power spectra of matter. Horizontal dashed lines are the linear biases associated to each MHIMh 
models. The Bagla10, AGN, and HOD A are the MHIMh relations from \citet{Bagla2010}, 
\citet{Villaescusa2016}, and \citet{Padmanabhan2017}, respectively. Adapted from \citet{Penin2017}. 
({\it Courtesy of Aur\'elie P\'enin})}
\label{fig:biasHI_for_report}
\end{figure}

Over the next decade, several intensity mapping experiments will probe unprecedented 
volumes of Universe, allowing these surveys to span ultra-large scales (provided foreground contamination can be mitigated) as well as non-linear ones. Our 
current approaches of modeling clustering will soon reach their limits. They rely on the 
assumption that structure formation is linear on large scales and non-linear on smaller 
scales, nevertheless this statement is only true for dark matter and not for biased tracers. 
Indeed, short and large scale modes of biased tracers are coupled: density fluctuations of 
the tracer are enhanced in more massive regions as compared to those in less massive 
regions. Using a perturbative approach, \citet{Penin2017} showed that this coupling 
translates into a significant contribution of non-linear terms to the power spectrum on large 
scales. Therefore, on linear scales the bias is not constant but scale-dependent. This 
effective bias depends on the relation between the tracer and the Large Scale Structure. 
In the case of HI, a simple relation between the HI mass and the halo mass (MHIMh) 
is assumed. The HI effective bias is, at most, 15\% lower than its linear counterpart at $z<1.5$ 
while it is higher at $z>1.5$. Nevertheless, the case of HI is peculiar as the HI linear bias is, 
uncommonly, below unity at low redshift and above at higher redshift \citep{2012ApJ...750...38M,Sarkar2016}. 
The effective bias is compared to the linear one in Fig.~\ref{fig:biasHI_for_report} for several MHIMh relations.

The scale dependence of the effective bias has a significant impact on the expected signatures 
of cosmological parameters. It alters the ratios of the BAO peaks, modifies the scale dependence 
of the power spectrum with $f_\mathrm{NL}$ on ultra large scales \citep{Umeh2016,Umeh2017}, 
and notably changes the shape of the 2-dimensional power spectrum for the estimation of the 
growth factor. Further work is required to quantify these modifications and to which extent they 
flaw the estimation of cosmological parameters.

However, the main source of uncertainty today is the MHIMh relation; it controls the effective bias, 
and more importantly, the temperature of HI which rules the overall amplitude of the power spectrum. 
Indeed, the actual observable is the product $T_\mathrm{HI}\, b_\mathrm{HI}$ and there is an order 
of magnitude difference between the HI temperatures predicted by current MHIMh prescriptions.
Lastly, the MHIMh relation rules the small scale behavior of the HI power spectrum. The fact that HI 
is less clustered than dark matter translates into a dip in the power spectrum on small scales as 
shown in Fig.~\ref{fig:biasHI_for_report}. 

From a broader point of view, these results hold for any intensity mapping (or galaxy) survey over 
several thousands of square degrees such as SPHEREX, among others. It is worth highlighting 
the fact that these results also apply to cross-correlations of two biased tracers. Finally, it is critical 
to improve our understanding of the bias of the tracers, namely how the line luminosity relates to dark matter.

\subsection{Semi-Analytic Models and More}

Existing simulations of the expected signal for line-intensity
mapping, particularly for lines other than 21-cm, have been carried
out using empirical scaling laws between dark matter halo properties
and line luminosity \citep[e.g.][]{Li:2015gqa, Sun2016}. This
approach is relatively simple to implement, and computationally
efficient, but has several disadvantages. First, line-intensity mapping is
expected to probe below the flux scales of objects that can be readily
detected individually at a given redshift, so it is inevitable that
these scaling relations must be extrapolated well outside of the
regime where they are well calibrated. Second, the emission luminosity
of optical and IR lines such as H$\alpha$, [OIII], CO, and [CII] are
sensitive to the detailed conditions in the sites of their production
in the interstellar medium (ISM) of galaxies, such as density,
temperature, metallicity, and local background radiation field
strength. The detailed physical connection between the efficiency of
production of these lines and dark matter halo properties is poorly
understood, and likely to be complex. Third, most of these empirical
approaches have so far not attempted to simultaneously model multiple
lines. As discussed further in the Techniques section, cross-correlating
intensity maps across multiple lines will have enormous
leverage. Finally, although line-intensity mapping experiments have great potential to
provide insights into the physics of galaxy formation and its
interplay with cosmology, the empirical approach is limited in its
ability to aid in this goal.

However, creating physics-based models for intensity mapping is an
extremely challenging problem, because of the very wide range of
scales and diversity of physical processes that come into play.  The
processes that determine the efficiency of star formation and
luminosity of lines such as CO and [CII] occur on sub-pc scales within
the ISM, while galaxy properties are known to depend on their large
scale environment on scales of Mpc. Line-intensity mapping experiments will map
significant fractions of the sky, requiring very large volume
simulations (tens of Gpc).

A promising approach for overcoming these challenges may be the use of
semi-analytic galaxy formation models \citep*[for a summary and
references, see][]{Somerville2015a}. Semi-analytic models
capture the cosmological formation history (including any correlation
with larger scale environment) using halo merger trees extracted
from dissipationless N-body simulations. Within these merger trees,
the approach applies simple but physically motivated recipes for a
broad variety of physical processes, which generally include cooling
and accretion of gas, conversion of cold gas into stars, feedback from
massive stars and supernovae, chemical enrichment, black hole
formation and feedback from accreting black holes. State-of-the art
semi-analytic models have been shown to produce consistent predictions
for global galaxy properties and their evolution compared with much
more computationally expensive numerical hydrodynamic simulations
\citep{Somerville2015a}.  In a recent generation of models the
cold ISM gas is partitioned into atomic, molecular, and ionized phases
according to recipes motivated by empirical relations or fitting
functions derived from numerical simulations \citep{Fu2010, Fu2012,
Lagos2011, Lagos2012, Popping2014, Somerville2015a}.

These models can then be coupled with radiative transfer and
Photodissociation Region (PDR) models to make predictions for the line
emission of rotational transitions of CO and of [CII] (\citet{Lagos2012}; 
\citet{Popping2014, Popping2016}). In the models developed by
G. Popping and collaborators, each galaxy is populated with ``clouds''
selected from a probability distribution function motivated by
observed molecular cloud mass functions, based on the average H$_2$
density in each annulus. The temperature of the gas and dust is then
calculated by computing the heating-cooling balance, taking into
account cosmic ray heating, photo-electric heating, gravitational
heating, and the exchange of energy between dust and gas. The primary
cooling mechanism for the gas is line radiation through CO, atomic
carbon [CI], and ionized carbon [CII]. Finally, the level populations of
the molecule or atom of interest, and the probability that a photon at
some position in the cloud can escape the system, are calculated \citep*[see
e.g.][]{Perez2011}.

To date, these studies have focused on comparing the model
predictions with observations of individually detected galaxies. P14
and P16 present a comparison of their model predictions with observed
scaling relations for CO and [CII] emission properties vs. galaxy
properties such as FIR luminosity, SFR, and stellar mass, as well as
CO and [CII] luminosity functions. They find very good agreement with
existing CO-galaxy scaling relations and LFs over a broad range of
redshifts. These models fare less well at reproducing the observed
scaling relations and LFs for [CII]. However, work in progress
(G. Popping et al. in prep) finds that implementing the approach
presented in \citet{Narayanan2017}, which takes into account
the radial dependence of conditions \emph{within} individual clouds,
will lead to improved predictions for CO and [CII] by these
semi-analytic models.

The approach described above can be used to create predictions for a
few times $\sim 10^4$ galaxies, but is probably still prohibitively
expensive for computing very large area maps. A promising approach
would be to use statistical techniques such as principal component
analysis or machine learning to develop computationally efficient
methods for assigning sets of galaxy properties (UV-FIR continuum plus
line emission) to dark matter halos in large volume cosmological
simulations. Models run on merger trees extracted from numerical
N-body simulations, as well as more detailed fully numerical
hydrodynamic simulations, should be used to study the ``second order''
dependence of line emission efficiency on large scale environment, and
to include this in these mappings.

Similarly, one can interface predictions from semi-analytic models
with semi-numerical models of radiative transfer and reionization
\citep[e.g.][]{Mesinger2007, Santos2010, Hassan2016} to create mock 21-cm maps. 
The semi-analytic model straightforwardly provides predictions for the ionizing 
emissivity of each halo, based on the modeled stellar population in each 
galaxy, thereby replacing the very simplified halo mass based
parameterizations previously utilized in these types of models. An
important unknown parameter is the escape fraction of ionizing
photons, which may vary depending on the instantaneous galaxy
properties such as star-formation rate, gas fraction, and metallicity.

\begin{figure}[h!]
\begin{center}
\includegraphics[scale=0.26]{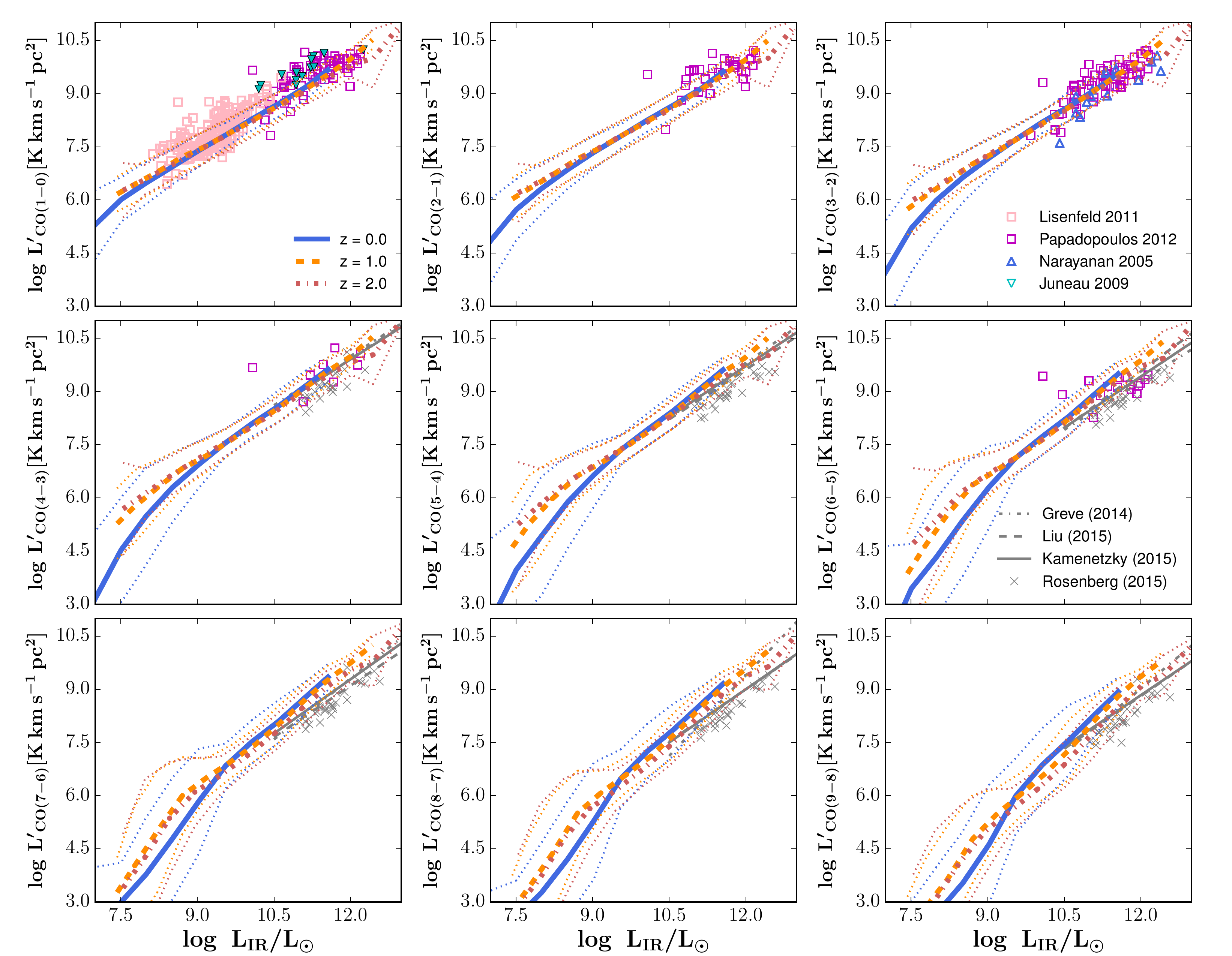}
\includegraphics[scale=0.29]{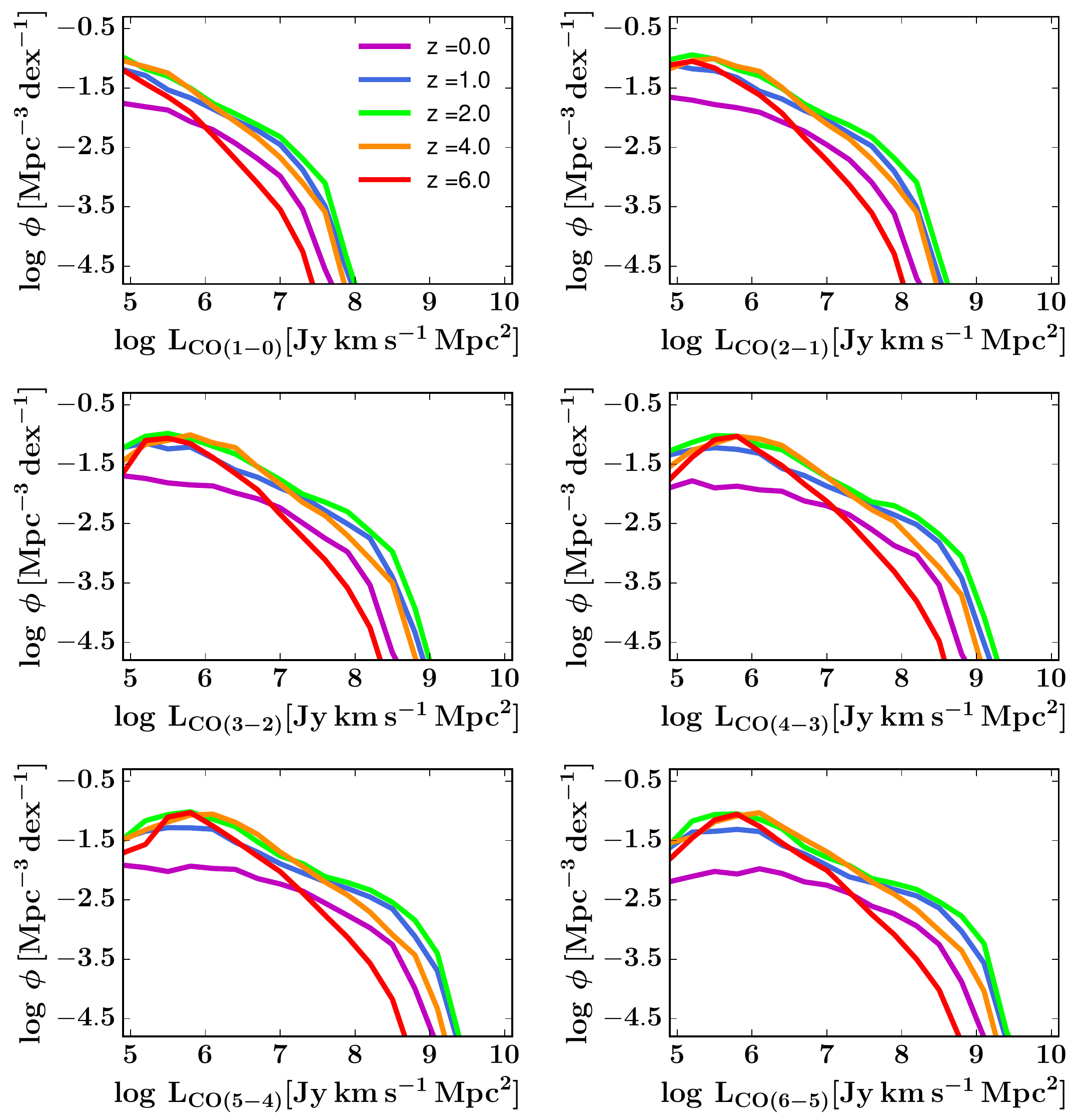}
\caption{{\bf Left} CO line-luminosity of CO J$=$1-0 up to CO J$=$9-8 as a function 
of FIR luminosity at redshifts $z=0$, $z=1$, and $z=2$. Model results are compared 
to observations taken from \citet{Narayanan2005}, \citet{Juneau2009}, 
\citet{Lisenfeld2011},\citet{Papadopoulos2012}, \citet{Greve2014}, \citet{Liu2015}, 
\citet{Rosenberg2015}, and \citet{Kamenetzky2016}. The thick lines show the median 
of the model predictions, whereas the dotted lines represent the two sigma deviation 
from the median. {\bf Right} Model predictions of the CO J$=$1--0 up to the CO J$=$6--5 
luminosity function of galaxies from $z=0$ out to $z=6$.  Figures from \citet{Popping2016} 
model. ({\it Courtesy of Gerg{\"o} Popping and Rachel Somerville})}
\label{fig:semian}
\end{center}
\end{figure}

In summary, galaxy formation models have traditionally focused on
predictions for the stellar properties of galaxies, such as UV-NIR
emission. Recently, they have been extended to make detailed
predictions for line emission such as CO and [CII]. It will be
important to continue to validate and refine these models by
confronting the predictions with deep observations of line properties
for individually detected galaxies, as well as with the predictions of
more detailed, spatially resolved numerical hydrodynamic simulations
\citep[e.g.][]{Pallottini2017}. With some additional effort, these
tools can be adapted to provide detailed forecasts for intensity
mapping, provide a testbed for developing new analysis tools, and to
aid in the interpretation of upcoming experiments in terms of the
constraints they can provide on the physics of galaxy formation and
evolution.

\subsection{High Resolution Numerical Hydrodynamic Simulations}

High resolution numerical hydrodynamic simulations can be used to reproduce the 
observed stellar mass-halo mass relation and Kennicutt-Schmidt relation.  
The FIRE simulations, for example, are cosmological zoom-in simulations which are being
used to make mock maps for the COMAP experiment.

\begin{figure}[!h]
\begin{center}
\includegraphics[width=0.7\columnwidth]{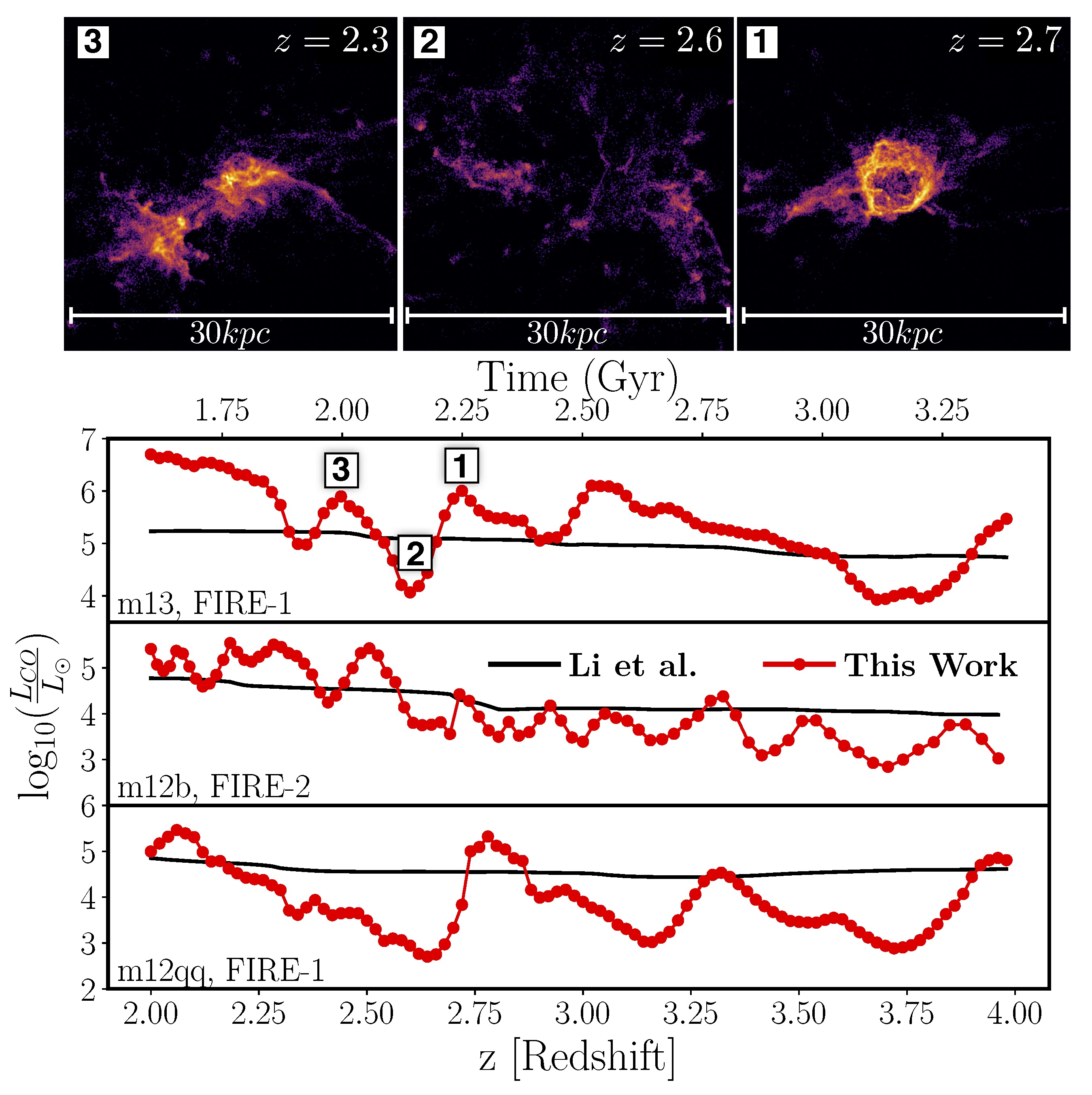}
\caption{CO Luminosity trajectories for three FIRE runs as a function of redshift. 
Top panel shows the ${\rm H}_2$ column density from a halo with mass $10^{12}\,M_\odot$ 
at different redshifts (the halo mass is $10^{13}\, M_\odot$ at $z=0$, hence the name "m13") 
 which correspond to the ${\rm H}_2$ emission before, during and 
after a gas blowout. FIRE-1 and FIRE-2 are different simulation runs. 
({\it Courtesy of Gunjan Lakhlani and Norman Murray})}
\label{fig:FIRE}
\end{center}
\end{figure}

Since CO gas is tightly correlated with molecular ${\rm H}_2$, the technique 
described in \citet{Krumholz2011} is first used to predict the molecular ${\rm H}_2$ fraction as a function 
of the local gas column density and metallicity of gas. Then the model described 
in \citet{Narayanan2012} is used to obtain CO luminosity as a function of the local molecular 
${\rm H}_2$ column density and metallicity of the gas. Contrary to previous studies in literature, 
the halo mass alone is found to not be a good indicator of the CO luminosity. As shown in 
Fig.~\ref{fig:FIRE}, the CO luminosity varies by 3 dex over a time in which the halo mass 
varies by less than a factor of two. This shows that there is at least  other physical 
parameter (besides halo mass) determining the CO luminosity of a given galaxy. The 
star-formation rate is strongly correlated with the CO luminosity. However, after accounting 
for the effect of the star-formation rate, one still observes 2 dex of fluctuations. Finding 
other hidden parameters which control the behavior of CO luminosity from a given 
halo is left to future work. 

\subsection{Modeling for Line-Intensity Mapping Cross-Correlations} 

\begin{figure}[h!]
\centering
\includegraphics[width=.7\columnwidth]{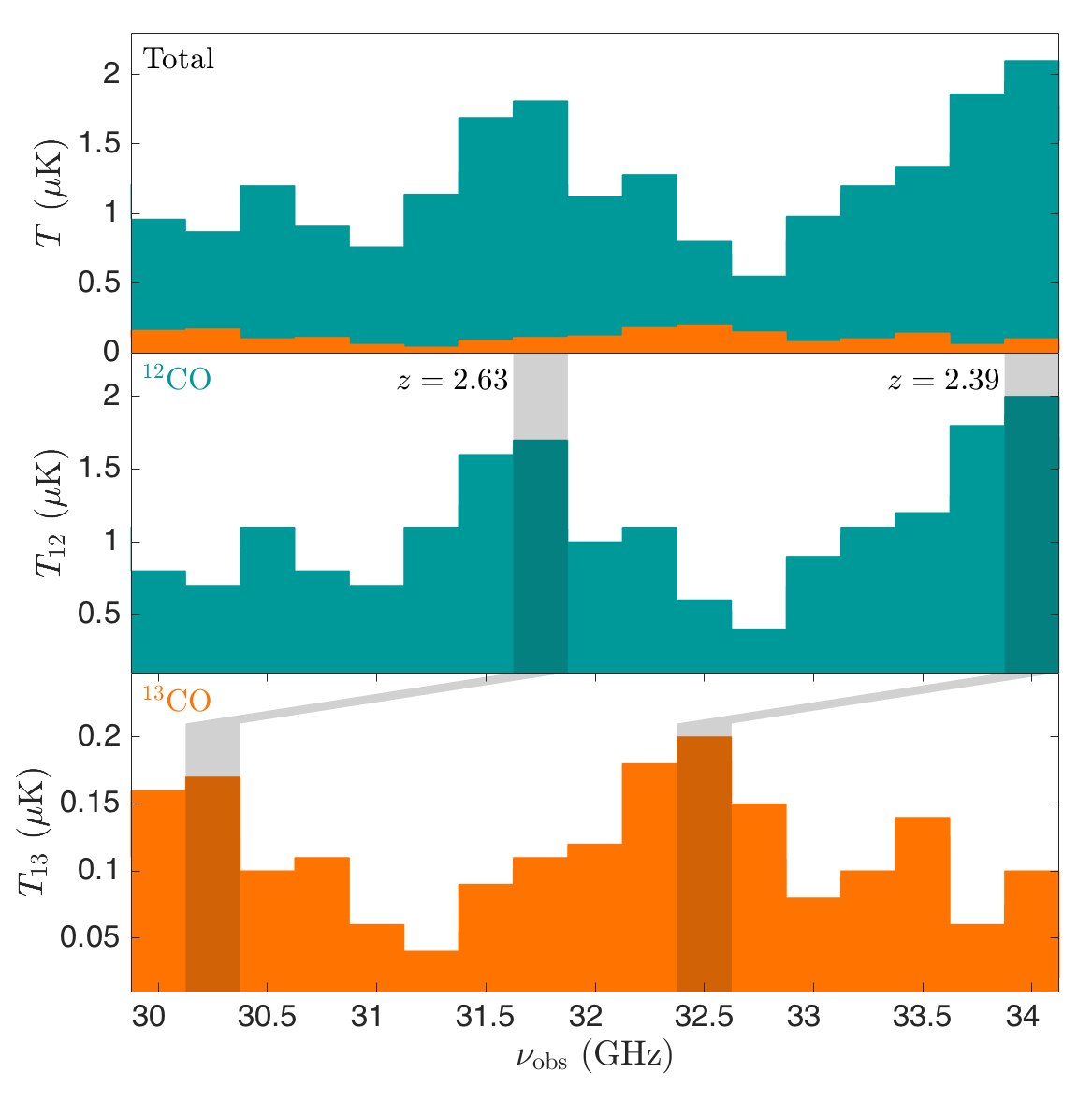}
\caption[Schematic view of $^{12}$CO/$^{13}$CO cross-correlation]{A schematic view of 
the contributions from $^{12}$CO and $^{13}$CO to a hypothetical intensity mapping survey.  
The top panel shows the total observed intensity in each frequency bin assuming that the 
observed $^{13}$CO intensity from all galaxies is 10\% of the $^{12}$CO intensity.  The 
middle and bottom panels show the contribution to the total signal from $^{12}$CO and 
$^{13}$CO emission respectively.  The shaded regions in these panels highlight emission 
that comes from the same slice of physical space. ({\it Courtesy of Patrick Breysse})}
\label{fig:CCvisual}
\end{figure}

Cross-correlations between different lines provide a number of exciting possibilities for 
intensity maps.  In addition to allowing tests of systematics and foreground cleaning, 
they also make it possible to obtain significantly more information about a target population 
than can be learned from a single line alone.  They can also allow access to lines far too 
faint to detect on their own.  For an example, consider the cross-correlation between the 
usual 115 GHz $^{12}$CO line and its 110 GHz $^{13}$CO isotopologue discussed in 
\citet{Breysse2016b}.  Figure \ref{fig:CCvisual} shows a schematic view of what the contributions 
to a survey from these two lines might look like.  A given set of galaxies will emit both 
$^{12}$CO and $^{13}$CO lines at a given position in physical space.  The two lines are then 
redshifted to different bands in frequency space as shown in the bottom two panels, then added 
together to produce the observed signal in the top panel.  For illustration purposes, Figure 
\ref{fig:CCvisual} assumes that the observed $^{13}$CO intensity from all galaxies is 10\% of the 
$^{12}$CO intensity.  

The cross-spectrum of a pair of lines takes a similar form to those in the Introduction:
\begin{equation}
P_{1\times2}(k,z)=\overline{I}_{1}(z)\overline{I}_{2}(z)\overline{b_{1}}(z)
\overline{b_{2}}(z)P_m(k,z)\\+P_{\rm{shot}}^{1\times2},
\end{equation}
where $\overline{I}_{1,2}$ and $\overline{b}_{1,2}$ are the mean intensities and biases of 
the two lines.  The shot noise amplitude becomes
\begin{equation}
P^{\rm{shot}}_{1\times2}\propto\int_0^\infty L_1 f(L_1)\frac{dn_{\rm{gal}}}{dL_{1}}dL_{1},
\end{equation}
where it was assumed that $L_2\equiv f(L_1)$.  The cross-spectrum thus allows a 
measurement of the intensity ratio of the lines as well as how they vary with respect 
to one another from source to source.  This gives it great potential as a tool to probe the 
high-redshift interstellar medium in great detail.





\newpage

\section{Techniques}
\label{sec:tech}

A range of theoretical work is underway to further develop line-intensity mapping analysis 
and simulation techniques. These efforts aim to: identify interesting science targets for 
line-intensity mapping surveys and produce methods for extracting this science 
(\S \ref{sec:gal_evolution} and \S \ref{sec:eor_lim}), to develop rapid techniques for generating mock
line-intensity mapping data cubes (\S \ref{sec:mocks}), to optimize analysis methodologies 
(\S \ref{sec:analysis_methods}), and to explore strategies for foreground mitigation 
(\S \ref{sec:foreground_mitigation}). This work is important for sharpening the science 
case for line-intensity mapping, for planning upcoming survey efforts, in developing 
end-to-end simulations of analysis pipelines, and ultimately for interpreting the actual 
survey data as it becomes available. The subsections below provide the flavor of some 
current work in this area. 

\subsection{Studying Galaxy Evolution with Intensity Mapping Cross-Correlations}
\label{sec:gal_evolution}

Line-intensity mapping data cubes may be cross-correlated with galaxy surveys, 
providing a reliable means of statistically detecting the cosmological matter distribution 
in datasets dominated by instrumental noise and systematic effects. The resulting cross-power 
spectrum, $P_{x}(k)$, depends on the bias factor of the optically selected galaxies,
$b_{\rm gal}$, the neutral hydrogen bias, $b_{\rm HI}$, and the average HI brightness
temperature, $\overline{T_{\rm HI}}$. It is therefore highly sensitive to the HI
distribution and to how the sample of optical galaxies is selected. 
The cross-correlation coefficient, $r(k)$, expresses the intrinsic correlation of the two 
probes and is defined as $ r(k) = P_x(k) \left[P_{\rm HI}(k) P_{\rm gal}(k) \right]^{-1/2}$.

As described in Section 3, the first cosmological detection of an HI intensity map was through cross-correlating
redshifted 21-cm data from the Green Bank Telescope (GBT) at $z \approx 0.8$ 
\citep{Masui:2012zc} with an overlapping optically-selected galaxy sample from the 
WiggleZ Dark Energy survey \citep{Drinkwater:2009sd}. Thus far, only upper
limits have been obtained on the HI intensity mapping auto power spectrum from the 
GBT data \citep{Switzer:2013ewa}, and so $r(k)$ has not yet been derived observationally. 
Nevertheless, the prospect of near-term measurements of $r(k)$ have motivated efforts 
to understand how this quantity may inform models of galaxy evolution.

First, simulations of the GBT-WiggleZ cross-correlation \citep{Wolz:2015ckn} found 
that the cross-correlation coefficient decreases on small scales around $k > 1.0$ Mpc$^{-1}$, 
and exhibits a varying shape, dependent on how the optical galaxies are selected. 
Specifically, the WiggleZ selected galaxies are highly star-forming and usually HI-rich 
and show a much higher correlation on small scales than quiescent, red galaxies. 
In further theoretical studies, this effect has been traced back to a previously 
unaccounted-for shot-noise contribution to the cross-correlation power spectrum which 
scales with the HI content of the optically selected sample \citep{Wolz:2017rlw}. The 
scale-independent shot-noise contribution is caused by Poisson noise in the discrete 
galaxy sample, which in cross-correlation with the HI maps is weighted by the average 
HI temperature of the sample, such that 
$P_{\rm shot}= \overline{T_{\rm HI,gal}}/N_{\rm gal}$.  The 
observed cross-correlation coefficient $\hat r(k)$ can be rewritten as 
$\hat r(k)=(P_{x}(k) + P_{\rm shot})\left(P_{\rm HI}(k) P_{\rm gal}(k)\right)^{-1/2}$. 
Measuring the cross shot-noise contribution either in $\hat r(k)$ or $P_x(k)$ allows 
one to determine the average HI mass of the selected galaxy sample from 
$\overline{T_{\rm HI, gal}}$. In other words, one can establish scaling relations between 
the selection criteria of the optical samples and their HI mass. 

\begin{figure}[h!]
\begin{center}
\includegraphics[width=0.65\textwidth]{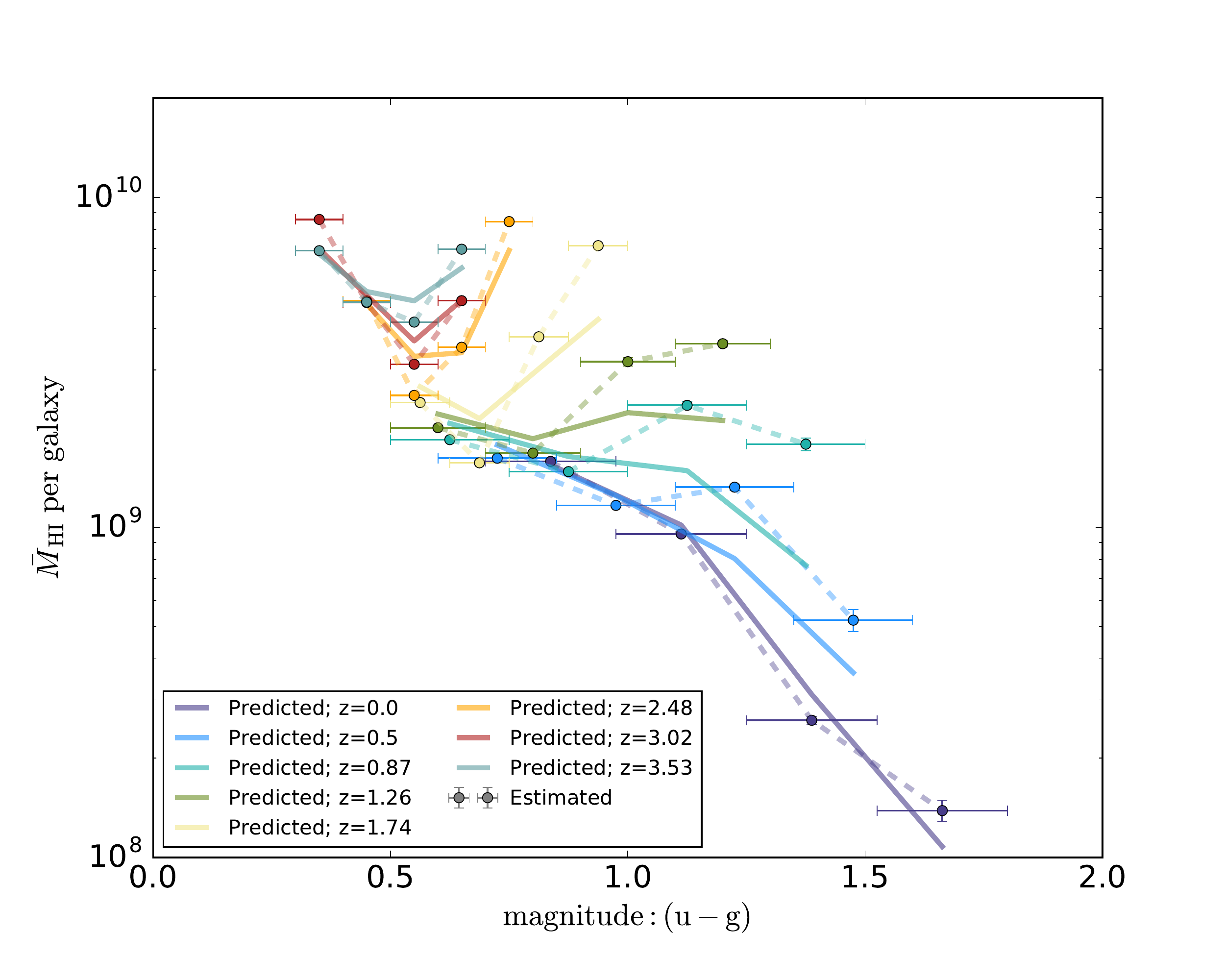}
\caption{Extracting galaxy properties from cross shot-noise measurements.
The average HI mass per galaxy sample, as selected by their color $(u-g)$ 
(moving from blue to red in the increasing direction along the x-axis) for redshifts 
from $0<z<3.5$. A $(100 {\rm Mpc}/h)^3$ volume EAGLE simulation box is used
with HI masses provided by \citet{Lagos:2015gpa}, based on the Gnedin-Krumholz 
scheme. The solid lines indicate the true scaling relation from the simulation output, 
while the dashed lines with circles show the values obtained through the cross 
shot-noise measurements. See \citet{Wolz:2017rlw} for a more detailed description 
of the method. ({\it Courtesy of Laura Wolz})}
\label{fig:HIscalingEAGLE}
\end{center}
\end{figure}

Fig.~\ref{fig:HIscalingEAGLE} shows an example of how the average HI mass relates 
to the optical color selection $(u-g)$, which indicates the level of star-formation
activity in a galaxy. These results are derived from the hydro-dynamical EAGLE simulation 
at redshifts $0<z<3.5$ \citep{Crain:2016ex}. The HI mass of each simulated galaxy is 
given in \citet{Lagos:2015gpa}. The solid lines indicate the true HI mass and color extracted 
directly from the simulation, while the circles mark the values derived from the shot-noise 
contribution to the simulated cross-correlation power spectrum. This illustrates how
the cross-correlation between intensity maps and optically selected galaxy samples
may be used to derive scaling relations between the gas contents and star-formation
properties of the samples, and thereby provide insight into universal laws governing
galaxy evolution.

\subsection{Complementing 21cm Probes of the EoR with other Line-Intensity Mapping Surveys}

\label{sec:eor_lim}

Line-intensity mapping observations may potentially be used to trace large scale 
structure in the galaxy distribution during the EoR, and thereby complement redshifted 
21-cm observations of intergalactic neutral hydrogen from the same epoch. Reionization 
involves the interplay between the ionizing sources -- which are likely abundant, low mass 
galaxies -- and intergalactic hydrogen. Stated generally, our understanding of reionization 
should therefore benefit from tracking both the intergalactic gas (through the 21-cm line) 
and the sources themselves (using other convenient emission lines).

More concrete synergies may also be identified. For example, cross-correlating a 21-cm 
data cube with emission in another line at the same redshift may help to verify a putative 
redshifted 21-cm detection \citep{Furlanetto:2006pg,Lidz:2008ry,Lidz:2011dx}. A significant 
challenge for 21-cm measurements from the EoR is to separate strong foreground contamination, 
and it may initially be tricky to distinguish a genuine cosmological signal from residual 
foreground contamination. A smoking-gun validation would be to show that the putative 21-cm 
signal correlates with another high redshift tracer of large-scale structure.  This, however, 
requires an additional probe which spans a large field of view, yet has accurate redshift 
information.\footnote{A survey with coarse photometric redshift estimates measures mainly
transverse modes, but these spectrally-smooth modes are hardest to separate from foregrounds 
in the 21-cm survey. In order to be well-matched to the 21-cm survey, accurate redshifts are 
required of the ``tracer'' survey.} Although high redshift galaxy surveys with the JWST or ALMA, 
for example, will provide valuable information about early galaxy populations, they are poorly 
suited for direct cross-correlation with 21-cm measurements given the $\sim$ arcminute fields 
of view of these instruments. A more promising approach for cross-correlation purposes is to 
give up on detecting individual sources, and to focus instead on measuring the spatial fluctuations 
in the combined emission from many individually unresolved galaxies in convenient tracer lines 
(e.g. \citet{Lidz:2011dx,Gong:2011ts}). In other words, one can perform line-intensity mapping 
observations -- using various emission lines --  to trace-out galaxy populations across the same 
large cosmic volumes in which the 21-cm surveys track neutral hydrogen. 

\begin{figure}[h!]
\begin{center}
\includegraphics[scale=0.5]{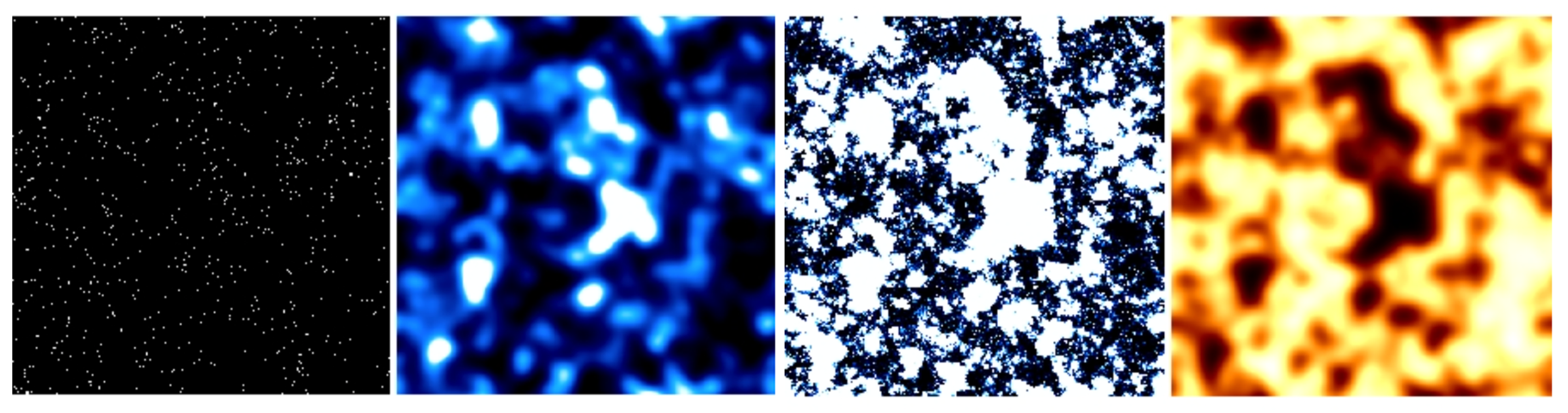}
\caption{\label{fig:21cm_co_illustrate} Complementing redshifted 21-cm observations 
of the EoR with intensity mapping observations in other lines. \textbf{Panel 1}: The 
simulated distribution of EoR galaxies. \textbf{Panel 2}: A model for the 
CO(2-1) line-intensity mapping signal from these galaxies. \textbf{Panel 3}: The 
ionization field from the simulation slice. The white regions are highly ionized, while 
the dark regions show neutral hydrogen. \textbf{Panel 4}: The redshifted 21-cm signal 
from the same region of the IGM. Each simulation slice is $130$ co-moving Mpc/$h$ 
on a side (spanning roughly a degree across on the sky), and $0.25$ Mpc/$h$ thick. 
The CO(2-1) and 21-cm maps are smoothed to $6$ arcminute spatial resolution and 
$0.035$ GHz spectral resolution. On large scales, the CO and 21-cm fields are 
anti-correlated. This large scale anti-correlation may be used to confirm a possible
redshifted 21-cm detection, and to help understand the properties of cosmic reionization.
({\it Courtesy of Adam Lidz})}
\end{center}
\end{figure}

\begin{figure}[h!]
\begin{center}
\includegraphics[scale=0.46]{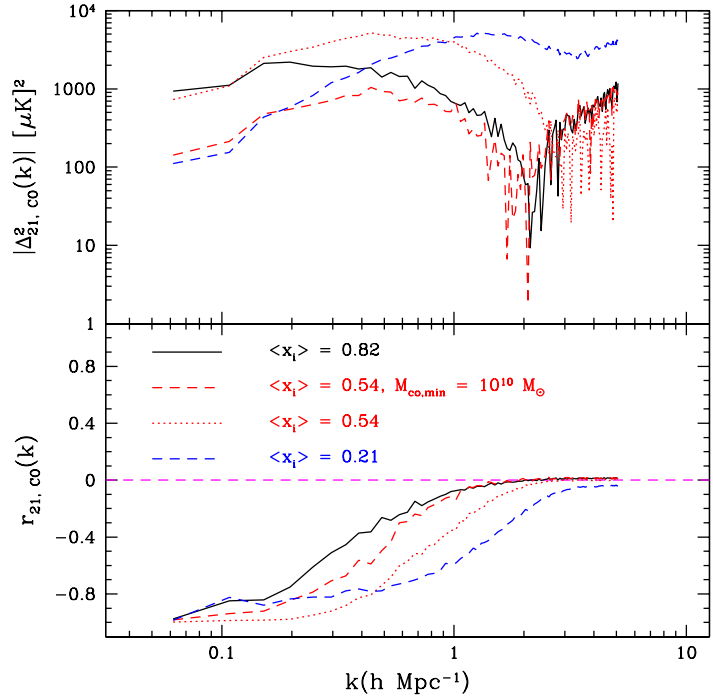}
\caption{\label{fig:21cm_co_xcorr} Cross-correlation between CO(2-1) and 21-cm brightness 
temperature fluctuations during the EoR. \textbf{Top Panel}: The absolute value of the cross 
power spectrum between the fluctuations in each line at different stages of the reionization process. 
The quantity $\avg{x_i}$ in the legend refers to the volume-averaged ionization fraction. 
\textbf{Bottom Panel}: The cross correlation coefficient between the two fields. On scales larger 
than the characteristic size of the ionized regions at the redshift of interest, the cross-correlation 
coefficient approaches $r_{\rm 21, CO} \sim -1$, while it goes to zero on significantly smaller scales. 
As reionization proceeds, a progressively larger fraction of the IGM volume is ionized, and the 
typical size of the ionized regions increases. The cross-correlation hence turns over at smaller 
and smaller wavenumber with increasing $\avg{x_i}$, charting the overall progress of reionization. 
From \citet{Lidz:2011dx}. ({\it Courtesy of Adam Lidz})}
\end{center}
\end{figure}

Fig.~\ref{fig:21cm_co_illustrate} illustrates how a hypothetical CO(2-1) line-intensity mapping 
survey might be fruitfully combined with 21-cm measurements.\footnote{Note that CO(2-1) 
provides just one illustrative example transition and other lines such as [CII] may, in fact, be 
superior for mapping galaxy populations during the EoR (e.g. \citet{Gong:2011mf}).} 
On large scales, the CO(2-1) emission should be anti-correlated with the 21-cm signal: bright 
areas in the CO map trace upward fluctuations in the galaxy abundance but these regions are 
ionized and dim in 21 cm. On the other hand, the 21-cm and CO fields should be roughly 
uncorrelated on scales much smaller than the size of the ionized regions. This occurs because 
the gas interior to ionized regions is highly ionized irrespective of the precise galaxy abundance, 
while completely neutral regions do not contain galaxies (see \citet{Lidz:2008ry,Lidz:2011dx} 
for further discussion and model variations). Fig.~\ref{fig:21cm_co_xcorr} shows this more 
quantitatively, plotting the cross-correlation coefficient between the model CO(2-1) and 21-cm 
fields. In short, the cross-correlation may be used to determine the typical size of ionized bubbles 
at different stages of the reionization process.

In addition, line-intensity mapping observations may provide valuable information about the 
bulk properties of galaxy populations during the EoR. This may be especially valuable given
recent observations,  which suggest that reionization is driven largely by abundant, low luminosity 
galaxies (e.g. \citet{Robertson:2015uda}). These faint sources are challenging to detect 
individually, but their collective impact may nevertheless be studied with line-intensity mapping 
observations. The key issue that requires further quantitative study, however, is just how luminous 
the bulk of the ionizing sources -- which may be quite metal poor -- are in convenient tracer lines. 
Targeted ALMA observations may provide valuable guidance here, which can then sharpen 
line-intensity mapping forecasts. If line-intensity mapping observations are possible in multiple
lines, and the relative strength of emission fluctuations in the different lines is measurable, this 
should help elucidate the bulk ISM properties in early galaxy populations.

\subsection{Developing Mock Line-Intensity Mapping Surveys}
\label{sec:mocks}

Mocking line-intensity mapping experiments requires accurate large scale structure simulations. 
N-body methods work well for a subset of these simulations \citep{Li:2015gqa}, but when larger 
cosmological volumes or many independent realizations are needed, more efficient methods are 
required. Examples include studying map-to-map covariances and cosmological parameter variations, 
especially for efforts to explore physics beyond the standard model of cosmology (e.g., primordial 
non-Gaussianity, dark energy, and modified gravity). 

To create mocks for COMAP the mass-peak-patch method for finding halos is used, based on 
ellipsoidal dynamics applied to Lagrangian (initial condition) regions defined by collapse along 
all three axes \citep{Bond:1993we}. The speed of this method allows an 
1140$^3$ Mpc$^3$, $2.4<z < 3.4$, n$_{cell}$ = 4096$^3$ simulation to be completed in 
only 580 CPU hours using 2.3Tb of memory on 2048 Intel Xeon E5540 2.53 GHz processors 
of the Scinet General Purpose Cluster. The resulting halo catalogs, with a minimum mass of 
$2.5\times 10^{10} M_{\odot}$, are used to generate a multi-frequency realization of the observed 
line intensity by assigning a CO(1-0) flux to each halo, using properties such as mass, redshift, 
and formation time. Given the COMAP resolution, equivalent maps would be obtained without 
using an HOD model for central and satellite galaxies in the halos. The final mock intensity map is 
created by binning the resulting halo fluxes by angular position and observed frequency, 
including redshift space distortions, using publicly available code\footnote{github.com/georgestein/limlam\_mocker}. 

\begin{figure}[h!]
\centering
\includegraphics[width=0.6\textwidth]{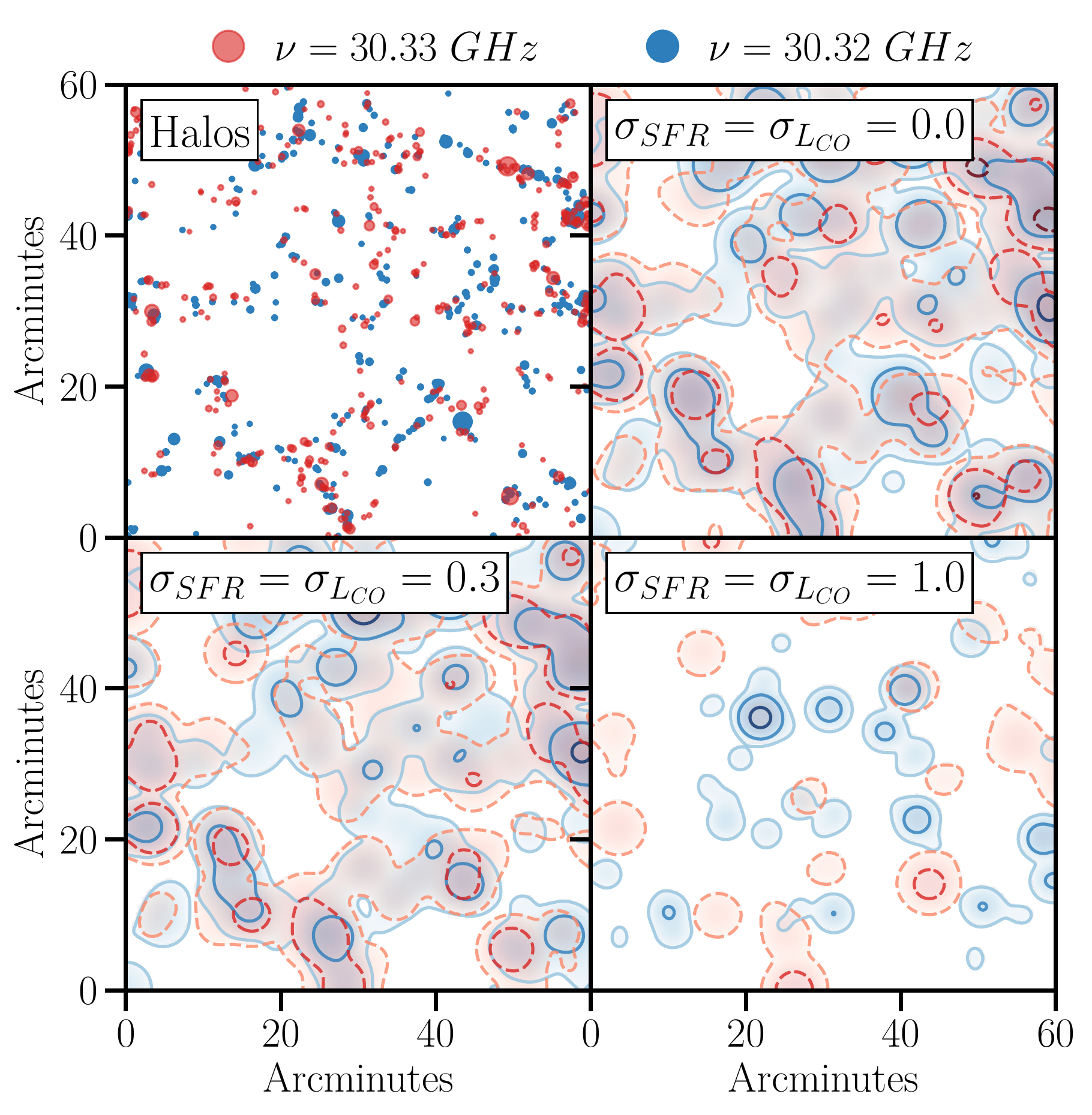}
\caption{\label{fig:peakpatch-COMAP} COMAP mocks created using a peak patch lightcone 
halo catalogue for the standard model of cosmology, $\Lambda$CDM. The cosmological 
parameters and $L_{CO}(M_{halo},z)$ model are the same as assumed in \citet{Li:2015gqa}. 
\textbf{Top Left}: Dark matter halos in two neighboring COMAP frequency slices 
(corresponding to CO(1-0) emission redshifts of $z=2.800, 2.801$) are shown to illustrate the 
level of slice-to-slice correlations. An HOD model is used to treat central and satellite galaxies. 
\textbf{Top Right}: CO(1-0) emission smoothed with a beam of FWHM=4$'$ . Contours 
correspond to brightness temperature thresholds of $T_{CO}$ = $0.1\mu K$, $1\mu K$, and 
$10 \mu K$. \textbf{Bottom Left}: A log-normal scatter of 0.3 dex on the $SFR(M_{halo},z)$ and 
$L_{CO}(L_{IR})$ relations is adopted, as in \citet{Li:2015gqa}. \textbf{Bottom Right}: The effect 
of increasing the log-normal scatter to 1.0 dex, more in line with the scatter measured in FIRE 
gasdynamical simulations. This scatter is primarily a consequence of the bursty nature of star 
formation, but a proper study of correlations among the variables defining $L_{CO}$ is still in 
progress. These maps were created with the publicly available code 
github.com/georgestein/limlam\_mocker. ({\it Courtesy of George Stein and J.~Richard Bond})}
\end{figure}

This technique has been used to create thousands of independent realizations covering the 
COMAP survey volume, which each consist of a $1140\,{\rm Mpc}$ comoving line of site depth, 
covering a $90$ deg$^2$ field of view. Initially COMAP plans to survey a few $\sim 2$ deg$^2$ 
regions, but the simulations cover a much larger region of the sky with the given minimum mass 
resolution. Example results are shown in Fig.~\ref{fig:peakpatch-COMAP}. One can see the 
cosmic web revealed, and so additional statistical measures beyond the traditional power spectrum 
are needed to fully describe the maps. For example, one can consider the $P(\{D,z\})$ statistic, 
a generalization of the classic radio astronomy one-point measure generalized to tomography, 
and applied to CO mapping by \citet{Breysse:2016szq} (which is discussed in more detail below), 
stacking on peaks, etc. This is especially 
important given the large sample variance for such small patches. Fig.~\ref{fig:peakpatch-COMAP} 
also illustrates the importance of (uncorrelated) scatter in the $L_{CO}(M,z)$ relationship. The 
same mock methods have been applied to HI intensity mapping for surveys such as CHIME and 
HIRAX, covering much larger volumes (although at much lower resolution), while [CII] intensity 
mapping mocks are underway.

\subsection{Line-Intensity Mapping Analysis Methods} 

\label{sec:analysis_methods}

\subsubsection*{Constructing an Optimal Observable}

It is interesting to consider in which regimes line-intensity mapping is advantageous to a 
``traditional’’ survey mode, where one considers catalogs of the individual objects that are
robustly detected in a survey. Of course upcoming survey data may ultimately be analyzed 
using an intensity mapping approach as well by cataloging individual objects, but it is still valuable to 
identify the optimal technique for achieving a given scientific goal.  

The intensity mapping method includes the information from all faint sources below the 
detection limit, by measuring the total emission from a (large) pixel (or voxel in 3D). In 
contrast, a traditional survey uses high angular resolution to pick out the bright sources, 
applying a thresholding operation to the pixels or voxels. These two distinct 
approaches can be described by a unified ``observable" function $\hat{O}(L)$. In both cases, a ``map" is created 
by applying $\hat{O}(L)$ on a voxel-by-voxel basis. In the galaxy detection methodology, the 
signal in a voxel is labeled as a ``detection" if it is brighter than a threshold luminosity $L_{th}$ 
(say 5 times the rms noise for a $5\sigma$ detection). The underlying density distribution is 
then encoded in this ``digital map", consisting of 1's (detections) and 0's (non-detections), and 
cosmological constraints can be extracted from the power spectrum of this map.  Therefore, in 
the galaxy detection case, $\hat{O}(L)$ is a step function with the step at $L_{th}$ . On the other 
hand, intensity mapping takes the power spectrum of the measured intensity (or luminosity) map 
directly, so the observable is a linear function of $L$, $\hat{O}(L)=L$.

\begin{figure}[h!]
\begin{center}
\includegraphics[scale=0.5]{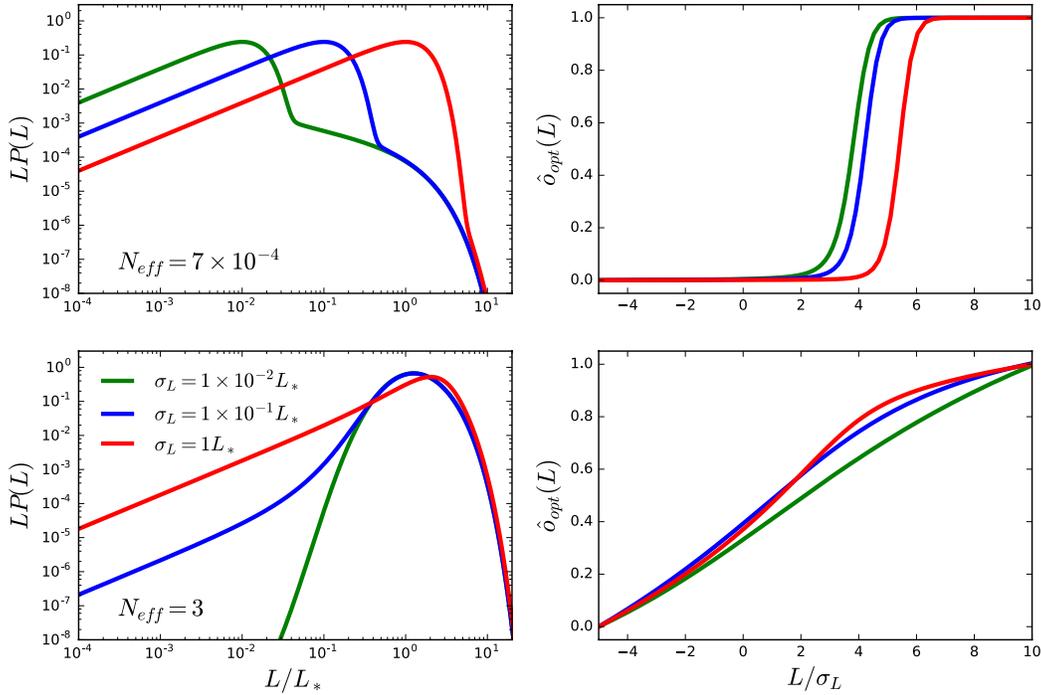}
\caption{Identifying the optimal observable. The probability distribution of total voxel 
luminosity (\textbf{Left Panels}) and the optimal observable/weight function (\textbf{Right Panels}) 
for a low (\textbf{Top Panels}) and high (\textbf{Bottom Panels}) effective number of sources 
per pixel, $N_{\rm eff}$. The colors indicate different levels of Gaussian instrumental noise, 
$\sigma_L$. In the low $N_{\rm eff}$ regime, the optimal observable is close to a step function, 
corresponding to the detection of bright sources, whereas in the high $N_{\rm eff}$ regime, 
the optimal observable is close to a linear function of L, corresponding to intensity mapping. 
({\it Courtesy of Yun-Ting Cheng})}
\label{fig:YunTingb}
\end{center}
\end{figure}

In Cheng et al. 2017 (in prep.), the authors generalize the line-intensity mapping formalism by 
seeking the ``optimal observable", $\hat{O}_{opt}(L)$, which is optimal for measuring the underlying 
density field for a given survey design and source population. The optimal observable can be 
derived from the probability distribution of the total luminosity in a voxel, $P(L)$, and its dependence 
on the underlying density. The key quantity distinguishing the line-intensity mapping and galaxy 
detection regimes is the effective number of sources per voxel, $N_{\rm eff}$. Fig.~\ref{fig:YunTingb} 
shows that the optimal observable is indeed very close to the line-intensity mapping and galaxy 
detection scenarios for $N_{\rm eff} > 1$ and $N_{\rm eff} < 1$, respectively. The source population 
is assumed to follow the Schechter form and different levels of Gaussian (instrumental) noise, 
$\sigma_L$, are considered. In the low $N_{\rm eff}$ regime (\textbf{Top Panels}, with 
$N_{\rm eff}=7\times 10^{-4}$), the optimal observables are similar to step functions with the steps 
at a few times $\sigma_L$, which is the traditional thresholding/galaxy detection  approach; while 
in the large $N_{\rm eff}$ scenario (\textbf{Bottom Panels}, with $N_{\rm eff}=3$), the optimal 
observable function is approximately linear over a wide range of $L$. This result justifies the 
usage of the line-intensity mapping approach in the highly confused limit. Moreover, 
Cheng et al. 2017 (in prep.) also identify scenarios where intermediate estimators are optimal; 
i.e., in these cases neither pure line-intensity mapping nor
galaxy detection analysis strategies are ideal.

\subsection{Multi-Tracer Analysis} 

A promising strategy to overcome the limits imposed by systematics and sample variance is to 
perform {\it multi-tracer} analyses, whereby two differently-biased tracers of large-scale structure 
are combined in such a way that sample variance is cancelled from the noise terms for some 
combinations of bias-dependent terms \citep{McDonald:2008sh}. The cross-correlation of multiple 
tracers should also significantly reduce many systematic effects, which tend to be uncorrelated 
between surveys. These methods should be fruitfully applied using various future line-intensity 
mapping survey data sets, both in conjunction with each other, and when combined with traditional 
galaxy surveys. For example, recent studies have found that post-reionization 21-cm intensity 
mapping surveys are well-matched to large optical surveys like LSST and Euclid  
\citep{Alonso:2015sfa, Fonseca:2015laa}. These analyses should help probe the power spectrum 
on ultra-large scales, which in turn may be used to test General Relativity and to search for possible 
signatures of primordial non-Gaussianity (see \S \ref{chap:goals} and references therein). 

Furthermore, on large scales a key quantity of interest for line-intensity mapping is
the product of the average specific intensity $\avg{I(z)}$ and a luminosity-weighted bias factor, 
$b(z)$ (see Eq.~(\ref{eq:PowSpec})). Determining this quantity from cross-correlating a 
line-intensity map with a traditional galaxy survey is more akin to template fitting than power 
spectral estimation, and so this can be done without sample variance \citep{Switzer:2017kkz}. 
In surveys of the cosmic microwave background, a common strategy is to map each mode to 
${\rm SNR}=1$ \citep{Knox:1996cd}. Time is better spent integrating a larger number of modes 
rather than having high signal-to-noise on a mode that is ultimately limited by sample variance. 
In contrast, determination of the amplitude through cross-correlation could pursue deeper surveys 
with fewer modes. Hence, the possibility of carrying out multi-tracer analyses has important 
implications for the design of line-intensity mapping surveys.

\subsubsection*{P(D) analysis}

\begin{figure}[h!]
\centering
\includegraphics[width=\textwidth]{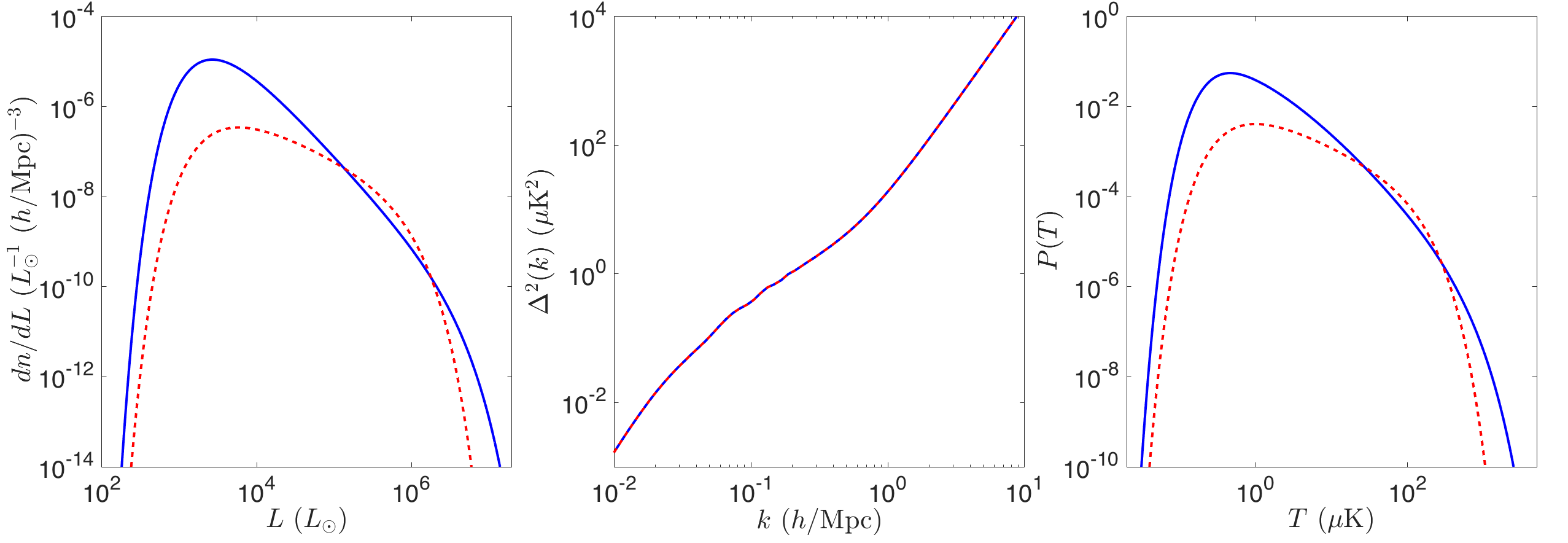}
\caption{Importance of statistics beyond the power spectrum and VIDs.
\textbf{Left Panel}: Two generic luminosity functions for a hypothetical line, each with the modified 
Schechter form from \citet{Breysse:2016szq}.  The red dashed curve is chosen to have five times 
fewer total emitters than the blue. \textbf{Center Panel}: Power spectra computed for the two 
luminosity functions in the left panel.  Parameters of the two Schechter functions were chosen 
so that these two power spectra are identical (for the purposes of this illustration, both spectra 
were assigned the same average bias factor).  An infinite number of other luminosity functions 
would also give the same spectrum.
 \textbf{Right Panel}:  VIDs for the two luminosity functions. Though the two functions have the 
 same power spectra, they can be distinguished by their VIDs. ({\it Courtesy of Patrick Breysse})}
\label{fig:PSlimits}
\end{figure}

The primary statistic used to date when discussing intensity maps is the power spectrum, $P(k)$.  
The power spectrum is the Fourier transform of the two-point correlation function, and is thus a 
two-point statistic.  As seen throughout this document, it is a very powerful tool for studying 
cosmological density fields.  However, it suffers from a key limitation: all of the information about 
a random field is contained within its power spectrum if and only if the field is perfectly Gaussian.  
The small-scale fluctuations in an intensity map are expected to be highly non-Gaussian, as the 
measured intensity is the product of highly nonlinear processes within the galaxy population.  
Thus, the power spectrum alone misses out on much of the information content of a map.  
This is illustrated in the left two panels of Fig.~\ref{fig:PSlimits}, where applying 
Eq.~(\ref{eq:PowSpec}) to two very different luminosity functions gives identical power spectra.

One possibility for accessing this extra information is to look additionally at the one-point 
statistics of a map.  \citet{Breysse:2015saa} and \citet{Breysse:2016szq} propose to do this 
using $P(D)$ analysis methods, which were originally developed for radio astronomy 
\citep{Scheuer1957} but have since been applied to observations across the electromagnetic 
spectrum (see, for example, \citet{Barcons1994, Windridge2000, Lee:2008fm, Patanchon:2009um}).  
$P(D)$ analysis provides a means of mapping the underlying luminosity function of a source 
population onto the observed histogram of voxel intensities (termed the Voxel Intensity Distribution, 
or VID by \citet{Breysse:2016szq}).  By doing so, one can measure the full shape of the luminosity 
function, rather than simply the first two moments (see the right-hand panel of Fig.~\ref{fig:PSlimits}).  
Intensity mapping is particularly well-suited to this type of analysis for two reasons.  First, the 
distances to the target sources are known, which allows one to infer the distribution of their 
intrinsic luminosities, while most $P(D)$ studies are limited to flux distributions.  In addition, 
because intensity maps are inherently three-dimensional, sources can be localized into small 
voxels rather than 2D pixels which span the full line-of-sight.  This lessens confusion caused by 
having many sources in each pixel.

\begin{figure}[h!]
\centering
\includegraphics[width=.7\columnwidth]{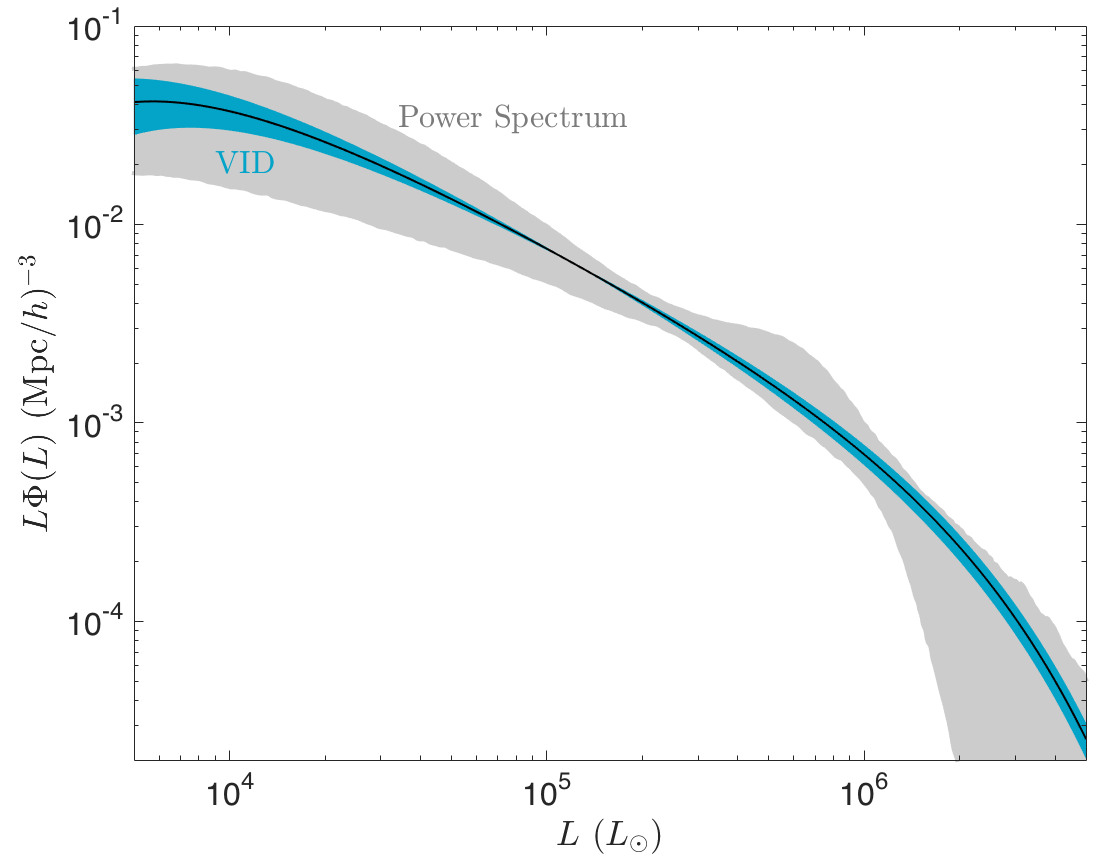}
\caption{Forecasted VID constraints on the CO(1-0) luminosity function. The shaded
regions show forecasted 95\% confidence regions around a model CO luminosity function
using the VID (blue) by \citet{Breysse:2016szq} and using the power spectrum (gray) by
\citet{Li:2015gqa}. ({\it Courtesy of Patrick Breysse})}
\label{fig:PDconstraints}
\end{figure}

\citet{Breysse:2016szq} applied $P(D)$ analysis to a CO(1-0) luminosity function model with 
a similar shape to that of \citet{Li:2015gqa}.  Fig.~\ref{fig:PDconstraints} compares the constraints 
on the luminosity function obtained using the power spectrum (shown in grey) to those using the 
VID (shown in blue).  The power spectrum estimates assume much stronger priors than the VID 
estimates, but the VID produces a considerably tighter constraint.  Both of these error bands make 
strong assumptions about the exact form of the luminosity function, and both neglect complications 
from foregrounds and systematics, and therefore should not be taken as an exact forecast of 
instrumental constraining power.  However, Fig.~\ref{fig:PDconstraints} does make clear the advantage 
of VID methods over the power spectrum when measuring luminosity functions.  \citet{Breysse:2016szq} 
go on to demonstrate the effectiveness of the VID in the presence of line and continuum foregrounds 
as well as gravitational lensing.  $P(D)$ techniques appear to be a promising tool for extracting 
astrophysical information from intensity maps.

\subsection{Foreground Contamination and Interloper Lines}
\label{sec:foreground_mitigation}

It is helpful to divide the foreground contamination problem for line-intensity mapping surveys into 
the case of continuum foregrounds from, e.g. synchrotron radiation or thermal emission from dust 
grains, and that of line emission foregrounds. Strategies for mitigating the continuum foregrounds 
have been well developed for the case of 21-cm intensity mapping, and these should carry over for 
surveys in other emission lines of interest as well. The main idea here is that the continuum foregrounds 
are spectrally smooth, while the line emission signal — although generally much lower in amplitude than 
the continuum foregrounds — has a great deal of spectral structure. Consequently, the continuum 
foreground contamination is confined to Fourier modes with small line-of-sight wavenumbers, unlike 
the signal itself, and this feature can be used to avoid or excise the contamination. 

One complication here is mode-mixing from the frequency dependence of the instrumental beam. 
The instrumental response inevitably moves spectrally smooth contamination into other regions of 
Fourier space. However, 21-cm studies have accounted for mode-mixing effects and shown that the 
continuum foregrounds nevertheless corrupt only a well-defined ``wedge" shaped region in the 
$k_\parallel-k_\perp$ plane. Here $k_\parallel$ denotes the line-of-sight component of the wavevector, 
$\k$, while $k_\perp$ labels the magnitude of components in the plane of the sky. A conservative analysis 
strategy is simply to excise the foreground corrupted wedge, and measure statistics in the remaining part 
of Fourier space. Although the details will depend on the specifics of each survey, this basic approach 
should be broadly applicable across different line-intensity mapping surveys.

A more challenging issue for surveys targeting emission lines such as CO, Ly-$\alpha$, and [CII], 
for example, is line confusion:  in general, multiple ``interloper'' line transitions from gas at various 
redshifts may overlap in observed frequency with that of the ``target'' line at the redshift of interest 
(e.g. \citet{Visbal:2010rz}). Consider the interesting example case of a [CII] intensity mapping survey 
at a target redshift of $z_t = 7$, corresponding to an observed frequency of $\nu_{\rm obs} = 238$ GHz. 
Here important interloper lines include CO(3-2) from $z=0.45$, CO(4-3) at $z=0.88$, CO(5-4) at $z=1.4$, 
and CO(6-5) at $z=1.8$.  Indeed, the spatial fluctuations in the combined emission from these CO 
transitions may exceed that from the [CII] emission of interest \citep{Silva:2014ira,Lidz:2016lub}. 
Unlike the continuum foregrounds, the interloper contamination will obviously have a great deal of 
spectral structure, and so a different approach is required to avoid or excise it. Note, however, that
interloper contamination is {\em not} expected to be an issue for redshifted 21-cm surveys,
simply because there are few conceivable emission lines at the low frequencies of
interest for these surveys. Below we discuss several different strategies for
removing or avoiding interloper contamination.

\subsubsection*{Blind Bright-Voxel Masking}

\begin{figure}
\begin{center}
\includegraphics[width=\textwidth]{MapsforMaskingNew.pdf}
\includegraphics[width=0.32\textwidth]{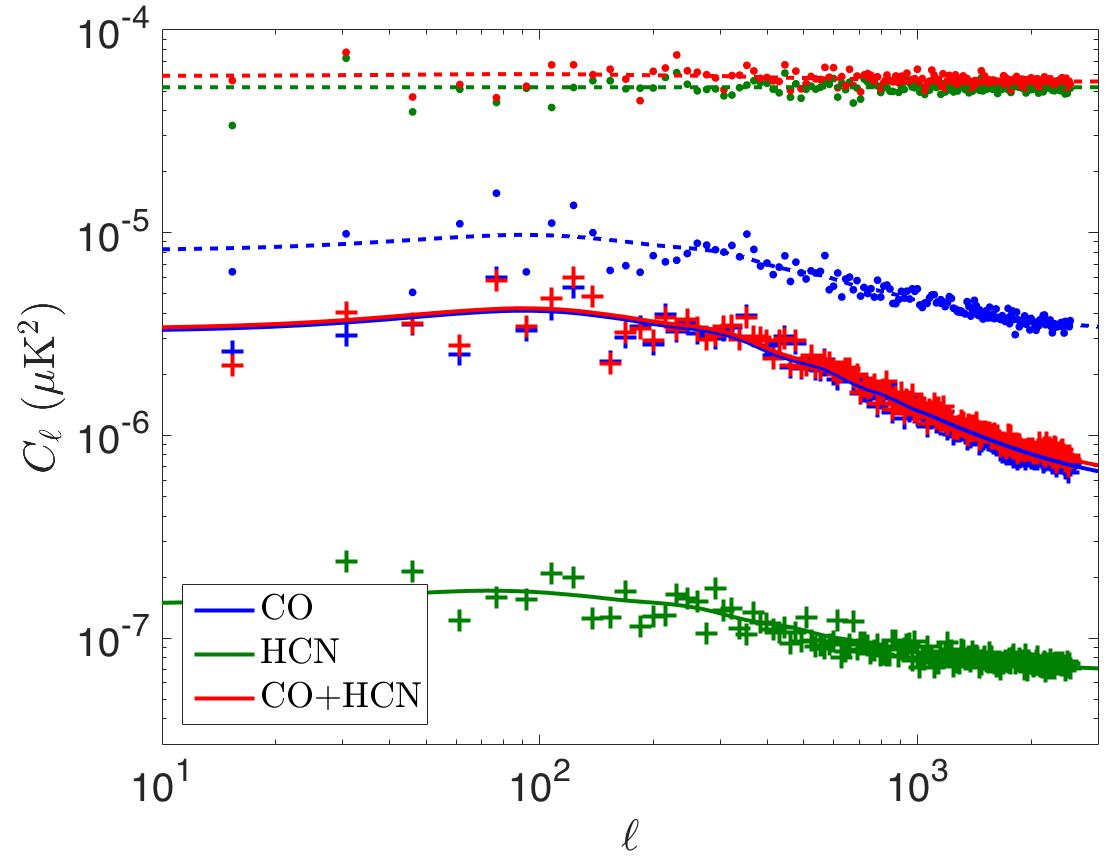}
\includegraphics[width=0.32\textwidth]{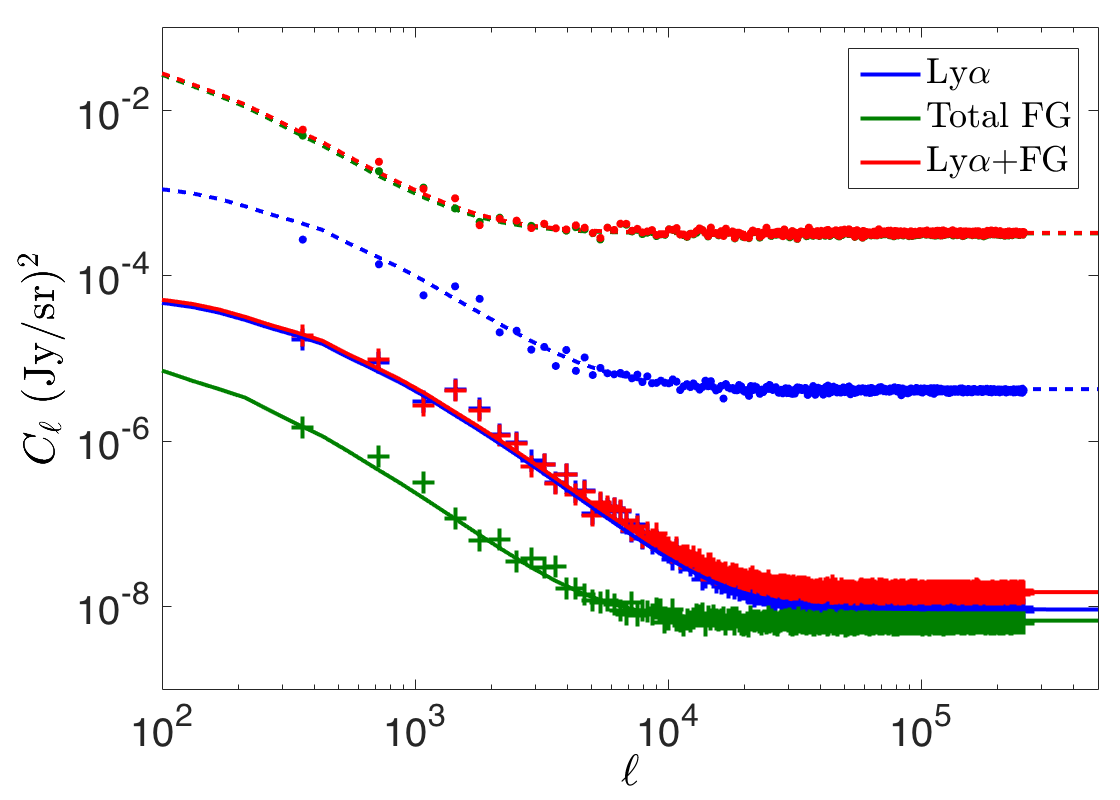}
\includegraphics[width=0.32\textwidth]{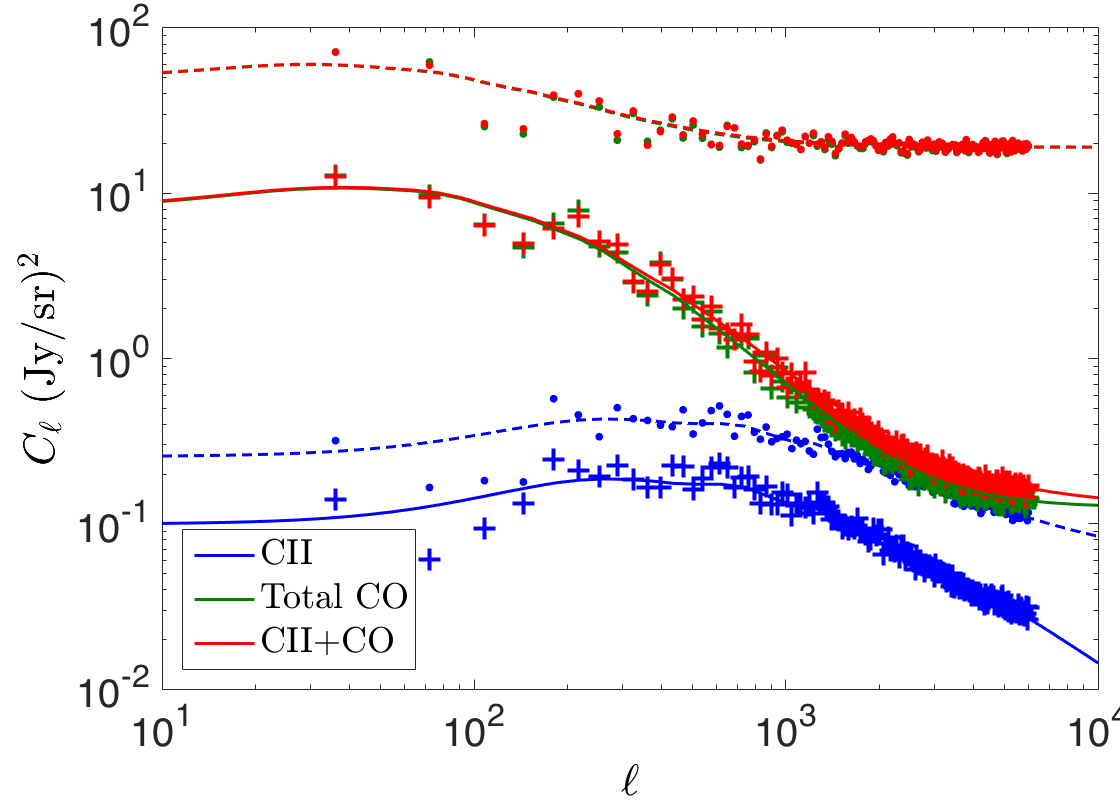}
\caption{Cleaning interlopers by bright voxel masking. \textbf{Top panels:} Simulated maps 
of CO(1-0) at $z=3$, Ly-$\alpha$ at $z=7$, and [CII] at $z=7$ (top row, left to right) along with 
their foregrounds (middle row). The CO simulations cover 550 deg$^2$ with 10 arcmin resolution, 
the Ly-$\alpha$ simulations cover 1 deg$^2$ with 0.1 arcmin resolution, and the [CII] simulations 
cover 100 deg$^2$ with 3.2 arcmin resolution. \textbf{Bottom panels:} Power spectra and best fit 
curves for each of the signals in the simulated maps above (blue), the total corresponding 
foreground contribution (green), and the total emission (red). Dashed curves/dots show the spectra 
before masking, solid curves/pluses show the results after masking all voxels brighter than 
$1\%, 3\%$ and $1\%$ of the voxels in the total emission map. For both CO and Ly-$\alpha$, 
the masking removes a large fraction of the foreground contamination. In the case of [CII] at EoR 
redshifts, however, the foregrounds continue to dominate over the signal regardless of the masking 
percentage. ({\it Courtesy of Ely Kovetz and Patrick Breysse})}
\label{fig:blindmasking}
\end{center}
\end{figure}

The first approach is to mitigate interloper contamination by masking out the
brightest voxels in a survey. This technique is motived, in part, by the fact that
important interloper lines generally come from gas at significantly lower redshifts
than the target lines. Since galaxy masses tend to grow with decreasing redshift,
one expects there to be more very bright sources of emission in the interloper lines
than in the target line. This means that the brightest voxels in a survey are often sourced by 
foreground galaxies; the foreground contamination can be removed or reduced by masking 
out the bright voxels.

This method, the blind masking of bright voxels to mitigate line foregrounds, is investigated 
in \cite{Breysse:2015baa}. Using empirical luminosity function models and simulated intensity 
maps these authors quantify how masking changes the simulated power spectra for three cases: 
CO(1-0) contaminated with HCN, Ly-$\alpha$ contaminated with various atomic lines, and [CII] 
contaminated with higher order CO lines. As demonstrated in Fig.~\ref{fig:blindmasking}, 
for the CO survey, the foreground line emission is faint enough on average that removing the 
brightest voxels significantly drops the amplitude of the foreground spectrum. In fact, this model 
makes pessimistic assumptions about the brightness of the HCN foreground (see also 
Fig.~\ref{fig:assumptions_interlopers}), yet bright pixel masking is still quite efficient. The high 
angular resolution in the Ly-$\alpha$ survey limits the foreground contamination to a few voxels, 
which can be easily masked. The [CII] survey, however, has both bright foregrounds and large 
voxels, and so blind masking is found to be ineffective.

For each emission line considered, masking bright voxels alters the recovered power spectrum
from its input, uncorrupted form. This means that masking removes some of the information in the 
spectrum. However, in the two surveys where masking is effective, the {\em shape} of the clustering 
component of the power spectrum is preserved after masking. Therefore, though much of the 
{\it astrophysical} content of the map is lost, the {\it cosmological} information contained in the shape 
of the clustering spectrum can still be recovered from a masked map. Thus, voxel masking seems 
to be a useful technique for obtaining information from even a highly contaminated CO or 
Ly-$\alpha$ map. If one is to obtain the remainder of the information in these surveys, it will be
 necessary to use some other foreground cleaning technique, such as cross-correlation, or to 
 augment it with a P(D) analysis of the progressively masked survey. If one is to fully unlock all 
 of the benefits of line-intensity mapping surveys, it is imperative that one utilizes these or other 
 methods to isolate the signal from the foregrounds.

\subsubsection*{Targeted Masking}

Another approach, ``targeted masking", uses existing (e.g. optically selected) galaxy catalogs 
at the redshifts of suspected interloper lines to identify which survey voxels to mask.
This strategy has been studied for use in [CII] and Ly$\alpha$ surveys (see \citet{Silva2013, Silva:2014ira}) 
and is being developed by the TIME project to mask-out CO interloper emission that 
lands in their observing band. The brightest CO lines in the TIME band come from CO transitions 
with upper rotational levels of $J_{\rm upper} = 3-6$ from
galaxies at $z < 3$. 

This approach requires catalogs of likely interloping galaxies with accurate redshift
information. TIME plans to observe patches of the sky that have good observational
coverage in multiple wavelength bands, and to supplement existing catalogs by obtaining additional 
spectroscopic redshifts using MOSFIRE or another similar instrument. Targeted masking should
be efficient provided the galaxy catalogs obtained contain good observable proxies for the 
CO interloper emission. The details of this methodology, including uncertainties
arising from scatter in the relation between CO emission and the properties of
catalog galaxies, are discussed in \citet{Sun16}.

\subsubsection*{Anisotropic Power Spectrum Method}

\begin{figure}
\begin{center}
\includegraphics[scale=0.452]{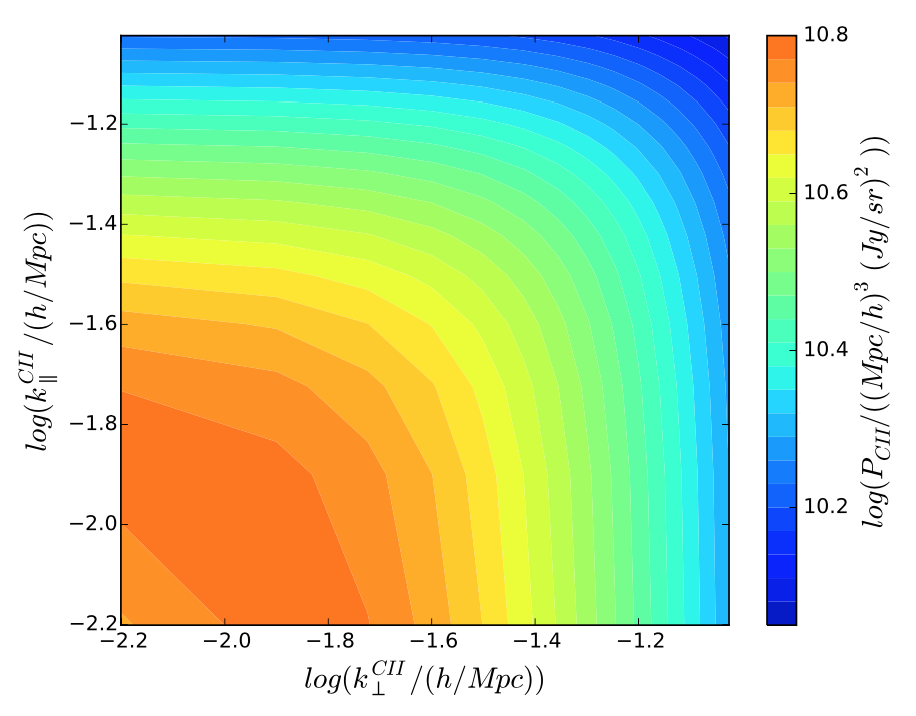}
\includegraphics[scale=0.45]{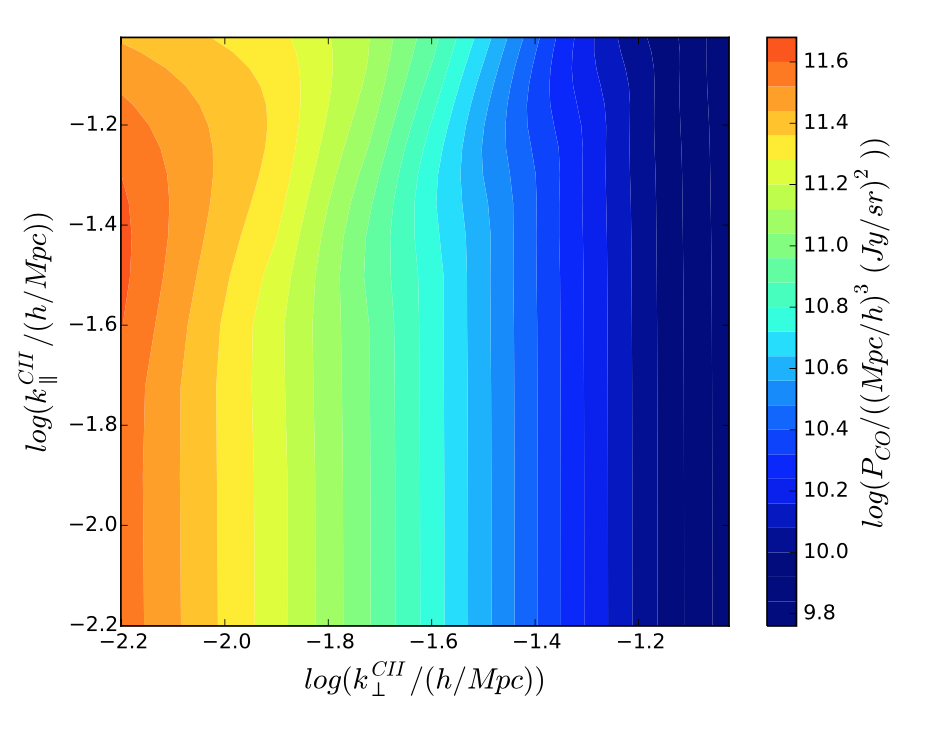}
\caption{Fitting-out interloper contamination using their anisotropic contribution 
to the line-intensity mapping power spectrum. \textbf{Left panel}: 3D Power spectrum of 
[CII] from $z=6$. \textbf{Right Panel} : The brightest CO interloper power spectrum, 
CO(3-2) from $z=0.27$, projected onto the [CII] co-moving frame. The CO interloper power 
spectrum contours are highly elongated as a consequence of the coordinate mapping 
distortion described in the text. ({\it Courtesy of Yun-Ting Cheng})}
\label{fig:YunTing_aniso}
\end{center}
\end{figure}

An additional technique for separating out the interloper contamination exploits the redshift 
dependence of the mapping between observed frequency and angular separation on the 
sky into co-moving spatial scales \citep{Visbal:2010rz,Gong:2013xda,Lidz:2016lub,Cheng:2016yvu}. 
If one assumes the target redshift in performing this mapping, then the interloper emission 
fluctuations are mapped to the wrong spatial scales. Specifically, consider interloper
emission at redshift $z_i$ contaminating a survey for line emission at redshift $z_t$. Further, 
let us denote the apparent co-moving line of sight coordinate of this interloper emission by 
$\tilde{x}_\parallel$ while the true coordinate is $x_\parallel$, and the apparent (true) transverse 
separation by $\tilde{\x}_\perp$ ($\x_\perp$). The apparent coordinates of the interloper emission
are related to the true coordinates by \citep{Lidz:2016lub}: 
$\tilde{x}_\parallel = \frac{H(z_i)}{H(z_t)} \frac{1+z_t}{1+z_i} x_\parallel$, and $\tilde{\x}_\perp 
= \frac{D_{\rm A, co}(z_t)}{D_{\rm A, co}(z_i)} \x_\perp$, where $H(z)$ is the Hubble parameter, 
and $D_{\rm A, co}(z)$ is the co-moving angular diameter distance. Under these coordinate 
transformations, the apparent power spectrum of the interloper emission becomes highly 
anisotropic owing to the different mis-mappings in the line of sight and transverse directions. 
This is essentially the Alcock-Paczynski effect \citep{Alcock:1979mp}, except here the warping 
arises from assuming an incorrect redshift rather than the wrong cosmology.

This interloper emission anisotropy is illustrated in Fig.~\ref{fig:YunTing_aniso} for
the case of a $z=6$ [CII] line-intensity mapping survey.  After projecting the observed 
volume onto the $z=6$ [CII] co-moving frame, the resulting total power spectrum is a 
superposition of [CII] and CO power spectra at different redshifts, each with a different 
but predictable 3D shape due to the projection. Fig.~\ref{fig:YunTing_aniso} shows the 
projected [CII] and brightest CO interloper -- CO (3-2) from $z=0.27$ — emission power 
spectra. \citet{Cheng:2016yvu} demonstrate, using an MCMC approach, that a sufficiently 
sensitive survey can use this distinctive anisotropy to fit out the interloper contamination. 
Although the method is illustrated here for the specific case of a $z=6$ [CII] survey, it 
should be broadly applicable to a range of different line-intensity mapping measurements.

\subsubsection*{Modeling Assumptions and Uncertainties}  

\begin{figure}\centering\includegraphics[height=0.42\linewidth]{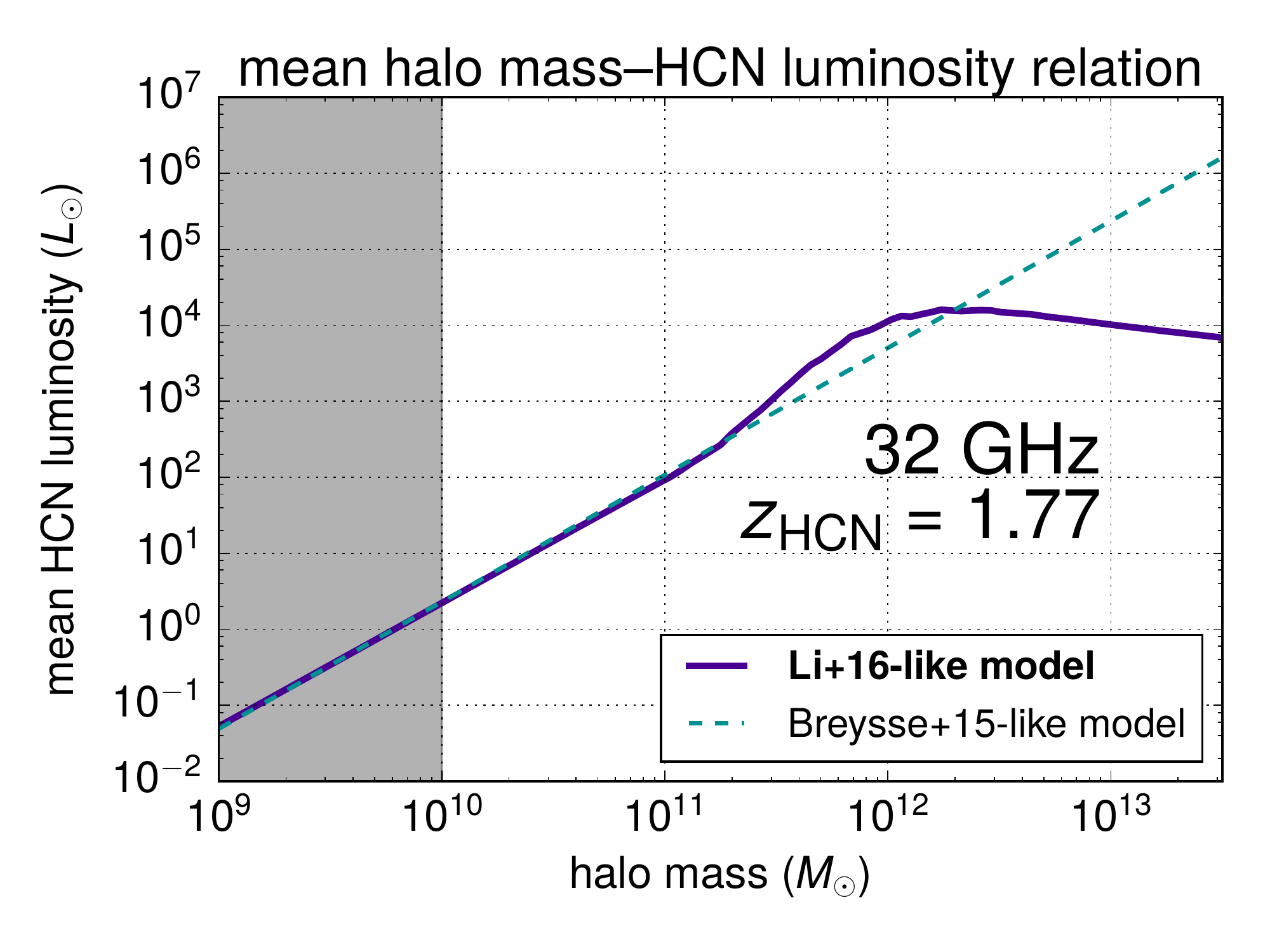}\qquad
\includegraphics[height=0.42\linewidth]{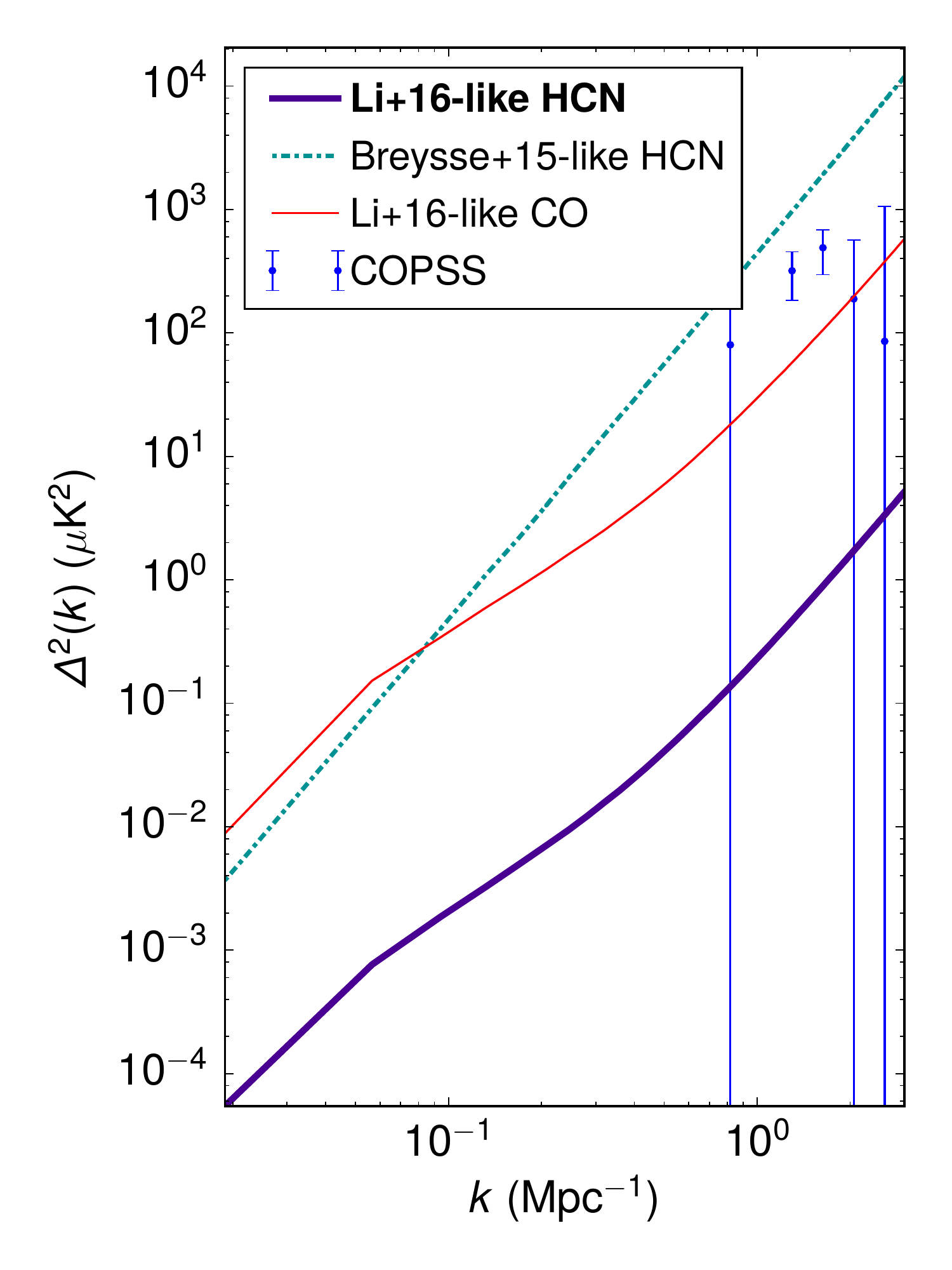}
\caption{Illustration of the difference in predicted HCN contamination in a CO survey
 based on the modeling approach taken, adapted from~\citet{Chung:2017uot}. 
 \textbf{Left Panel:} the mean halo mass--HCN line-luminosity relation, given a 
 power-law model following~\citet{Breysse:2015baa} and a model following~\citet{Li:2015gqa}, 
 which incorporates a declining star-formation efficiency at high-mass. \textbf{Right Panel:} 
 the resulting HCN auto power spectrum, plotted as $\Delta^2(k)$ against the COPSS 
 tentative detections and the CO power spectrum expected from the model in~\citet{Li:2015gqa}. 
 The power-law model predicts high shot-noise contamination from HCN, but the more 
 complex model does not, since it lacks rare, extreme-luminosity ($\gtrsim10^4\,L_\odot$) 
 HCN emitters. ({\it Courtesy of Dongwoo Chung})}
\label{fig:assumptions_interlopers}
\end{figure}

It is worth emphasizing explicitly that interloper contamination predictions are subject 
to significant uncertainties in the modeling and sensitive to the precise assumptions 
made in these calculations. One illustration of this is provided by Fig.~\ref{fig:assumptions_interlopers} 
for the specific case of HCN(1-0) interloper contamination for CO(1-0) line-intensity mapping surveys. 

If a power-law luminosity halo mass relation is assumed, as in \citet{Breysse:2015baa}, 
then simulated HCN emission may seriously impact CO power spectra and
detection significances in a COMAP-like survey, if steps are not made to remove
this contamination (e.g. Fig.~\ref{fig:blindmasking}). However, a more realistic
model informed by empirical constraints on high mass galaxies likely includes
a turnover at high mass owing to declining star-formation efficiency. 
Incorporating this turnover in both CO and HCN modeling, as in~\citet{Li:2015gqa}, 
\citet{Chung:2017uot} find that the simulated HCN auto spectrum lies several orders of 
magnitude below the CO auto spectrum, across all spatial modes (see Figure~\ref{fig:assumptions_interlopers}). 
The resulting contamination in total detection significance is small, and while several other 
interloper lines emit at luminosities similar to HCN(1-0), one nonetheless expects the total effect 
to be subdominant compared to other systematics and foregrounds expected in CO surveys~\citep{Chung:2017uot}.
More generally, it is important to bear in mind that power-law models often over-predict the 
relative contribution of high-mass emitters (with halo masses greater than $\sim {10}^{12}\,M_\odot$) 
to the line-intensity spectrum, especially to the shot-noise component. The precise impact, 
however, depends on, for example,  the relationship between line luminosity and 
star-formation rate.
Future refinements in modeling efforts (such as those described in \S \ref{sec:model}), studies 
based on semi-numerical simulations \cite{Silva:2014ira} and additional empirical constraints should 
help improve efforts to forecast the impact of interlopers and techniques to mitigate them.


\chapter{Conclusion}
\bigskip
%

This Status Report has summarized the recent ideas in intensity mapping theory and experiment 
presented at the first two annual workshops held at Stanford and Johns Hopkins in 2016 and 2017. 
After providing a short introduction to the field, various high-profile science goals that have been identified were
laid out above, followed by a description of the first detections made in this new but rapidly advancing field. 
A separate section was dedicated to the experimental frontier, covering the different instruments that are planned 
to harvest the enormous volume between redshifts of unity to those of reionization and beyond.
The final chapter discussed the methodologies being pursued as part of the effort to model and simulate the signals
being sought, as well as the techniques that have been devised to efficiently analyze the data statistically,
separate out foreground contamination and make optimal use of cross correlations between measurements at
different frequencies. 
As this report demonstrates, the intensity mapping field is standing on firm ground, supported globally by the efforts 
of many dozens of scientists, and is set to make giant leaps forward over the next few years. 
The motivation for continued investment in this research, both in experiments and theory, is as strong as can be.

\bibliography{ScienceReport}

\end{document}

%% file: author_list_indexed.tex
\newcounter{affilcount}



\bigskip

\begin{raggedright}

E.~D.~Kovetz\textsuperscript{1,*},
M.~P.~Viero\textsuperscript{2},
A.~Lidz\textsuperscript{3},
L.~Newburgh\textsuperscript{4}, 
M.~Rahman\textsuperscript{1},
E.~Switzer\textsuperscript{5},
M.~Kamionkowski\textsuperscript{1},
J.~Aguirre\textsuperscript{3},
M.~Alvarez\textsuperscript{6},
J.~Bock\textsuperscript{7},
J.~R.~Bond\textsuperscript{6},
G.~Bower\textsuperscript{8},
C.~M.~Bradford\textsuperscript{9},
P.~C.~Breysse\textsuperscript{6},
P.~Bull\textsuperscript{9},
T.~C.~Chang\textsuperscript{9},
Y.~T.~Cheng\textsuperscript{7},
D.~Chung\textsuperscript{2},
K.~Cleary\textsuperscript{7},
A.~Cooray\textsuperscript{10},
A.~Crites\textsuperscript{7},
R.~Croft\textsuperscript{11},
O.~Dor\'{e}\textsuperscript{7,9},
M.~Eastwood\textsuperscript{7},
A.~Ferrara\textsuperscript{12},
J.~Fonseca\textsuperscript{13},
D.~Jacobs\textsuperscript{14},
G.~Keating\textsuperscript{15},
G.~Lagache\textsuperscript{16},
G.~Lakhlani\textsuperscript{6},
A.~Liu\textsuperscript{17,18},
K.~Moodley\textsuperscript{19},
N.~Murray\textsuperscript{6},
A.~Penin\textsuperscript{19},
G.~Popping\textsuperscript{20},
A.~Pullen\textsuperscript{21},
D.~Reichers\textsuperscript{22},
S.~Saito\textsuperscript{23},
B.~Saliwanchik\textsuperscript{19},
M.~Santos\textsuperscript{13,24},
R.~Somerville\textsuperscript{25,26},
G.~Stacey\textsuperscript{22},
G.~Stein\textsuperscript{6},
F.~Villaescusa-Navarro\textsuperscript{26},
E.~Visbal\textsuperscript{26},
A.~Weltman\textsuperscript{27},
L.~Wolz\textsuperscript{28},
M.~Zemcov\textsuperscript{29}


\bigskip
\footnotesize

\textsuperscript{1}{Department of Physics and Astronomy, Johns Hopkins University, 3400 N. Charles St., Baltimore, MD 21218, USA}\\
\textsuperscript{2}{Kavli Institute for Particle Astrophysics and Cosmology, Stanford University, 382 Via Pueblo Mall, Stanford, CA 94305}\\
\textsuperscript{3}{Department of Physics and Astronomy, University of Pennsylvania, 209 South 33rd Street, Philadelphia, PA 19104, USA}\\
\textsuperscript{4}{Department of Physics, Yale University, New Haven, CT 06520}\\
\textsuperscript{5}{NASA Goddard Space Flight Center, Greenbelt, MD, USA}\\
\textsuperscript{6}{Canadian Institute for Theoretical Astrophysics, University of Toronto, 60 St. George st., Toronto, ON, M5S 3H8, Canada}\\
\textsuperscript{7}{Division of Physics, Math and Astronomy, California Institute of Technology, 1200 E. California Blvd. Pasadena, CA 91125}\\
\textsuperscript{8}{Academia Sinica Institute of Astronomy and Astrophysics, 645 N. A'ohoku Place, Hilo, HI 96720, USA}\\
\textsuperscript{9}{Jet Propulsion Laboratory, California Institute of Technology, 4800 Oak Grove Drive, Pasadena, CA 91109, USA}\\
\textsuperscript{10}{Kavli Institute for Particle Astrophysics and Cosmology, Physics Department, Stanford University, Stanford, CA 94305, USA}\\
\textsuperscript{11}{McWilliams Center for Cosmology, Department of Physics, Carnegie Mellon University, Pittsburgh, PA 15213, USA}\\
\textsuperscript{12}{Normale Superiore, Piazza dei Cavalieri 7, 56126 Pisa, Italy}\\
\textsuperscript{13}{Department of Physics and Astronomy, University of the Western Cape, Cape Town 7535, South Africa}\\
\textsuperscript{14}{School of Earth and Space Exploration, Arizona State University, 781 E Terrace Mall, Tempe, AZ 85287}\\
\textsuperscript{15}{Harvard-Smithsonian Center for Astrophysics, 60 Garden Street, Cambridge, MA 02138, USA}\\
\textsuperscript{16}{Aix Marseille Univ, CNRS, LAM, Laboratoire d'Astrophysique de Marseille, Marseille, France}\\
\textsuperscript{17}{Department of Astronomy and Radio Astronomy Laboratory, University of California, Berkeley, CA 94720, USA}\\
\textsuperscript{18}{Department of Physics and McGill Space Institute, McGill University, Montreal, QC H3A 2T8, Canada}\\
\textsuperscript{19}{Astrophysics and Cosmology Research Unit, University of KwaZulu-Natal, Durban, 4041, South Africa}\\
\textsuperscript{20}{European Southern Observatory, Karl-Schwarzschild-Strasse 2, 85748, Garching, Germany}\\
\textsuperscript{21}{Center for Cosmology and Particle Physics, Department of Physics, New York University, New York, NY, 10003, USA}\\
\textsuperscript{22}{Department of Astronomy, Cornell University, Ithaca, NY 14853, USA}\\
\textsuperscript{23}{Max-Planck-Institut fuer Astrophysik, Karl-Schwarzschild-Str.\ 1, 85748, Garching, Germany}\\
\textsuperscript{24}{SKA South Africa, 3rd Floor, The Park, Park Road, Pinelands, 7405, South Africa}\\
\textsuperscript{25}{Department of Physics and Astronomy, Rutgers University, USA}\\
\textsuperscript{26}{Center for Computational Astrophysics, Flatiron Institute, 162 5th Ave, 10010, New York,NY, USA}\\
\textsuperscript{27}{Department of Mathematics and Applied Mathematics, University of Cape Town, Rondebosch, 7700, South Africa}\\
\textsuperscript{28}{School of Physics, University of Melbourne, Parkville, VIC 3010, Australia}\\
\textsuperscript{29}{School of Physics and Astronomy, Rochester Institute of Technology, Rochester, NY 14623, USA}\\

\vspace{0.1in}

\textsuperscript{*}{email: \email{ekovetz1@jhu.edu}}

\end{raggedright}

%% file: endorser_list_indexed.tex



The following people have endorsed this document as a 2017 Status Report for the Intensity Mapping field:
\bigskip

\begin{raggedright}

James Aguirre\textsuperscript{1},
Matthieu Bethermin\textsuperscript{2},
James Bock\textsuperscript{3},
Geoffrey C.~Bower\textsuperscript{4},
Charles M.~Bradford\textsuperscript{3},
Patrick C.~Breysse\textsuperscript{5},
Philip Bull\textsuperscript{6},
Tzu-Ching Chang\textsuperscript{6},
Yun-Ting Cheng\textsuperscript{3},
Dongwoo Chung\textsuperscript{10},
Sarah Church\textsuperscript{10},
Kieran Cleary\textsuperscript{3},
Asantha Cooray\textsuperscript{11},
Rupert A.~C.~Croft\textsuperscript{12},
Clive Dickinson\textsuperscript{13},
Joshua S.~Dillon\textsuperscript{14},
Olivier Dor\'e\textsuperscript{3,6},
Michael W.~Eastwood\textsuperscript{3},
Andrea Ferrara\textsuperscript{15},
Pedro G.~Ferreira\textsuperscript{16},
Anastasia Fialkov\textsuperscript{17},
Jos\'e Fonseca\textsuperscript{18},
Steven R.~Furlanetto\textsuperscript{19},
Brandon Hensley\textsuperscript{6},
Daniel Jacobs\textsuperscript{20},
Marc Kamionkowski\textsuperscript{21},
Garrett K.~Keating\textsuperscript{22},
Ely D.~Kovetz\textsuperscript{21},
Elisabeth Krause\textsuperscript{10},
Guilaine Lagache\textsuperscript{23},
Daniel Lenz\textsuperscript{3,6},
Adam Lidz\textsuperscript{1},
Adrian Liu\textsuperscript{24,25},
Abraham Loeb\textsuperscript{17,26},
Tobias Marriage\textsuperscript{21},
Daniel P.~Marrone\textsuperscript{27},
Kiyoshi Masui\textsuperscript{28},
Norman Murray\textsuperscript{5},
Laura Newburgh\textsuperscript{29}, 
Gergo Popping\textsuperscript{30},
Alkistis Pourtsidou\textsuperscript{31},
Anthony R.~Pullen\textsuperscript{32},
Mubdi Rahman\textsuperscript{21},
J.~Richard Bond\textsuperscript{5},
Dominik A.~Riechers\textsuperscript{33},
Brant Robertson\textsuperscript{34},
Shun Saito\textsuperscript{35},
Mario G.~Santos\textsuperscript{18},
Marta B.~Silva\textsuperscript{36},
Rachel S.~Somerville\textsuperscript{37,38},
Gordon J.~Stacey\textsuperscript{33},
George Stein\textsuperscript{5},
Guochao Sun\textsuperscript{3},
Eric Switzer\textsuperscript{39},
Joaquin D.~Vieira\textsuperscript{40},
Matteo Viel\textsuperscript{41},
Marco P.~Viero\textsuperscript{10},
Francisco Villaescusa-Navarro\textsuperscript{38},
Eli Visbal\textsuperscript{38},
Amanda Weltman\textsuperscript{42}

\bigskip
\tiny

\textsuperscript{1}{Department of Physics and Astronomy, University of Pennsylvania, Philadelphia, PA 19104, USA}\\
\textsuperscript{2}{Laboratoire d'astrophysique de Marseille, Joliot-Curie 13388 Marseille cedex 13, France}\\
\textsuperscript{3}{Division of Physics, Math and Astronomy, California Institute of Technology, 1200 E. California Blvd. Pasadena, CA 91125}\\
\textsuperscript{4}{Academia Sinica Institute of Astronomy and Astrophysics, 645 N. A'ohoku Place, Hilo, HI 96720, USA}\\
\textsuperscript{5}{Canadian Institute for Theoretical Astrophysics, University of Toronto, 60 St. George st., Toronto, ON, M5S 3H8, Canada}\\
\textsuperscript{6}{Jet Propulsion Laboratory, California Institute of Technology, 4800 Oak Grove Drive, Pasadena, CA 91109, USA}\\
\textsuperscript{7}{That Institution, Another University, Country zipcode}\\
\textsuperscript{8}{Those Departments, The University, State, Country zipcode}\\
\textsuperscript{9}{Department, University, State, Country zipcode}\\
\textsuperscript{10}{Kavli Institute for Particle Astrophysics and Cosmology and Physics Department, Stanford University, Stanford, CA 94305, USA}\\
\textsuperscript{11}{Department of Physics and Astronomy, University of California, Irvine CA 92697}\\
\textsuperscript{12}{McWilliams Center for Cosmology, Department of Physics, Carnegie Mellon University, 5000 Forbes Avenue, Pittsburgh, PA 15213, USA}\\
\textsuperscript{13}{Jodrell Bank Centre for Astrophysics, School of Physics and Astronomy, The University of Manchester, Oxford Road, Manchester, M13 9PL, U.K.}\\
\textsuperscript{14}{NSF AAPF Fellow, Department of Astronomy, University of California, Berkeley, 501 Campbell Hall, Berkeley, CA 94720-3411}\\
\textsuperscript{15}{Scuola Normale Superiore, Piazza dei Cavalieri 7, 56126 Pisa, Italy}\\
\textsuperscript{16}{Astrophysics, University of Oxford ,Keble Road ,Oxford OX1 3RH, UK}\\
\textsuperscript{17}{Institute for Theoretical Computation, Harvard University, Cambridge, MA 02138, USA}\\
\textsuperscript{18}{Department of Physics and Astronomy, University of the Western Cape, Cape Town 7535, South Africa}\\
\textsuperscript{19}{Department of Physics and Astronomy, University of California Los Angeles, Los Angeles, CA 90095}\\
\textsuperscript{20}{School of Earth and Space Exploration, Arizona State University, 781 E Terrace Mall, Tempe, AZ 85287}\\
\textsuperscript{21}{Department of Physics and Astronomy, Johns Hopkins University, 3400 N. Charles St., Baltimore, MD 21218, USA}\\
\textsuperscript{22}{Harvard-Smithsonian Center for Astrophysics, 60 Garden Street, Cambridge, MA 02138, USA}\\
\textsuperscript{23}{Aix Marseille Univ, CNRS, LAM, Laboratoire d'Astrophysique de Marseille, Marseille, France}\\
\textsuperscript{24}{Department of Astronomy and Radio Astronomy Laboratory, University of California, Berkeley, CA 94720, USA}\\
\textsuperscript{25}{Department of Physics and McGill Space Institute, McGill University, Montreal, QC H3A 2T8, Canada}\\
\textsuperscript{26}{Astronomy department, Harvard University, 60 Garden street, Cambridge, MA 02138, USA}\\
\textsuperscript{27}{Steward Observatory, University of Arizona, 933 N. Cherry Ave., Tucson, AZ 85716 USA}\\
\textsuperscript{28}{Department of Physics and Astronomy, University of British Columbia, 6224 Agricultural Road, Vancouver, BC ,V6T 1Z1, Canada}\\
\textsuperscript{29}{Department of Physics, Yale University, New Haven, CT 06520}\\
\textsuperscript{30}{European Southern Observatory, Karl-Schwarzschild-Strasse 2, 85748, Garching, Germany}\\
\textsuperscript{31}{School of Physics and Astronomy, Queen Mary University of London, Mile End Road, London, E1 4NS, United Kingdom}\\
\textsuperscript{32}{Center for Cosmology and Particle Physics, Department of Physics, New York University, New York, NY, 10003, USA}\\
\textsuperscript{33}{Department of Astronomy, Cornell University, 220 Space Sciences Building, Ithaca, NY 14853, USA}\\
\textsuperscript{34}{Department of Astronomy and Astrophysics, University of California, Santa Cruz, 1156 High Street, Santa Cruz, CA 95064}\\
\textsuperscript{35}{Max-Planck-Institut fuer Astrophysik, Karl-Schwarzschild-Str. 1, 85748 Garching bei Muenchen, Germany}\\
\textsuperscript{36}{Kapteyn Astronomical Institute, University of Groningen, Landleven 12, 9747 AG, Groningen, The Netherlands}\\
\textsuperscript{37}{Department of Physics and Astronomy, Rutgers University, USA}\\
\textsuperscript{38}{Center for Computational Astrophysics, Flatiron Institute, 162 5th Ave, 10010, New York,NY, USA}\\
\textsuperscript{39}{NASA Goddard Space Flight Center, Greenbelt, MD, USA}\\
\textsuperscript{40}{Department of Astronomy, The University of Illinois at Urbana-Champaign, 1002 W. Green Street, Urbana, IL 61801, USA}\\
\textsuperscript{41}{SISSA, via Bonomea, 265 , I-34136 Trieste , Italy}\\
\textsuperscript{42}{Department of Mathematics and Applied Mathematics, University of Cape Town, Rondebosch, 7700, South Africa}\\

\end{raggedright}